\documentclass[12pt]{article}

\input{packagePreamble.sty}
\usepackage{endnotes}
\usepackage{sansmath}
\usepackage{amssymb}
\usepackage{enumitem}
\usepackage{scalerel}

\usepackage{hyperendnotes}

\usepackage[tocindentauto]{tocstyle}
\usetocstyle{standard}

\usepackage{array,longtable}
\newcolumntype{C}{>{$}c<{$}} 
\usepackage[page,toc]{appendix}

\input{commands.sty}
\input{formulae.sty}

\renewcommand{\Id}{\operatorname{Id}}
\newcommand{\paragraphsection}[1]{\paragraph{#1}\mbox{}\\}
\newcommand{\sets}{\mathsf{Set}}
\newcommand{\setfin}{\mathsf{Fin}\mathsf{Set}}
\newcommand{\finvect}{\mathsf{Fin}_{\mathbb{R}}\mathsf{Vect}}
\newcommand{\fintensor}[1]{\mathsf{Fin}_{\mathbb{R}}\mathsf{Tens}_{#1}}
\newcommand{\morphisms}{\mathsf{Mor}}
\newcommand{\multiindex}{\bm{\mathsf{MltIdx}}}
\newcommand{\V}{\mathbb{V}}
\newcommand{\W}{\mathbb{W}}
\newcommand{\B}{\mathscr{B}}
\renewcommand{\L}{\mathcal{L}}
\newcommand{\dual}[1]{\left(#1 \right)^\ast}

\NewDocumentCommand{\unitvector}{om}{\mathbf{e}_{#2}^{\smallsuper{ \IfNoValueTF{#1}{\MakeUppercase{#2}}{#1}    }  }}

\newcommand{\bT}{\mathbb{T}}
\newcommand{\unittensor}[1]{\mathbfcal{I}_{#1}}
\newcommand{\T}{\mathbfcal{T}}

\newcommand{\genMltidx}[2]{[#1_{\scaleto{#2\mathstrut}{0.5em} }]}
\newcommand{\Mltidx}[2][O]{\genMltidx{#2}{[#1]}}
\newcommand{\genmltidx}[2]{#1_{\scaleto{#2\mathstrut}{0.5em}}}
\newcommand{\mltidx}[2][O]{\genmltidx{#2}{[#1]}}
\newcommand{\M}[1][O]{\Mltidx[#1]{M}}
\newcommand{\m}[1][O]{\mltidx[#1]{m}}
\newcommand{\N}[1][O_2]{\Mltidx[#1]{N}}
\newcommand{\n}[1][O_2]{\mltidx[#1]{n}}
\newcommand{\smallsuper}[1]{\scriptsize\scaleto{(#1)\mathstrut}{0.7em} }
\newcommand{\smallone}{\smallsuper{1}}
\newcommand{\smalltwo}{\smallsuper{2}}
\newcommand{\MI}{ \Mltidx[O_1]{M^{\smallone}}  }
\newcommand{\mi}{\mltidx[O_1]{m^{\smallone}}  }
\newcommand{\MII}{\Mltidx[O_2]{M^{\smalltwo}}   }
\newcommand{\mii}{\mltidx[O_2]{m^{\smalltwo}} }
\newcommand{\Md}{ \Mltidx[O_d]{M^{\smallsuper{d}}}  }
\newcommand{\md}{\mltidx[O_d]{m^{\smallsuper{d}}}  }
\newcommand{\MD}{ \Mltidx[O_D]{M^{\smallsuper{D}}}  }
\newcommand{\mD}{\mltidx[O_D]{m^{\smallsuper{D}}}  }
\newcommand{\NI}{\Mltidx[D_1]{N^{\smallone}  } }
\newcommand{\NII}{\Mltidx[D_2]{N^{\smalltwo}  }   }
\renewcommand{\P}[1][O_3]{\Mltidx[#1]{P}}
\newcommand{\LL}{\mathbfcal{L}}
\newcommand{\summing}[1]{\sum_{#1 =1}^{\MakeUppercase{#1}}}

\renewcommand{\u}[1]{\mathbf{u}^{\smallsuper{#1}}}
\renewcommand{\v}[1]{\mathbf{v}^{\smallsuper{#1}}}
\newcommand{\w}[1]{\mathbf{w}^{\smallsuper{#1}}}
\newcommand{\x}[1]{\mathbf{x}^{\smallsuper{#1}}}
\newcommand{\lex}{\operatorname{vec}_{\operatorname{lex}}}
\newcommand{\shape}{\mathfrak{F}}
\newcommand{\dualshape}{\mathfrak{F}^\ast}
\renewcommand{\hom}{\operatorname{Hom}}
\newcommand{\fotimes}{\otimes_{func}}
\newcommand{\bigfotimes}{{\bigotimes_{func\ }}}
\newcommand{\idendual}[1]{\psi^{\smallsuper{#1}}}
\newcommand{\dualfunc}[2]{\varphi_{#1}^{\smallsuper{#2}}}
\newcommand{\isodual}[1]{\Phi^{\smallsuper{#1}}}

\renewcommand*\theendnote{\alph{endnote}}
\renewcommand{\makeenmark}{\textsuperscript{\theenmark}} 
\renewcommand\enoteformat{
  \makebox[0pt][r]{\theenmark.\rule{0pt}{\dimexpr\ht\strutbox+\baselineskip}}
}

\title{Tensor Algebra \\and its \\Applications to Data Science and Statistics}
\author{By \\William E. Krinsman}
\date{A thesis submitted in partial satisfaction of the \\requirements for the degree of \\Master of Arts \\in \\Biostatistics \\in the \\Graduate Division \\of the \\University of California, Berkeley\\\phantom{t}\phantom{t}\phantom{t}
Committee in charge:\\\phantom{t} Professor Lexin Li, Chair \\Professor Mark van der Laan \\Professor Rajarshi Mukherjee}

\DeclareMathAlphabet\mathbfcal{OMS}{cmsy}{b}{n}

\begin{document}

\pagenumbering{roman}
\setcounter{page}{-1}

\maketitle

\break
\phantom{t}
\break

\renewcommand{\baselinestretch}{0.9975}\normalsize
\setcounter{tocdepth}{5}
\tableofcontents
\renewcommand{\baselinestretch}{1.0}\normalsize

\break

\pagenumbering{arabic}
\setcounter{page}{1}

\section{Multi-Indexes}
\label{sec:multi-index}

An \textbf{index set} is any set of the form $[M] \defequals\{1 , \dots, M \}$ for some $M \ge 1, M \in \mathbb{N}$.\\ 

An element $m$ of an index set $[M]$ is called an \textbf{index} (plural \textbf{indexes} or \textbf{indices}) \cite{uschmajew_pde}.\\

Any set which can be written as the Cartesian product of a finite number of index sets\footnote{Here $\bigtimes$ denotes the Cartesian product of sets.}\\
${\M \defequals \bigtimes_{o=1}^O [M_o]}$ (where for each $o \in [O]$, $M_o \ge 1$, $M_o \in \mathbb{N}$) will be called a \textbf{multi-index set}.\\ 

An element $\m \defequals (m_1, \dots, m_o, \dots, m_O)$ of a multi-index set ${\M \defequals \bigtimes_{o=1}^O [M_o]}$ is called a \textbf{multi-index}\footnote{This should not be confused with ``multi-index'' as it is defined with respect to e.g. partial derivatives or polynomials. That notion of multi-index corresponds to equivalence classes of elements in special (``cubical'') multi-index sets.}\footnote{The definition of ``multi-index'' given in \cite{cichocki_tensor_networks} corresponds to elements of an index set which is bijective with a ``multi-index set'' as defined in this review, along with a fixed choice of bijection. Compare appendix \ref{sec:order-multi-index}.} (plural \textbf{multi-indexes} or \textbf{multi-indices}).\\ 

Given a multi-index set ${\M \defequals \bigtimes_{o=1}^O [M_o]}$, for each $o \in [O]$ the corresponding index set $[M_o]$ is called the $o$th \textbf{mode} (or sometimes \textbf{way})\cite{chemometrics}\cite{generic_typical_ranks}\cite{Acar2009} of the multi-index set, and $M_o$ is its \textbf{size}\\

In a data science context, each mode usually corresponds to a ``group of variables'' or ``factor'', e.g. the $M_o$ variables in the $o$'th group could be points in time or space at which the data was sampled\cite{NIPS2014_5429}\cite{1910.09499}. Compare this interpretation with chapter 1 of \cite{chemometrics}. Identifying the $o$'th group of variables with the set $[M_o]$ amounts to assigning some order to the variables (see appendix \ref{sec:equiv-char-total}).\\

For a given multi-index set ${\M \defequals \bigtimes_{o=1}^O [M_o]}$, the number of modes, i.e. the natural number $O$, is called the \textbf{order} of the multi-index set. For example, an index set is a multi-index set of order $1$. The index set $[O]$ will be called the \textbf{mode index set} of the multi-index set $\M$.\\

The order of the multi-index set thus corresponds to the number of factors in an experiment. The entire multi-index set $\M$ corresponds to an experiment for which a ``(full) factorial design'' is used, i.e. for which a value is measured for each possible combination of variables. Thus each multi-index can be interpreted as representing a particular combination of variables.\\

One may think of there as being a function from the mode index set:
\[
 [O] \xrightarrow{M} \mathbb{N} 
\]
such that $M(o) \defequals M_o$ is the size of the $o$'th mode of the multi-index set. Each such \textbf{mode function} defines a unique multi-index set, and each multi-index set defines a unique mode function.\\

A multi-index set will be called \textbf{degenerate} if any of its modes has size $1$ (i.e. if any of its modes is equal to the singleton set $[1]\defequals\{1\}$). A multi-index set which is not degenerate, equivalently for which \textit{none} of the modes has size one, will be called \textbf{non-degenerate}. Equivalently, a non-degenerate multi-index set corresponds to a mode function whose image is contained in $\mathbb{N} \setminus \{1\}$.\\

For example, $[1]\times [2] \times [3], [4] \times [1] \times [4],$ and $[3] \times[1] \times [1] \times [5] \times [1]$ are all degenerate multi-index sets, whereas $[2] \times [3]$, $[2] \times [3] \times [4]$, $[4] \times [4]$, and $[3] \times [5]$ are all non-degenerate multi-index sets.

%%% Local Variables:
%%% mode: latex
%%% TeX-master: "../new_notes_draft"
%%% End:

\section{Tensors}
\label{sec:tensors}

A (real-valued) \textbf{tensor} is\endnote{
That this definition essentially corresponds to the nebulous notion of a ``multidimensional array'' is somewhat inherently obvious, but see e.g. the comment after equation (5) in Section 2.2. of \cite{mohlenkamp} for precedent for this definition. This definition can also be found in equation (2.3) of \cite{uschmajew_pde}. See also equation (2.2) of \cite{uschmajew_pde} regarding indices. For the sake of simplicity, in this review, using the notation of \cite{uschmajew_pde}, the Hilbert spaces $\mathscr{V}_{\mu}$ will always be $\R^M$ for some $M\in\mathbb{N}$ endowed with the standard inner product i.e. $\ell^2 (\mathscr{I}_\mu)$, the orthonormal basis for each $\mathscr{V}_\mu$ will always be the standard basis of unit vectors for $\R^M$, and thus the orthonormal basis for $\mathscr{V}$ will always be the unit tensors defined below and $\mathscr{V} = \ell^2(\mathscr{I})$. Thus the unitary isomorphism $\ell^2(\mathscr{I})$ to $\mathscr{V}$ will always be the identity. Comparing with example 2 of section 15-10 of \cite{lim_hla}, again the vector spaces $V_o$ will always be $\R^{M_o}$ for some $M_o$ and the basises $\mathfrak{B}_o$ for each $\R^{M_o}$ will always be the corresponding standard basises of unit vectors. Cf. appendix \ref{sec:simpl-assumpt}.
} 
a function $\mathbfcal{T} : \bigtimes_{o=1}^O [M_o] \to \mathbb{R}$\cite{uschmajew_pde}\cite{mohlenkamp}. Some sources call these functions\endnote{
The argument presented for this in \cite{lim_nonnegative} seems to be that ``tensor'' can refer to an element of any tensor product of general vector spaces (see \cite{atiyah-macdonald} or \cite{hackbusch} for a precise definition), and thus using it to refer to these functions, which themselves are elements of a vector space which is the tensor product of other vector spaces (see below), would be ambiguous. The same reasoning would also seem to suggest that, since elements of a general vector space are also referred to as ``vectors'', in order to avoid ambiguity elements of $\R^M$ should not be referred to as ``vectors'', even though $\R^M$ itself is a vector space, and should instead be referred to as ``hypomatrices''. Thus perhaps it is not altogether surprising that the authors of \cite{lim_nonnegative} seem to characterize adhering to such a naming convention as ``perverse''.
} 
\textbf{hypermatrices} instead\cite{Cooper2012}\cite{banerjee_spectral}\cite{uschmajew_pde}\cite{lim_nonnegative}\cite{liqun2017tensor}\cite{Gelfand1994}. Another common name, more often found in older literature, is \textbf{(multidimensional/multiway) arrays}\cite{chemometrics}\cite{Acar2009}\cite{Comon_symmetric}\cite{Cooper2012}\cite{generic_typical_ranks}\cite{Zhang2016}.\\ 

For a given tensor $\mathbfcal{T}: \M \to \R$, the order of the multi-index set $\M$ which is its domain is called the \textbf{order} of the tensor\cite{kolda_bader}\cite{sidiropoulos_review}. Similarly, the modes $[M_1], \dots, [M_o], \dots, [M_O]$ of the multi-index set $\M$ are also often referred to as the \textbf{modes} of the tensor $\mathbfcal{T}$.\\ 

The multi-index set which is the domain of the tensor will be called the \textbf{shape}\footnote{Cf. p. 491 of \cite{Janzamin2019}. This is a common term in programming libraries, but no terminology for this concept seems to be established or even commonly used in the technical literature.} of the tensor\cite{Janzamin2019}.\\ 

An \textbf{element}, also known as an \textbf{entry}, of a tensor is the value of the tensor evaluated at a given multi-index. For a given tensor $\mathbfcal{T} : \M \to \R$, the element $\mathbfcal{T}(\m)$ corresponding to the multi-index $\m \in \M$ will usually be written in shorthand as $\mathcal{T}_{\m}$.\\

A tensor can be interpreted as a function which, given the factorial design for an experiment with multiple groups of variables, returns the data resulting from the measurements made during the experiment. Each entry of a tensor can be interpreted as a measurement, since the function returns the value (a real number) measured or observed for every possible combination of variables.\\

A tensor of order $2$ is called a \textbf{matrix}. Given a matrix $\mathbf{A}: [H] \times [W] \to \mathbb{R}$, the natural number $H$ is called the \textit{height} of the matrix, and the natural number $W$ is called the \textit{width} of the matrix. The $(h,w)$'th element $A_{(h,w)}$  of the matrix $\mathbf{A}$ will be written for convenience as $A_{hw}$. A matrix can be depicted using a table of $H W$ numbers, with $H$ rows and $W$ columns, as follows: 
\[ 
 \mathbf{A} =
  \begin{bmatrix}
    A_{11} & A_{12} & \dots & A_{1W} \\
A_{21}  & \ddots & & \vdots \\
\vdots & & \ddots & \vdots \\

A_{H1} & A_{H2} & \dots & A_{HW}
  \end{bmatrix}
\]
Data from experiments is most commonly represented using a matrix. In order to do this for experiments with more than two groups of variables, it is necessary to recode combinations of variables as ``pseudo-variables''. In other words, representing data from such an experiment as a matrix instead of as a higher-order tensor involves ignoring some of the structure inherent in the data. This can often lead to one drawing incorrect or unnecessarily ambiguous conclusions from the data.\\

A \textbf{degenerate tensor} is a tensor whose shape is a degenerate multi-index set, cf. Definition 3.24 of \cite{hackbusch}. For example, all row vectors (matrices with shape $[1] \times [W]$) are degenerate tensors, as are all column vectors (matrices with shape $[H] \times [1]$) degenerate tensors.\\

Given sets $X$ and $Y$, the set of all functions $f: X \to Y$ will be denoted $Y^X$.\\ 

Given two index sets $[M]$ and $[N]$, the number of functions $f: [M] \to [N]$, i.e. the number of elements of the set $[N]^{[M]}$, is equal to the natural number $N^M$.\\

Given an arbitrary set $X$, verify that the set $\mathbb{R}^X$ of all functions $X \to \mathbb{R}$ is a real vector space (i.e. $\mathbb{R}$-module) with addition and scalar multiplication defined ``element-wise'', i.e. addition for given $f_1, f_2 \in \mathbb{R}^X$ is defined such that for all $x \in X$, $(f_1+f_2)(x) \defequals f_1(x) + f_2(x)$, and scalar multiplication for given $r \in \mathbb{R}, f \in \mathbb{R}^X$ is defined such that for all $x \in X$, $(r \cdot f)(x) \defequals r f(x)$.\\

Given a multi-index set $\M = \bigtimes_{o=1}^O [M_o]$, the set of all tensors $\mathbfcal{T}: \M \to \mathbb{R}$ of shape $\M$ can be denoted $\mathbb{R}^{\M}$. This is a real vector space of dimension: 
\[
{ \left|\M\right|=M_1 \cdots  M_o \cdots  M_O}\,.
\] 
A space of the form $\R^{\M}$, for any multi-index set $\M$, will be called a \textbf{coordinate space}.\\ 

Every vector space is isomorphic as a vector space to $\R^{[M]}$ for some $M \in \mathbb{N}$, and similarly every tensor space is isomorphic as a tensor space\footnote{For definitions of ``tensor space'' and ``tensor space isomorphism'', please see appendix \ref{sec:categ-vect-spac}.} to $\R^{\M}$ for some multi-index set $\M$. These isomorphisms amount to specifying ``coordinates'' for the vector or tensor space, whence the name.\\

Given a coordinate space $\R^{\M}$, one possible choice of basis is always given by the \textbf{unit tensors}. For a given multi-index $\m \in \M$, the unit tensor $\mathbfcal{I}_{\m}$ is defined as the following:
\[ 
\forall \mu_{[O]} \in \M \,, \quad \mathbfcal{I}_{\m}(\mu_{[O]}  )  \defequals
  \begin{cases}
    0 &  \mu_{[O]} \not= \m \\
    1 & \mu_{[O]} = \m
  \end{cases} \,.
\]
These tensors correspond to the \textbf{indicator functions}\footnote{Sometimes also called \textbf{characteristic functions}. Compare this with section 3.1.2 of \cite{hackbusch}.} of the multi-indices $\m$ in the multi-index set $\M$. As such, clearly there is a bijection between the set of all such tensors and $\M$.\\

It is straightforward to see that every $\mathbfcal{T} \in \mathbb{R}^{\M}$ can be written as a linear combination of these unit tensors, so that they span all of $\mathbb{R}^{\M}$, and it is also not hard to see that they are linearly independent of one another, therefore they are a maximal linearly independent set of elements of $\mathbb{R}^{\M}$, thus a basis\cite{lim_hla}. There are $|\M| = M_1 \dots M_o \dots M_O$ such ``unit tensors'' (the cardinality of $\M$), explaining why that is the dimension of $\R^{\M}$ \cite{hackbusch}. In other words, the dimension of the coordinate space $\R^{\M}$ always equals the cardinality of the multi-index $\M$.\\

Given a tensor $\mathbfcal{T} \in \mathbb{R}^{\M}$, it can be written as a linear combination of unit tensors as follows:
\[ 
\mathbfcal{T} = \sum_{\m \in \M} \mathcal{T}_{\m} \mathbfcal{I}_{\m} \,. 
 \]
Note that the $\mathcal{T}_{\m}$ above are scalars, whereas the $\mathbfcal{I}_{\m}$ are tensors (of the same shape and thus also order as $\mathbfcal{T}$). As a simple special case\footnote{Compare this with example 1. from section 14.2 of \cite{silva_hla}.}, consider $\mathbb{R}^{[2]\times[2]}$. Then the unit tensors are
\[
  \begin{bmatrix}
    1 & 0 \\ 0 & 0
  \end{bmatrix},
  \begin{bmatrix}
    0 & 1 \\ 0 & 0
  \end{bmatrix},
  \begin{bmatrix}
    0 & 0 \\ 1 & 0
  \end{bmatrix},
  \begin{bmatrix}
    0 & 0 \\ 0 & 1
  \end{bmatrix} \,.
\]
The above claims should be easy to verify in this special case, while at the same time the substance of the arguments involved differs little from the general case.
\[
  \begin{bmatrix}
    A_{11} & A_{12} \\ A_{21} & A_{22}
  \end{bmatrix} = A_{11}   \begin{bmatrix}
    1 & 0 \\ 0 & 0
  \end{bmatrix} + A_{12} \begin{bmatrix}
    0 & 1 \\ 0 & 0
  \end{bmatrix} + A_{21}  \begin{bmatrix}
    0 & 0 \\ 1 & 0
  \end{bmatrix} + A_{22} \begin{bmatrix}
    0 & 0 \\ 0 & 1
  \end{bmatrix}
\]
Given a natural number $L \ge 1$, any element of the set $\mathbb{R}^{[L]}$ is a tensor of order $1$ and will be called a \textbf{vector} (of \textit{length} $L$). Given a vector $\mathbf{v}: [L] \to \mathbb{R}$, i.e. $\mathbf{v} \in \R^{[L]}$, its $\ell$'th element $\mathbf{v}(\ell)$ will be denoted for convenience $v_{\ell}$. This is ``essentially the same'' space as $\bigtimes_{\ell=1}^L \mathbb{R}$ (see appendix \ref{sec:vector-identification}) (where $\bigtimes_{\ell=1}^L$ denotes the $L$-fold Cartesian product), so as a result in the sequel both spaces will be denoted interchangeably as $\mathbb{R}^L$, and elements of both spaces will be denoted by tuples, a.k.a. ordered lists, of real numbers $(v_1, \dots, v_{\ell}, \dots, v_L)$ of length $L$. Observe that $\mathbb{R}^L$ has dimension $L$, and that the unit vectors (a.k.a. unit tensors which, as always, form a basis) in this case correspond to the standard basis of $\R^L$, namely $(1, 0, \dots, 0), (0,1,0, \dots, 0), \dots, (0, \dots, 0, 1)$.\\

A tensor of order $O$ whose shape is the $O$-fold Cartesian product of the same index set, i.e. $\bigtimes_{o=1}^O [M]$, is called a \textbf{cubical tensor}\footnote{Also sometimes called a \textbf{hyper-cubical tensor}. For brevity cubical tensor will be used in what follows.}. A tensor is cubical if and only if all of its modes have the same size\cite{lim_hla}, or equivalently the function ${M:[O] \to \mathbb{N}}$ defined in section \ref{sec:multi-index} has a single point in its image. Square matrices and (trivially) all vectors are examples of cubical tensors.

%%% Local Variables:
%%% mode: latex
%%% TeX-master: "../new_notes_draft"
%%% End:

\theendnotes
\setcounter{endnote}{0}

\section{The Tensor Product}
\label{sec:tensor-product}

For two arbitrary sets $X$, $Y$, and functions $f: X \to \mathbb{R}$ and $g: Y \to \mathbb{R}$, their \textbf{tensor product} $f \fotimes g: X \times Y \to \mathbb{R}$ is the function defined (cf. \cite{mohlenkamp} or section 1.1.3 of \cite{hackbusch}) using the rule:
\[ 
 f \fotimes g : (x,y) \mapsto f(x)g(y) \,.  
\]
Given tensors $\mathbfcal{T}: \M[O_1] \to \mathbb{R}$ and $\mathbfcal{U}: \N \to \mathbb{R}$, their tensor product $\mathbfcal{T}\fotimes \mathbfcal{U}$ as functions is almost exactly the same\footnote{Here $\times$ does not refer to the Cartesian product of the multi-index sets, but rather to a closely related operation. See Appendix \ref{product-pedantry} for a clarification, if necessary. Thus this tensor product is not \textit{exactly} the same as $\fotimes$.} as their tensor product as tensors $\mathbfcal{T}\otimes \mathbfcal{U}$, which is defined to be the function $\mathbfcal{T} \otimes \mathbfcal{U} : \M[O_1] \times \N \to \mathbb{R}$ defined using the rule:
\[ 
\mathbfcal{T} \otimes \mathbfcal{U}: \m[O_1] \times \n \defequals \m[O_1]\n \defequals (m_1, \dots, m_{O_1}, n_1, \dots, n_{O_2}) \mapsto \mathcal{T}_{\m[O_1]} \mathcal{U}_{\n} \,.
 \] 
Therefore, given a tensor $\mathbfcal{T}$ of order $O_1$ and a tensor $\mathbfcal{U}$ of order $O_2$, their tensor product $\mathbfcal{T} \otimes \mathbfcal{U}$ is a tensor of order $O_1 + O_2$ (cf. section 1.1.1 of \cite{hackbusch} or section 2. of \cite{Comon_symmetric}).\\

This operation on tensors is sometimes also called the \textbf{(Segre) outer product} \cite{lim_nonnegative}\cite{lim_hla}\cite{Comon_symmetric}\cite{liqun2017tensor}\cite{Zhang2016} and is denoted differently in some sources (cf. \cite{kolda_bader}\cite{li_sparse}\cite{qi_survey}\cite{comon_2002}\cite{DeLathauwer2000}\cite{chemometrics}).\\

Given an integer $D \ge 2$, and $D$ arbitrary sets $X_1, \dots, X_D$, the \textbf{$D$-fold tensor product} 
\[
 f_1 \fotimes \dots \fotimes f_d \fotimes \dots \fotimes f_D \defequals \bigfotimes_{d=1}^D f_d 
\] 
of functions $f_1: X_1 \to \mathbb{R}, \dots, f_D: X_D \to \mathbb{R}$ is defined (c.f. section 1.1.3 of \cite{hackbusch}) using the rule:
\[ 
 \bigfotimes_{d=1}^D f_d : (x_1, \dots, x_d, \dots, x_D) \mapsto f_1(x_1) \dots f_d(x_d) \dots f_D(x_D) = \prod_{d=1}^D f_d(x_d) \,. 
 \]
Given an integer $D \ge 2$, and $D$ tensors $\mathbfcal{T}^{\smallone} \in \mathbb{R}^{\MI}, \dots, \mathbfcal{T}^{\smallsuper{D}} \in \mathbb{R}^{\MD}$, their \textbf{D-fold tensor product} as tensors $\bigotimes_{d=1}^D \mathbfcal{T}^{\smallsuper{d}} \defequals \mathbfcal{T}^{\smallone} \otimes \dots \otimes \mathbfcal{T}^{\smallsuper{D}} \in \mathbb{R}^{ \bigtimes_{d=1}^D \Md }$ is the tensor:
\[  
\bigotimes_{d=1}^D \mathbfcal{T}^{\smallsuper{d}}: \bigtimes_{d=1}^D \md \defequals \mi \mii\dots  \mD \defequals (m_1^{(1)}, \dots, m_{O_D}^{(D)}) \mapsto \mathcal{T}^{\smallone}_{\mi} \cdots \mathcal{T}^{\smallsuper{D}}_{\mD} = \prod_{d=1}^D  \mathcal{T}^{\smallsuper{d}}_{\md} \,.
 \]
In particular, the $D$-fold tensor product (as tensors) of a tensor of order $O_1$, a tensor of order $O_2$, $\dots$, and a tensor of order $O_D$ is a tensor of order $\sum_{d=1}^D O_d$.\\

Given an integer $D \ge 2$, the \textbf{tensor product} of the vector spaces $\R^{\MI}$, \dots, $\R^{\MD}$ is defined to be the set $\bigotimes_{d=1}^D \R^{\Md} \defequals \R^{ \MI  \times \cdots \times \MD }$. Note in particular that 
\[ 
\mathbfcal{T}^{\smallone} \in \R^{\MI}, \dots, \mathbfcal{T}^{\smallsuper{D}} \in \R^{\MD} \quad \implies \quad \bigotimes_{d=1}^D \mathbfcal{T}^{\smallsuper{d}} \in \bigotimes_{d=1}^D \R^{\Md} = \R^{\MI \times \cdots \times \MD } \,.
 \]
Consider any unit tensor $\mathbfcal{I}_{\m} \in \mathbb{R}^{\M}$, where $\M = \bigtimes_{o=1}^O [M_o]$ and ${\m = (m_1, \dots, m_O)}$. When $\unitvector[M_o]{m_o}$ denotes the $m_o$'th unit vector in $\mathbb{R}^{M_o}$ for all $o \in [O]$(cf. Fact 3 of 15-4 in \cite{lim_hla}),
\[ 
\unittensor{\m}= \unitvector{m_1} \otimes \unitvector{m_2} \otimes \dots \otimes \unitvector{m_O}= \bigotimes_{o=1}^O \unitvector[M_o]{m_o} \,.  
\]
A tensor which can be written as the outer product of vectors is called an \textbf{elementary tensor}\cite{hackbusch} or a \textbf{rank-one tensor}\cite{kolda_bader}. The above shows that all unit tensors are elementary tensors. Since the unit tensors of $\mathbb{R}^{\M}$ form a basis for $\mathbb{R}^{\M}$, it follows that \textbf{\textit{every tensor can be written as a linear combination of elementary tensors}}. However in general there is no unique\endnote{To clarify, given a specific set of elementary tensors which form a basis, then of course there is a unique way to write the given tensor as a linear combination of those specific, given elementary tensors. But for writing the tensor as a sum of \textit{some} arbitrary, as of yet unknown, or not chosen, set of elementary tensors, there is no unique way to do so.} way to do so.\\

One can verify that the tensor product as defined above satisfies the following properties\footnote{Compare fact 6 and fact 12 of section 14.2 of \cite{silva_hla}.}:

\begin{itemize}
\item Given $\mathbfcal{T} \in \mathbb{R}^{\M[O_1]}, \mathbfcal{U} \in \mathbb{R}^{\N}$, for any $r \in \mathbb{R}$ one has that:
\[  
(r \cdot \mathbfcal{T}) \otimes \mathbfcal{U}  =  r \cdot (\mathbfcal{T} \otimes \mathbfcal{U}) = \mathbfcal{T} \otimes (r \cdot \mathbfcal{U})  \,. 
 \]
\item Given $\mathbfcal{T}^{\smallone}, \mathbfcal{T}^{\smalltwo} \in \mathbb{R}^{\M[O_1]}, \mathbfcal{U} \in \mathbb{R}^{\N}$, one has that:
\[ 
 (\mathbfcal{T}^{\smallone} + \mathbfcal{T}^{\smalltwo}) \otimes \mathbfcal{U} = \mathbfcal{T}^{\smallone} \otimes \mathbfcal{U} + \mathbfcal{T}^{\smalltwo} \otimes \mathbfcal{U} \,.
  \]
\item Given $\mathbfcal{T} \in \mathbb{R}^{\M[O_1]}, \mathbfcal{U}^{\smallone}, \mathbfcal{U}^{\smalltwo} \in \mathbb{R}^{\N}$, one has that:
\[ 
 \mathbfcal{T} \otimes (\mathbfcal{U}^{\smallone} + \mathbfcal{U}^{\smalltwo}) = \mathbfcal{T} \otimes \mathbfcal{U}^{\smallone} + \mathbfcal{T} \otimes \mathbfcal{U}^{\smalltwo} \,.
  \]
\item Given $\mathbfcal{S} \in \mathbb{R}^{\Mltidx[O_1]{L}}, \mathbfcal{T} \in \mathbb{R}^{\M[O_2]}, \mathbfcal{U} \in \mathbb{R}^{\N[O_3]}$, one has that:
\[ 
\mathbfcal{S} \otimes ( \mathbfcal{T} \otimes \mathbfcal{U}) = \mathbfcal{S} \otimes \mathbfcal{T} \otimes \mathbfcal{U} = (\mathbfcal{S} \otimes \mathbfcal{T}) \otimes \mathbfcal{U} \,.
 \]
\end{itemize}
It suffices to show that the tensors in question have the same shape, and that all of their entries are equal to one another, since two tensors of the same shape\footnote{And of course if two tensors do not have the same shape then they are automatically unequal.} are equal if and only if all of their entries are the same. Thus the verification comes down to basic properties of real numbers.\\

The first three properties above show that the tensor product is a bilinear function\\
 ${\mathbb{R}^{\M[O_1]} \times \mathbb{R}^{\N} \to \mathbb{R}^{\M[O_1] \times \N}}$. It can in fact be considered the\textit{ quintessential} such bilinear function, because \textit{every} bilinear function ${\mathbb{R}^{\M[O_1]} \times \mathbb{R}^{\N} \to \mathbb{R}^{\M[O_1] \times \N}}$ can be written in terms of the tensor product in a unique way, since it satisfies the following important property:

\begin{quote}
For any real vector space $\V$, and any bilinear function $\Phi: {\mathbb{R}^{\M[O_1]} \times \mathbb{R}^{\N} \to \V}$, there always exists a unique \textit{linear}\footnote{This is not a typo, it is very much supposed to say linear. The function being linear, as opposed to bilinear, is precisely what makes the tensor product so important: it can reduce ``multilinear problems'' to ``linear problems''.} function $\phi: \mathbb{R}^{\M[O_1] \times \N} \to \V$ such that 
\[
\Phi(\mathbfcal{T}, \mathbfcal{U}) = \phi(\mathbfcal{T} \otimes \mathbfcal{U})
\] 
for any tensors $\mathbfcal{T} \in \mathbb{R}^{\M[O_1]}, \mathbfcal{U} \in \mathbb{R}^{\N}$.  
 \end{quote}

The above is called the \textbf{universal property of the\endnote{\label{tensor-universal-property-uniqueness}Or better perhaps, ``of \textit{\textbf{a}} tensor product'', since any real vector space $\W$ such that there is a vector space isomorphism ${\psi:\mathbb{R}^{\M[O_1] \times \N} \to \W}$ also satisfies the same universal property (cf. fact 3. of section 14.2 \cite{silva_hla}), using the bilinear product $\mathbfcal{T} \otimes_{\psi} \mathbfcal{U} \defequals \psi(\mathbfcal{T} \otimes \mathbfcal{U})$. Conversely, given a real vector space $\W$ such that there is a bilinear product $\mathbb{R}^{\M[O_1]} \times \mathbb{R}^{\N} \to \W$ satisfying the universal property, then there is a unique vector space isomorphism ${\psi:\mathbb{R}^{\M[O_1] \times \N} \to \W}$ (cf. fact 1. of section 14.2 of \cite{silva_hla}), whose existence is a result of the universal property.} tensor product}\cite{silva_hla}.\\ 

Similarly, the $D$-fold tensor product is a multilinear function, and satisfies the analogous universal property for multilinear functions\footnote{Cf. chapter 2 of \cite{atiyah-macdonald} or fact 1 of section 15.2 of \cite{lim_hla}.} with domain in $\bigtimes_{d=1}^D \mathbb{R}^{\Md}$, namely:

\begin{quote}
 Given any multilinear function ${\Phi: \bigtimes_{d=1}^D \mathbb{R}^{\Md} \to \V}$, there always exists a unique \textit{linear} function ${\phi: \mathbb{R}^{\bigtimes_{d=1}^D \Md} \to \V}$ such that:
\[
 \Phi(\mathbfcal{T}^{\smallone}, \dots, \mathbfcal{T}^{\smallsuper{D}}) = \phi(\mathbfcal{T}^{\smallone} \otimes \dots \otimes \mathbfcal{T}^{\smallsuper{D}}) \,.
\]
for any tensors $\mathbfcal{T}^{\smallone} \in \R^{\MI}$, \dots, $\mathbfcal{T}^{\smallsuper{D}} \in \R^{\MD}$.
\end{quote}

Analogous to how knowing the values of a linear function on the elements of a basis is sufficient to characterize the function entirely\endnote{Because any (finite-dimensional) vector space satisfies the universal property of the free module generated by the elements of any of its basises. See \cite{rotman} and/or \cite{hilton-stammbach} for details on the universal property of free modules. The book \cite{roman} also introduces the concept of free modules, but not in terms of their universal property. \cite{roman} only has linear algebra as a prerequisite, while the treatment in \cite{rotman} or \cite{hilton-stammbach} requires prior familiarity with abstract algebra.}, the universal property of the tensor product allows one to define a linear function on $\mathbb{R}^{\M}$ while specifying its values only for elementary tensors (cf. section 3.3.1 of \cite{hackbusch}). Again, every tensor can be written as a linear combination of elementary tensors, but not uniquely. So in general if the values of a function are given only for the elementary tensors, attempting to extend that function by linearity to all tensors could fail, since one way of writing a given tensor as a linear combination of elementary tensors might imply a different value of the function evaluated at that tensor from another way of writing the tensor as a linear combination of elementary tensors. However, if the value of the function for any given elementary tensor, $\phi(v_1 \otimes \dots \otimes v_O)$, equals the value $\Phi(v_1, \dots, v_O)$ of a \textit{multilinear} function ${\Phi: \bigtimes_{o=1}^O \mathbb{R}^{M_o}} \to \V$, then the universal property of the tensor product guarantees that extending $\phi$ by linearity to all tensors will succeed. In other words, in that case the value of $\phi$ for any given tensor will not depend on the chosen decomposition of that tensor into a linear combination of elementary tensors.\\

Notice that the multiplication of polynomials also satisfies all four of the properties above:

\begin{itemize}
\item $(r \cdot f) g = r \cdot (fg) = f (r\cdot g) $.
\item $(f_1 + f_2) g = f_1 g + f_2 g$.
\item $f (g_1 + g_2) = f g_1 + fg_2$.
\item $f(gh) = fgh = (fg)h$.
\end{itemize}

In fact, the fundamental difference between the product of polynomials and the tensor product is that the former is commutative while the latter is not. In other words, given two polynomials $f,g$, one always has that $fg = gf$, but given two tensors $\mathbfcal{T}, \mathbfcal{U}$, in general one has that $\mathbfcal{T} \otimes \mathbfcal{U} \not= \mathbfcal{U} \otimes \mathbfcal{T}$. \\

This is obviously the case when the shapes of $\mathbfcal{T}$ and $\mathbfcal{U}$ are different, since (if $\M[O_1]$ is the shape of $\mathbfcal{T}$ and $\N$ is the shape of $\mathbfcal{U}$) then $\mathbfcal{T} \otimes \mathbfcal{U}$ has shape $\M[O_1] \times \N$ but $\mathbfcal{U} \otimes \mathbfcal{T}$ has shape $\N \times \M[O_1]$, and two tensors with different shapes clearly cannot be equal.\\

However, this is true even when $\mathbfcal{T}$ and $\mathbfcal{U}$ do have the same shape. As an example, consider $\mathbfcal{T} = (1,0) \in \mathbb{R}^2$, $\mathbfcal{U} = (0,1) \in \mathbb{R}^2$, then one has that:
\[ 
 \mathbfcal{T} \otimes \mathbfcal{U} =
  \begin{bmatrix}
    0 & 1 \\ 0 & 0
  \end{bmatrix} \not=
  \begin{bmatrix}
    0 & 0 \\ 1 & 0
  \end{bmatrix} = \mathbfcal{U} \otimes \mathbfcal{T} \,.
\]
While extremely simple, this example is actually typical. Very loosely speaking\endnote{Speaking less loosely, going from $\mathbfcal{T} \otimes \mathbfcal{U}$ to $\mathbfcal{U} \otimes \mathbfcal{T}$ amounts to applying a linear isomorphism which sends the standard basis of unit tensors in the space of tensors with the shape of $\mathbfcal{T} \otimes \mathbfcal{U}$ to the standard basis of unit tensors in the space of tensors with the shape of $\mathbfcal{U} \otimes \mathbfcal{T}$. This is true regardless of whether $\mathbfcal{T}$ and $\mathbfcal{U}$ have the same shape. This isomorphism is also a result of applying the tensor shape functor (defined in section \ref{sec:tens-shape-funct}) to a bijection between $\M[O_1]$ and $\N$, corresponding to a permutation of (a contiguous partition of) modes. Compare this with appendix \ref{sec:kron-prod-coord}.}., $\mathbfcal{U} \otimes \mathbfcal{T}$ will always have the same entries as $\mathbfcal{T} \otimes \mathbfcal{U}$ but in different places.\\

In any case, the above illustrates that the property which distinguishes the tensor product from the product of polynomials is the non-commutativity\footnote{Compare this claim with fact 8. and example 1. of section 14.9 of \cite{silva_hla}.}. Therefore (loosely speaking), \textit{tensors are non-commutative polynomials}. There are several, inter-related ways of making this statement precise, but giving the background to explain them is outside the scope of this review. Here is a dictionary:\\

\begin{tabular}{|l|l|}
  \hline
  \textbf{Polynomials}& \textbf{Tensors}\\
  \hline
  degree & order \\
  \hline
  monomial & elementary tensor \\
  \hline
  commutative & non-commutative \\
  \hline
\end{tabular}\\

It is possible to make this dictionary more extensive, see appendix \ref{sec:append-non-comm} for a brief, non-rigorous overview. (Compare this with section 14.9 of \cite{silva_hla}, and in particular see example 1 therein.)\\ 

Observe that the fact that elementary tensors correspond to non-commutative monomials means that the statement: ``Every tensor can be written as a linear combination of elementary tensors'' is the direct analogue of the statment that: ``Every polynomial can be written as a linear combination of monomials''.  The constraint that two tensors can be added together only if they have the same shape\footnote{Without defining some notion of ``formal addition''.}, and the fact that two tensors with the same shape have the same order, means that all tensors are linear combinations of ``non-commutative monomials'' (elementary tensors) with the same ``non-commutative degree'' (order), i.e. to be precise tensors don't correspond to the non-commutative version of arbitrary polynomials, \textit{tensors correspond to the non-commutative version of homogeneous\footnote{A polynomial is called homogeneous when it is a linear combination of monomials who all have the same degree.} polynomials}. (Non-commutative general polynomials correspond to taking formal linear combinations of tensors of different shapes and orders, but that isn't useful in practice.)\\ 

The algebraic relationship between tensors and homogeneous polynomials has made it possible to apply tools from algebraic geometry and commutative algebra to the study of tensors in order to make discoveries which would have been much more difficult otherwise. For some examples, consult any of e.g. \cite{landsberg}\cite{Oeding2016}\cite{Pfeffer2019}\cite{seigal_thesis}\cite{robeva_thesis}\cite{nonneg_uniqueness}. In particular, such methods have been fruitful for studying the geometry of spaces of tensors\cite{Sturmfels2012}\cite{10.5555/2230976.2230988}\cite{Garcia2005}\cite{semialg_nonneg}\cite{Seigal2017}; generalizations of eigenvalues, singular values, and decompositions based on them to tensors\cite{Comon1996}\cite{Robeva2016}\cite{Abo2017}\cite{Chang2013}\cite{Boralevi2017}\cite{Robeva2017}\cite{Cartwright2013}\cite{Nie2009}; and studying the properties of various notions of tensor rank\cite{Bernardi2011}\cite{Shitov2018}\cite{semialg_nonneg}\cite{comon:hal-02361504}\cite{Diaz2016}\cite{Seigal2017}\cite{seigal_thesis}\cite{Comon_symmetric}\cite{Allman2015}\cite{Carlini2011}. \\

On a more mundane level, the fact that tensors are essentially ``non-commutative homogeneous polynomials'' also means that algebraically manipulating expressions involving the outer product is very easy: as long as one doesn't add together two tensors with different shapes, nor change the order of multiplication, anything which one would otherwise do with polynomials is allowed.\\

For a both elementary\endnote{Elementary in the sense that the characterization of the tensor product is properties-based, requiring only an understanding of the terms ``vector space'', ``basis'', and ``bilinear'', but not phrased explicitly in terms of the universal property (although in a way equivalent to the characterization using the universal property), and thus doesn't need to introduce either free vector spaces or quotient vector spaces in order to carry out the standard construction of a vector space satisfying the universal property. It is also possibly more accessible for being phrased in terms of basises.} and rigorous introduction to both the tensor product, and the algebra of tensor products, the reader is encouraged to consult Chapter 8 of \cite{vinberg}.

%%% Local Variables:
%%% mode: latex
%%% TeX-master: "../new_notes_draft"
%%% End:

\subsection{Some Applications of the Tensor Product to Statistics}
\label{sec:moments-cumulants}

Starting with a random vector $\mathbf{x}$, one can take the $O$-fold tensor product of $\mathbf{x}$ with itself, denoted compactly by $\mathbf{x}^{\otimes O}$, to produce a random order-$O$ tensor. Taking the expectation of the resulting random tensor, i.e. considering the (non-random) order-$O$ tensor:
\[ 
 \mathbb{E}[ \mathbf{x}^{\otimes O}  ] 
  \]
(where the expectation is to be understood entrywise), the result is the tensor containing all of the $O$'th moments of the distribution of the random vector $\mathbf{x}$.\\

As a special case, if $\boldsymbol{\mu}$ denotes the (non-random) vector $\mathbb{E}[\mathbf{x}]$, then of course $\mathbf{x} - \boldsymbol{\mu}$ is also a random vector. Consequently, one can also define the ``centered'' random order-$O$ tensor $(\mathbf{x} - \boldsymbol{\mu})^{\otimes O}$, as well as the following (non-random) order-$O$ tensor:
\[ 
\mathbb{E} [  (\mathbf{x} - \boldsymbol{\mu})^{\otimes O}  ] \,,
  \]
which consists of all of the $O$'th \textit{central} moments of the distribution of the random vector $\mathbf{x}$. The most commonly encountered example is the \textit{covariance matrix} of the distribution:
\[ 
\mathbb{E}[ (\mathbf{x} - \boldsymbol{\mu} )\otimes (\mathbf{x} - \boldsymbol{\mu})  ] \,. 
 \]
The (higher-order) \textit{cumulants} of a distribution are also best understood in terms of tensors, see e.g. \cite{mccullagh} or \cite{cumulant_book}, although a detailed discussion of cumulants of probability distributions is outside the scope of the review. The interested reader is encouraged to also consult Section 12.1 of \cite{landsberg} or in particular Chapter 2 of \cite{mccullagh}, although both the notation and the conceptual focus in the latter work are quite different from that found in this review, obscuring their mathematical relatedness. See appendix \ref{sec:tens-in-tens-out} for an attempt to bridge that gap. At the very least, the above equations are to be compared with equation (2.2) in \cite{mccullagh}. It should also be noted that the scope of this work is limited entirely to what would be called in \cite{mccullagh} contravariant tensors, since the full range of issues raised by considering dual spaces (needed to understand covariant tensors and so-called mixed tensors) is also beyond the scope of this review\endnote{Limiting the scope of the review this way has precedent for example in \cite{lim_hla}: ``A perhaps unconventional aspect of our approach is that for clarity we isolate the notion of covariance and contravariance... from our definition of a tensor. We do not view this as an essential part of the definition but a source of obfuscation''.} (although see appendix \ref{sec:repr-dual-spac}).\\

For a reference discussing in depth the relationship between dual vector spaces and the tensor product, and in a relatively elementary but nevertheless rigorous manner, the reader is again encouraged to consult Chapter 8 of \cite{vinberg} (although the notation also differs from the notation used in \cite{mccullagh}). For a somewhat unconventional and terse reference using notation ``intermediate'' between that of \cite{mccullagh} and \cite{vinberg}, and which also discusses (theoretical) applications to machine learning, the reader is encouraged to consult \cite{smolensky}. The book\cite{amari} gives a gentler introduction to the same physics notation used by \cite{mccullagh}, and also discusses the use of both covariant and contravariant tensors in (Riemannian) geometry and its applications to statistics and machine learning, called ``information geometry''. For more about information geometry see e.g. \cite{BarndorffNielsen1986},\cite{info-geo}, \cite{jost},\cite{arwini}, \cite{amari-nagaoka}, \cite{murray-rice}, etc.

%%% Local Variables:
%%% mode: latex
%%% TeX-master: "../new_notes_draft"
%%% End:

\theendnotes
\setcounter{endnote}{0}

\section{Tensor Shape Functors}
\label{sec:tens-shape-funct}

Define a function\endnote{From the objects of the category $\multiindex$, i.e. multi-index sets, to the objects of the category $\finvect$, i.e. finite-dimensional real vector spaces. Technically there are issues with the foundations of mathematics which must first be addressed to justify calling this a ``function'', but those are entirely outside of scope and will be ignored in this review.}:
\[ 
 \mathfrak{F}: \operatorname{Obj}(\multiindex) \to \operatorname{Obj}(\finvect) \,,
  \]
which, given a multi-index set $\M$, returns the coordinate space $\R^{\M}$ whose elements are tensors of shape $\M$, in other words for every multi-index set $\M$ one has the identity:
\[ 
 \mathfrak{F} (\M) = \R^{\M}\,. 
\]
Also, define a function:
\[
  \mathfrak{F}: \operatorname{Mor}(\multiindex) \to \operatorname{Mor}(\finvect) \,, 
\]
from functions between multi-index sets to linear functions between finite-dimensional real vector spaces which, given a function:
\[ 
\M[O_1] \overset{f}{\longrightarrow} \N \,, 
\]
returns the unique linear function:
\[  
\mathfrak{F}(\M[O_1]) = \R^{\M[O_1]} \xrightarrow{\mathfrak{F}(f)} \R^{\N} = \mathfrak{F}(\N) 
  \]
such that for each $\m[O_1] \in \M[O_1]$ the $\m[O_1]$'th unit tensor of $\R^{\M[O_1]}$ is sent to the $f(\m[O_1])$'th unit tensor of $\R^{\N}$, i.e.:
\[ 
\mathfrak{F}(f): \unittensor{\m} \mapsto \unittensor{f(\m)} \,.
  \]
One then has the following properties:
\begin{itemize}
\item For any multi-index set $\M$, $\mathfrak{F}$ always sends the identity of $\M$ to the identity of $\R^{\M}$:
  \[ 
\mathfrak{F}(\Id_{\M}) = \Id_{\R^{\M}} = \Id_{\mathfrak{F}(\M)} \,.
 \]
\item Given two functions
  \[  
\M[O_1] \overset{f}{\longrightarrow} \N \quad \text{and} \quad \N \overset{g}{\longrightarrow} \P  \quad \implies \quad \M[O_1] \xrightarrow{g \circ f} \Mltidx[O_3]{P}  \,,
 \]
the following is always true:
  \[  
\mathfrak{F}(g \circ f) = \mathfrak{F}(g) \circ \mathfrak{F}(f) \,.
 \]
  \end{itemize}
Because these two properties are satisfied, one may say that $\mathfrak{F}$ is a (\textbf{covariant}) \textbf{functor}\endnote{
    \label{free-functor-endnote}$\shape$ can be considered the restriction to the subcategory $\multiindex$ of $\setfin$ of functors:
    \[ 
\mathfrak{F}: \setfin \to \finvect
 \]
    (called ``free functors'') which, given any finite set $S$, return a finite-dimensional vector space $\V$ and an inclusion map $\iota: S \to \V$ which together satisfy the universal property of the free (real) vector space generated by $S$ (note that $\iota(S)$ is always a basis for $\V$, in some sense a ``standard'' one, at least relative to $\iota$), and which, given any function $S \xrightarrow{f} T$ between finite sets, returns the unique linear function $\mathfrak{F}(S) \to \mathfrak{F}(T)$ determined by the universal property. Any two such functors are naturally equivalent, thus often one speaks suggestively, albeit technically speaking arguably inaccurately, of ``the'' free functor. So $\mathfrak{F}$ can be interpreted as standing for ``\textbf{f}ree'' as well as for ``\textbf{f}unctor''.
  }:
  \[
 \shape : \multiindex \to \finvect \,, 
\]
  from the category of multi-index sets to the category of finite-dimensional real vector spaces.\\

  The above functor $\shape: \multiindex \to \finvect$ will be called the \textbf{tensor shape functor}, since when given a shape (i.e. a multi-index set), the functor returns the space of all tensors with that given shape, and when given a function between shapes (i.e. multi-index sets), the functor returns a linear function between spaces of tensors with the corresponding given shapes.\\

  It is also possible to define another functor:
  \[ 
 \dualshape : \multiindex \to \finvect
  \]
  such that for every multi-index set $\M$:
  \[ 
 \dualshape(\M) = \shape(\M) = \R^{\M} \,, 
 \]
  and for every function:
  \[ 
\M[O_1] \xrightarrow{f} \N  
\]
  between multi-index sets, $\mathfrak{F}^\ast$ returns the linear function\footnote{It is left as a verification for the reader to check that the function $\mathfrak{F}^\ast (f)$ so defined really is a linear function.}:
  \[ 
 \mathfrak{F}^\ast (f): \R^{\N} \to \R^{\M[O_1]} \,, 
\]
  which is defined for every $\U \in \R^{\N}$ via the rule of assignment:
  \[ 
\dualshape(f) : \U \mapsto \U \circ f \,. 
\]
  In particular, note that for each $\n \in \N$:
  \[
 \dualshape (f) : \unittensor{\n} \mapsto \sum_{\substack{\m[O_1]:\\ f(\m[O_1]) = \n } } \unittensor{\m[O_1]} \,, 
 \]
  and observe that, when $\M[O_1] \xrightarrow{f} \N$ is invertible with inverse $\N \xrightarrow{f^{-1}} \M[O_1]$, then:
  \[  
 \shape(f) = \dualshape(f^{-1}) \,, \quad \dualshape(f) = \shape(f^{-1}) \,.
 \]
  As before, one always has the relationship:
  \[  
\dualshape (\Id_{\M}) = \Id_{\R^{\M}} = \Id_{\dualshape(\M)} \,,
 \]
  but given two functions $\M[O_1] \xrightarrow{f} \N$, $\N \xrightarrow{g} \P$, and their composite $\M[O_1] \xrightarrow{g \circ f} \P$, instead of the relationship analogous to that for $\shape$, the following is true:
  \[ 
\dualshape ( g \circ f) = \dualshape (f) \circ \dualshape (g) \,.  
\]
In comparison to $\shape$, one sees that $\dualshape$ composes morphisms in the \textit{opposite} order, which means that it is a so-called \textbf{contravariant functor}\endnote{
    Strictly speaking, a contravariant functor 
\[
\multiindex \to \finvect
 \] 
is actually a functor 
\[ 
\multiindex^{\operatorname{op}} \to \finvect  
\] 
and thus in some sense not a functor $\multiindex \to \finvect$ at all. Of course this all depends on the precise definition of functor one is using. The older convention defined ``covariant functors'' and ``contravariant functors'', while the newer convention defines only ``functor'' (thus the term ``covariant functor'' being redundant) with contravariant functors just being functors from the opposite category, see e.g. section 10.3 of \cite{ghrist2014elementary}.
  }\endnote{Covariant tensors correspond to dual spaces (cf. appendix \ref{sec:coord-vector-spaces}), which correspond to the \textit{contravariant} dual space functor (cf. appendix \ref{sec:interpr-dual-tens}), while contravariant tensors actually correspond to \textit{covariant} functors. The reason for the opposite order of the two terminologies is that the ``variance'' with respect to which tensors are defined actually refers to the coordinates (cf. appendix \ref{sec:coord-isom-tens}) of the tensors, not the tensors themselves, even though their coordinates correspond to elements of the dual space and therefore have the opposite ``variance''. (The coordinates of a covariant tensor \textit{do} correspond to a contravariant tensor, see appendix \ref{sec:repr-dual-spac} for an explanation.)} $\multiindex \to \finvect$. To contrast it with the tensor shape functor, this contravariant functor $\dualshape$ will be called the \textbf{dual tensor shape functor}.\\

When both multi-index sets are of order $1$, then given any morphism $f: [M] \to [N]$ in $\multiindex$, it is always true that the matrix representing the morphism $\dualshape(f): \R^N \to \R^M$ in $\finvect$ with respect to the standard basises is the transpose of the $[N] \times [M]$ matrix representing the morphism $\shape: \R^M \to \R^N$ in $\finvect$ with respect to the standard basises.\\

  For example, given $f:[2] \to [3]$ such that $f(1) = 1$ and $f(2)=2$, then the matrix representing $\shape(f)$ with respect to the standard basises is:
  \[
    \begin{bmatrix}
      1 & 0 \\ 0 & 1 \\ 0 & 0
    \end{bmatrix} \,,
  \]
  while the matrix representing $\dualshape(f)$ with respect to the standard basises is:
  \[
    \begin{bmatrix}
      1 & 0 & 0 \\
      0 & 1 & 0
    \end{bmatrix} = \left(   \begin{bmatrix}
      1 & 0 \\ 0 & 1 \\ 0 & 0
    \end{bmatrix}  \right)^\top \,.
\]
Similarly, given $g:[3]  \to [2]$ such that $g(1)=1$, $g(2)=2$, $g(3)=1$, the matrix representing $\shape(g)$ with respect to the standard basises is:
\[
  \begin{bmatrix}
    1 & 0 & 1 \\ 0 & 1 & 0
  \end{bmatrix} \,,
\]
while the matrix representing $\dualshape (g)$ with respect to the standard basises is:
\[
  \begin{bmatrix}
    1 & 0 \\ 0 & 1 \\ 1 & 0
  \end{bmatrix} =
  \left(  \begin{bmatrix}
    1 & 0 & 1 \\ 0 & 1 & 0
  \end{bmatrix} \right)^\top \,.
\]
The above seems to suggest that $\dualshape$ is related to the dual functor\footnote{This sends every vector space $\V$ to its dual, $\L(\V; \R)$, and sends every linear function to its adjoint \cite{mccullagh-functor}.} $\mathfrak{D}: \finvect \to \finvect$. This is actually true; see appendix \ref{sec:interpr-dual-tens} for the details of the relationship. \\

The most important properties of the tensor shape functors are the following:
\begin{itemize}
\item For any two objects $\M[O_1]$, $\N$ in $\multiindex$, the following is always true:
  \[
    \begin{array}{rcl}
      \shape\phantom{^\ast}( \M[O_1] \times \N ) & = & \shape\phantom{^\ast}(\M[O_1]) \otimes \shape\phantom{^\ast}(\N) \,, \\
      \dualshape(\M[O_1] \times \N) & = & \dualshape(\M[O_1]) \otimes \dualshape(\N) \,.
    \end{array}
  \]
\item For any two morphisms $\MI \xrightarrow{f_1} \NI$, $\MII \xrightarrow{f_2} \NII$ in $\multiindex$, the following is always true (the definition of tensor product of linear functions is found in section \ref{sec:tens-prod-line}):
  \[
    \begin{array}{rcl}
      \shape\phantom{^\ast}(f_1 \times f_2) &= & \shape\phantom{^\ast}(f_1) \otimes \shape\phantom{^\ast}(f_2) \,,\\
      \dualshape(f_1 \times f_2) & = & \dualshape(f_1) \otimes \dualshape(f_2) \,.
    \end{array}
  \]
\end{itemize}
In other words the tensor shape functors transform\footnote{
This property can be used to explain why the incidence matrix of the categorical product (sometimes also called ``tensor product'') of graphs is the Kronecker product of the incidence matrices of the original graph\cite{imrich2008topics}.
} Cartesian products in $\multiindex$ to tensor products in $\finvect$. Tensor product structure in $\finvect$ is related to Cartesian product structure in $\multiindex$ via the tensor shape functors, or in yet other words functoriality\endnote{
  Both the Cartesian product in $\setfin$ and thus in $\multiindex$, and the tensor product in $\finvect$, endow their respective categories with the structure of a ``closed, symmetric monoidal category''. (This is also true of the category of (finite) pointed sets using the modified Cartesian product, also known as the ``smash product'', and thus also for the full subcategory of pointed multi-index sets.) Thus both the Cartesian product in $\multiindex$ and the tensor product in $\finvect$ are called ``monoidal products''. The above results say that both tensor shape functors transform the monoidal product of $\multiindex$ to the monoidal product of $\finvect$ of $\finvect$. They therefore may be described as ``monoidal functors''.\\

The tensor shape functors also transform the categorical sum (a.k.a. coproduct) $+$/$\sqcup$ of $\setfin$ (disjoint union) to the categorical sum (a.k.a. coproduct) $\oplus$ of $\finvect$ (direct sum or equivalently Cartesian product). These operations both endow $\setfin$ and $\finvect$ with alternative monoidal product structures respectively. The tensor shape functors also transform the disjoint unions in $\setfin$ to direct sums in $\finvect$, which means that there is another sense in which the tensor shape functors transform a monoidal product of $\multiindex$ to a monoidal product of $\finvect$. (Making this sense precise is tedious and probably requires redefining the category $\multiindex$ to be closed under sums of arbitrary multi-index sets, not just sums of index sets. Such would probably correspond to allowing the formal addition of tensors of different shapes, which generally is not useful in practice.) Thus the tensor shape functors allow one to represent a lot of structure from $\multiindex$ inside $\finvect$, and could likely even be called ``doubly monoidal functors''.
} makes it possible to reduce problems involving tensor products of coordinate spaces to problems involving Cartesian products of multi-index sets. This aids e.g. the study of unfoldings and of permutations of modes. Compare e.g. Definition 3.3 from \cite{Wang2017}, where the use of a function between multi-index sets to define the unfolding map between tensors implicitly uses the tensor shape functor.\\

The above two properties mean that the tensor shape functors relate not only the category of multi-index sets with the category of finite-dimensional vector spaces: they also induce functors from the categories of factorized multi-index sets to the categories of finite-dimensional real tensor spaces\footnote{See appendix \ref{sec:categ-multi-index} for the definition of the categories of factorized multi-index sets, and appendix \ref{sec:categ-vect-spac} for the definition of the categories of finite-dimensional real tensor spaces.}. More specifically, one has that, for every $D>1$, there exist analogous tensor shape functors:
\[
  \begin{array}{rcl}
    \multiindex_D & \xrightarrow{ \shape }& \fintensor{D} \,, \\
     \multiindex_D^{\operatorname{op}} & \xrightarrow{\dualshape} & \fintensor{D} \,,
  \end{array}
\]
not just for the case of $D=1$ where $\multiindex_1 \defequals \multiindex$ and $\fintensor{1} \defequals \finvect$. This says that the tensor shape functors transform Cartesian product structure inside of $\multiindex$ to tensor product structure inside of $\finvect$ in a reliable and straightforward manner.\\

Note that the tensor shape functors also transform the disjoint unions (i.e. ``sums'') of index sets to the direct sums of vector spaces, and combining this with the properties above means that the Cartesian products of sums of index sets are transformed to tensor products of direct sums of vector spaces by the tensor shape functors. In other words, the tensor shape functors also enable the study of block structure of tensors through the study of Cartesian products of partitioned index sets. Much more can be said about this, the details of which are outside the scope of this review.\\

The above properties also mean that the tensor shape functors can be said to transfer the arithmetic of cardinalities of index sets and multi-index sets to the arithmetic of dimensions of vector spaces (cf. Examples 10.18 and 10.19 of \cite{ghrist2014elementary}). Specifically one has the relationships:
\[ 
|[M] + [N]| = |[M]| + |[N]| \quad \text{and since} \quad \shape([M] + [N]) = \shape([M]) \oplus \shape([N])\,,
\]
\[
\quad \dim(\shape([M] + [N])) = \dim(\shape([M]) \oplus \shape([N]) ) = \dim (\shape([M])) + \dim(\shape( [N] )) \,. 
 \]
\[
 | \M[O_1] \times \N| = | \M[O_1] |\cdot |\N| \quad \text{and since} \quad \shape(\M[O_1] \times \N) = \shape(\M[O_1]) \otimes \shape(\N) \,, 
\]
\[ 
\quad \dim(\shape(\M[O_1] \times \N)) = \dim(\shape(\M[O_1]) \otimes \shape(\N)) = \dim(\shape(\M[O_1]))\cdot\dim(\shape(\N)) \,.
 \]
More generally, the tensor shape functors enable one to transfer the study of combinatorics (finite sets and functions between them) to the study of linear algebra (vector spaces and linear functions between them), and thus reduce linear algebraic questions to combinatorial questions.\\

For other instances of functoriality hidden within statistics and data science, consult e.g. \cite{carlsson} for how functoriality manifests in the study of clustering and related problems, and \cite{mccullagh-functor} for how functoriality manifests in the definition and study of statistical models.

%%% Local Variables:
%%% mode: latex
%%% TeX-master: "../new_notes_draft"
%%% End:

\theendnotes
\setcounter{endnote}{0}

\section{Tensor Product Decompositions}
\label{sec:outer-prod-decomp}

Every order $O$ tensor can be written as the sum of elementary tensors (i.e. $O$-fold tensor products of vectors), as the decomposition of any tensor into a linear combination of the unit tensors (with the tensor's entries as coefficients) attests. However, such a representation is more useful if there are fewer terms in the sum. The extreme example of writing a tensor as a linear combination of the unit tensors also demonstrates this, since such a decomposition is no more and no less useful than directly knowing the values of all of the entries of the tensor. \\

For examples of more useful decompositions, consider a given $H \times W$ matrix $\mathbf{A}$, let $\mathbf{A}_{:w}$ denote the $w$'th column for any $w \in [W]$, and let $\mathbf{A}_{h:}$ denote the $h$'th row for any $h \in [H]$. Then it is always possible to write $\mathbf{A}$ using both of the following tensor product decompositions:
\[  
\mathbf{A} = \sum_{w \in [W]} \mathbf{A}_{:w} \otimes \unitvector{w} \quad \text{and} \quad \mathbf{A} = \sum_{h=1}^H  \unitvector{h} \otimes \mathbf{A}_{h:} \,. 
 \]
The first of the two tensor product decompositions above will be called the \textbf{standard column-wise decomposition}, and the second will be called the \textbf{standard row-wise decomposition}.\\

Analogously for any higher-order ($O \ge 3$) tensor $\mathbfcal{T}$, given a fixed mode $o \in [O]$, let 
\[  
 { \genmltidx{m}{[O] \setminus o } \defequals (m_1, \dots, m_{o - 1}, m_{o +1}, \dots, m_O) }
\] 
denote a multi-index omitting the $o$'th index. These are elements of: 
\[  
\genMltidx{M}{[O]\setminus o} \defequals\bigtimes_{\omega \not= o}^O [M_\omega] \,,
 \] 
the corresponding multi-index set formed from the shape of $\mathbfcal{T}$ by omitting the $o$'th mode. Then it is always possible to write $\mathbfcal{T}$ using $| \genMltidx{M}{[O]\setminus o}| = \prod_{\omega \not= o}^O M_\omega$ terms as:
\[  
\mathbfcal{T} = \sum_{  \genmltidx{m}{[O]\setminus o}  \in  \genMltidx{M}{[O]\setminus o} }  \unitvector{m_1} \otimes \dots \otimes  \unitvector[M_{o -1}]{m_{o-1}} \otimes \mathbfcal{T}_{\genmltidx{m}{[O]\setminus o} } \otimes \unitvector[M_{o +1}]{m_{o+1}} \otimes \dots \otimes  \unitvector{m_O}
 \]
where $\mathbfcal{T}_{\genmltidx{m}{[O]\setminus o} }$ denotes the \textbf{fiber} of $\mathbfcal{T}$ corresponding to $\genmltidx{m}{[O]\setminus o}$, i.e. the vector formed by using the indices in the multi-index $\genmltidx{m}{[O]\setminus o}$ to fix all of the indices of $\mathbfcal{T}$ except for the $o$'th index,. (Compare section 2 of \cite{kolda_bader}.) For $O=3$ one can write more explicitly (where $:$'s represent an omitted index):
\[
 \begin{array}{rcl}
    \mathbfcal{T}    & & \\

   (\text{for }o = 1) &=& \displaystyle  \sum_{\substack{(m_2, m_3) \\\in [M_2] \times [M_3]}} \! \mathbfcal{T}_{: m_2 m_3} \otimes \unitvector{m_2}  \otimes \unitvector{m_3}  \\
   (\text{for }o=2) &=& \displaystyle\sum_{\substack{(m_1, m_3) \\\in [M_1] \times [M_3]}} \! \unitvector{m_1}  \otimes \mathbfcal{T}_{m_1 : m_3} \otimes \unitvector{m_3} \\
   (\text{for }o=3) & = &\displaystyle \sum_{\substack{(m_1, m_2) \\ \in [M_1] \times [M_2]}} \! \unitvector{m_1} \otimes \unitvector{m_2}  \otimes \mathbfcal{T}_{m_1 m_2 :} \,.
  \end{array}
  \]
Using the terminology of \cite{kolda_bader}, the fibers $\mathbfcal{T}_{: m_2 m_3}$ are called \textbf{column} fibers, the fibers $\mathbfcal{T}_{m_1 : m_3}$ are called \textbf{row} fibers, and the fibers $\mathbfcal{T}_{m_1m_2:}$ are called \textbf{tube} fibers. Thus in analogy with the matrix case, the first of the three outer product decompositions above will be called the \textbf{standard column-wise decomposition}, the second will be called the \textbf{standard row-wise decomposition}, and the third will be called the \textbf{standard tube-wise decomposition}. Notice in particular that the rows $\mathbf{A}_{h:}$ and columns $\mathbf{A}_{:w}$ of a matrix $\mathbf{A}$ are also fibers using the definition given above.

\subsection{Tensor Rank}
\label{sec:tensor-rank}

The smallest possible number of non-redundant (i.e. linearly independent) terms in any such sum (exactly) equalling the tensor is known as the \textbf{rank}\footnote{This is the usage found in \cite{kolda_bader} and elsewhere, but some sources refer to the order of the tensor as its ``rank''.} of the tensor\cite{silva_hla}\cite{lim_hla}. This is the motivation for why some sources preferentially call elementary tensors ``rank-one tensors''. If the rank of a given tensor $\mathbfcal{T}$ is the (positive integer) $R$, then (see \cite{kolda_bader}) any decomposition of $\mathbfcal{T}$ into a linear combination of $R$ elementary tensors is called a \textbf{rank decomposition} of $\mathbfcal{T}$\cite{lim_hla}. \\

When the tensor is a matrix, this notion of rank coincides exactly with the rank of the linear transformation represented by the matrix. In short this is because\footnote{Compare this with example 4 of section 14.2 of \cite{silva_hla}.} any rank-one matrix $\mathbf{v} \otimes \mathbf{w}$ corresponds to a linear function which takes the dot product of the input vector with $\mathbf{w}$ and then multiplies $\mathbf{v}$ by the resulting scalar (which is a rank-one linear transformation), see appendix \ref{sec:repr-line-funct}.\\ 

Techniques like the singular value decomposition (SVD) allow one to find the rank of any matrix in polynomial time. This is not the case for tensors of order $O \ge 3$. For such tensors, finding the exact rank is known to in general be NP-hard (see \cite{hillar_lim}\cite{hastad}). In the case of an $H \times W$ matrix $\mathbf{A}$, it follows from the two tensor product decompositions given above that an upper bound on its rank is $\min \{ H, W\}$. It is well-known that this upper bound is tight. In general, the analogous upper bound for higher-order tensors, $\min_{\omega \in [O]} \{ \prod_{o \not= \omega}^O M_O  \}$, e.g. for an order-three tensor $\min \{ M_1 M_2, M_1 M_3, M_2 M_3  \}$, is also the tightest upper bound on the rank known to hold generally. (The upper bound follows from the fiber-wise tensor product decompositions given above.) However, tighter upper bounds for the largest possible rank of a tensor of a given shape (the \textit{maximum rank}) are known for e.g. special shapes of order-three tensors, compare section 3.1 of \cite{kolda_bader}.\\

Almost all\footnote{Both in the sense that the set of matrices which do not have full rank has Lebesgue measure zero and in the sense that the set of matrices which do is an open set in the Zariski topology (and thus dense).} matrices in $\R^{H \times W}$ attain the upper bound on their rank of $\min \{H,W \}$. Analogous statements are \textit{not} true for higher-order tensors. Given a fixed tensor shape $\M$, it is not necessarily the case that ``most'' tensors achieve the maximum possible rank, and in fact it is not necessarily the case that ``most'' tensors have any given rank (called a ``generic rank''), necessitating the notion of ``typical rank'' for higher-order tensors, see for example \cite{generic_typical_ranks}.

\subsection{Tensor Product Decompositions for a Fixed Rank}
\label{sec:outer-prod-decomp-2}

Given a rank $R$ matrix $A$, the ``best''\footnote{With respect to either Frobenius or spectral norm.} rank $r$ approximation (for $r \le R$) of $A$ is known\footnote{The result is known as the Eckart-Young theorem \cite{Strang2018}.} to be the first $r$ terms of the singular value decomposition of $A$ (when the terms in the SVD are ordered by decreasing size of the singular values). Unfortunately, for higher-order tensors, no analogous result exists. In fact, for a given integer $r > 1$, a best rank-$r$ approximation of a given tensor $\mathbfcal{T}$ might not even exist, since the set of all tensors with rank equal to $r$ is not closed \cite{kolda_bader}\cite{hillar_lim}, or equivalently the tensor rank function is not upper semicontinuous\cite{desilva_lim}, necessitating the concept of ``border rank''. For example, in $\mathbb{R}^{2 \times 2 \times 2}$, \textit{no} rank-three tensor has a best rank-two approximation\cite{hillar_lim}. This phenomenon doesn't depend upon the choice of norm or Bregman divergence used, and occurs with positive probability\cite{desilva_lim}. For more details regarding such difficulties, consult e.g. section 3.3 of \cite{kolda_bader} or section 7 of \cite{hillar_lim}. Even in the case of $r=1$, where a best approximation is guaranteed to exist\footnote{The set of all rank-one tensors is closed, corresponding to the Segre variety from algebraic geometry\cite{landsberg}.}, the problem of finding it is in general NP-hard\cite{hillar_lim}.\\ 

Thus, in general, not only is the problem of finding the exact rank of a given tensor $\mathbfcal{T}$ computationally intractable, so is the problem of finding an optimal low rank approximation for $\mathbfcal{T}$ for a fixed rank $r$. Methods for finding a rank $r$ approximation for a fixed rank $r$ in polynomial time do exist, but they either come with guarantees of only ``weak optimality'' which usually hold under strong and/or difficult to verify assumptions, or have no optimality guarantees of any sort at all and are completely heuristic. A detailed discussion is outside the scope of this review. The interested reader is urged to consult for example section 3.4 of \cite{kolda_bader} for an introduction. In practice, methods for finding such non-optimal approximate decompositions are usually what is being referred to when the term \textbf{CP decomposition}\footnote{``CP'' is an acronym which stands for two acronyms: ``CANDECOMP'' (canonical decomposition) and ``PARAFAC'' (parallel factors), see Table 3.1 of \cite{kolda_bader} for more names.} is used, since both finding exact decompositions and optimal approximate decompositions are usually intractable. Moreover, since one typically chooses $r$ to be (relatively) small, conceptually CP decompositions usually also correspond in practice to ``low rank (non-optimal) approximate outer product decompositions''.\\

For overviews of some of the results about the computational hardness of tensor problems, consult e.g. \cite{hillar_lim} or section \textbf{H.} of \cite{curtensor1}. At the very least, outer product decompositions of higher-order tensors benefit from uniqueness properties comparable to and in some cases better than the counterpart matrix results, see e.g. \cite{Kruskal1976}\cite{Kruskal1977}\cite{Kruskal1989}\cite{Sidiropoulos2000} for some basic results about uniqueness.

%%% Local Variables:
%%% mode: latex
%%% TeX-master: "../new_notes_draft"
%%% End:

\subsection{Applications of Tensor Product Decompositions to Statistics}
\label{apps-outer-prod}

For a statistician, one of the most important benefits of calculating tensor product decompositions of tensor-valued data is arguably psychological: it allows one to understand a complicated tensor, and thus interpret it statistically, via a divide-and-conquer approach:

\begin{enumerate}
\item Understand the tensor as a sum of elementary (i.e. rank one) tensors. This reduces the problem of understanding arbitrary tensor data to the simpler problem of understanding data formatted as elementary tensors.
\item Understand each elementary tensor in terms of the vectors in its tensor product factorization. This reduces the problem of understanding data formatted as elementary tensors to the problem of understanding vector-valued data.
\end{enumerate}

Thus, heuristically speaking, tensor product decompositions allow the statistician to reduce the (potentially very complicated) problem of understanding tensor-valued data to (multiple instances of) the problem of understanding vector-valued data. This can be very helpful inasmuch as most statisticians have more training and experience doing the latter than doing the former.\\

Note also that, to a statistician, the fact that these tensor product
decompositions are usually approximate is most likely not so
important: the data is already being interpreted as a noisy
approximation of an unobservable ``ground truth'' anyway. Replacing the original tensor-valued data with approximate
tensor product decomposition versions of that data is not a major issue
inasmuch as the original data is already being interpreted as something which is not itself ``the truth''.\\

Moreover, using approximate tensor product decompositions is analogous
to assuming the ``manifold hypothesis'', that for real-world high-dimensional
data the ``data tend to lie near a low
dimensional manifold'' \cite{manifold_hypothesis}. Thus even for
very high-dimensional (e.g. corresponding to the number of entries in
a tensor of a given shape) data, one should still expect to be able to
faithfully represent the phenomena of interest by considering a much
lower-dimensional (e.g. corresponding to the number of entries in all
of the vectors constituting a rank-$R$ approximate tensor product
decomposition of a tensor of a given shape) space. Thus using
(approximate) tensor decompositions, in particular tensor product
decompositions, corresponds to a specific instance of a general
philosophy widely employed to make problems in high-dimensional
statistics tractable.

\subsubsection{Special Case: Matrices}
\label{sec:spec-case:-matr}

The product of two matrices can be written as the sum of the tensor products of the columns of the first matrix with the rows of the second matrix. In contrast to the more typical entrywise ``row-column'' way of understanding matrix multiplication, the tensor product ``column-row'' is considered by some to be the ``right way''\cite{Strang2018} of understanding matrix multiplication.\\ 

More specifically, given\footnote{$H$ stands for ``height'', $W$ stands for ``width'', and $M$ stands for both ``middle'' and ``mode''.} an $H \times M$ matrix $\mathbf{A} \in \mathbb{R}^{[H] \times [M]}$ and an $M \times W$ matrix $\mathbf{B} \in \mathbb{R}^{[M] \times [W]}$, letting $\mathbf{A}_{:m} \in \mathbb{R}^H$  for all $m \in [M]$ denote the columns of $\mathbf{A}$, and letting $\mathbf{B}_{m:} \in \mathbb{R}^W$ for all $m \in [M]$ denote the rows of $\mathbf{B}$, one always has the following relationship:
\[ 
\mathbf{A B} = \sum_{m=1}^M \mathbf{A}_{:m} \otimes \mathbf{B}_{m:}  \,. 
\]
It is easy to show that both the left and right hand sides of the
above equation are matrices in $\mathbb{R}^{[H] \times [W]}$, and that
all of their entries are the same. Therefore one sees that all
factorizations of a matrix into the (matrix) product of two matrices
are tensor product decompositions. Thus tensor product decompositions
are the generalization of a very familiar concept.\\

Moreover, the
Eckart-Young theorem\cite{Strang2018} guarantees that for tensors of order two, the best\footnote{In terms of both
  Frobenius norm or spectral norm.} rank $R$ approximate tensor product decomposition can be found
in polynomial time by taking the first $R$ terms of that matrix's
singular value decomposition (SVD), which can be written as
\[ 
 \sum_{r=1}^R (\sigma_r \mathbf{u}_r) \otimes \mathbf{v}_r \,, 
 \]
where $\sigma_r$ denotes singular values, $\mathbf{u}_r$ denotes left
singular vectors, and $\mathbf{v}_r$ denotes right singular
vectors. Applications of the SVD to statistics are well-known\cite{Elden}\cite{Gentle}\cite{Gentle2017}, for
example principal components analysis (PCA) can be considered as a special case\footnote{Since the
  decompositions given by the spectral theorem and the singular value
  decomposition coincide for a positive semidefinite matrix (PSD), and
  a matrix is PSD if and only if it is the covariance matrix of some
  distribution.} and is used extensively in the analysis of data\cite{pca_review}. SVD is perhaps the most widely used and widely known low-rank approximation method\cite{extraordinary_svd}\cite{Golub2013}\cite{Kalman1996}, e.g. as it is used in PCA,
and its role in calculating Moore-Penrose pseudoinverses\cite{pseudoinverse_review} also makes it relevant to solving ordinary least squares problems as well. Pseudoinverses and the SVD can be used for the analysis of genomics data, see \cite{Alter2004}.\\

Two other very widely used examples of two-term matrix factorizations are the Gram-Schmidt process implemented via QR factorizations \cite{rank_revealing_QR}\cite{Gentle}\cite{Gentle2017}\cite{Elden}\cite{Golub2013}, and Gaussian elimination implemented via LU decompositions\cite{Gentle}\cite{Gentle2017}\cite{Elden}\cite{Golub2013}. The numerous applications of matrix factorizations to data science and statistics overall are already a very well-studied subject, see for example any of \cite{Elden}\cite{Gentle}\cite{Gentle2017}\cite{Dunn} for introductions. Matrix factorizations also have many applications to subfields of data science, such as bioinformatics, see e.g.  \cite{bioinformatics_matrix_factorization}\cite{bioinformatics_matrix_factorization2}\cite{bioinformatics_matrix_factorization3}\cite{bioinformatics_matrix_factorization4}. \\

Since no matrix factorization can improve upon the SVD in terms of accuracy with respect to any unitarily invariant norm\cite{eck-young-unitarily-invariant1}\cite{eck-young-unitarily-invariant2} (again this is the statement of the Eckart-Young Theorem), proposed alternate matrix factorizations almost always emphasize improvements with regards to SVD's greatest potential weakness, namely interpretability. Considering the ``tradeoff curve'' for accuracy vs. interpretability, modest sacrifices of accuracy are often worth substantial gains in interpretability. And in many cases the required sacrifice of accuracy is actually small\cite{jordan_nmf}.\\

In semi-nonnegative matrix factorization (NMF), an arbitrary matrix $\mathbf{M} \in \R^{I \times J}$ is decomposed as
\[  
\mathbf{M} \approx  \mathbf{C}\mathbf{W}^\top \,, 
   \]
where the entries of the weight matrix $\mathbf{W} \in \R^{J \times K}$ are constrained to be nonnegative\cite{gillis_seminmf}\cite{jordan_nmf}. This has the effect of representing the columns of the original matrix $\mathbf{M}$ as \textit{conic} combinations:
\[   
\mathbf{M}_{:j} = \sum_{k=1}^K W_{jk} \mathbf{C}_{:k} 
 \]
of the columns of the (usually substantially lower-rank) matrix $\mathbf{C}\in \R^{I \times K}$. The columns of the matrix $\mathbf{C}$ can be interpreted as \textit{components} or \textit{patterns} and the matrix $\mathbf{C}$ itself can be interpreted as a ``dictionary''\cite{Hamon}, thus the goal of the semi-NMF problem can be interpreted as finding an accurate approximate representation of all of the columns of $\mathbf{M}$ as conic combinations of a small number of dictionary elements. The semi-NMF problem can also be interpreted as a relaxation of the $K$-means clustering problem \cite{jordan_nmf}. The idea is that in the $K$-means clustering problem\footnote{This formulation of $K$-means shows that it is a more difficult problem than semi-NMF, which is known to be NP-hard\cite{gillis_seminmf}, and thus one sees that the problem of finding an exact solution of the $K$-means objective is NP-hard.}, all of the weights would be constrained to be either exactly $1$ or exactly $0$, i.e. indicators variables for cluster membership. Thus the columns of $\mathbf{C}$ are analogous to cluster centroids. \\

Moreover, just like the unconstrained CP decomposition for higher-order tensors\cite{hillar_lim}, the constrained CP decomposition for second order tensors represented by the semi-NMF procedure is known to be NP-hard already for the case of finding a best rank-$1$ approximation, and to sometimes not have any well-defined solution for higher-rank approximations\cite{gillis_seminmf}.\\

In the case of non-negative matrix factorization\cite{Lee_1999}\cite{wang2013nonnegative}, the original matrix $\mathbf{M}$ is assumed to be nonnegative, and now, like $\mathbf{W}$, the matrix $\mathbf{C}$ is constrained to be non-negative. This ensures that the nonnegative columns of $\mathbf{M}$ are written as conic combinations of \textit{nonnegative} vectors, increasing interpretability compared to e.g. SVD where they might be written instead as a linear combination with negative coefficients of columns with negative entries. In certain contexts, e.g. like signal processing, such negative coefficients, or vectors with negative coefficients, correspond to ``unphysical'' scenarios and thus lack any useful interpretation, making the SVD wholly unsuitable. For an introduction to the applications of NMF to signal processing see e.g. \cite{Fu2019} or \cite{amari_nonnegative}.\\

NMF can be considered the special case of non-negative CP decomposition for order two tensors. While finding an optimal unconstrained CP decompositions for a second order tensor can be done in polynomial time (again by the Eckart-Young theorem this is just computing the SVD), the corresponding problem with non-negativity constraints is already NP-hard\cite{vavasis2009}\cite{gillis_nmf} for second order tensors, not just for higher order tensors\cite{amari_nonnegative}. One way to see this is by relating NMF to the NP-hard\cite{DasG95} Nested Polytope Problem\cite{gillis_nmf} or to the special case and also NP-hard Intermediate Simplex Problem\cite{vavasis2009}. Another way is to relate the NMF problem to the problem of testing the copositivity of a matrix\cite{copositive1}\cite{copositive2}, which is NP-hard\cite{copositiveNPHard}.

%%% Local Variables:
%%% mode: latex
%%% TeX-master: "../new_notes_draft"
%%% End:

\subsubsection{Outer Product Decompositions of Higher-Order Tensors}
\label{sec:outer-prod-decomp-1}

Applications of outer product decompositions to higher-order tensors
are still an active area of research, so a comprehensive overview
is impossible. As was implied above, in any application where
tensor-valued data occurs, outer product decompositions are useful for
increased interpretability. At a high level, using (low
rank) approximate outer product decompositions of higher order tensors
often makes it possible to drastically decrease the number of free
parameters in the problem. Consider a cubical order $O$ tensor where
the dimension of each mode is $M$. Then $M^O$ parameters are required
to specify each entry of the tensor directly, which is exponential in
the order of the tensor. In contrast, a rank $R$
(approximate) outer product decomposition of the tensor only requires
$RMO$ parameters to specify the tensor. Depending on the value of $R$,
this can lead to drastic reductions in: (i) the space in memory
required to store the tensor, important for memory-bound applications,
(ii) the number of computations required to operate on the tensor,
import for compute-bound applications, (iii) the amount of
data/observations required to make meaningful statistical
inferences. Thus outer product decompositions
enable much more efficient algorithms and much more powerful
statistical procedures for working with tensor-valued data.\\

The
book\footnote{Unlike in \cite{hackbusch}, no distinction between ``tensor decompositions'' and
  ``tensor representations'' is made in this review.}\cite{hackbusch} is a comprehensive reference
regarding efficient computations with tensors. An example of
powerful statistical procedures using outer-product decompositions can
be found in the work on tensor regression for neuroimaging data done
by Professor Lexin Li.

%%% Local Variables:
%%% mode: latex
%%% TeX-master: "../new_notes_draft"
%%% End:

\paragraphsection{Regression with Tensor Predictors}
\label{CP-tensor-predictor}

For example,
consider the problem examined in \cite{li_glm} of regressing a
scalar response variable $y$ (e.g. a binary indicator of disease
status) against both vector-valued ``ancillary''\footnote{This is not
  the exact language used in the paper, merely this author's interpretation.} covariates (e.g. age
and sex) $\mathbf{z}$ and a tensor-valued ``primary'' covariate
$\mathbfcal{X}$ (e.g. an fMRI image). Such a setup corresponds to using
fMRI scans of the brain, accounting for the patient's age and sex, to
predict incidence of neurological disorders like ADHD or autism. Any
patterns uncovered from the best-fitting parameters might then be used
to examine existing hypotheses or develop new hypotheses about which
parts of the brain are relevant to the onset of these diseases.\\

In the classical GLM, all covariates are vector-valued, so
``collapsing'' the ``ancillary'' and ``primary'' covariates into a
single vector $\mathbf{x}$, the predicted values of $Y$, $\mu =
\mathbb{E}[y|\mathbf{x}]$, are found via the relationship involving a
strictly increasing (thus invertible) link function $g$:
\[ 
 g(\mu) = \alpha + \boldsymbol{\beta} \bullet \mathbf{x}  \,, 
 \]
where $\alpha$ is an intercept coefficient, and $\boldsymbol{\beta}$ is a
coefficient (column) vector to be solved for.\\

To generalize to the case of a
tensor-valued covariate $\mathbfcal{X}$, the naive approach is to replace $\boldsymbol{\beta}$ with a
coefficient tensor $\mathbfcal{B}$ and take the entrywise dot product\footnote{In other words, this is a contraction along all $O$ modes corresponding to the mode index set $[O]$ of the multi-index set $\M$. Compare this with Fact 14 of section 15-1 in \cite{lim_hla}.} (denoted by $\bullet_{[O]}$) of
$\mathbfcal{B}$ and $\mathbfcal{X}$:
\[  
g(\mu) = \alpha + \boldsymbol{\gamma} \bullet \mathbf{z} +  \mathbfcal{B} \bullet_{[O]}
  \mathbfcal{X}  \,, 
 \]
where the extra vector of coefficients $\boldsymbol{\gamma}$ needs to be introduced
to account for the ``ancillary'' vector-valued covariates $\mathbf{z}$
which can no longer be ``collapsed'' into the same array as the
``primary'' tensor-valued covariate $\mathbfcal{X}$. Under this model the
number of parameters corresponding to the coefficient tensor
$\mathbfcal{B}$ is equal to the number of entries in the covariate tensor
$\mathbfcal{X}$, which in general can be extremely large. Li et al give
the example of a $256 \times 256 \times 256$ MRI image where this
corresponds to $256^3 \approx 16,000,000$ parameters. Not only is this
in general computationally intractable, but moreover most real-world
applications will not have sample sizes anywhere near large enough for
this many parameters to be fit with any remotely meaningful
statistical guarantees.\\

Moreover, this approach corresponds to essentially ``vectorizing'' the
tensor $\mathbfcal{X}$, and thus completely ignoring ``the inherent
spatial structure of the image that possesses a wealth of
information'' \cite{li_glm}. Especially since one is actually interested
in ``regressing against brain regions'', rather than against individual voxels
of the MRI image which are by themselves mostly meaningless, this is a
very serious disadvantage. A space corresponding to relevant brain
regions would of course be much lower-dimensional than the space
corresponding to all of the individual voxels in the image, which
suggests that something like the ``manifold hypothesis'' should be applicable to this
problem, and that it should be possible to replace the coefficient
tensor $\mathbfcal{B}$ with a much lower-dimensional estimand and still
retain statistical fidelity. This is exactly what the authors of \cite{li_glm} did as well as what they found. More specifically, they
replaced the above naive GLM with:
\[  
 g(\mu) = \alpha + \boldsymbol{\gamma} \bullet \mathbf{z}  + \left(
    \sum_{r=1}^R \boldsymbol{\beta}_1^{\smallsuper{r}} \otimes \boldsymbol{\beta}_2^{\smallsuper{r}} \otimes \cdots
    \otimes \boldsymbol{\beta}_O^{\smallsuper{r}} \right) \bullet_{[O]} \mathbfcal{X}  \,,
  \]
i.e. they replaced the coefficient tensor $\mathbfcal{B}$ with a rank-$R$
tensor product decomposition. When $O=3$, like with MRI images, this
changes the number of $\boldsymbol{\beta}$ parameters to be fit for the
tensor-valued covariates to be $R(256+256+256)
= 768R$, a drastic reduction from the $16$ million for the naive
approach. Again, this is important not just for computational
feasibility, interpretability, and retaining spatial information, but
it is also necessary for MRI studies where the available sample size
is usually at most only on the order of several hundred subjects. The paper \cite{li_glm} contains many more salient details, as well as a description of the method's promising results. This work has also recently been extended by the authors to incorporate a generalized estimating equation (GEE) approach to account for ``intra-subject correlation of responses''\cite{tensor_gee}.

%%% Local Variables:
%%% mode: latex
%%% TeX-master: "../new_notes_draft"
%%% End:

\paragraphsection{Regression with Tensor Responses}
\label{CP-tensor-response}

Such an approach is useful not only for regression with scalar
responses and tensor
covariates, but also regression with tensor responses and vector
covariates. The latter situation is also applicable to neuroimaging, where now one
attempts to predict MRI or fMRI scans based on disease status and
controlling for additional covariates such as age and sex, instead of
using the MRI or fMRI scans to predict disease status\cite{li_sparse}. In this case one models the order-$O$ tensor-valued responses
$\mathbfcal{Y}_i$ as being generated from observations of i.i.d. random variables:
\[  
\mathbfcal{Y}_i = \mathbfcal{B} \bullet_{O+1} \mathbf{x}_i +
  \mathbfcal{E}_i \,, 
\]
where $\mathbfcal{B}$ is an order-$(O+1)$ coefficient tensor, the
$\mathbf{x}_i$ are the vector-valued covariates, and the $\mathbfcal{E}_i$ are order-$O$
tensor-valued errors (independent of the covariates
$\mathbf{x}_i$).\\

The notation requires $\bullet_{O+1}$ some
explanation. It is the same concept which is denoted by the symbol $\times_o$ in \cite{kolda_bader} and
corresponds to the following operation. Given any, it doesn't matter which\footnote{Compare this claim with Fact 12 of section 15-1 in \cite{lim_hla}.}, \textit{exact} tensor product
decomposition of $\mathbfcal{B}$ into a sum of
elementary tensors:
\[ 
  \mathbfcal{B} =  \sum_{r=1}^R \boldsymbol{\beta}_1^{\smallsuper{r}} \otimes \cdots \otimes
  \boldsymbol{\beta}_O^{\smallsuper{r}} \otimes \boldsymbol{\beta}_{O+1}^{\smallsuper{r}} \,, 
  \]
then essentially by definition the following relationship holds:
\[  
\mathbfcal{B} \bullet_{O+1} \mathbf{x}_i = \sum_{r=1}^R  \left(
   \boldsymbol{\beta}_{O+1}^{\smallsuper{r}} \bullet \mathbf{x}_i  \right) \boldsymbol{\beta}_1^{\smallsuper{r}} \otimes
  \cdots \otimes \boldsymbol{\beta}_O^{\smallsuper{r}} \,. 
\]
In other words, one essentially ``takes the dot product of
$\mathbf{x}_i$ along the $(O+1)$'th mode of the coefficient tensor
$\mathbfcal{B}$''. This is a well-defined operation, since the function:
\[
  \begin{array}{rcl}
  \R^{M_1} \times \cdots \times \R^{M_O}
    \times \R^{M_{O+1}}  & \to & \R^{[M_1] \times \cdots \times [M_O]} \\
    (\boldsymbol{\beta}_1, \dots, \boldsymbol{\beta}_O, \boldsymbol{\beta}_{O+1}) &  \mapsto &
 \left\langle
    \mathbf{x}_i, \boldsymbol{\beta}_{O+1} \right\rangle \boldsymbol{\beta}_1 \otimes \cdots
  \otimes \boldsymbol{\beta}_O                                                                                               
  \end{array}
 \,, 
  \]
is multilinear, and thus by the universal property of the tensor
product extends from a function defined on elementary tensors to a
well-defined linear function ${\R^{[M_1] \times \cdots \times [M_O] \times
    [M_{O+1}]}} \to {\R^{[M_1] \times \cdots \times [M_O]}}$.\\

An explicit formula for this operation in terms of the input and output
tensors is given in \cite{kolda_bader} which can be thought of as using a rank-one decomposition of the input tensor in
terms of its mode-$(O+1)$ fibers. In other words, writing $\mathbfcal{B}$
using the tensor product decomposition\footnote{Observe that $[O+1]\setminus(O+1) = [O]$, so $\genmltidx{m}{[O+1]\setminus(O+1)} = \m:$ and $\genMltidx{M}{[O+1]\setminus(O+1)} = \M$. }:
\[ 
 \mathbfcal{B} = \sum_{ \m \in  \M }
  \unitvector{m_1} \otimes \cdots \otimes
  \unitvector{m_O} \otimes \mathbfcal{B}_{\m : } \,, 
 \]
then via the definition given above one has that the following relationship is true:
\[ 
 \mathbfcal{B} \bullet_{O+1} \mathbf{x}_i = \sum_{ \m  \in
   \M }  \left( \mathbfcal{B}_{\m : } \bullet   \mathbf{x}_i   \right)
  \unitvector{m_1}  \otimes \cdots \otimes
   \unitvector{m_O}   
\]
which, up to differences in notation, is the same formula as the one
found in \cite{kolda_bader} using the fact that, when a tensor is written as a
linear combination of the unit tensors, the coefficients correspond
exactly to the entries of the tensor. So the above formula says that
the $\m$'th entry of the tensor $\mathbfcal{B} \bullet_{O+1}
\mathbf{x}_i$ is $ \mathbfcal{B}_{\m:} \bullet \mathbf{x}_i  $, the dot product of
$\mathbf{x}_i$ with the $\m$'th mode-$(O+1)$ fiber of $\mathbfcal{B}$, exactly the same as the
formula given in \cite{kolda_bader}. This operation can
be thought of as a special case\endnote{Specifically the linear
  transformation in question is a linear functional $\R^{M+1} \to \R$,
given by taking the dot product with $\mathbf{x}_i$. The resulting
order-$O+1$'th tensor is degenerate since the dimension
of the last mode is $1$. Applying the standard isomorphism found in
Remark 3.25 of \cite{hackbusch}, the result is the non-degenerate
order-$O$ tensor defined above.} of the ``mode-$o$'' product which will
be discussed later in connection with Tucker decompositions.\\

Compared to the case of regression with tensor-valued predictors, the number of parameters using a naive approach to this problem is even more extreme. Using again the example of $256 \times 256 \times 256$ MRI images, the number of parameters with the naive approach is equal to the number of entries of the coefficient tensor $\mathbfcal{B}$. In this case $\mathbfcal{B}$ is a $4$th-order tensor with $256 \times 256 \times 256 \times d$ where $d$ is the length of the covariates vectors $\mathbf{x}_i$. In \cite{li_glm} the value of $d$ is $5$, so in that case the number of parameters to be estimated would be $5 \times 256^3 \approx 83,000,000$. \\

Even using a slightly less naive approach, where the parameters corresponding to the tensor-valued responses are ``separated'' from the parameters corresponding to the covariates, i.e.
\[  
 \mathbfcal{B} = \mathbfcal{B}_O \otimes \boldsymbol{\beta}_{O+1} \,, 
 \]
where $\mathbfcal{B}_O$ is an order-$O$ tensor with the same shape as the tensor-valued responses (and thus also the same shape as the error tensors $\mathbfcal{E}_i$) and $\boldsymbol{\beta}_{O+1}$ is a vector with the same shape as the covariates, the number of parameters to fit would still be the same as the number of parameters for the completely naive approach mentioned above in the case of tensor-valued predictors ($256^3 + 5 \approx 16,000,000$). Therefore the case of regression with tensor-valued responses benefits even more from utilizing full tensor product decompositions than the case of regression with tensor-valued predictors, which itself already benefitted substantially from such an approach.\\

The approach used successfully in \cite{li_sparse} is to instead use a low-rank coefficient tensor $\mathbfcal{B}$:
\[  
\mathbfcal{B} = \sum_{r=1}^R w_r \boldsymbol{\beta}_1^{\smallsuper{r}} \otimes \cdots \otimes \boldsymbol{\beta}_O^{\smallsuper{r}} \otimes \boldsymbol{\beta}_{O+1}^{\smallsuper{r}} \,, 
 \]
where the weights $w_r$ are assumed to be bounded away from both $0$ and $\infty$ to avoid pathological behavior, and the vectors $\boldsymbol{\beta}_o^{\smallsuper{r}}$ are assumed to all be sparse and lie in the unit sphere.\\

While this is technically a constrained tensor product decomposition (as opposed to a ``pure'' one), the same motivations for using tensor product decompositions hold for both the case of regression with tensor-valued predictors and regression with tensor-valued responses. Computational cost is reduced, interpretability is increased, statistical efficiency is increased, spatial structure is preserved, and the data is expected a priori to be low-rank (e.g. regions in the brain are important features, and individual voxels aren't) in any case.

%%% Local Variables:
%%% mode: latex
%%% TeX-master: "../new_notes_draft"
%%% End:

\paragraphsection{Other Applications of Tensor Product Decompositions of Higher-Order Tensors}

Similar benefits have been found in applications to other areas. The book \cite{cumulant_book} describes the use of CP decompositions to define generalizations of the Schur complement and Cholesky factorization applicable to (symmetric) higher-order tensors and uses them to study cumulant tensors.

%%% Local Variables:
%%% mode: latex
%%% TeX-master: "../new_notes_draft"
%%% End:

\theendnotes
\setcounter{endnote}{0}

\section{Tucker Decompositions}
\label{sec:tuck-decomp-1}

\subsection{Mode-$o$ Products}
\label{sec:mode-o-products}

Any introduction to Tucker decompositions requires an explanation of the mode-$o$ product\footnote{Called the mode-$m$ product in \cite{kolda_bader}, but the terminological difference is unimportant, because both $o$ and $m$ are meant to denote some arbitrary index corresponding to one of the modes of the tensor.}, since that operation is at the core of the Tucker decomposition's definition.
As usual with operations on tensors, it is easiest to define the mode-$o$ product on elementary tensors and then use the universal property of the tensor product to extend the definition by
linearity to all tensors.\\

Consider an order-$O$ elementary tensor $\mathbfcal{T}\in \R^{\M}$,
\[  
\mathbfcal{T} = \mathbf{v}_1 \otimes \cdots \otimes \mathbf{v}_{o-1} \otimes \mathbf{v}_o \otimes \mathbf{v}_{o+1} \otimes \cdots \otimes \mathbf{v}_O \,. 
 \]
Then given a linear transformation ${\mathcal{L}}: \R^{M_o} \to \R^{N_o}$, of course the vector $\mathcal{L} (\mathbf{v}_o) \in \R^{N_o}$, and using that new vector one can define a new elementary tensor $\mathbfcal{T} \bullet_o \mathcal{L} \in \R^{ \Mltidx{\tilde{M}}  }$
\[
 \mathbfcal{T} \bullet_o \mathcal{L} \defequals \mathbf{v}_1 \otimes \cdots \otimes  \mathbf{v}_{o-1} \otimes \mathcal{L}(\mathbf{v}_o) \otimes \mathbf{v}_{o+1} \otimes \cdots \otimes \mathbf{v}_O \,, 
 \]
where $\Mltidx{\tilde{M}} \defequals [M_1] \times \cdots\times [M_{o-1}] \times [N_o] \times [M_{o+1}] \times \dots \times [M_O]$.\\

As a consequence of the linearity of $\mathcal{L}$, one has that the function $\bigtimes_{o=1}^O \R^{M_o} \to  \R^{\Mltidx{\tilde{M}}}$,
\[ 
  (\mathbf{v}_1, \cdots, \mathbf{v}_{o-1}, \mathbf{v}_o, \mathbf{v}_{o+1}, \dots, \mathbf{v}_O) \mapsto \mathbf{v}_1 \otimes \dots \otimes \mathbf{v}_{o-1} \otimes \mathcal{L}(\mathbf{v}_o) \otimes \mathbf{v}_{o+1} \otimes \dots \otimes \mathbf{v}_O \,,  
 \]
is multilinear. Therefore the universal property of the tensor product says that extending the above operation on elementary tensors via linearity to arbitrary tensors leads to a well-defined linear function $\R^{ \M } \to \R^{\Mltidx{\tilde{M}}}$. Specifically, the \textbf{mode-$o$ product} is defined as follows: given an arbitrary tensor $\mathbfcal{T} \in \R^{ \M }$, and \textit{any} (exact) tensor product decomposition of $\mathbfcal{T}$:
\[ 
\mathbfcal{T} = \sum_{k=1}^K  \mathbf{v}_1^{ \smallsuper{k} } \otimes \cdots \otimes \mathbf{v}_{o-1}^{\smallsuper{k}  } \otimes \mathbf{v}_o^{ \smallsuper{k}  } \otimes \mathbf{v}_{o+1}^{ \smallsuper{k} } \otimes \cdots \otimes \mathbf{v}_O^{ \smallsuper{k}  } \,,     
\]
then the \textbf{mode-$o$ product} of the tensor $\mathbfcal{T}$ with the linear transformation $\mathcal{L}: \R^{M_o} \to \R^{N_o}$ is the following tensor in $\R^{\Mltidx{\tilde{M}}}$ (where $\Mltidx{\tilde{M}}$ is the same multi-index set as before), denoted by ${\mathbfcal{T} \bullet_o \mathcal{L} }$:
\[ 
\mathbfcal{T} \bullet_o \mathcal{L} = \sum_{k=1}^K  \mathbf{v}_1^{  \smallsuper{k}  } \otimes \cdots \otimes \mathbf{v}_{o-1}^{\smallsuper{k} } \otimes \mathcal{L} (\mathbf{v}_o^{ \smallsuper{k} }) \otimes \mathbf{v}_{o+1}^{ \smallsuper{k}  } \otimes \cdots \otimes \mathbf{v}_O^{ \smallsuper{k}  } \,.     
\]
Again, while it may look like the above definition depends crucially on the chosen tensor product decomposition for $\mathbfcal{T}$, in fact the universal property of the tensor product guarantees\footnote{Compare this claim with Fact 12 of section 15-1 in \cite{lim_hla}.} that the resulting tensor will be the same regardless of which tensor product product decomposition was originally chosen for $\mathbfcal{T}$. Therefore the mode-$o$ product is well-defined using this definition.\\

The review \cite{kolda_bader} gives another definition of mode-$o$ product, one using matrices and the elements of the tensors in question. Here an explanation for how the two definitions are equivalent is given.\\

Given a tensor $\mathbfcal{T} \in \R^{\M}$, if one takes its standard tensor product decomposition in terms of its mode-$o$ fibers, denoted as before for each $\genmltidx{m}{[O]\setminus o} \in \genMltidx{M}{[O]\setminus o}$ by $\mathbfcal{T}_{\genmltidx{m}{[O]\setminus o}  } \in \R^{M_o}$, namely:
\[   
\mathbfcal{T} = \sum_{ \genmltidx{m}{[O]\setminus o} \in \genMltidx{M}{[O]\setminus o} } \unitvector{m_1} \otimes \ldots \otimes  \unitvector[M_{o-1}]{m_{o-1}} \otimes \mathbfcal{T}_{\genmltidx{m}{[O]\setminus o} } \otimes \unitvector[M_{o+1}]{m_{o+1}}  \otimes \ldots \otimes \unitvector{m_O} \,,  
 \]
then the definition given above for the mode-$o$ product implies that:
\[ 
 \mathbfcal{T} \bullet_{o} \mathcal{L} =  \sum_{\genmltidx{m}{[O]\setminus o} \in \genMltidx{M}{[O]\setminus o}  } \unitvector{m_1}\otimes \ldots \otimes \unitvector[M_{o-1}]{m_{o-1}}  \otimes \mathcal{L} \left(  \mathbfcal{T}_{ \genmltidx{m}{[O]\setminus o} }  \right) \otimes \unitvector[M_{o+1}]{m_{o+1}} \otimes \ldots \otimes \unitvector{m_O} \,. 
 \]
The above equation says that each mode-$o$ fiber has the linear transformation $\mathcal{L}$ applied to it\cite{kolda_bader}. To find the $\mltidx{\tilde{m}}$'th entry of $\mathbfcal{T} \bullet_o \mathcal{L}$, one can look at the\endnote{Since the multi-index sets $\M$ and $\Mltidx{\tilde{M}}$ differ at most only in their $o$'th mode, $\genMltidx{M}{[O]\setminus o} = \genMltidx{\tilde{M}}{[O]\setminus o}$. In particular $\mathbfcal{T}$ and $\mathbfcal{T} \bullet_o \mathcal{L}$ are guaranteed to have the same number of mode-$o$ fibers, with a one-to-one correspondence given by $\mathbfcal{T}_{\genmltidx{m}{[O]\setminus o} } \leftrightarrow \mathcal{L} \left( \mathbfcal{T}_{ \genmltidx{m}{[O]\setminus o}}\right)$, as hopefully the equation above makes clear (since $\genmltidx{\tilde{m}}{[O]\setminus o} = \genmltidx{m}{[O]\setminus o}$).} $\genmltidx{\tilde{m}}{[O]\setminus o}$'th mode-$o$ fiber $\mathbfcal{T}_{\genmltidx{\tilde{m}}{[O]\setminus o} }$ of $\mathbfcal{T}$, apply $\mathcal{L}$ to it, then take the $\tilde{m}_o$'th entry of $\mathcal{L}\left(\mathbfcal{T}_{\genmltidx{\tilde{m}}{[O]\setminus o}}\right)$. One gets the same entrywise formula given in \cite{kolda_bader}, if one replaces $\mathcal{L}$ with a matrix $\mathbf{L}$ representing it with respect to the standard basises of $\R^{M_o}$ and $\R^{N_o}$ (see appendix \ref{sec:repr-line-funct}) and represents the mode-$o$ fibers with column vectors\endnote{Which are technically matrices, not vectors. They are ``degenerate'' in the sense of Definition 3.2.4 \cite{hackbusch}, since one of their modes has dimension $1$. The standard isomorphism found in Remark 3.25 of \cite{hackbusch} is usually used to identify vectors with column vectors, but the existence of a conventional linear isomorphism between the two spaces doesn't mean that they are literally equal, just that they may be considered equivalent for the purposes of linear algebra.}.

%%% Local Variables:
%%% mode: latex
%%% TeX-master: "../new_notes_draft"
%%% End:

\subsection{Tensor Product of Linear Transformations}
\label{sec:tens-prod-line}

The mode-$o$ product can be understood as a special case of a more general construction, namely the tensor product of linear transformations. Given $O$ linear transformations, ${\mathcal{L}_1 : \R^{M_1} \to \R^{N_1}  }$, ${\mathcal{L}_2 : \R^{M_2} \to \R^{N_2}  }$, $\dots$, ${\mathcal{L}_O : \R^{M_O} \to \R^{N_O}  }$, verify that the following function is multilinear:
\[ 
 \bigtimes_{o=1}^O \R^{M_o} \to \R^{\N[O]} \,, \quad (\mathbf{v}_1, \mathbf{v}_2, \dots, \mathbf{v}_O) \mapsto \mathcal{L}_1 (\mathbf{v}_1) \otimes \mathcal{L}_2(\mathbf{v}_2) \otimes \cdots \otimes \mathcal{L}(\mathbf{v}_O) \,.   \]
Therefore, as a consequence of the universal property of the tensor product, the following operation is well-defined and results in a linear function\cite{silva_hla}\cite{atiyah-macdonald}\cite{hackbusch}. The \textbf{tensor product} of linear transformations ${\mathcal{L}_1 : \R^{M_1} \to \R^{N_1}  }$, ${\mathcal{L}_2 : \R^{M_2} \to \R^{N_2}  }$, $\dots$, ${\mathcal{L}_O : \R^{M_O} \to \R^{N_O}  }$ is a linear function $\bigotimes_{o=1}^O \mathcal{L}_o: \R^{\M} \to \R^{\N[O]}$ defined as follows: given a tensor $\mathbfcal{T} \in \R^{\M}$ and \textit{any} (exact) tensor product decomposition of the order-$O$ tensor $\mathbfcal{T}$, for example:
\[ 
 \mathbfcal{T} = \sum_{k=1}^K \v{k}_1 \otimes\v{k}_2 \otimes \cdots \otimes \v{k}_O \,, 
 \]
then the value of $\bigotimes_{o=1}^O \mathcal{L}_o$ evaluated at $\mathbfcal{T}$ is the following\footnote{Compare this with Fact 7. from \cite{lim_hla}.}:
\[
 \left(\bigotimes_{o=1}^O \mathcal{L}_o \right) (\mathbfcal{T}) = \sum_{k=1}^K  \mathcal{L}_1(\v{k}_1) \otimes \mathcal{L}_2 (\v{k}_2) \otimes \cdots \otimes \mathcal{L}_O(\v{k}_O) \,. 
\]
Given $O$ linear transformations, ${\mathcal{L}_1 : \R^{M_1} \to \R^{N_1}  }$, ${\mathcal{L}_2 : \R^{M_2} \to \R^{N_2}  }$, $\dots$, ${\mathcal{L}_O : \R^{M_O} \to \R^{N_O}  }$, as well as $O$ linear transformations, ${\mathcal{K}_1 : \R^{N_1} \to \R^{P_1}  }$, ${\mathcal{K}_2 : \R^{N_2} \to \R^{P_2}  }$, $\dots$, ${\mathcal{K}_O : \R^{N_O} \to \R^{P_O}  }$, it follows from the definitions that for any tensor $\mathbfcal{T} \in \R^{\M}$ the following relationship holds:
\[
  \left(\bigotimes_{o=1}^O (\mathcal{K}_o \circ \mathcal{L}_o) \right) (\mathbfcal{T}) = \sum_{k=1}^K \mathcal{K}_1(\mathcal{L}_1(\v{k}_1)) \otimes \cdots \otimes \mathcal{K}_O(\mathcal{L}_O(\v{k}_O)) = \left(\left( \bigotimes_{o=1}^O \mathcal{K}_o \right) \circ \left( \bigotimes_{o=1}^O \mathcal{L} \right) \right) (\mathbfcal{T}) \,,
  \]
or in other words, the ``tensor product distributes over composition"\footnote{\label{kronecker_footnote}After vectorization, this corresponds to the ``mixed product'' property of the Kronecker product. See \ref{sec:spec-cases:-vect} below. It also corresponds to the tensor product being functorial, as does the identity $\bigotimes_{o=1}^O \Id_{\R^{M_o}} = \Id_{\R^{\M}}$. }:
\[ 
 \bigotimes_{o=1}^O (\mathcal{K}_o \circ \mathcal{L}_o) = \left( \bigotimes_{o=1}^O \mathcal{K}_o \right) \circ \left( \bigotimes_{o=1}^O \mathcal{L}_o  \right) \,. 
\]
Letting $\Id_M$ denote the identity (linear) transformation $\operatorname{Id}_M: \R^M \to \R^M$, it immediately follows from the above definitions that for any $o \in [O]$, any linear transformation $\mathcal{L}: \R^{M_o} \to \R^{N_o}$, and any order-$O$ tensor $\mathbfcal{T} \in \R^{\M}$, the following identity holds:
\[ 
 \mathbfcal{T} \bullet_o \mathcal{L} = (\Id_{M_1} \otimes \cdots \Id_{M_{o-1}} \otimes \mathcal{L} \otimes \Id_{M_{o+1}} \otimes \cdots \otimes \Id_{M_O}) (\mathbfcal{T}) \,.
 \]
Therefore the mode-$o$ product with the linear transformation $\mathcal{L}$ can be thought of as ``padding'' $\mathcal{L}$ via tensor products with the identity linear transformations of $\R^{M_1}, \dotsm \R^{M_{o-1}}, \R^{M_{o+1}}, \dots, \R^{M_O}$.\\

Since the identity transformation following or preceding any other function is equal to the original function ($\Id \circ f = f \circ \Id = f$), the above makes it possible to conclude that, for any $o_1 \not= o_2$ with $o_1, o_2 \in [O]$, and any order-$O$ tensor $\mathbfcal{T} \in \R^{\M}$ and linear transformations $\mathcal{L}_{o_1}: \R^{M_{o_1}} \to \R^{N_{o_1}}$, $\mathcal{L}_{o_2}: \R^{M_{o_2}} \to \R^{N_{o_2}}$, the following property of mode-$o$ products is valid:
\[ 
  \mathbfcal{T} \bullet_{o_1} \mathcal{L}_{o_1} \bullet_{o_2} \mathcal{L}_{o_2} = \mathbfcal{T} \bullet_{o_2} \mathcal{L}_{o_2} \bullet_{o_1} \mathcal{L}_{o_1} \,. 
\]
In other words, mode-$o$ products applied to \textit{distinct} modes ``commute'' with each other. \\

The \textbf{Kronecker product}\footnote{It is this operation which is represented using the symbol $\otimes$ in \cite{kolda_bader}, \cite{li_glm}, \cite{li_sparse}, and others.} $\mathbf{L}_1 \otimes_{\operatorname{Kr}} \mathbf{L}_2$ of matrices is related to the coordinate representation of the tensor product $\L_1 \otimes \L_2$ of the linear transformations $\L_1, \L_2$ represented by the two matrices $\mathbf{L}_1, \mathbf{L}_2$, respectively. (See appendix \ref{sec:coord-isom} for what is meant here by ``coordinate representation''.) While both the tensor $\mathbf{L}_1 \otimes \mathbf{L}_2$ and the coordinate representation of $\L_1 \otimes \L_2$ are fourth order tensors, they are not quite the same tensor (but are related). See appendix \ref{sec:kron-prod-coord} for a detailed explanation.

%%% Local Variables:
%%% mode: latex
%%% TeX-master: "../new_notes_draft"
%%% End:

\subsection{Definition of Tucker Decomposition}
\label{sec:defin-tuck-decomp}

Given an order-$O$ tensor $\mathbfcal{T} \in \R^{\M}$, a \textbf{Tucker decomposition} of $\mathbfcal{T}$ writes $\mathbfcal{T}$ in the form:
\[  
\mathbfcal{T} = \mathbfcal{G} \bullet_1 \mathcal{L}_1 \cdots \bullet_O \mathcal{L}_O \,, 
 \]
where $\mathbfcal{G}$ is an order-$O$ tensor called the \textit{core tensor}\cite{kolda_bader}, with $\mathbfcal{G} \in \R^{\Mltidx{I}}$, and where for each $o \in [O]$ one has that $\mathcal{L}_o: \R^{I_o} \to \R^{M_o}$ is a linear function. The following definitions are equivalent to the one given above due to the properties of tensor products of linear transformations:
\[ 
\mathbfcal{T} = \mathbfcal{G} \bullet_{\pi(1)} \mathcal{L}_{\pi(1)} \cdots \bullet_{\pi(O)} \mathcal{L}_{\pi(O)} \,,  
\]
where $\pi: [O] \to [O]$ is any permutation (see appendix \ref{sec:equiv-char-perm}) of the index set $[O]$, and:
\[ 
 \mathbfcal{T} = \left(\bigotimes_{o=1}^O \mathcal{L}_o \right) (\mathbfcal{G}) \,.  
\]
Note that in practice one usually does not demand that the decomposition is exact. In other words, for applications the data scientist is usually content with approximate Tucker decompositions:
\[   
\mathbfcal{T} \approx \left( \bigotimes_{o=1}^O \mathcal{L}_o \right) (\mathbfcal{G}) \,. 
\]
From a data compression standpoint, it is obvious that such a decomposition of $\mathbfcal{T}$ is only likely to be useful if the core tensor $\mathbfcal{G}$ has substantially fewer non-zero entries than $\mathbfcal{T}$, e.g. because for every $o \in [O]$ one has that $I_o << M_o$ so that the dimensions of the core tensor $\mathbfcal{G}$ are much smaller, or e.g. because $\mathbfcal{G}$ is extremely sparse. It is clear that Tucker decompositions of any given tensor $\mathbfcal{T}$ always exist: for example, assume that for each $o \in [O]$ the linear function $\mathcal{K}_o: \R^{M_o} \to \R^{I_o}$ is injective, thus post-invertible. Then the functions $\mathcal{K}_o^\dagger : \R^{I_o} \to \R^{M_o}$ corresponding to the Moore-Penrose pseudoinverses\footnote{Specifically, $\mathcal{K}_o^\dagger \circ \mathcal{K}_o \circ \mathcal{K}_o^\dagger  = \mathcal{K}_o^\dagger $, $\mathcal{K}_o \circ \mathcal{K}_o^\dagger  \circ \mathcal{K}_o = \mathcal{K}_o$, the kernel of $\mathcal{K}_o^\dagger$ is the orthogonal complement of the image of $\mathcal{K}_o$, and the image of $\mathcal{K}_o^\dagger$ is the orthogonal complement of the kernel of $\mathcal{K}_o$.} are post-inverses, i.e. $\mathcal{K}_o^\dagger  \circ \mathcal{K}_o = \operatorname{Id}_{M_o}$, and therefore one can write:
\[ 
 \mathbfcal{G} \defequals \left(\bigotimes_{o=1}^O \mathcal{K}_o \right) (\mathbfcal{T})  \text{ and for all $o\in [O]$, } \mathcal{L}_o \defequals \mathcal{K}_o^\dagger \quad\implies \quad \mathbfcal{T} = \left( \bigotimes_{o=1}^O \mathcal{L}_o \right) (\mathbfcal{G}) \,. 
 \]
However, such a Tucker decomposition, while mathematically valid, is not guaranteed to be useful. In particular, one has that $I_o \ge M_o$ for every $o \in [O]$, such that the number of entries of the core tensor $\mathbfcal{G}$ is no smaller than, and in general larger than, that of $\mathbfcal{T}$, and in general $\mathbfcal{G}$ is not sparse.\\

Therefore the challenge in applying Tucker decompositions to data analysis problems is not necessarily that of finding \textit{any} Tucker decomposition of the given tensor $\mathbfcal{T}$, but that of finding (approximate) Tucker decompositions of $\mathbfcal{T}$ which are actually useful. Heuristic ``usefulness'' corresponds in practice to some choice of possible constraints on either the core tensor $\mathcal{G}$, the linear transformations\footnote{Which in practice are usually replaced with their representations as matrices, $\mathbf{L}_1, \dots, \mathbf{L}_O$, which according to \cite{kolda_bader} are usually referred to as the ``factor matrices'' of the Tucker decomposition.} $\mathcal{L}_1, \dots, \mathcal{L}_O$, or both. This is because, without any constraints, the problem of finding a ``best Tucker decomposition'' lacks any unique solution, and those solutions which do exist and can easily be found are likely to not be useful, as explained implicitly above. For example, consider the case where all of the linear transformations $\mathcal{K}_o$ are simply the identity transformations: this is clearly an exact Tucker decomposition with $0$ error, and yet also completely useless. Of course, as for any computational problem, which constraints are chosen can have dramatic effects on the computational complexity, as well as on the choices of algorithms available.

%%% Local Variables:
%%% mode: latex
%%% TeX-master: "../new_notes_draft"
%%% End:

\subsection{Generalization of CP Decomposition}
\label{sec:gener-cp-decomp}

Assume one has a (useful) Tucker decomposition for the tensor $\mathbfcal{T}$ in terms of a core tensor $\mathbfcal{G}$:
\[ 
  \mathbfcal{T} = \left( \bigotimes_{o=1}^O \mathcal{L}_o  \right) (\mathbfcal{G}) \,.  \]

Like any tensor, one can express $\mathbfcal{G}$ as a linear combination of unit tensors using its entries:
\[   
\mathbfcal{G} = \sum_{\mltidx{i} \in \Mltidx{I}  } \mathcal{G}_{\mltidx{i}} \unittensor{\mltidx{i}} \,,  
\]
where, to recall from section \ref{sec:tensors}, the notation $\unittensor{\mltidx{i}}$ denotes the unit tensor all of whose entries are $0$ except for the $\mltidx{i}$'th, which equals one. As explained in \ref{sec:tensor-product}, each of these unit tensors has a standard tensor product decomposition in terms of unit vectors, such that one can write $\mathbfcal{G}$ as
\[ 
 \mathbfcal{G} = \sum_{\mltidx{i} \in \Mltidx{I}} \mathcal{G}_{\mltidx{i}} \left( \unitvector{i_1} \otimes \unitvector{i_2} \otimes \cdots \otimes \unitvector{i_O} \right) \,.  
\]
Therefore the above Tucker decomposition of the tensor $\mathbfcal{T}$ is equivalent to writing $\mathbfcal{T}$ as
\[ 
 \mathbfcal{T} = \left( \bigotimes_{o=1}^O \mathcal{L}_o \right) \left(  \sum_{\mltidx{i} \in \Mltidx{I}} \mathcal{G}_{\mltidx{i}} \left(  \unitvector{i_1} \otimes \unitvector{i_2} \otimes \cdots \otimes \unitvector{i_O} \right)   \right) \,. 
 \]
Since $\bigotimes_{o=1}^O \mathcal{L}_o$ is a linear function $\R^{\Mltidx{I}} \to \R^{\M}$, it distributes over scalar multiplication and addition, meaning that the Tucker decomposition of $\mathbfcal{T}$ also allows one to write it as
\[ 
 \mathbfcal{T} = \sum_{\mltidx{i} \in \Mltidx{I}} \mathcal{G}_{\mltidx{i}} \left( \bigotimes_{o=1}^O \mathcal{L}_o \right) \left(  \unitvector{i_1}  \otimes \unitvector{i_2} \otimes \cdots \otimes \unitvector{i_O} \right) \,.  
\]
Finally, using the definition of tensor product of linear transformations, one arrives at the conclusion that the Tucker decomposition of $\mathbfcal{T}$ is also equivalent to the following expression\cite{lim_hla}:
\[
  \mathbfcal{T} = \sum_{\mltidx{i} \in \Mltidx{I}} \mathcal{G}_{\mltidx{i}} \left(  \mathcal{L}_1 ( \unitvector{i_1} ) \otimes \mathcal{L}_2 (\unitvector{i_2} ) \otimes \cdots \otimes \mathcal{L}_O( \unitvector{i_O} )   \right) \,. 
 \]
Notice how this would be exactly the same as the tensor product decomposition for $\mathbfcal{G}$ in terms of unit tensors if everywhere $\unitvector[I_o]{i_o}$ were replaced with $\mathcal{L}_o (\unitvector[I_o]{i_o})$. Therefore Tucker decompositions can also be thought of as ``$\mathbfcal{G}$-weighted\footnote{This interpretation of the Tucker decomposition is referred to as using $\mathbfcal{G}$ as a ``kernel'' for the decomposition in \cite{cumulant_book}.} tensor product decompositions'' or of as a ``fancy\footnote{I.e. instead of ``non-fancy'' unit tensors which are tensor products of $\unitvector[I_o]{i_o}$, ``fancy unit tensors'' which are tensor products of $\mathcal{L}_o (\unitvector[I_o]{i_o})$ are used instead. Of course these ``fancy unit tensors'' in general aren't actual unit tensors.} version of the standard tensor product decomposition in terms of unit tensors''.\\

The above suggests that regular, ``non-fancy'', tensor product decompositions can be written as Tucker decompositions, and indeed they can. The key to doing this is choosing the correct core tensor $\mathbfcal{G}$. Specifically if the core tensor is the order-$O$ ``hyper-diagonal''\footnote{The authors of \cite{kolda_bader} call it the ``superdiagonal'' instead of the ``hyper-diagonal''. The latter seems to correspond better to other terms for higher-dimensional generalizations of familiar concepts, such as ``hyper-plane'' or ``hyper-cube''.} tensor of size $R$:
\[  
 \sum_{r=1}^R \underbrace{ \unitvector{r}  \otimes \cdots \otimes \unitvector{r}  }_{\text{$O$ times}} = \sum_{r=1}^R \bigotimes_{o=1}^O \unitvector{r} = \sum_{\mltidx{i} \in \Mltidx{I}} \delta_{\mltidx{i}} \unittensor{\mltidx{i}}  \,, 
\]
where $\delta_{\mltidx{i}}$ denotes the Kronecker delta which equals $1$ only if all of the entries of the multi-index $\mltidx{i}$ are the same and which equals $0$ otherwise, and where $I_1 = I_2 = \cdots = I_O \defequals R$, i.e. ``hyper-diagonal'' tensors are always cubical tensors. (In particular $\Mltidx{I} = [R]^O$ in this case.) Then a Tucker decomposition for $\mathbfcal{T}$ using this core tensor has the following form:
\[ 
 \begin{array}{rcl}
\mathbfcal{T} &=& \displaystyle \sum_{\mltidx{i} \in \Mltidx{I}} \delta_{\mltidx{i}} \left( \bigotimes_{o=1}^O \mathcal{L}_o \right) \left( \unitvector{i_1} \otimes \unitvector{i_2} \otimes \cdots \otimes \unitvector{i_O} \right) \\
& = &\displaystyle \summing{r}    \mathcal{L}_1 ( \unitvector{r}  ) \otimes \mathcal{L}_2 ( \unitvector{r} ) \otimes \cdots \otimes \mathcal{L}_O( \unitvector{r} )   \,.
\end{array}
\]
The above is clearly an tensor product decomposition of $\mathbfcal{T}$. In particular, if $\mathbf{L}_1, \dots, \mathbf{L}_O$ denote the matrices representing the linear transformations $\mathcal{L}_1, \dots, \mathcal{L}_O$, then for every $r \in [R]$ and $o \in [O]$, one has that $\mathcal{L}_o( \unitvector{r}  ) = (\mathbf{L}_o)_{:r}$, i.e. the $r$'th column\footnote{Represented as a vector, i.e. with one mode, and not as a column vector (with two modes, one of which is degenerate).} of $\mathbf{L}_o$, so the above is equivalent to:
\[ 
\mathbfcal{T} = \sum_{r=1}^R (\mathbf{L}_1)_{:r} \otimes (\mathbf{L}_2)_{:r} \otimes \cdots \otimes (\mathbf{L}_{O})_{:r} \,, \]
which is the same as the form of the CP decomposition in terms of factor matrices described in section 3 of \cite{kolda_bader} (which in this case would be written as $\mathcal{T}= [[ \mathbf{L}_1, \mathbf{L}_2, \cdots, \mathbf{L}_O ]]$ in the notation used by \cite{kolda_bader}, \cite{kolda}, and \cite{Kruskal1977}). Therefore CP decompositions are just special Tucker decompositions where the core tensor is constrained to be a hyper-diagonal tensor.\\

When a CP decomposition is written with scalar coefficients (denoted $[[\boldsymbol{\lambda}; \mathbf{V}_1,\dots, \mathbf{V}_O]]$ in \cite{kolda_bader}):
\[
\summing{k} \lambda_k \v{k}_1 \otimes \cdots \otimes \v{k}_O \,,
\]
this also corresponds to a Tucker decomposition where the core tensor is hyper-diagonal. The only difference is that the $k$'th element of the hyper-diagonal of the core tensor $\mathbfcal{G}_{k\dots k}$ in this case will be $\lambda_k$ instead of $1$. This form of CP decomposition is equivalent to the $[[\mathbf{W}_1, \dots, \mathbf{W}_O]]$ form:
\[
\summing{k} \w{k}_1 \otimes \cdots \otimes \w{k}_O 
\]
due to the multilinearity of the tensor product. Specifically, when $\lambda_1 = \dots = \lambda_K = 1$, then $[[\boldsymbol{\lambda}; \mathbf{V}_1,\dots, \mathbf{V}_O]]$ obviously reduces to $[[\mathbf{V}_1,\dots, \mathbf{V}_O]]$, so the $[[\mathbf{W}_1, \dots, \mathbf{W}_O]]$ form is a special case of the $[[\boldsymbol{\lambda}; \mathbf{V}_1,\dots, \mathbf{V}_O]]$ form. Conversely, given $[[\boldsymbol{\lambda}; \mathbf{V}_1,\dots, \mathbf{V}_O]]$, it can always be written in the $[[\mathbf{W}_1, \dots, \mathbf{W}_O]]$ form by ``spreading the scalar coefficient among the tensor factors'' (which is always possible and always works due to the multilinearity of the tensor product). For example, by setting $\w{k}_1 = \lambda_k \v{k}_1$ for all $k$ and $\w{k}_o = \v{k}_o$ for all $k$ when $o > 1$. Thus, by using a ``weighted'' hyper-diagonal core tensor, ``weighted'' CP decompositions (i.e. ones with scalar coefficients) can also be written as special cases of Tucker decompositions. That it should be possible to write ``weighted'' CP decompositions \textit{somehow} as a special case of Tucker decompositions should also be immediately obvious from the equivalence of weighted and unweighted CP decompositions.\\

Note the implications this has for the computational complexity of most problems involving Tucker decompositions of higher-order tensors. Any algorithm for solving such a problem which would be guaranteed to find in polynomial time an exact Tucker decomposition, including in the case that the core tensor is constrained to be a hyper-diagonal tensor, would be capable of finding an exact CP decomposition in polynomial time. Thus it is perhaps not surprising that the algorithms for finding most types of useful (i.e. constrained) Tucker decompositions suffer the same limitations as those for finding CP decompositions, namely that they only offer approximate solutions, not exact ones, rely on heuristics, or are only guaranteed to work effectively for a limited class of tensors.

%%% Local Variables:
%%% mode: latex
%%% TeX-master: "../new_notes_draft"
%%% End:

\subsection{Applications of Tucker Decompositions}
\label{sec:appl-tuck-decomp}

The same heuristic motivations for CP decompositions discussed in \ref{apps-outer-prod} are also relevant for Tucker decompositions, and therefore won't be discussed again in great detail here. It is perhaps worth mentioning though that, unlike the CP decomposition which decomposes a tensor of order $O$ ``into $O$-tuples of vectors'', a Tucker decomposition will return a core tensor of the same order $O$ as well as $O$ matrices. While the matrices returned by the Tucker decomposition may not be much more difficult to interpret than the vectors returned from a CP decomposition, the core higher-order tensor returned by the Tucker decomposition potentially could be much more challenging to interpret.\\

Therefore the benefit to interpretability from a Tucker decomposition might not be as great as from a CP decomposition. Even in the cases when the resulting core tensor is substantially smaller than the original tensor, thus making the Tucker decomposition helpful from a data compression standpoint, the resulting order $O$ core tensor could still be difficult to interpret. This potential disadvantage might be minimal in the case where the interpretation of the core tensor is mostly unimportant or irrelevant, and all that matters is that it be small, or it might not even be present in cases where the core tensor is easy to interpret (e.g. for a CP decomposition when the core tensor is just hyper-diagonal, and thus also extremely sparse).\\ 

This challenge is sometimes circumvented by applying a CP decomposition to the core tensor resulting from the Tucker decomposition. It is also important for the linear transformations represented by the matrices resulting from the Tucker decomposition to be relatively easily interpretable. For example, if all of the factor matrices are invertible, then the original tensor has the same tensor rank as the core tensor\cite{lim_hla} (and even the same multilinear ranks, see section \ref{sec:mult-mult-rank}), and even when the factor matrices are not invertible, the tensor rank of the core tensor is still an upper bound for the tensor rank of the original tensor (and again the same is true for multilinear ranks) \cite{desilva_lim}.\\

Often it is possible to interpret the modes of the core tensor in a Tucker decomposition as corresponding to groups of ``hidden variables''. In this case the factor matrices represent linear transformations which transform the hidden variables to the variables which are actually observed in the experiment. For example, when the factor matrices are invertible (and thus square, so the shape $\Mltidx{I}$ of the core tensor equals the shape $\M$ of the original tensor), then the Tucker decomposition corresponds to a change of basis of $\R^{M_o}$ for each $o\in [O]$\cite{lim_hla}, and the elements of the old basises can be interpreted as the ``hidden variables'' while the elements of the new basises can be interpreted as the ``observed variables''. (Cf. the discussion in section \ref{sec:multi-index}. There is a bijection between the elements of $[M_o]$ and the elements of any basis of $\R^{M_o}$, allowing them to be identified. Since the elements of $[M_o]$ can be identified with a ``group of variables'' as discussed in section \ref{sec:multi-index}, by transitivity the elements of the chosen basis of $\R^{M_o}$ can also be so identified.)  \\

In what follows, the manner in which special cases of Tucker decompositions manifest all of the time in data science will be discussed, and subsequently some case studies of how Tucker decompositions of higher-order tensors have been used in statistics will be reviewed.

\subsubsection{Special Case: Vectors}
\label{sec:spec-cases:-vect}

The following problem is ubiquitous in data science and statistics, and it may not even be an exaggeration to say that its study was what first led to the birth of linear algebra as an area of research: given a column vector $\mathbf{Y} \in \R^{N \times 1}$, a way of expressing it as
\[ 
 \mathbf{Y} = \mathbf{A X}
  \]
for a fixed matrix $\mathbf{A} \in \R^{N \times M}$, and \textit{some} column vector $\mathbf{X} \in \R^{M \times 1}$, is to be found. This is the problem of solving linear equations, and it is actually a special case of a Tucker decomposition. Or, it could be argued, Tucker decompositions are a special case of solving linear equations\endnote{It depends on how one defines what it means to ``solve a linear equation'': the specific problem described above is equivalent to a special case of finding a Tucker decomposition, while if one defines solving a linear equation to be finding an $\vec{x} \in V$ (for some abstract vector space $V$) such that $\vec{y} = \mathcal{L}(\vec{x})$ for a fixed/given $\vec{y} \in W$ and fixed/given linear $\mathcal{L}: V \to W$, then Tucker decompositions are a special case of that more general problem. Of course, that more general problem also includes problems like solving linear operator equations between function spaces.}.\\

To clarify, one usually knows that the column vector $\mathbf{Y}$ is the coordinate representation of a known vector\footnote{Or more generally in some finite-dimensional vector space linearly isomorphic to $\R^N$, i.e. with dimension $N$.} $\mathbf{y} \in \R^N$ with respect to some known basis $\mathscr{C}$, and that the matrix $\mathbf{A}$ is the coordinate representation of a known linear transformation $\mathcal{L}: \R^M \to \R^N$ with respect to the known basis $\mathscr{C}$ of $\R^N$ and some known basis $\mathscr{B}$ of\footnote{Or again more generally some finite-dimensional vector space linearly isomorphic to $\R^M$, i.e. with dimension $M$.} $\R^M$. Therefore solving the above problem is equivalent to solving the problem of finding some vector $\mathbf{x} \in \R^M$ which is a solution for the equation
\[ 
 \mathbf{y} = \mathcal{L} ( \mathbf{x}) \,,  
\]
where again, for the sake of emphasis, $\mathbf{y}$ and $\mathcal{L}$ are already considered known and fixed.\\

The problem of finding a Tucker decomposition for a given tensor $\mathbfcal{T}$ was defined above as finding a way to write it as the result of applying some tensor product of linear transformations $\bigotimes_{o=1}^O \mathcal{L}_o$ to some core tensor $\mathbfcal{G}$. In other words, if one writes ${\mathbfcal{Y} \defequals \mathbfcal{T}}$, ${\mathcal{L} \defequals \bigotimes_{o=1}^O \mathcal{L}_o}$, and ${\mathbfcal{X} \defequals \mathbfcal{G}}$, then the problem of finding a Tucker decomposition for $\mathbfcal{T}$ is the same as finding a way to write it as
\[ 
 \mathbfcal{Y} = \mathcal{L} ( \mathbfcal{X}) \,.
  \]
The only way this is different from the general problem of writing $\mathbfcal{Y} = \mathcal{L}( \mathbfcal{X})$, for \textit{some} linear transformation $\R^{\M} \to \R^{\N[O]}$, is the additional assumption that the linear transformation $\mathcal{L}$ factorizes as the tensor product $\bigotimes_{o=1}^O \mathcal{L}_o$ of $O$ linear transformations, where $O$ is the common order of both $\mathbfcal{Y}$ and $\mathbfcal{X}$. In other words, finding a Tucker decomposition for $\mathbfcal{Y}$ differs from the more general situation of writing $\mathbfcal{Y}$ as $\mathcal{L} (\mathbfcal{X})$ by the additional assumption that the linear transformation $\mathcal{L}$ must ``respect'' or ``preserve'' the ``multilinear structure'' of $\mathbfcal{X}$ and $\mathbfcal{Y}$.\\

Of course, in the case that $\mathbfcal{Y}$ (and thus also $\mathbfcal{X}$) is a tensor of order $1$, i.e. a vector, this additional restriction amounts to no restriction at all\footnote{In other words, one trivially has that $\bigotimes_{o=1}^1 \mathcal{L}_o = \mathcal{L}_1$, so the assumption that $\mathcal{L}$ ``factorizes'' as the tensor product of a single linear transformation is the same as saying that $\mathcal{L}$ is an arbitrary linear transformation.}. Therefore, when $\mathbfcal{Y}$ has a single mode, and the linear transformation $\mathcal{L}$ is fixed/known, the problem of finding a Tucker decomposition for $\mathbfcal{Y}$ is the same as the problem of solving the linear equation $\mathbfcal{Y} = \mathcal{L} (\mathbfcal{X})$. Thus the quintessential problem of linear algebra is a special case of the problem of finding Tucker decompositions for tensors.\\

One might reasonably ask why solving linear equations is so much easier\footnote{One can determine in polynomial time whether a solution to the problem exists, and if a solution to the problem does exist, then the solution can be found in polynomial time.} than, e.g. finding CP decompositions of higher-order tensors, even though both are special cases of Tucker decompositions. The distinction between the two problems is more subtle than that between higher-order and lower-order tensors. The different constraints in the two problems appear to be what actually distinguish them from the viewpoint of computational complexity.\\

The first issue to note is that, even in the case of order one tensors, i.e. solving ``matrix-vector'' linear equations, the addition of constraints on the core tensor (i.e. the solution of the linear equation) are often already sufficient to transform that problem into an NP-hard one. For example, when solving an underdetermined system of linear equations (i.e. one for which, even though there is a single fixed linear transformation, there are infinitely many possible core tensors), requiring that the core tensor be sparse (finding the core tensor which minimizes the $L_0$ ``norm'') is an NP-hard problem. (For more information about the NP-hard ``Sparsest Solution Problem'' (SSP), which has been studied for its applications to signal processing, see e.g. \cite{donoho1}, \cite{donoho2}, \cite{Qu2016}, \cite{pmlr-v23-spielman12}, \cite{Demanet2014}, or \cite{cotter}.)\\ 

The constraints on the core tensor when finding an (optimal) CP decomposition for a higher-order tensor are similar, since requiring the core tensor to be hyper-diagonal not only amounts to a very severe sparsity constraint on the core tensor, but also a very specific sparsity pattern. Thus, in light of the NP-hardness of the SSP even for order $1$ tensors, it is perhaps less surprising that finding (optimal) CP decompositions for higher-order tensors is also NP-hard. The SSP, also known as the ``Sparsest Vector Problem'', has been found to be a special case of another problem, the ``Low-Rank Basis Problem'', and in fact another special case of the ``Low-Rank Basis Problem'' was shown (perhaps not coincidentally) to be solvable using CP decompositions\cite{lowranksubspace}.\\

The second issue to note is that, when operating under the same constraints as for solving linear equations (i.e. single, fixed linear transformation, no constraints on the core tensor), the problem of finding a Tucker decomposition for a higher-order tensor is \textit{exactly as easy}\footnote{I.e. the problem can also be resolved in polynomial time.}. This is because both the tensor and the linear transformation can both be ``vectorized'', turning the problem into one involving a core tensor of order one, and thus one resolvable in polynomial time.\\

The number of entries of the ``vectorized'' tensor is the same as the number of entries of the original tensor, so one might object that this isn't actually a polynomial time problem, since the number of entries of a tensor grows exponentially, $\Omega(M_{\min}^O)$, with the dimension $M_{\min}$of its smallest mode as the order $O$ increases. However, that's not relevant, since the situation under discussion corresponds to a fixed tensor, and thus a tensor of a fixed order. Therefore, in terms of the computational complexity analysis, either the number of entries of the tensor, or the dimension of its largest mode should be considered variable, and not the order. For a \textit{fixed} order $O$, the number of entries of a tensor is polynomial in the dimension $M_{\max}$ of its largest mode, specifically\endnote{Admittedly using a variable $O$ in an expression using Big-Oh notation is probably not a great choice. In this case it seemed like the best way to maintain consistency with the notation used elsewhere throughout this review however. Hence $\mathscr{O}$ will be used in this paper to represent Big-Oh notation, for greater unambiguity.} $\mathscr{O}(M_{\max}^O)$.\\

For a fixed order $O$ tensor $\mathbfcal{T}$, the problem of finding a Tucker decomposition $\mathbfcal{T} = \left(  \bigotimes_{o=1}^O \mathcal{L}_o  \right)( \mathbfcal{G})$ for fixed $\mathcal{L}_1, \dots, \mathcal{L}_O$, i.e. solving for $\mathbfcal{G}$, is equivalent\cite{cumulant_book}\cite{kolda} to solving the linear equation:
\[
\vectorize{\mathbfcal{T}} = (\mathbf{L}_O \otimes_{Kr} \mathbf{L}_{O-1} \otimes_{Kr} \cdots \otimes_{Kr} \mathbf{L}_1)\vectorize{\mathbfcal{G}} \,,
 \]
where $\operatorname{vec}_{\operatorname{col}}$ denotes \textbf{col}exicographical \textbf{vec}torization\endnote{Corresponding to \textbf{col}umn-major \textbf{vec}torization of a matrix. Note that although \cite{cumulant_book} also appears to use colexicographical vectorization like \cite{kolda} and \cite{kolda_bader} do, the formula given in Example 4 uses the reverse order Kronecker product ${\mathbf{L}_1 \otimes_{Kr} \mathbf{L}_2 \otimes_{Kr} \cdots \otimes_{Kr} \mathbf{L}_O}$, even though that order actually corresponds to lexicographical vectorization.} of a tensor, $\otimes_{Kr}$ denotes the Kronecker product\cite{kronecker} of matrices, and as before $\mathbf{L}_o$ denotes the matrix representation of $\mathcal{L}_o$. (Cf. Proposition 3.7 of \cite{kolda}.) This equivalence owes to the fact that $\R^{\Mltidx{I}}$ and $\R^{\prod_{o=1}^O I_o}$ are vector spaces of the same dimension, and (colexicographical) vectorization is linear and sends a basis of $\R^{\Mltidx{I}}$ (the unit tensors) to a basis of $\R^{\prod_{o=1}^O I_o}$ (the unit vectors). The matrix in the above linear equation is effectively the representation of ${\bigotimes_{o=1}^O \mathcal{L}_o}$ with respect to this new basis. See \cite{pereyra} and \cite{hjj} for more about how this linear equation relates to multilinear algebra. The core tensor of the Tucker decomposition $\mathbfcal{G}$ is recovered by ``tensorizing''\footnote{The function $\R^{\Mltidx{I}} \to \R^{\prod_{o=1}^O I_o}$ given by colexicographical vectorization is a linear isomorphism, and thus has a linear inverse, $\R^{\prod_{o=1}^O I_o} \to \R^{\Mltidx{I}}$. Applying this inverse to $\vectorize{\mathbfcal{G}}$ is what ``tensorizes'' it and recovers $\mathbfcal{G}$.} the solution of the linear equation, $\vectorize{\mathbfcal{G}}$.\\

Since the time complexity of solving a linear equation $\mathbf{Y} = \mathbf{A}\mathbf{X}$ (where again $\mathbf{Y}$ is a fixed column vector, $\mathbf{A}$ is a fixed matrix, and $\mathbf{X}$ is an unknown column vector to be solved for) is $\mathscr{O}(M^3)$ using an LU decomposition, where $M$ is the dimension of $\mathbf{Y}$, it follows that the time complexity of finding a Tucker decomposition $\mathbfcal{T} = \left(  \bigotimes_{o=1}^O \mathcal{L}_o  \right)( \mathbfcal{G})$ for a fixed $\mathbfcal{T}$ whose largest mode's dimension is $M_{\max}$ and fixed $\mathcal{L}_1, \dots, \mathcal{L}_O$ is $\mathscr{O}( (M_{\max}^O)^3  ) = \mathscr{O}(M_{\max}^{3O})$, since the dimension of the (column) vector $\vectorize{\mathbfcal{T}}$ is $M_{\max}^O$, and finding a Tucker decomposition for $\mathbfcal{T}$ under the given constraints is equivalent to solving the linear equation involving $\vectorize{\mathbfcal{T}}$ as described above. Note in particular that the time complexity $\mathscr{O}(M_{\max}^{3O})$ is polynomial in $M_{\max}$. Therefore, inasmuch as finding Tucker decompositions for higher-order tensors, under the same constraints as those usually found in the problem of solving linear equations, is also an ``easy'' problem, it might be less surprising that solving linear equations is actual a special case of finding Tucker decompositions.\\

Directly forming the matrix $\mathbf{L}_O \kronecker \mathbf{L}_{O-1} \kronecker \cdots \kronecker \mathbf{L}_1$ in computer memory, and then performing Gaussian elimination with it, is not likely to be a good idea in practice\footnote{Polynomial time complexity does not inherently make a problem ``easy'' for practical purposes, especially in cases like these where the degree of the polynomial involved is large.}, since the matrix is usually enormous. This also fails to exploit the available ``multilinear structure'': the matrix is not the flattening of an arbitrary linear transformation $\R^{\Mltidx{I}} \to \R^{\M}$, it is the flattening of one with a tensor product structure, $\bigotimes_{o=1}^O \mathcal{L}_o$. Extra structure means that algorithms can make extra assumptions, which often makes it possible to improve their efficiency. The problem of efficiently solving linear equations like this has been studied for decades\cite{pereyra}\cite{hjj}, due to its applications to numerical analysis\cite{pereyra} (e.g. Lagrange interpolation, Hermite interpolation, multidimensional numerical integration a.k.a. ``hyper-cubature'', finite element methods), fast matrix multiplication\cite{hjj}\cite{granata}, signal processing\cite{granata} (e.g. discrete Fourier transforms, Walsh-Hadamard transforms, discrete Hartley transforms, discrete cosine transforms, and linear convolutions), etc.\\

Incidentally, blockwise recursion can be used to create fast\footnote{I.e. ``sub-cubic'', or in yet other words, $\mathscr{O}(M^d)$ for $d<3$.} algorithms for LU decomposition (thus for solving linear equations) from fast algorithms for matrix multiplication\endnote{See, for example, \cite{dumasGPRS16}, \cite{bailey1991}, for papers doing this with Strassen's algorithm. For a basic introduction to the ``algorithmic equivalence'' of matrix multiplication and matrix inversion (which subsequently can be used to explain the ``algorithmic equivalence'' between matrix multiplication and LU decomposition), see section 28.2 of \cite{cormen}. Chapter 16 of \cite{burgisser} explains comprehensively why all of these problems (matrix multiplication, matrix inversion, LU decomposition) have the same complexity, albeit in a much less basic and much less introductory manner than \cite{cormen}.}, such as Strassen's algorithm which is $\mathscr{O}(M^{\log_2(7)}) \approx \mathscr{O}(M^{2.8})$. These fast matrix multiplication algorithms can be understood in terms of tensor product decompositions of third order tensors, see\footnote{For a detailed explanation, cf. applications 2. and 3. of section 15.3 of \cite{lim_hla}. Perhaps less widely accessible treatments include e.g. Sections 1.1., 1.2., 3.8. of \cite{landsberg}, and Chapters 15-16 of \cite{burgisser}.} for example \cite{benson} and \cite{legall}. Classically this formulation has often been expressed in terms of Kronecker products, see e.g. \cite{hjj}\cite{granata}. So both multiplying matrices and solving linear equations, which are workhorses of statistics and data science, are also special cases of Tucker decompositions.

%%% Local Variables:
%%% mode: latex
%%% TeX-master: "../new_notes_draft"
%%% End:

\subsubsection{Special Case: Matrices}
\label{sec:spec-cases:-matrix}

Consider the problem of Tucker decomposition for an order-$2$ tensor, i.e. for a matrix. Given a matrix $\mathbf{M} \in \R^{H \times W}$, one can use the standard row-wise outer product decomposition\footnote{Due to the universal property of the tensor product, the result will not depend on which outer product decomposition is chosen, so the freedom exists to use whichever outer product decomposition simplifies the analysis.} to show that the mode-$1$ product corresponds to matrix multiplication\footnote{Remember from the first example of section \ref{sec:spec-case:-matr} that the product of two matrices is equal to the sum of the outer products of the columns of the first matrix with the rows of the second matrix.} from the left, specifically:
\[ 
 \mathbf{M} \bullet_1 \mathcal{L}_1 = \left(  \sum_{h=1}^H \unitvector{h} \otimes \mathbf{M}_{h:}  \right) \bullet_1 \mathcal{L}_1 = 
\sum_{h=1}^H  \mathcal{L}_1 (\unitvector{h} ) \otimes \mathbf{M}_{h:} = \mathbf{L}_1 \mathbf{M} = \mathbf{M} \bullet_1 \mathbf{L}_1  \,,  
  \]
where $\mathbf{L}_1$ is the matrix representing the linear transformation $\mathcal{L}_1$ in the standard way. Completely analogously, one can also use the standard column-wise outer product decomposition\footnote{Here also the result will not depend on the choice of outer product decomposition due to the universal property.} to show that the mode-$2$ product corresponds to matrix multiplication from the right: 
\[  
\mathbf{M} \bullet_2 \mathcal{L}_2 = \left( \sum_{w=1}^W  \mathbf{M}_{:w} \otimes \unitvector{w} \right) \bullet_2 \mathcal{L}_2 = \sum_{w=1}^W \mathbf{M}_{:w} \otimes \mathcal{L}_2 ( \unitvector{w}  ) = \mathbf{M} \mathbf{L}_2^\top = \mathbf{M} \bullet_2 \mathbf{L}_2 \,, 
 \]
where similarly $\mathbf{L}_2$ is the matrix representing the linear transformation $\mathbf{L}_2$ in the standard way.\\

Thus every Tucker decomposition of a matrix $\mathbf{T}$ is\cite{cumulant_book} a three-term factorization of the form:
\[ 
 \mathbf{T} = \mathbf{L}_1 \mathbf{G} \mathbf{L}_2^\top = \mathbf{G} \bullet_1 \mathbf{L}_1 \bullet_2 \mathbf{L}_2 \,, 
 \]
where the matrix $\mathbf{G}$ is the ``core tensor'', the matrix $\mathbf{L}_1$ represents the linear transformation which is applied to the first mode of the core tensor, and finally the matrix $\mathbf{L}_2$ represents the linear transformation which is applied to the second mode of the core tensor. Conversely, every three-term factorization of a matrix is a Tucker decomposition of that matrix (``in disguise'').\\

When the core tensor is the identity matrix, as a special case of the result from section \ref{sec:gener-cp-decomp} one gets
\[
\mathbf{T} = \mathbf{L}_1 \ \mathbf{I}\  \mathbf{L}_2^\top = \mathbf{L}_1 \mathbf{L}_2^\top\,,
\]
which of course is also the same as the result mentioned in section \ref{sec:spec-case:-matr} (take $\mathbf{A}= \mathbf{L}_1$ and $\mathbf{B}=\mathbf{L}_2^\top$). Here one is taking the outer products of the columns of the first matrix with the columns of the second matrix. In contrast the way two-term matrix factorizations are written in section \ref{sec:spec-case:-matr}, one takes the outer products of the columns of the first matrix with the \textit{rows} of the second matrix. Of course these two ways of writing two-term matrix factorizations are equivalent to the extent that the rows of a matrix are the same as the columns of its transpose.\\

When the core matrix is diagonal $\boldsymbol{\Lambda} = \operatorname{Diag}(\lambda_1, \dots, \lambda_K)$, then the result of section \ref{sec:gener-cp-decomp} gives
\[
\mathbf{T} = \mathbf{L}_1 \boldsymbol{\Lambda} \mathbf{L}_2^\top = \summing{k} \lambda_k (\mathbf{L}_1)_{:k} \otimes (\mathbf{L}_2)_{:k} \,.
\]
Thus three-term matrix factorizations of the form $\mathbf{L}_1 \boldsymbol{\Lambda} \mathbf{L}_2^\top $ where $\boldsymbol{\Lambda} $ is diagonal amount to ``weighted'' CP decompositions. Of course, as discussed in section \ref{sec:gener-cp-decomp}, ``weighted'' CP decompositions are equivalent to ``unweighted'' ones, which in the matrix case means that these special three-term matrix factorizations can also easily be written as two-term matrix factorizations, something implicitly discussed already in section \ref{sec:spec-case:-matr} with reference to the singular value decomposition.

\paragraphsection{Singular Value Decompositions}
\label{tucker-svds}

Such three-term matrix factorizations are ubiquitous in data science. Of course the most prominent and important examples are the singular value decomposition (SVD) of an arbitrary matrix $\mathbf{M}$:
\[  
\mathbf{M} = \mathbf{U} \boldsymbol{\Sigma} \mathbf{V}^\top \,,  
\]
and the eigenvalue decomposition\footnote{Whose existence for symmetric matrices is guaranteed by the spectral theorem.} of a symmetric matrix $\mathbf{S}$:
\[ 
 \mathbf{S} = \mathbf{V} \boldsymbol{\Lambda} \mathbf{V}^\top \,.
 \]
In the case that the symmetric matrix is positive semi-definite, the eigenvalue decomposition and the singular value decomposition coincide. This is the situation for principal components analysis (PCA), since the covariance matrix of any distribution is positive semi-definite.\\ 

Therefore, inasmuch as SVD and PCA are fundamental tools in data science\cite{Elden}\cite{Gentle}\cite{Gentle2017}\cite{pca_review}\cite{extraordinary_svd}\cite{Golub2013}\cite{Kalman1996}, so are Tucker decompositions. For instance, SVD and PCA have both been used numerous times for the analysis of gene expression data\cite{genome_svd}\cite{genome_svd2}\cite{genome_svd3}\cite{genome_svd4}\cite{genome_svd5}. \\ 

Moreover, the Tucker decomposition interpretation also motivates why the form of both of these factorizations is usually written $\mathbf{A B C}^\top$ instead of $\mathbf{DEF}$: the former corresponds to applying the linear transformation represented by the last matrix to the rows of the second matrix, whereas the second corresponds to the more roundabout process of applying the adjoint of the linear transformation represented by the last matrix to the rows of the second matrix.\\ 

It was also mentioned in section \ref{sec:spec-case:-matr} how, by distributing the scalars found in the central matrix, both SVD and PCA can be considered special cases of CP decompositions. This makes sense in light of how, as was shown above, CP decompositions are a special case of Tucker decompositions when the core tensor is constrained to be hyper-diagonal. Indeed, the central matrix (i.e. core tensor) in both the SVD ($\boldsymbol{\Sigma}$) and in the eigenvalue decomposition ($\boldsymbol{\Lambda}$) are both diagonal matrices, which is the same thing as hyper-diagonal tensors of order $2$.

\paragraphsection{CUR Decompositions}
\label{cur-matrix}

One example of another three-term matrix factorization besides SVD or PCA which is used in statistics and data science is CUR decompositions. There are technically speaking many different forms of CUR decompositions\cite{cur1}\cite{cur2}\cite{cur3}\cite{cur4}\cite{cur5}\cite{cur6}, which differ in whether they are exact or approximate, which optimization problems they are (approximate) solutions to, and which algorithms are used to find them. However, in all cases the general idea is the same. Given a matrix $\mathbf{M}$, the goal is to write it (approximately) as a three-term factorization:
\[ 
\mathbf{M} \approx \mathbf{CUR} \,, 
 \]
where the columns of $\mathbf{C}$ are some subset of the columns of $\mathbf{M}$, the rows of $\mathbf{R}$ are some subset of the rows of $\mathbf{M}$, and the core matrix $\mathbf{U}$ is the ``glue'' which allows one to reconstruct all or ``most'' of $\mathbf{M}$ from the subset of columns found in $\mathbf{C}$ and the subset of rows found in $\mathbf{R}$. As in any Tucker decomposition, the ideal is to have that $\mathbf{U}$ is as small as possible, for the purposes of both data compression and ease of interpretation. Smaller core matrices $\mathbf{U}$ correspond to matrices $\mathbf{C}$ with fewer columns and matrices $\mathbf{R}$ with fewer rows, or in other words the ability to meaningfully interpret the matrix $\mathbf{M}$ using as few of its original columns and rows as possible.\\

The intuition behind this decomposition, regardless of the algorithm used to compute it, is that for many matrices which arise in practice it is often the case that only a small subset of the rows and/or columns of the original matrix are necessary to capture the ``vast majority'' of the information contained in the data. The method proposed for computing this decomposition in \cite{cur_mahoney} belongs to a class of randomized algorithms motivated by the same intuition. For many problems these new methods enable one to perform highly accurate matrix calculations while only using a randomly selected subset of the entries or columns, which can greatly decrease the computational time or space required. For an introduction to these methods, see any of \cite{randnla2}\cite{randnla3}\cite{randnla4}\cite{mahoney-notes}.\\

As a result of the Eckart-Young theorem, it is clear that an approximate CUR decomposition will never be as close to the original matrix in Frobenius or spectral norm as the corresponding truncated SVD of the same rank. However, one of the primary reasons for applying any Tucker decomposition, beyond just data compression, is often interpretability, and as argued convincingly in \cite{cur_mahoney} this is in many cases a decisive strength of CUR decompositions over the SVD.\\ 

Consider an example where the rows of the original matrix $\mathbf{M}$ correspond to distinct genes, and the columns of $\mathbf{M}$ correspond to distinct patients. In this case, while the left singular vectors correspond to linear combinations of patients (or ``eigenpatients'' in the terminology of \cite{cur_mahoney}), the columns of $\mathbf{C}$ correspond to actual patients. Similarly, while the right singular vectors correspond to linear combinations of genes (or ``eigengenes''), the rows of $\mathbf{R}$ correspond to actual genes.\\ 

This makes it not only easier to correctly interpret the data (e.g. the columns in $\mathbf{C}$ may correspond to ``representative'' patients, while the rows in $\mathbf{R}$ may correspond to ``important'' genes), but also to avoid incorrect conclusions based on invalid reification\cite{cur_mahoney} of the left or right singular vectors, i.e. mistakes based on treating ``eigenpatients'' as if they were actual patients or ``eigengenes'' as if they were actual genes. In other words, CUR decompositions address a common criticism of SVD when used for the analysis of gene expression data, that ``the [singular] vectors themselves are completely artificial and do not correspond to actual [DNA expression] profiles'' \cite{genome_svd}.\\

One potential disadvantage of the CUR decomposition compared to the SVD with respect to interpretability is the core tensor $\mathbf{U}$. Specifically, while the core matrix in an SVD is always guaranteed to be diagonal, and thus extremely sparse with a very predictable sparsity pattern, there are no guarantees for the structure of the core matrix of a CUR decomposition (at least using the algorithm described in \cite{cur_mahoney}). Moreover, the entries of the core matrix in the SVD decomposition are highly interpretable as singular values, whereas there may not be any meaningful interpretation of the entries of the core matrix in a CUR decomposition. So there is an ``interpretability'' tradeoff between the two decompositions: the SVD sacrifices interpretability of the left and right factor matrices for interpretability of the core matrix, while CUR decompositions sacrifice interpretability of the core matrix in favor of interpretability of the left and right factor matrices.\\

While in general the approximation error in terms of Frobenius or spectral norm of a CUR decomposition will be higher than that for the SVD, the algorithm proposed in \cite{cur_mahoney} selects the columns in $\mathbf{C}$ and the rows in $\mathbf{R}$ in a way which is designed to heavily utilize the information contained in the rank-$k$ (truncated) SVD. Specifically, the columns of $\mathbf{C}$ are chosen so that they are most likely highly correlated with the span of the top $k$ right singular vectors, and the rows of $\mathbf{R}$ are chosen so that they are most likely highly correlated with the span of the top $k$ left singular vectors. Then the core matrix $\mathbf{U}$ is defined to be $\mathbf{C}^\dagger \mathbf{M} \mathbf{R}^\dagger$, where $^\dagger$ denotes the Moore-Penrose pseudoinverse\cite{pseudoinverse_review}.

%%% Local Variables:
%%% mode: latex
%%% TeX-master: "../new_notes_draft"
%%% End:

\paragraphsection{Sparse SVD and Sparse PCA}

Inasmuch as it is intended to be a more interpretable replacement for the truncated SVD and it is designed to approximate the SVD's most desirable properties, the CUR decomposition of \cite{cur_mahoney} can be considered analogous to ``sparse SVD'' or ``sparse PCA'' methods such as those from \cite{sparse_svd1}, \cite{sparse_svd2}, or \cite{sparse_svd3}. Yet those methods still require the practitioner to examine and analyze the coefficients of the resulting ``singular vectors''\footnote{In other words, columns of the left factor matrix and rows of the right factor matrix.} to identify which rows or columns of the original matrix correspond to important features, the way normal SVD does. (That or they still somewhat encourage researchers to make the mistake of ``the reification of the artificial singular directions that offer little insight into the underlying biology'' \cite{cur_mahoney}.) This analysis is prone to errors and misunderstandings, see e.g. \cite{coefficient_interpretation} for a detailed discussion. These ``sparse SVD'' methods only make the analysis easier, but they do not obviate it. In contrast, the CUR decomposition identifies important rows and columns automatically, completely eliminating the need for any such analysis.\\ 

At the same time, the core matrix of these ``sparse SVD'' methods is usually diagonal and thus more easily interpretable than the core matrix of the CUR decomposition. Therefore these methods can perhaps be considered to inhabit a point of the ``core-matrix-/factor-matrix-interpretability tradeoff curve'' between the positions of the regular truncated SVD and of CUR decompositions. Just like the CUR decomposition method from \cite{cur_mahoney}, all of the ``sparse SVD'' and ``sparse PCA'' methods from \cite{sparse_svd1}, \cite{sparse_svd2}, and \cite{sparse_svd3} have all been shown to be applicable to the analysis of gene expression data, as well as a possible replacement for truncated SVD in that context.

%%% Local Variables:
%%% mode: latex
%%% TeX-master: "../new_notes_draft"
%%% End:

\paragraphsection{Convex Nonnegative Matrix Factorization}
\label{tucker-cnmf}

Convex NMF is a constrained case of semi-NMF as defined in \ref{sec:spec-case:-matr}. Unlike in NMF, but like in semi-NMF, in convex NMF the original data matrix $\mathbf{M}$ is not assumed to have only nonnegative entries. Instead, the matrix $\mathbf{C}$ is assumed to factorize as $\mathbf{C} = \mathbf{AL}$, where $\mathbf{A}$ is a matrix whose columns are ``atoms'', and the columns of the nonnegative matrix $\mathbf{L}$ are constrained\footnote{In the case the $\mathbf{L}$ were also a square matrix, then it would be a stochastic matrix a.k.a. a left stochastic matrix.} to sum to one\cite{jordan_nmf}. This constraint removes the scale indeterminacy between $\mathbf{A}$ and $\mathbf{H}$ \cite{Hamon}, and ensures that the columns of $\mathbf{C}$ are \textit{convex} combinations of the columns of $\mathbf{A}$. The matrix $\mathbf{L}$ can be thought of as the ``labeling'' matrix, since it gives the ``labels'' (coefficients) allowing one to write the dictionary elements (columns of $\mathbf{C}$) as convex combinations of ``atoms'' (columns of $\mathbf{A}$) \cite{Hamon}.\\

In summary, convex NMF writes an arbitrary\footnote{Actually \cite{Hamon} considers only the special case where the original matrix $\mathbf{M}$ is assumed to be non-negative and as a result the matrix $\mathbf{A}$ in the decomposition is constrained to be non-negative.} matrix $\mathbf{M}$ as a Tucker decomposition:
\[   
 \mathbf{M} \approx \mathbf{A} \mathbf{L} \mathbf{W}^\top \,,   
\]
where the core tensor $\mathbf{L}$ contains labels which allows one to construct a small dictionary $\mathbf{C}$ from convex combinations of the atoms in $\mathbf{A}$, and then the non-negative weights in $\mathbf{W}$ allow one to express all of the data points (columns of $\mathbf{M}$) as conic combinations of the dictionary elements. The paper \cite{jordan_nmf} considers the special case where the atoms $\mathbf{A}$ are constrained to equal the data points themselves, i.e. where $\mathbf{A} = \mathbf{M}$. This case leads to a matrix factorization conceptually similar to the CUR decomposition. Like the CUR and Tucker decompositions in general, ``convex-NMF stands out for its interpretability''\cite{jordan_nmf}. Depending on the context, other possible choices of atoms $\mathbf{A}$ can also lead to easily interpretable dictionary elements\cite{Hamon}. The choice of $\mathbf{A}$ can be seen as a ``source of supervision that guides learning'' \cite{Hamon}, or in other words a way to make the matrix factorization a semi-supervised learning method. Another reason why such matrix factorizations tend to be interpretable is because the constraints, while not explicitly requiring sparsity of the core or right factor matrices, tend to encourage such sparsity in practice\cite{jordan_nmf}.

%%% Local Variables:
%%% mode: latex
%%% TeX-master: "../new_notes_draft"
%%% End:

\subsubsection{Tucker Decompositions of Higher Order Tensors}
\label{sec:tucker-hot-apps}

Tucker decompositions of higher-order tensors have been used in numerous applications, for example the analysis of genetic data \cite{gsvd1}\cite{gsvd2}. Below some specific cases are described in detail.

\paragraphsection{Tensor Versions of CUR Decomposition}
\label{curt}

There are also (unsurprisingly) tensor versions of CUR, see for example \cite{curtensor2} or \cite{curtensor1}. In particular, the CURT decomposition as defined in \textbf{Defintion B.6} of \cite{curtensor1} is a Tucker decomposition of the original tensor, where $\mathbfcal{U}$ is the core tensor, and the matrices \textbf{C} (containing columns of the original tensor), \textbf{R} (containing rows of the original tensor), \textbf{T} (containing tubes of the original tensor), represent the three linear transformations generating the original tensor from the core tensor $\mathbfcal{U}$. (Review section \ref{sec:outer-prod-decomp} for a reminder of what these terms mean for a third-order tensor.) The paper \cite{curtensor1} also proposes a randomized algorithm for computing the approximate CURT decomposition. This is interesting because randomization algorithms might be a promising method for efficiently generating approximate solutions for tensor decomposition problems like this for which exact solutions are computationally prohibitive, based on their previous success in quickly finding approximate solutions to problems in linear algebra. See \cite{mahoney-notes} for an introduction to the principles and some of the results of this field for linear algebra problems. Section 3.5 of \cite{cichocki_tensor_networks} gives an overview of such randomized selection or ``sketching'' methods for tensor decompositions.

\paragraphsection{Discrete Fourier Transform}
\label{dfts}

A so-called ``two dimensional discrete Fourier transform'' which transforms a given $H \times W$ matrix $\mathbf{M}$ along both of its modes can be written\cite{Tolimieri1997}:
\[  
\mathbf{F}_H \mathbf{M} \mathbf{F}_W \,, 
 \]
where the matrices $\mathbf{F}$ correspond to the matrices of the one-dimensional discrete Fourier transforms of the appropriate dimensions, and $\mathbf{F}_W^\top = \mathbf{F}_W$ so that one could also express this as
\[ 
 \mathbf{F}_H \mathbf{M} \mathbf{F}_W^\top 
\]
since these matrices are symmetric. Comparing this with the formula from \ref{sec:spec-cases:-matrix}, one can see that the Fourier transformed matrix may in fact be written as a Tucker decomposition that has the original matrix as the core tensor and the discrete Fourier transforms as the factor matrices. This makes sense intuitively: in order to apply a discrete Fourier transform to a matrix, one needs to apply a discrete Fourier transform to its columns, and a discrete Fourier transform to its rows.\\

The formula for the discrete Fourier transforms of so-called ``multidimensional arrays'' (i.e. tensors), called the ``multidimensional discrete Fourier transform'' or the ``$N$-dimensional Fourier transform'', is often given in terms of the individual entries of the transformed array\cite{Nussbaumer1982}\cite{Tolimieri1997}\cite{Nechepurenko1990}, or a formula is given in terms of a vectorized form of the original array\cite{Tolimieri1997}\cite{Nechepurenko1990}\cite{DFT_kronecker}\cite{granata}. Close examination of these formulae reveals that they are actually describing the resulting matrix as a Tucker decomposition whose core tensor is the original array, and with factor matrices representing the discrete Fourier transforms applied to each of the modes of the array:
\[ 
 \left(  \bigotimes_{o=1}^O \mathcal{F}_{I_o}  \right) \mathbfcal{A} = \mathbfcal{A} \bullet_1 \mathbf{F}_{I_1} \cdots \bullet_O \mathbf{F}_{I_O} \,,  
\]
where $\mathbfcal{A}$ denotes the original untransformed array, $\mathcal{F}$ denotes the various linear transformations corresponding to the discrete Fourier transforms of the appropriate dimensions, and $\mathbf{F}$ denotes the corresponding matrix representations (i.e. the same type of matrices as in the formula above).\\

As an aside, it perhaps also interesting to note that the multidimensional \textit{continuous} Fourier transform also corresponds to the tensor product of functions\cite{osgood_fourier}.

\begin{quote}
  If you really want to impress your friends and confound your enemies, you can invoke \textit{tensor products} in this context. In mathematical parlance the separable~signal~$f$ [${f(x_1, \dots, x_n) = f_1(x_1)f_2(x_2)\dots f_n(x_n))}$] is the tensor product of the functions $f_i$ and one writes
  \[ 
f = f_1 \otimes f_2 \otimes \cdots \otimes f_n \,, 
 \]
and the formula for the Fourier transform as
\[ 
 \mathcal{F}(f_1 \otimes f_2 \otimes \cdots \otimes f_n) = \mathcal{F}f_1 \otimes \mathcal{F}f_2 \otimes \cdots \otimes \mathcal{F}f_n \,. 
 \]
People run in terror from the $\otimes$ symbol. Cool.
\end{quote}

The relevance of Tucker decompositions (in their vectorized form corresponding to Kronecker products) to discrete Fourier transforms has been studied for decades, see e.g. \cite{DFT_kronecker}\cite{granata}. Tucker decompositions are also natural for expressing other discrete transforms of ``multidimensional arrays'' such as Walsh-Hadamard transforms\cite{DFT_kronecker}\cite{granata}. In \cite{anandkumar_fourier}, Discrete Fourier transforms were used as subroutines for efficiently computing approximate CP decompositions via randomization.

%%% Local Variables:
%%% mode: latex
%%% TeX-master: "../new_notes_draft"
%%% End:

\paragraphsection{Regression with Tensor Predictors}
\label{tucker-tensor-predictor}

In \cite{tucker-predictor}, the authors apply Tucker decompositions to the same GLM-type model used for regression with tensor covariates that was already discussed in \ref{CP-tensor-predictor}. Specifically, consider again the scenario where one wants to use a  regress a scalar response variable $y$ against both ``ancillary'' vector-valued covariates $\mathbf{z}$ and ``primary'' tensor-valued covariates $\mathbfcal{X}$ (e.g. medical imaging data). The naive way to do this would be to solve the GLM:
\[ 
 g(\mu) = \alpha + \boldsymbol{\gamma} \bullet \mathbf{z}  +  \mathbfcal{B} \bullet_{[O]} \mathbfcal{X}  \,,
 \]
where $\bullet_{[O]}$ as before denotes entrywise dot product, so that the coefficient tensor $\mathbfcal{B}$ corresponds to as many parameters as there are entries in $\mathbfcal{X}$. As explained in section \ref{CP-tensor-predictor}, in most applications of interest this is obviously an intractable statistical problem, and so the authors of \cite{li_glm} proposed a vastly improved regression method using CP decomposition.\\

However, as the authors of \cite{tucker-predictor} point out, the fact that CP decomposition corresponds to a Tucker decomposition with a hyperdiagonal core tensor means that modeling the coefficient tensor $\mathbfcal{B}$ via a CP decomposition affords far less flexibility than using an arbitrary Tucker decomposition to model the coefficient tensor $\mathbfcal{B}$ would. This has practical consequences inasmuch as the hyperdiagonal core tensor must always be hypercubical, since in cases where one of the multilinear rank components of the ``true'' coefficient tensor $\mathbfcal{B}$ was substantially larger than the others, using a CP decomposition would force either (i) a model with many more parameters than necessary for all but one of the modes\footnote{The mode with the substantially larger multilinear rank component.}, or (ii) a model with many fewer parameters which is relatively accurate except for one mode, for which it is substantially inaccurate\footnote{Again that mode being the one with the substantially larger multilinear rank component.}.\\

In contrast, using a general Tucker decomposition model for the coefficient tensor $\mathbfcal{B}$ removes the need for any such tradeoff, by making it possible to use a non-cubical core tensor, in this case one for which all of the modes were the same dimension, except for one whose dimension was substantially larger. Thus one can recover a model with no more parameters than necessary which nevertheless is still accurate for all modes of the coefficient tensor $\mathbfcal{B}$, in other words a model whose parsimony enables it to utilize a limited sample size as much as possible, while retaining the accuracy of a CP model with more parameters. While the scenario described above might sound artificial to an external observer, it actually is not uncommon in practice\cite{tucker-predictor}.\\

At the same time, this Tucker regression model retains all of the advantages of the model proposed in \cite{li_glm}, which makes sense inasmuch as the model from \cite{li_glm} can be shown to be a special case. More specifically, for both regression models it is true that ``it exploits the special structure of the tensor data, reduces the dimensionality to enable efficient model estimation, and provides a sound low-rank approximation to a potentially high-rank signal'' \cite{tucker-predictor}. The authors also explain multiple benefits this model has compared to the CP regression model with regards to interpretation. The interested reader is urged to consult \cite{tucker-predictor} for full details about the Tucker regression model:
\[
  \begin{array}{rcl}
    g(\mu)    & =& \alpha + \boldsymbol{\gamma} \bullet \mathbf{z}  +\left(\mathbfcal{G}  \bullet_1 \mathbf{B}_1 \cdots \bullet_O \mathbf{B}_O  \right) \bullet_{[O]} \mathbfcal{X}   \\
\smallskip   & = & \displaystyle \alpha +  \boldsymbol{\gamma} \bullet \mathbf{z}  + \left(\sum_{ \mltidx{r}  \in \Mltidx{R}  }\mathcal{G}_{\mltidx{r}} \left( \boldsymbol{\beta}_1^{\smallsuper{r_1}} \otimes \cdots \otimes \boldsymbol{\beta}_O^{\smallsuper{r_O}}  \right) \right) \bullet_{[O]} \mathbfcal{X}  \,,
   \end{array}
 \]
where, for a coefficient tensor $\mathbfcal{B} \in \R^{\M}$ and a core tensor $\mathbfcal{G} \in \R^{\Mltidx{R}}$, the factor matrices $\mathbf{B}_o$ are such that $\mathbf{B}_o \in \R^{M_o \times R_o }$ and have columns $\boldsymbol{\beta}_o^{r_o} \in \R^{M_o}$ for all $1 \le r_o \le R_o$.

%%% Local Variables:
%%% mode: latex
%%% TeX-master: "../new_notes_draft"
%%% End:

\paragraphsection{Regression with Tensor Responses}
\label{tucker-tensor-response}

The paper \cite{tucker-response} applies Tucker decompositions to a model for regression with tensor response variables which is very similar to the one that was already discussed in \ref{CP-tensor-response}. Specifically, one again assumes that the order-$O$ tensor responses $\mathbfcal{Y} \in \R^{\M}$ may be represented in the form:
\[ 
 \mathbfcal{Y}_i = \mathbfcal{B} \bullet_{O+1} \mathbf{x}_i + \mathbfcal{E}_i \,,  
\]
where the coefficient tensor $\mathbfcal{B}$ is a tensor of order $O+1$, the $\mathbf{x}_i$ are vector-valued covariates, and the $\mathbfcal{E}_i$ are order-$O$ tensor-valued errors assumed to be independent of the covariates $\mathbf{x}_i$. However, while in \cite{li_sparse} the error tensors could have any distribution\endnote{Since the analysis in that paper provides a finite sample bound which only uses the $s$-sparse spectral norm of the $\mathbfcal{E}_i$ and which otherwise does not need to use any further properties of the error tensors, although the bound could be simplified further in the case that the entries of the error tensors $\mathbfcal{E}_i$ were assumed to be i.i.d. Gaussian.}, in \cite{tucker-response} the error tensors are assumed to further have a \textit{separable Kronecker covariance structure}. See section \ref{sec:separ-kron-covar} about this term.\\

Building from the work on envelope methods for regression with vector-valued responses found in \cite{cook_envelope}\cite{cook2}, the authors of \cite{tucker-response} propose an analogous \textit{generalized sparsity principle}. The main idea behind this assumption is that "whereas the usual sparsity principle focuses on individual variables, [the generalized sparsity principle] permits linear combinations of the variables to be irrelevant"\cite{tucker-response}. The statement of the principle in terms of subspaces is also reminiscent to the idea behind hierarchical tensor decompositions as phrased in \cite{hierarchical_tucker}, but the model for the coefficient tensor $\mathbfcal{B}$ in \cite{tucker-response} is not a hierarchical Tucker decomposition (see section \ref{sec:tensor-networks}). Instead, the model from \cite{tucker-response} assumes that $\mathbfcal{B}$ admits the following Tucker decomposition:
\[ 
  \mathbfcal{B} = \boldsymbol{\Theta} \bullet_1 \mathbf{U}_1 \bullet_2 \mathbf{U}_2 \cdots \bullet_O \mathbf{U}_o \bullet_{O+1} \mathbf{I}_{M_{O+1}} \,,  
\]
where $\boldsymbol{\Theta} \in \R^{\Mltidx[O+1]{I}}$ is an order $(O+1)$ tensor (and $I_{O+1} = M_{O+1}$), for all $o \in [O]$ the $\mathbf{U}_o \in \R^{M_o \times I_o}$ are orthogonal matrices (i.e. $\mathbf{U}_o^\top \mathbf{U}_o = \mathbf{I}_{I_o}$), and $\mathbf{I}_D$ denotes the $D \times D$ identity matrix.\\

The generalized sparsity principle corresponds to assuming that for each mode $o \in [O]$, only a proper subspace of $\R^{M_o}$ is likely to be relevant for predicting the response. The columns of $\mathbf{U}_o$ for each $o \in [O]$ constitute an orthonormal basis for the subspace of $\R^{M_o}$ assumed to be relevant to predicting the response from the covariates. Thus the Tucker decomposition with orthogonal factor matrices directly encodes the generalized sparsity principle assumption used by the authors.\\

The assumption that for each mode $o \in [O]$, only a strict subspace of $\R^{M_o}$ is relevant for predicting the response, also allows one to decompose the covariance matrix $\boldsymbol{\Sigma}_o$ for each mode as follows:
\[ 
\boldsymbol{\Sigma}_o =  \boldsymbol{\Omega}_o \bullet_1 \mathbf{U}_o \bullet_2 \mathbf{U}_o +  \boldsymbol{\Omega}_{0o} \bullet_1 \mathbf{U}_{0o} \bullet_2 \mathbf{U}_{0o} \,, 
 \]
where the columns of $\mathbf{U}_{0o} \in \R^{M_o \times (M_o - I_o)}$ are an orthonormal basis for the orthogonal complement of the space spanned by the columns of $\mathbf{U}_o$, and thus correspond to the subspace of $\R^{M_o}$ irrelevant for predicting the response. The matrix $\boldsymbol{\Omega}_o \in \R^{I_o \times I_o}$ represents the ``part of the covariance of the $o$-th mode in the subspace relevant to predicting the response'', and the Tucker decomposition applied to it amounts to embedding $\R^{I_o}$ into $\R^{M_o}$ as the subspace relevant to predicting the response. Completely analogously, $\boldsymbol{\Omega}_{0o} \in \R^{(M_o-I_o)\times(M_o - I_o)}$ represents the ``part of the covariance of the $o$-th mode in the subspace \textit{ir}relevant to predicting the response'', and again the Tucker decomposition applied to it amounts to embedding $\R^{(M_o - I_o)}$ into $\R^{M_o}$ as the subspace irrelevant to the regression (on other words the orthogonal complement of the relevant subspace).\\

The paper \cite{tucker-response} contains many more details. For example, the fact that the factor matrices in this Tucker decomposition are constrained to be orthogonal means that the relevant optimization problems for a blockwise coordinate descent approach is over the space of orthogonal matrices, also known as the Grassmannian manifold. Therefore Grassmannian optimization methods become relevant\cite{Ishteva_2008}\cite{Eld_n_2009}\cite{Borckmans_2010}, and the authors of \cite{tucker-response} discuss how to apply them here. Other estimation methods are considered, consistency and asymptotic normality of the method are proved, and superior performance on real datasets over other methods widely used in practice is demonstrated.

%%% Local Variables:
%%% mode: latex
%%% TeX-master: "../new_notes_draft"
%%% End:

%%% Local Variables:
%%% mode: latex
%%% TeX-master: "../new_notes_draft"
%%% End:

\theendnotes
\setcounter{endnote}{0}

\section{Tensor Contractions}
\label{sec:tensor-contractions}

Tensor contractions are the natural generalization of the dot product and matrix multiplication, with mode-$o$ products as defined in section \ref{sec:mode-o-products} being a special case. (Cf. Facts 10, 11, and 13 of section 15-1 of \cite{lim_hla}.) This review follows the convention of some other sources in denoting contractions with a $\bullet$ symbol, cf. e.g. \cite{qi_survey}. Roughly speaking, tensor contractions can also be understood as the reformulation of tensor-valued linear functions of tensors as multilinear functions, cf. appendix \ref{sec:tens-in-tens-out} for a slightly less rough description of this idea or section 2.3 of \cite{uschmajew_pde}. For example, Tucker decompositions are linear in the core tensor for fixed factor matrices, but multilinear in the factor matrices for fixed core tensors (cf. section 2 of \cite{learning_latent_variable_models} or the description of Tucker decompositions as ``multilinear matrix multiplication'' in \cite{lim_hla}, \cite{Hackbusch2016}, \cite{desilva_lim}, and other sources). Tensor contractions have historically been an intense topic of study in mathematical physics and differential geometry\cite{Qi2005}, but their pertinence to statistics seems to be only more recently recognized (cf. \cite{mccullagh}). For a review of properties satisfied by tensor contractions, see \cite{lim_hla}, in particular section 15-1. Tensor contractions are useful e.g. for constructing neural networks\cite{8014977}, but (besides matrix multiplication) perhaps they are most important because of tensor networks, see section \ref{sec:tensor-networks}.\\

There are essentially two ways to think about tensor contractions: as an ``external'' operation (between two tensors), or as an ``internal'' operation (within a single tensor). Both ways are related to the other. The ``external'' way can be considered a generalization of the dot product of two vectors, while the ``internal'' way can be considered a generalization of the trace of a matrix. Matrix multiplication and mode-$o$ products\endnote{It might be necessary to allow permuting modes of tensors in order to recover mode-$o$ products as a special case of the precise definition from section 15-1 of \cite{lim_hla}, except in the case when $o=O$. Given a tensor $\T$ and matrix $\mathbf{M}$, the mode-$o$ product $\T \bullet_o \mathbf{M}$ of $\T$ by $M$ is the result of contracting the first mode of $\mathbf{M}$ with the $o$'th mode of $\T$ in a way such that the second mode of $\mathbf{M}$ becomes the $o$'th mode of the resulting tensor, cf. section \ref{sec:mode-o-products}. This seems to differ from the definition of contraction from \cite{lim_hla}, where the second $\mathbf{M}$ would always become the last ($O$'th) mode of the resulting tensor. In particular, Fact 13 of section 15-1 of \cite{lim_hla} seems to be incorrect using the definition of contraction from \cite{lim_hla}, although the analogous order-invariance property (see section \ref{sec:defin-tuck-decomp}) is true for the definition from section \ref{sec:mode-o-products}. This becomes a non-issue if one is allowed to transpose the $o$'th and $O$'th modes of the resulting tensor.} are important examples of tensor contractions. Below some of these most basic examples of tensor contractions are discussed to provide intuition. For a precise and general definition of tensor contraction, see e.g. section 15-1 of \cite{lim_hla}, in particular Fact 12.

\subsection{The Dot Product}
\label{sec:dot-product}

Given two vectors $\mathbf{x}, \mathbf{y} \in \R^M$ (observe how they must have the same dimension/length, i.e. be in the same space $\R^M$), their \textbf{(standard) dot product} is defined as:
\[
\mathbf{x} \bullet \mathbf{y} = \summing{m} x_m y_m = \summing{m} \pi_m(\mathbf{x}) \pi_m(\mathbf{y}) =\summing{m} (\mathbf{x} \bullet \unitvector{m})(\mathbf{y} \bullet \unitvector{m}) \in \R \,,
\]
where for all $m \in [M]$ the symbol $\pi_m$ denotes the $m$'th standard coordinate projection $\R^M \to \R$ corresponding to the (dual basis of) the standard basis of $\R^M$, i.e. the linear function(al) $\R^M \to \R$ which returns the $m$'th entry of the given vector. (Cf. appendix \ref{sec:vector-identification} as well as appendix \ref{sec:coord-vector-spaces}.)\\

Due to the distributive property it can be readily seen that the dot product is a bilinear function $\R^M \times \R^M \to \R$. In particular, fixing the vector $\mathbf{x} \in \R^M$, the function $\R^M \to \R$ with rule of assignment $\mathbf{y} \mapsto \mathbf{x} \bullet \mathbf{y}$ is a linear function, as is the function $\R^M \to \R$ with rule of assignment $\mathbf{x} \mapsto \mathbf{x} \bullet \mathbf{y}$ derived from fixing the vector $\mathbf{y} \in \R^M$ instead. Moreover, every linear function(al) $\R^M \to \R$ can be written using the dot product in a unique way (a fact which is essentially the finite-dimensional version of the Riesz Representation Theorem). Specifically, given any linear function(al) $\varphi: \R^M \to \R$, from linearity it follows that its rule of assignment equals:
\[
\mathbf{x} \mapsto \mathbf{x} \bullet \mathbf{f} = \mathbf{f} \bullet \mathbf{x} \,,
\]
where $\mathbf{f} \in \R^M$ is the vector whose $m$'th entry $f_m$ equals $\varphi(\unitvector{m})$. (Cf. appendices \ref{sec:repr-dual-spac} and \ref{sec:repr-line-funct}.)

\subsection{The Trace}
\label{sec:trace}

Since the dot product is a multilinear function $\R^M \times \R^M \to \R$, the universal property of the tensor product satisfied by the outer product and $\R^{[M] \times [M]}$, the space of $M \times M$ matrices, guarantees that there is a \textit{unique} linear function $\Phi: \R^{[M] \times [M]}$ such that for any two vectors $\mathbf{x}, \mathbf{y} \in \R^M$ :
\[
\mathbf{x} \bullet \mathbf{y} = \Phi ( \mathbf{x} \otimes \mathbf{y}) \,.
\]
As might have been guessed already from the section heading, that linear function $\Phi: \R^{[M] \times [M]}$ happens to be the trace operator $\trace$. Given the standard column-wise tensor product decomposition of a matrix $\mathbf{A} \in \R^{[M] \times [M]}$ (remember that the universal property of the tensor product guarantees that the value of $\Phi$ will not depend on the chosen tensor product decomposition), consider how:
\[
\Phi(\mathbf{A}) = \Phi \left( \summing{m} \mathbf{A}_{:m} \otimes \unitvector{m}  \right) = \summing{m} \Phi(\mathbf{A}_{:m} \otimes \unitvector{m}) = \summing{m} \mathbf{A}_{:m} \bullet \unitvector{m} = \summing{m} A_{mm} = \trace(\mathbf{A}) \,.
\]
The standard row-wise tensor product decomposition of $\mathbf{A}$ can also demonstrate this result:
\[
\Phi(\mathbf{A}) = \Phi \left( \summing{m} \unitvector{m} \otimes \mathbf{A}_{m:} \right) = \summing{m} \Phi (\unitvector{m} \otimes \mathbf{A}_{m:}) = \summing{m} \unitvector{m} \bullet \mathbf{A}_{m:} = \summing{m} A_{mm} = \trace(A) \,.
\]
It is a possibly counterintuitive fact that the trace $\trace(A)$, the sum of the diagonal values of $\mathbf{A}$, is also equal to the sum of the eigenvalues of $\mathbf{A}$ when $\mathbf{A}$ is a diagonalizable matrix, i.e. $\mathbf{A} = \mathbf{P} \boldsymbol{\Lambda} \mathbf{P}^{-1}$ for a diagonal matrix $\boldsymbol{\Lambda}$ whose diagonal entries are the eigenvalues of $\mathbf{A}$ (with multiplicity). However, this fact can be explained readily using the characterization of the trace via the universal property of the tensor product, since the matrix factorization $\mathbf{A} = \mathbf{P} \boldsymbol{\Lambda} \mathbf{P}^{-1}$ corresponds to the tensor product decomposition (where in what follows $\lambda_m \defequals \boldsymbol{\Lambda}_{mm}$, the $m$'th eigenvalue of $\mathbf{A}$):
\[
\mathbf{A} = \summing{m}\lambda_m \mathbf{P}_{:m} \otimes (\mathbf{P}^{-1})_{m:} = (\lambda_m \mathbf{P}_{:m}) \otimes (\mathbf{P}^{-1})_{m:}  = \mathbf{P}_{:m} \otimes (\lambda_m (\mathbf{P}^{-1})_{m:}  ) \,.
\]
More specifically, as a result of this tensor product decomposition one concludes that:
\[
  \begin{array}{rcl}
\trace(\mathbf{A}) = \Phi(\mathbf{A})    & = & \displaystyle \Phi \left( \summing{m} (\lambda_m \mathbf{P}_{:m}) \otimes (\mathbf{P}^{-1})_{m:} \right) = \summing{m} \Phi( (\lambda_m \mathbf{P}_{:m}) \otimes (\mathbf{P}^{-1})_{m:} ) \smallskip \\
& = & \displaystyle \summing{m} (\lambda_m \mathbf{P}_{:m}) \bullet (\mathbf{P}^{-1})_{m:} = \summing{m} \lambda_m (\mathbf{P}_{:m} \bullet (\mathbf{P}^{-1})_{m:} \smallskip \\
& = & \displaystyle \summing{m} \lambda_m (1) = \summing{m} \lambda_m \,.
  \end{array}
\]
The second to last equality follows from the fact that a matrix $\mathbf{Y} \in \R^{[N] \times [M]}$ is a right inverse of the matrix $\mathbf{X} \in \R^{[M] \times [N]}$, if and only if $\mathbf{X}$ is a left inverse of $\mathbf{Y}$, if and only if $\mathbf{XY}$ equals the $M \times M$ identity matrix $\mathbf{I}$, if and only if for all $(m_1, m_2) \in [M] \times [M]$:
\[
\mathbf{X}_{m_1:} \bullet \mathbf{Y}_{:m_2} = \delta_{m_1 m_2} =
\begin{cases}
  1 & m_1 = m_2 \\
0 & m_1 \not= m_2 
\end{cases} \,.
\]
This is just the standard ``row then column'' definition of matrix multiplication. (It also explains why the ``columns of $\mathbf{Q}$ are orthonormal'' and the ``$\mathbf{Q}^\top \mathbf{Q} = \mathbf{I}$'' characterizations of orthogonal matrices $\mathbf{Q}$ are equivalent.) When $N=M$, $\mathbf{X}=\mathbf{P}^{-1}$, and $\mathbf{Y}=\mathbf{P}$, the desired conclusion immediately follows. i.e. that $(\mathbf{P}^{-1})_{m:} \bullet \mathbf{P}_{:m} = \mathbf{P}_{:m} \bullet (\mathbf{P}^{-1})_{m:} = 1$ for all $m \in [M]$.\\

While the dot product is a bilinear operation between two distinct vectors, using one mode each from two different tensors, the trace is a linear operation ``within'' a single tensor, using two modes of the same tensor. Both of these operations characterize the same ``essential notion'' of contraction in different ways, with both being related to the other via the universal property of the tensor product. In particular, all of these properties generalize to contractions of higher-order tensors.

\subsection{Matrix Multiplication}
\label{sec:matr-mult}

Given two matrices $\mathbf{X} \in \R^{[H] \times [M]}$ and $\mathbf{Y} \in \R^{[M] \times [W]}$, their product $\mathbf{XY} \in \R^{[H] \times [W]}$ is equal to the contraction of the second mode of the first matrix with the first mode of the second matrix, $\mathbf{X} \bullet_2 \mathbf{Y}$. As a convention, given a contraction symbol $\bullet$, unless otherwise specified it will be assumed to refer to the contraction of the first/left tensor's mode which corresponds to the subscript of $\bullet$, with the first mode of the second/right tensor. Specifically, given outer product decompositions $\mathbf{X} = \summing{i} \mathbf{a}_i \otimes \boldsymbol{\alpha}_i$ and $\mathbf{Y} = \summing{j} \boldsymbol{\beta}_j \otimes \mathbf{b}_j$, their product is the matrix:
\[
\mathbf{XY} \defequals \mathbf{X} \bullet_2 \mathbf{Y} = \summing{i} \summing{j} (\boldsymbol{\alpha}_i \bullet \boldsymbol{\beta}_j ) \mathbf{a}_i \otimes \mathbf{b}_j \,.
\]
Observe that this corresponds to a bilinear function $\R^{[H] \times [M]} \times \R^{[M] \times [W]} \to \R^{[H] \times [W]}$ in $\mathbf{X}$ and $\mathbf{Y}$, and thus in particular linear in $\mathbf{X}$ for any fixed $\mathbf{Y}$ as well as linear in $\mathbf{Y}$ for any fixed $\mathbf{X}$.\\

Because this operation is bilinear, the value of the above definition does \textit{not} depend on the chosen outer product decompositions of $\mathbf{X}$ and $\mathbf{Y}$. For example, by choosing the standard row-wise decomposition for $\mathbf{X}$ and the standard column-wise decomposition for $\mathbf{Y}$ one gets the result:
\[
\mathbf{XY} = \left( \summing{h} \unitvector{h} \otimes \mathbf{X}_{h:}  \right) \bullet_2 \left( \summing{w} \mathbf{Y}_{:w} \otimes \unitvector{w} \right) = \summing{h} \summing{w} (\mathbf{X}_{h:} \bullet \mathbf{Y}_{:w}) \unitvector{h} \otimes \unitvector{w} \,.
\]
This says that the $(h,w)$'th entry of $\mathbf{XY}$ is the dot product of the $h$'th row of $\mathbf{X}$ with the $w$'th column of $\mathbf{Y}$, exactly the same as the typical definition of matrix multiplication usually given.\\

Similarly, by choosing the standard column-wise decomposition for $\mathbf{X}$ and the standard row-wise decomposition for $\mathbf{Y}$, one can derive this other useful identity:
\[
  \begin{array}{rcl}
   \mathbf{XY} & = &\displaystyle \left( \summing{m} \mathbf{X}_{:m} \otimes \unitvector{m}   \right) \bullet_2 \left(  \summing{m} \unitvector{m} \otimes \mathbf{Y}_{m:}   \right) = \sum_{m_1=1}^M \sum_{m_2=1}^M (\unitvector[M]{m_1} \bullet \unitvector[M]{m_2} ) \mathbf{X}_{:m_1} \otimes \mathbf{Y}_{m_2 :}\smallskip \\
& = &\displaystyle  \sum_{m_1=1}^M \sum_{m_2=1}^M \delta_{m_1 m_2} \mathbf{X}_{:m_1} \otimes \mathbf{Y}_{m_2:} = \summing{m} \mathbf{X}_{:m} \otimes \mathbf{Y}_{m:} \,.
  \end{array}
\]
This is exactly the result mentioned in section \ref{sec:spec-case:-matr}, that the product of two matrices is equal to the sum of the outer products of the columns of the first matrix with the rows of the second matrix.\\

Both of the above results can be generalized\endnote{
Using the language of section \ref{sec:subt-rearr-tens}, the generalizations of these two results are one expression of the ``duality'' which for each tensor exists between the family of subtensors indexed by its contracted modes and the family of subtensors indexed by its uncontracted modes. For each tensor the subtensors corresponding to the contracted modes have multi-indices from the multi-index set formed by the uncontracted modes, and the subtensors corresponding to the uncontracted modes have multi-indices from the multi-index set formed by the contracted modes. For the contraction to be well-defined the multi-index set formed by the contracted modes must be the same for both tensors. The multi-index set of the tensor resulting from the contraction is the product of the multi-index set formed by the uncontracted modes of the first tensor with the multi-index set formed by the uncontracted modes of the second tensor.
} to arbitrary tensor contractions. Like the first result above, choosing outer product decompositions of the tensors which utilize the subtensors corresponding to the contracted modes and unit tensors corresponding to the uncontracted modes leads to a formula\footnote{Such a formula is the definition usually given for tensor contraction, compare e.g. the definitions from \cite{mccullagh} or \cite{lim_hla}.} for the entries of the resulting tensor as dot products of those subtensors indexed by the uncontracted modes. Like the second result above, choosing outer product decompositions of the tensors which utilize unit tensors corresponding to the contracted modes and the subtensors corresponding to the uncontracted modes leads to a formula\footnote{Compare Fact 12 of section 15-1 of \cite{lim_hla}. In this case the dot products will reduce to Kronecker deltas.} expressing the tensor resulting from the contraction as the sum of outer products of subtensors indexed by the contracted modes.\\ 

The product of two matrices has a straightforward relationship with the fourth order tensor $\mathbf{X} \otimes \mathbf{Y}$. Specifically, since matrix multiplication is an ``external'' contraction of two tensors, i.e. it is a bilinear function $\R^{[H] \times [M]} \times \R^{[M] \times [W]} \to \R^{[H] \times [W]}$ and generalizes the dot product, the universal property of the tensor product leads to the existence of a corresponding ``internal contraction'' of a single tensor, i.e. a linear function $\R^{[H] \times [M] \times [M] \times [W]} \to \R^{[H] \times [W]}$ which generalizes the trace.\\

Given any outer product decomposition of a fourth order tensor $\T \in \R^{[H] \times [M] \times [M] \times [W]} \to \R^{[H] \times [W]}$, one can use it to define the following linear function $\R^{[H] \times [M] \times [M] \times [W]} \to \R^{[H] \times [W]} \to \R^{[H] \times [W]}$:
\[
  \begin{array}{rcl}
\trace_{(2)(3)}(\T)    &= & \displaystyle \trace_{(2)(3)} \left( \summing{k} \v{k}_1 \otimes \v{k}_2 \otimes \v{k}_3 \otimes \v{k}_4  \right) \smallskip \\
& \defequals & \displaystyle \summing{k} \trace(\v{k}_2 \otimes \v{k}_3) \v{k}_1 \otimes \v{k}_4 = \summing{k} (\v{k}_2 \bullet \v{k}_3) \v{k}_1 \otimes \v{k}_4 \,.
  \end{array}
\]
It follows from the above definitions that $\trace_{(2)(3)}$ is the linear function corresponding to matrix multiplication given by the universal property of the tensor product. More concretely, it is easy to see that for any two given matrices $\mathbf{X} \in \R^{[H] \times [M]}$ and $\mathbf{Y} \in \R^{[M] \times [W]}$, one has the relationship:
\[
\mathbf{X} \bullet_2 \mathbf{Y} = \trace_{(2)(3)}(\mathbf{X} \otimes \mathbf{Y}) \,.
\]
As implied already above, this pattern continues to generalize to arbitrary tensor contractions.

%%% Local Variables:
%%% mode: latex
%%% TeX-master: "../new_notes_draft"
%%% End:

\theendnotes
\setcounter{endnote}{0}

\section{Tensor Generalizations of Eigen- and Singular Values and Vectors}
\label{sec:gener-eigenv-eigenv}

Two different papers, \cite{Qi2005} and \cite{LekHengLim}, published in 2005 independently defined generalizations of eigenvectors and eigenvalues for symmetric tensors. The first\cite{Qi2005} used an algebraic approach while the second\cite{LekHengLim} used a variational approach (generalizing the characterization of eigenvalues of symmetric matrices using the Rayleigh quotient), but arrived at similar results.\\ 

For a given symmetric tensor $\mathbfcal{S} \in \R^{[M]^O}$, what \cite{Qi2005} calls an $Z$-eigenvalue/$Z$-eigenvector\footnote{``$Z$'' evidently stands for ``Zhou'', who first suggested the definition to the author of \cite{Qi2005}.} pair of $\mathbfcal{S}$, and what \cite{LekHengLim} calls an $\ell^2$-eigenvalue/$\ell^2$-eigenvector pair of $\mathbfcal{S}$, is any pair $(\lambda, \mathbf{x})$, where $\lambda \in \R$, and $\mathbf{x} \in \R^M$ is such that $\mathbf{x} \bullet \mathbf{x} = ||\mathbf{x}||^2_2=1$, satisfying the equation:
\[
\mathbfcal{S} \bullet_2 \mathbf{x} \cdots \bullet_O \mathbf{x} = \lambda \mathbf{x} \,.
\]
Note that, given a solution $(\lambda, \mathbf{x})$ of the form above, for any real number $t \in \R$ one has that $(t^{O-2}\lambda, t\mathbf{x})$ also satisfies the above equation\footnote{Although technically speaking the only $t$ such that $(t^{O-2}\lambda, t \mathbf{x})$ still satisfies the constraints of the original optimization problem from which this equation was derived in \cite{LekHengLim} is $t=-1$, as noted in \cite{Cartwright2013}.}. Thus for tensors of order $O \ge 3$, this definition of eigenvalue and eigenvector leads to a situation where the value of the eigenvalue depends on the magnitude of the eigenvector, very much unlike the matrix case. However, it still makes sense to call two $Z$-eigenpairs\footnote{Which for convenience is now redefined to be any pair satisfying the above equation, regardless of the value of $||\mathbf{x}||_2$ (as long as $\mathbf{x}$ is not the all-zeros vector), following \cite{Cartwright2013}.} $(\lambda, \mathbf{x})$ and $(\lambda', \mathbf{x}')$ equivalent if $\lambda' = t^{O-2}\lambda$ and $\mathbf{x}'=t\mathbf{x}$ for some non-zero $t \in \R$. The variational approach from \cite{LekHengLim} lends itself only to real-valued eigenvectors and eigenvalues, but the algebraic characterization given above also works when $\lambda$ and $\mathbf{x}$ are allowed to be complex-valued\cite{Qi2005} and in that case even $\mathbfcal{S}$ can be complex-valued too\cite{Cartwright2013}. In the complex case, both the normalizations $\mathbf{x} \bullet \mathbf{x} = 1$ and\footnote{Where $\bar{\mathbf{x}}$ denotes the complex conjugate.} $\mathbf{x} \bullet \bar{\mathbf{x}} = 1$ can be considered, or the same equivalence relation as above except that now $t$ can be any non-zero number in $\mathbb{C}$\cite{Cartwright2013}. Then in the complex case \cite{Qi2005} calls the above $E$-eigenvalues and $E$-eigenvectors ($E$ for Euclidean).\\

For any vector $\mathbf{x} \in \R^M$, define $\mathbf{x}^n \in \R^M$ to be the ``entrywise $n$-th power of $\mathbf{x}$'', i.e. for all $m \in [M]$, $(\mathbf{x}^n)_m = (x_m)^n$. Then for a given symmetric tensor $\mathbfcal{S} \in \R^{[M]^O}$, what \cite{Qi2005} calls an $H$-eigenvalue/$H$-eigenvector pair of $\mathbfcal{S}$, and what \cite{LekHengLim} calls\footnote{When $O$ is odd, the definition from \cite{LekHengLim} (derived from the variational formulation of the problem) differs slightly. However the same author uses this definition in later work\cite{hillar_lim}, which matches that found in \cite{Qi2005} and \cite{chang2008}.} an $\ell^O$-eigenvalue/$\ell^O$-eigenvector pair of $\mathbfcal{S}$ is any pair $(\lambda, \mathbf{x})$ with $\lambda \in \R$ and $\mathbf{x } \in \R^M$ satisfying the equation:
\[
\mathbfcal{S} \bullet_2 \mathbf{x} \cdots \bullet_O \mathbf{x} = \lambda \mathbf{x}^{O-1} \,.
\]
Notice that for this equation, when $\mathbf{x}$ is replaced with $\mathbf{x}' = t \mathbf{x}$ (for any $t \in \R$), both the left and right-hand sides of this equation scale by $t^{O-1}$, in other words the equation is homogeneous. Thus like the eigenvalues of a matrix, and unlike $Z$-eigenvalues, the value of an $H$-eigenvalue does \textit{not} depend on the magnitude of the eigenvector. One consequence of this scale-invariance is that both \cite{Qi2005} and \cite{LekHengLim} consider any norm constraint on $\mathbf{x}$ to be unnecessary\footnote{Although again technically speaking only those $\mathbf{x}$ such that $||\mathbf{x}||_O=1$ satisfy the original optimization problem from which \cite{LekHengLim} derives this equation.}. Finally \cite{Qi2005} calls the complex analogues of these just ``eigenvalues'' and ``eigenvectors''. Note that while sources directly following the conventions and terminology of \cite{Qi2005} (such as \cite{banerjee_spectral}\cite{Cooper2012}\cite{Pearson2013}) will use ``tensor eigenvalue'' and ``tensor eigenvector'' to mean ``complex analogue of (H/$\ell^O$)-eigenvector'' and ``complex analogue of (H/$\ell^O$)-eigenvalue'' respectively, other sources have different conventions regarding the default meaning of ``tensor eigenvalue'' and ``tensor eigenvector''. For example, in \cite{hillar_lim} the default meaning of ``tensor eigenvalue'' and ``tensor eigenvector'' is ($Z$/$\ell^2$)-eigenpairs, while in \cite{robeva_thesis} and \cite{Cartwright2013} the default meaning is (equivalence classes of unconstrained) $E$-eigenpairs.\\

Notice why the assumption that $\mathbfcal{S}\in \R^{[M]^O}$ is symmetric becomes relevant in the above definitions:
\[
\mathbfcal{S} \bullet_2 \mathbf{x} \cdots \bullet_O \mathbf{x} = \mathbfcal{S} \bullet_1 \mathbf{x} \bullet_3 \mathbf{x} \cdots \bullet_O \mathbf{x} = \cdots = \mathbfcal{S} \bullet_1 \mathbf{x} \bullet_2 \mathbf{x} \cdots \bullet_{O-1} \mathbf{x} \,.
\]
Thus if one wants to extend any one of the above two (or four, depending on how one counts) eigenpair definitions to arbitrary cubical tensors $\mathbfcal{T} \in \R^{[M]^O}$, for which all of the $O$ values:
\[
\mathbfcal{T} \bullet_2 \mathbf{x} \cdots \bullet_O \mathbf{x}, \quad \dots, \quad \mathbfcal{T} \bullet_1 \mathbf{x} \bullet_2 \mathbf{x} \cdots \bullet_{O-1} \mathbf{x}
\]
can in general be distinct, one needs to do one of the following: (1) arbitrarily choose one of the $O$ possible definitions above, or (2) have $O$ different types of eigenpairs for non-symmetric cubical tensors, one type corresponding to each possible definition. Both \cite{Qi2007} and \cite{Cartwright2013} implement the first approach by always contracting $\mathbf{x}$ along the \textit{last} $O-1$ modes of $\mathbfcal{T}$, whereas \cite{hillar_lim} implements the first approach by always contracting $\mathbf{x}$ along the \textit{first} $O-1$ modes of $\T$. The second approach is used in \cite{LekHengLim}, where for each $o \in [O]$ the eigenpairs following the definition corresponding to contracting $\mathbf{x}$ along every mode of $\T$ \textit{except} for the $o$'th mode are called mode-$o$ eigenvectors with associated mode-$o$ eigenvalues. This is analogous to the definition from section \ref{CP-tensor-response}, except that involves a contraction on the $o$'th mode only and no others, while this involves contractions on every other mode besides the $o$'th mode. As observed in \cite{LekHengLim}, the second approach generalizes the concepts of right eigenvectors (and right eigenvalues) of matrices, corresponding to the equation $\mathbf{M} \bullet_2 \mathbf{x} = \lambda \mathbf{x}$, and left eigenvectors (and left eigenvalues) of matrices, corresponding to the equation $\mathbf{M} \bullet_1 \mathbf{x} = \lambda \mathbf{x}$, so the second approach is no less reasonable than something which is commonly done in the matrix case (e.g. when studying Markov chains and the Perron-Frobenius theorem). Of course the most common convention in the matrix case is to automatically assume that ``eigenvector'' and ``eigenvalue'' mean ``right eigenvector'' and ``right eigenvalue'', i.e. that $\mathbf{x}$ is contracted on all but the first mode, $\mathbf{M} \bullet_2 \mathbf{x} = \lambda \mathbf{x}$, so the first approach as used in \cite{Qi2007} and \cite{Cartwright2013} is also the generalization of a common convention from the special matrix case.\\

One reason there is not so much tension in the matrix case between the convention of always assuming that ``eigenvector'' always refers to ``right eigenvector'' and the convention of considering both left and right eigenvectors explicitly is the fact that the left eigenvectors of $\mathbf{M}$ are always the same as the right eigenvectors of $\mathbf{M}^\top$. The analogous fact is true also for higher-order tensors: if one interchanges the first mode and the $o$'th mode of $\T$ before contracting $\mathbf{x}$ on all but the first mode of the resulting tensor, the result will be the same as if one had left the modes in place and originally contracted $\mathbf{x}$ on all of the modes except the $o$'th mode. However perhaps an even more important reason is that, for a matrix, the left eigenvalues always equal the right eigenvalues. (Equivalently the (right) eigenvalues of $\mathbf{M}^\top$ always equal the (right) eigenvalues of $\mathbf{M}$.) In this case, the analogous fact is \textit{not} true for higher-order tensors. Counterexamples occur even with some of the simplest possible higher-order tensors. Consider $\T \in \R^{[2]^3}$ with entries:
\[  \T_{::1} =
  \begin{bmatrix}
    1 & 1 \\ 1 & 1
  \end{bmatrix} \,, \quad \T_{::2} =
  \begin{bmatrix}
    2 & 2 \\ 2 & 2 
  \end{bmatrix} \,.
\]
Because $\T_{1::}=\T_{:1:}$ and $\T_{2::}=\T_{:2:}$, or in other words since $\T$ is invariant under interchange of its first two modes, both the mode-$1$ and the mode-$2$ $Z$-eigenvalues and $Z$-eigenvectors are the same, namely $(6t, (t,t))$ for every $t \in \R$, with the normalized $Z$-eigenpairs being $(3 \sqrt{2}, (\frac{1}{\sqrt{2}}, \frac{1}{\sqrt{2}} ))$ and $(-3 \sqrt{2}, (-\frac{1}{\sqrt{2}}, -\frac{1}{\sqrt{2}}))$. In contrast, the mode-$3$ $Z$-eigenvalues (unlike the matrix case) and the mode-$3$ $Z$-eigenvectors (like the matrix case) are not the same, namely the $Z$-eigenpairs are $(9t, (t, 2t))$ for every $t \in \R$, with normalized $Z$-eigenpairs $(\frac{9}{\sqrt{5}}, (\frac{1}{\sqrt{5}}, \frac{2}{\sqrt{5}}))$ and $(-\frac{9}{\sqrt{5}}, (-\frac{1}{\sqrt{5}}, -\frac{2}{\sqrt{5}}))$. Moreover the counterexample does not rely on using $Z$-eigenpairs instead of $H$-eigenpairs. Again both the mode-$1$ and mode-$2$ $H$-eigenvalues and $H$-eigenvectors are the same, namely $(6, (t,t))$ for every $t \in \R$ and $(0, (t,-t))$ for every $t \in \R$ (since these are $H$-eigenvectors there is no need to normalize). In contrast again one has that the mode-$3$ $H$-eigenpairs are distinct, namely ${(3+2\sqrt{2}, (t, \sqrt{2}t))}$ for all $t \in \R$ and $(3-\sqrt{2}, (t,-\sqrt{2}t))$ for all $t \in \R$.\\

One benefit of the variational approach from \cite{LekHengLim} is that it readily extends to also provide generalizations of singular values and singular vectors to tensors. The framework in \cite{LekHengLim} allows one to define a distinct type of eigenpair for each natural number $p$ corresponding to the $\ell^p$ norm, with the special cases of $p=2$ (mostly) coinciding with $Z$-eigenvalues and $p=O$ coinciding with $H$-eigenvalues. Given an arbitrary, not even necessarily cubical, real-valued tensor $\T \in \R^{\M}$, one can extend that variational approach to define distinct notions of singular values and singular vectors by choosing a (potentially) different norm for each $\R^{M_o}$. However for simplicity one usually considers the case where the same $p$-norm is chosen for every $\R^{M_o}$. The tensor versions of singular values and singular vectors considered in \cite{Robeva2017} and \cite{robeva_thesis} correspond to the $\ell^2$-singular values and $\ell^2$-singular vectors from \cite{LekHengLim}. Specifically, given a tensor $\T \in {\R^{\M}}$, an $\ell^2$-singular value of $\T$ corresponding to an $\ell^2$-singular vector tuple of $\T$ are a real number $\sigma \in \R$ and $O$ vectors such that $\mathbf{x}_o \in \R^{M_o}$ for each $o \in O$ which together are a solution to the system of equations:
\[
  \begin{array}{rcl}
   \T \bullet_2 \mathbf{x}_2 \cdots \bullet_O \mathbf{x}_O & = & \sigma \mathbf{x}_1 \,,\\
\T \bullet_1 \mathbf{x}_1 \bullet_3 \mathbf{x}_3 \cdots \bullet_O \mathbf{x}_O & = & \sigma \mathbf{x}_2 \,,\\
& \vdots & \\
\T \bullet_1 \mathbf{x}_1 \cdots \bullet_{O-1} \mathbf{x}_{O-1} & = & \sigma \mathbf{x}_O \,,
  \end{array}
\]
or equivalently are, for each $o \in [O]$, a solution to the equation:
\[
\T \bullet_1 \mathbf{x}_1 \cdots \bullet_{o-1}\mathbf{x}_{o-1} \bullet_{o+1} \mathbf{x}_{o+1} \cdots \bullet_O \mathbf{x}_O = \sigma \mathbf{x}_o \,.
\]
The optimization problem in \cite{LekHengLim} from which one derives the above equations has the additional constraints that $\sigma \ge 0$ and that for every $o \in [O]$ one has $||\mathbf{x}_o||_2 = 1$. If one is willing to consider any $(\sigma, \mathbf{x}_1, \dots, \mathbf{x}_O)$ satisfying the above equations an $\ell^2$-singular value with associated $\ell^2$-singular vector tuple (regardless of whether they satisfy the constraints $\sigma \ge 0$ and $||\mathbf{x}_o||_2 =1 $ for all $o \in [O]$), then starting from any $(\sigma, \mathbf{x}_1, \dots, \mathbf{x}_O)$ which satisfies the above equations (as long as the $\mathbf{x}_o$'s are all non-zero), all other $\ell^2$-singular values and associated $\ell^2$-singular vector tuples can be generated by some combination of the following operations on $\sigma$ and $\mathbf{x}_1$, \dots, $\mathbf{x}_O$:
\begin{itemize}
\item Given a non-zero $t \in \R$, multiply every $\mathbf{x}_o$ by $t$, and multiply $\sigma$ by $t^{O-2}$, i.e. $(t^{O-2}\sigma, t\mathbf{x}_1, \dots, t\mathbf{x}_O)$ will always also be a solution to the above equations.
\item Flip the sign of an odd number of the $\mathbf{x}_o$'s and replace $\sigma$ with $-\sigma$.
\item Flip the sign of an even number of the $\mathbf{x}_o$'s and leave $\sigma$ unchanged.
\end{itemize}
Leaving aside the somewhat subtle sign indeterminacy issues which appear now that one must consider $O$ distinct vectors simultaneously instead of just a single vector, this is the same scale indeterminacy problem one also has for ($\ell^2$/Z)-eigenpairs. (Note that the sign indeterminacy issue is still relevant even when $O=2$ even though the scale indeterminacy issue is not.)\\

Also mentioned explicitly in \cite{LekHengLim} and \cite{hillar_lim} are the $\ell^O$-singular values of $\T$ each corresponding to an $\ell^O$-singular vector tuple of $\T$, which together for each $o \in [O]$ are a solution to the equation:
\[
\T \bullet_1 \mathbf{x}_1 \cdots \bullet_{o-1}\mathbf{x}_{o-1} \bullet_{o+1} \mathbf{x}_{o+1} \cdots \bullet_O \mathbf{x}_O = \sigma \mathbf{x}_o^{O-1} \,,
\]
or, equivalently stated, are a solution to the sytem of equations:
 \[
  \begin{array}{rcl}
   \T \bullet_2 \mathbf{x}_2 \cdots \bullet_O \mathbf{x}_O & = & \sigma \mathbf{x}_1^{O-1} \,,\\
\T \bullet_1 \mathbf{x}_1 \bullet_3 \mathbf{x}_3 \cdots \bullet_O \mathbf{x}_O & = & \sigma \mathbf{x}_2^{O-1} \,,\\
& \vdots & \\
\T \bullet_1 \mathbf{x}_1 \cdots \bullet_{O-1} \mathbf{x}_{O-1} & = & \sigma \mathbf{x}_O^{O-1} \,.
  \end{array}
\]
Just like before, the optimization problem used in \cite{LekHengLim} to derive these equations also requires that $\sigma \ge 0$ and $||\mathbf{x}_o||_O = 1$ for every $o \in [O]$, i.e. every $\mathbf{x}_o$ has unit $\ell^O$-norm. For any value of $O$, given a solution ${(\sigma, \mathbf{x}_1, \dots, \mathbf{x}_O)}$ to the above equations, for every non-zero $t \in \R$, ${(\sigma, t\mathbf{x}_1, \dots, t\mathbf{x}_O)}$ is also a solution to the equations. When $O$ is even, given any solution $(\sigma, \mathbf{x}_1, \dots, \mathbf{x}_O)$ to the above equations, the result of flipping the sign of an odd number of the $\mathbf{x}_o$'s and replacing $\sigma$ with $-\sigma$ is still a solution, as is flipping the sign of an even number of the $\mathbf{x}_o$'s and leaving $\sigma$ unchanged. When $O$ is odd, there is no sign indeterminacy whatsoever. Leaving aside the sign indeterminacy issues which can occur when $O$ is even, the invariance of $\ell^O$-singular values under rescalings of the $\ell^O$-singular vectors mirrors the situation with ($\ell^O$/Z)-eigenvalues while being unlike the behavior of either the $\ell^2$-singular values or the ($\ell^2$/Z)-eigenvalues (for $O \ge 3$). \\

Relationships between eigenvalues, eigenvectors and singular values, singular vectors for higher-order tensors are potentially much more complex for higher-order tensors than for matrices. For some musings on the subject, see appendix \ref{sec:relat-betw-eigen}. However, for any tensor there is a close relationship between its best rank-one approximation in Frobenius norm and its largest $\ell^2$-singular value and corresponding $\ell^2$-singular vector tuple. (The largest $\ell^p$-singular value is also the norm of the tensor induced by the $p$-norms of the $\R^{M_o}$, just like for matrices, a fact utilized in \cite{Wang2017}.) Similarly, for any symmetric tensor there is a close relationship between its best rank-one symmetric approximation in Frobenius norm and its largest in magnitude $\ell^2$-eigenvalue and corresponding eigenvector (generalizing the variational characterization of eigenvalues via the Rayleigh quotient for matrices). In fact, due to a theorem of Banach, the best rank-one symmetric approximation always equals the best rank-one approximation, and so just as for symmetric matrices for higher-order symmetric tensors there is always a close relationship between the largest in magnitude $\ell^2$-eigenvalue and corresponding eigenvector and the largest $\ell^2$-singular value and corresponding $\ell^2$-singular vector tuple. Details for all of these claims are discussed in \cite{hillar_lim} and to some extent also \cite{LekHengLim}. It also seems a direct consequence of the definitions that (for any $\ell^p$ norm) a vector is a mode-$o$ eigenvector simultaneously for all $O$ modes and always corresponding to the same eigenvalue if and only if that eigenvalue is (up to signs) also a singular value of the tensor corresponding to a singular vector tuple consisting of $O$ copies (up to signs) of the vector.\\

The paper \cite{hillar_lim} demonstrated that for real tensors of order 3, the problems of finding an eigenvalue, approximating an eigenvector, finding a singular value, and approximating a singular vector are all NP-hard, and that all of this remains true even when restricting to the case of symmetric tensors. This is true for both the $\ell^2$/$Z$ and the $\ell^O$/$H$ versions, and it also implies the analogous statements for tensors of higher order. These facts are also shown to be connected with the result that finding the best (in Frobenius norm) rank-one approximation of any tensor is NP-hard. As was mentioned in section \ref{sec:odeco-tensors}, these problems are tractable at least for odeco tensors\cite{Zhang2001}\cite{1503.01375}\cite{learning_latent_variable_models}, due to the close relationships between the singular vectors of an odeco tensor and the vectors in its orthogonal CP decomposition and between the eigenvectors of a symmetrically odeco tensor and the vectors in its orthogonal Waring decomposition\cite{1103.0203}\cite{robeva_thesis}\cite{Robeva2016}. Besides those facts mentioned already above, generalizations of eigenpairs and singular values and vector tuples to higher-order tensors have many surprising properties compared to the matrix case. Several such useful facts are developed in the paper \cite{Cartwright2013} for $E$-eigenvalues and $E$-eigenvectors (of complex-valued tensors) using algebro-geometric methods. Algorithms for computing (approximate) eigenvalues for special classes of tensors, for example non-negative tensors\cite{Ng2010} (see section \ref{sec:non-negative-tensors}) and symmetric tensors\cite{Kolda2011} (see section \ref{sec:symm-antisymm-tens}).  The recent monograph\cite{liqun2017tensor} provides a very comprehensive introduction to the recent developments, e.g. even including a chapter about applications to spectral hypergraph theory. For more about sepctral hypergraph theory and other applications, see section \ref{sec:appl-netw-with}.

%%% Local Variables:
%%% mode: latex
%%% TeX-master: "new_notes_draft"
%%% End:

% \input{symmetric_tensors.tex}

\section{Conclusion}
\label{sec:conclusion}

To give some perspective about the numerous other tensor-related topics which are relevant in the context of data science and statistics, but which were not described or mentioned at all above, below some of those other important topics are briefly described and references are given.

\subsection{Subtensors: Fibers, Slices, and Hyperslices}
\label{sec:subt-rearr-tens}

Fibers, slices\cite{kolda_bader}, hypersplices, and general subtensors\footnote{Given two multi-index sets $M_1$ and $M_2$, a \textbf{subtensor} of $\T: M_2 \to \R$ is any tensor of the form $\T \circ \phi: M_1 \to \R$ for any injective function $M_1 \xrightarrow{\phi} M_2$. General subtensors are not necessarily as useful as special cases.}, can be given a more precise definition and description. Given a proper non-empty subset $S$ of the mode index set $[O]$, there is always a family of hyperslices corresponding to the modes in $S$ as well as another family of hyperslices to the modes in $[O] \setminus S$. The simplest example is the case of a matrix, where these are the columns of the matrix and the rows of the matrix. For an order $3$ tensor, using the terminology from section 2 of \cite{kolda_bader}, the dual family of the columns are the horizontal slices (and vice versa), the dual family of the rows are the lateral slices (and vice versa), and the dual family of the tubes are the frontal slices (and vice versa). These pairs of families of subtensors are ``dual'' in the sense that it is possible to generalize to arbitrary higher-order tensors statements such as ``the row rank always equals the column rank'', and ``the matrix whose entries are the contractions of the rows of a first matrix with the columns of a second matrix equals the matrix which is the sum of the outer products of the columns of the first matrix with the rows of the second matrix''.\\

Besides (hyper)slices, the most important types of subtensors are likely the blocks\footnote{Related, although not to be confused with, the blocks of a set partition, cf. appendix \ref{sec:part-finite-sets}.} of a block tensor, which generalize the notion of block of a block matrix. Blocks and block tensors can be understood using direct sums of vector spaces, and thus as hinted already in section \ref{sec:tens-shape-funct}, are related via the tensor shape functor to the disjoint unions (categorical sums/coproducts) of index sets. Developing these ideas in detail is left to future work. However, notions of block tensor have already been defined less precisely in previous work and used to study, among other subjects, decompositions of block tensors\cite{DeLathauwer2008} and clusters defined via tensor blocks\cite{Wang_2019}.

%%% Local Variables:
%%% mode: latex
%%% TeX-master: "../new_notes_draft"
%%% End:

\subsection{Tensor Unfoldings and Rearranging Tensor Entries}
\label{sec:tens-unfold-rearr}

Also overlooked was a precise description of tensor \textit{unfoldings}\cite{Wang2017}, the most important special cases of which are vectorizations and matricizations, see e.g. \cite{kolda_bader}. These are sometimes also referred to as tensor \textit{flattenings}\cite{lim_hla}. For the sake of simplicity, what is usually done is to restrict attention to those unfoldings of an order $O$ tensor into an order $D$ tensor which can be written as a morphism in $\fintensor{D}$ consisting of the $D$-fold tensor product of the same type of vectorization (usually lexicographical or colexicographical, see appendix \ref{sec:lexic-colex-order}). These unfoldings can also be described as the result of the tensor shape functor (see section \ref{sec:tens-shape-funct}) applied to bijections of multi-index sets which have $D$-fold Cartesian product structure, i.e. which are morphisms in $\multiindex_D$. One benefit to this approach is that certain subtensors of the original tensor can more easily be identified with subtensors of the unfolded tensor, a fact utilized in \cite{Wang2017}.\\

Matricizations are an essential sub-routine of the HOSVD heuristic algorithm\cite{HOSVD} for finding an approximate Tucker decomposition\cite{Hackbusch2016}. The HOSVD algorithm is used so often that some sources use the term ``Tucker decomposition'' to refer specifically to solutions of the HOSVD optimization problem, rather than in the far broader sense used in this review. (That broader sense is referred to elsewhere, e.g. \cite{hackbusch}, as ``Tucker representation'' or ``Tucker format'' instead).\\

Tensor unfoldings as defined above are in turn special cases of tensor \textit{reshufflings}\cite{reshuffling}, which intuitively are those vector space isomorphisms which only rearrange the entries of a tensor without altering its values, and which can more precisely be described as the set of all vector space isomorphisms between coordinate spaces arising from the application of the tensor shape functor to a bijection of multi-index sets. Restricting focus to only reshufflings greatly reduces the size of the class of potential vector space isomorphisms under consideration, which is helpful for clarity of theoretical analysis. That these isomorphisms also correspond to essentially ``re-indexing'' the entries of a tensor also has the practical benefit that their computer implementations correspond to simply converting between data structures, without needing to calculate new values.

%%% Local Variables:
%%% mode: latex
%%% TeX-master: "../new_notes_draft"
%%% End:

\subsection{Multilinear and Multiplex Rank}
\label{sec:mult-mult-rank}

Unfoldings, and in particular matricizations, are a useful tool for studying multilinear rank and multiplex rank\cite{HOSVD}\cite{lim_hla}. These concepts are generalizations of the notions of ``column rank'' and ``row rank'' to tensors of arbitrary order. (See \cite{Hitchcock_1927}\cite{Hitchcock_1928} for one definition of multiplex rank, and \cite{Diaz2016} for a more modern definition.) Unlike in the matrix case, these notions of rank usually do not coincide with the tensor rank. (See \cite{desilva_lim} or section 15-7 of \cite{lim_hla} for an overview of some of the properties of multilinear rank.) Even when the multilinear ranks are the same, the tensor rank can still be strictly greater. The tensor rank can also be strictly greater than all of the multiplex ranks even when they are all the same value as well. Consider any of the following eight order $3$ tensors:
\begin{longtable}{CCCCC}
    \unitvector[2]{1} \otimes \unitvector[2]{1} \otimes \unitvector[2]{1} & + & \unitvector[2]{1} \otimes \unitvector{2} \otimes \unitvector{2} & + & \unitvector{2} \otimes \unitvector[2]{1} \otimes \unitvector{2} \,, \\
\unitvector[2]{1} \otimes \unitvector[2]{1} \otimes \unitvector[2]{1} & + & \unitvector[2]{1} \otimes \unitvector{2} \otimes \unitvector{2} & + & \unitvector{2} \otimes \unitvector{2} \otimes \unitvector[2]{1} \,,\\
\unitvector[2]{1} \otimes \unitvector[2]{1} \otimes \unitvector[2]{1} & + & \unitvector{2} \otimes \unitvector[2]{1} \otimes \unitvector{2} & + & \unitvector{2} \otimes \unitvector{2} \otimes \unitvector[2]{1} \,, \\
\unitvector[2]{1} \otimes \unitvector[2]{1} \otimes \unitvector{2} & + & \unitvector[2]{1} \otimes \unitvector{2} \otimes \unitvector[2]{1} & + & \unitvector{2} \otimes \unitvector[2]{1} \otimes \unitvector[2]{1} \,, \\
\unitvector[2]{1} \otimes \unitvector[2]{1} \otimes \unitvector{2} & + & \unitvector[2]{1} \otimes \unitvector{2} \otimes \unitvector[2]{1} & + & \unitvector{2} \otimes \unitvector{2} \otimes \unitvector{2} \,, \\
\unitvector[2]{1} \otimes \unitvector[2]{1} \otimes \unitvector{2} & + & \unitvector{2} \otimes \unitvector[2]{1} \otimes \unitvector[2]{1} & + & \unitvector{2} \otimes \unitvector{2} \otimes \unitvector{2} \,, \\
\unitvector[2]{1} \otimes \unitvector{2} \otimes \unitvector[2]{1} & + & \unitvector{2} \otimes \unitvector[2]{1} \otimes \unitvector[2]{1} & + & \unitvector{2} \otimes \unitvector{2} \otimes \unitvector{2} \,,\\
\unitvector[2]{1} \otimes \unitvector{2} \otimes \unitvector{2} & + & \unitvector{2} \otimes \unitvector[2]{1} \otimes \unitvector{2} & + & \unitvector{2} \otimes \unitvector{2} \otimes \unitvector[2]{1} \,.
\end{longtable}
All of these tensors have tensor rank $3$, however all of their multilinear ranks are $(2,2,2)$. Since the multiplex ranks for slices for an order $3$ tensor are the same as the multilinear ranks, their slice-multiplex ranks are also all $(2,2,2)$. (This is an instance of the ``duality'' for families of subtensors mentioned above in section \ref{sec:subt-rearr-tens}.) Thus the tensor ranks of these tensors are all strictly greater than any of their multiplex ranks or multilinear ranks, even though those are all equal.\\

However, for any tensor both the multilinear ranks and the multiplex ranks are always all lower bounds for the tensor rank\cite{desilva_lim}\cite{Diaz2016} (as above sometimes strict lower bounds) and relatively easy to compute lower bounds at that. In some cases they can be used to calculate upper bounds for the tensor rank as well\cite{Diaz2016}. As for tensor rank, methods from algebraic geometry can also be fruitful for studying questions about the multilinear and multiplex ranks of tensors, see e.g. \cite{Carlini2011} or \cite{Diaz2016}. These concepts are often used for the analysis of the Tucker decomposition (see e.g. the discussion in section \ref{tucker-tensor-predictor} or section 4.1 of \cite{kolda_bader}). Multilinear rank is also often used as an alternative to the tensor rank for constraints defining the optimization problems corresponding to approximate tensor decompositions, see e.g. \cite{NIPS2016_6302}, \cite{Eld_n_2009}, \cite{Ishteva_2008}, \cite{Borckmans_2010}, or \cite{Gandy_2011}. The most well-known such algorithm seems to be the HOOI (higher-order orthogonal iterations) algorithm\cite{HOOI}. However, it has been shown in some cases (see e.g. \cite{Yuan_2015}) that optimizing in terms of the multilinear rank can lead to poorer inference than if the tensor rank had been used instead.

%%% Local Variables:
%%% mode: latex
%%% TeX-master: "../new_notes_draft"
%%% End:

\subsection{Algorithms for (Approximate) Tensor Decompositions}
\label{sec:algorithms}

There are several common classes\footnote{To clarify, none of the classes of algorithms listed below are necessarily meant to be mutually exclusive.} of algorithms for (approximately) solving the optimization problems corresponding to finding (approximate) tensor decompositions. Having a higher-level, even if extremely crude, understanding of these common types of algorithms can be quite helpful for navigating the literature. See e.g. chapter 3 of \cite{Janzamin2019} or section 5 of \cite{1711.10781} for an overview.

\subsubsection{Alternating Least Squares}
\label{sec:altern-least-squar}

Perhaps the simplest, or at least the most well-known, is the family of alternating least squares (ALS) algorithms. As the name suggests, they are a special case of the wider class of alternating minimization algorithms (which are widely used in statistics\cite{de_Leeuw_1994}) where each step solves a least squares problem. For example, these types of alternating minimization algorithms are also often commonly used for constrained matrix factorizations like NMF and semi-NMF\cite{jordan_nmf}\cite{Hamon}, not just for their generalizations to constrained tensor decompositions such as semi-NTF\cite{Wang_2019}. A heuristic explanation for the class of alternating least squares algorithms can be found in the survey papers\cite{sidiropoulos_review}\cite{kolda_bader}, see also section 3.7 of \cite{Janzamin2019}. Each ``block'' of the alternating steps is usually reduced to solving a least squares problem for matrices by matricizing the tensors in question. This has allowed randomized methods for the solution of the least squares problem for matrices to be extended to create randomized ALS algorithms for CP decompositions\cite{Battaglino_2018}\cite{1703.09074} (see section \ref{sec:rand-tens-decomp}). Although the approach is entirely heuristic, it does seem to work fairly successfully in practice\cite{Comon_2009}\cite{NIPS2015_5920}. Attempts to provide theoretical explanations for the observed effectiveness of ALS algorithms in practice include the papers \cite{10.5555/3294996.3295123}, \cite{mohlenkamp}, \cite{Espig}, and \cite{Oseledets_2018}. (\cite{Rohwedder2013} focuses on explaining ALS for the tensor train format.) The article \cite{10.1145/1102351.1102451} uses an ALS approach for non-negative tensor factorization and, highlighting the relationship between non-negative tensors and undirected graphical models (see section \ref{sec:non-negative-tensors}), compares the approach to the EM algorithms often used for the analogous problems in statistics relating to graphical models. The book \cite{amari_nonnegative} also discusses several ALS algorithms relevant for non-negative matrix factorizations (NMF) and non-negative tensor factorizations (NTF) (discussing in part the hierarchical ALS algorithms from \cite{10.5555/1776684.1776708}), while the paper\cite{Kim_2013} attempts to provide a unified theoretical framework for describing the application of such algorithms to NMF and NTF problems. Since ALS algorithms are heuristic, their effectiveness often depends on their initializations\cite{HOOI}\cite{Comon_2009}. One work\cite{Wang_2019} attempting to address this sensitivity of ALS algorithms to initial conditions initialized their ALS algorithm for semi-non-negative tensor factorization using the HOSVD\cite{HOSVD}, which they then used to study genetic data.

\subsubsection{Tensor Power Iteration}
\label{sec:tens-power-iter}

Another important class of algorithms for tensor problems are the so-called tensor power methods, an example of which is the HOOI algorithm\cite{HOOI} mentioned in \ref{sec:mult-mult-rank}. One principal reason these methods are important is because they can be used to find exact tensor decompositions for a special sub-class of tensors\cite{Zhang2001}\cite{learning_latent_variable_models}, odeco tensors, which are discussed further in section \ref{sec:odeco-tensors}. In particular, this property makes these algorithms relevant to the use of tensor decomposition problems in independent components analysis (ICA) and the analysis of latent variable models\cite{pmlr-v40-Anandkumar15}. They are also frequently used for the tensor PCA problem\cite{pmlr-v65-anandkumar17a}\cite{10.5555/2969033.2969150}\cite{1702.07449}, see section \ref{sec:tensor-clustering}, but they have also been used for different problems, e.g. discriminative learning\cite{1412.2863}. Section 3.5 of the review book \cite{Janzamin2019} focuses on the class of tensor power iteration algorithms in detail.

\subsubsection{Randomized Tensor Decompositions}
\label{sec:rand-tens-decomp}

A class of algorithms which seems to have recently grown in popularity is randomized tensor decomposition algorithms, perhaps inspired by the previous success of similar methods for more computationally tractable problems in linear algebra\cite{randnla2}\cite{randnla3}\cite{randnla4}\cite{mahoney-notes}. One reason many researchers like these methods is that they can drastically reduce the number of required computations, while often leading to little, or even negligible, decreases in accuracy. For example, work on the randomized least squares problem for matrices has been extended to create a randomized algorithm for the CP decomposition\cite{Battaglino_2018}\cite{1703.09074} (cf. section \ref{sec:altern-least-squar}). Another approach to randomized CP decompositions has made use of the fast Fourier transform\cite{anandkumar_fourier}. Randomized methods for computing generalizations of the CUR decomposition to tensors\cite{curtensor1}\cite{curtensor2} were discussed in section \ref{curt}. The randomized algorithm for computing the matrix CUR decomposition was also used successfully in \cite{perros} as a subroutine for computing a sparsified hierarchical Tucker decompositions (see section \ref{sec:tensor-networks}). Randomized methods for the Tucker decomposition\cite{Tsourakakis_2010}\cite{Che_2018}\cite{Minster_2020} and the tensor train decomposition\cite{Huber_2017}\cite{Che_2018} are also active areas of research.

%%% Local Variables:
%%% mode: latex
%%% TeX-master: "../new_notes_draft"
%%% End:

\subsection{Non-Negative Tensors}
\label{sec:non-negative-tensors}

Non-negative tensors are tensors all of whose entries are non-negative. Because the space of non-negative tensors of a given rank is closed, optimization problems restricted to the space of non-negative tensors have many beneficial properties compared to the correspond optimization problems for arbitrary tensors, see \cite{lim_nonnegative} or \cite{nonneg_uniqueness} for details. Moreover, non-negative tensors have a deep connection with undirected probabilistic graphical models, also known as Markov random fields, see section 4 of \cite{lim_nonnegative} for the most basic case, or \cite{hypergraph-duality} for a complete explanation. In the simplest case of order two tensors, this connection reduces to the independently described\cite{10.1145/1076034.1076148}\cite{10.5555/1597538.1597594} equivalence of non-negative matrix factorization (NMF) and probabilistic latent semantic analysis.\\

For an example of the ``naive Bayes interpretation'' mentioned in \cite{lim_nonnegative}, say that discrete/categorical random variables $X,Y$ are conditionally independent given discrete/categorical random variable $Z$, written as $X \independent Y | Z$, i.e. for each possible state $z$ of $Z$, and for all possible states $x, y$ of $X$ and $Y$ respectively, one has that $\mathbb{P}(X=x, Y=y|Z=z) = \mathbb{P}(X=x|Z=z) \mathbb{P}(Y=y|Z=z)$. So if\footnote{$\boldsymbol{\pi}$ for Greek ``p'', ``p'' for ``\textbf{p}robability''.} $\boldsymbol{\pi}_X^{\smallsuper{Z=z}}$ denotes the probability vector\footnote{Since $X$, $Y$, $Z$ are discrete their distributions can be encoded in vectors whose lengths are the numbers of their possible states, and whose entries are the probabilities that the random variables assume the corresponding states.} for $X$ given $Z=z$, and correspondingly if $\boldsymbol{\pi}_Y^{\smallsuper{Z=z}}$ denotes the probability vector for $Y$ given $Z=z$, then the conditional independence statement says that, for every state $z$ of $Z$, the joint \textit{conditional} distribution of $X$ and $Y$ given $Z=z$ is:
\[
 \mathbf{P}_{X.Y}^{\smallsuper{Z=z}} = \boldsymbol{\pi}_X^{\smallsuper{Z=z}} \otimes \boldsymbol{\pi}_Y^{\smallsuper{Z=z}}\,.
\]
Moreover, for the \textit{unconditional} joint probability distribution $\mathbf{P}_{X,Y}$ of $X$ and $Y$, one has by the law of total probability the following expression for each $x$ and $y$:
\[
\begin{array}{rcl}
  \mathbb{P}(X=x, Y=y) &=& \sum_z \mathbb{P}(X=x, Y=y|Z=z)\mathbb{P}(Z=z) \\
&=& \sum_z \mathbb{P}(X=x|Z=z)\mathbb{P}(Y=y|Z=z)\mathbb{P}(Z=z) \,.
\end{array}
\]
The law of total probability thus implies that the joint distribution of $X$ and $Y$ can be written as:
\[
   \mathbf{P}_{X,Y} = \summing{z} \mathbb{P}(Z\!=\!z) \boldsymbol{\pi}_X^{\smallsuper{Z=z}} \otimes \boldsymbol{\pi}_Y^{\smallsuper{Z=z}} \,.
\]
In other words, stating $X \independent Y |Z$ is equivalent to specifying a tensor product decomposition for the second order tensor describing the joint distribution of $X$ and $Y$, or equivalently (see sections \ref{sec:gener-cp-decomp} and \ref{sec:spec-cases:-matrix}) by describing the joint distribution of $X$ and $Y$ via the  Tucker decomposition:
\[
 \mathbf{P}_{X,Y} = \mathbf{P}_{X|Z} \operatorname{Diag}(\boldsymbol{\pi}_Z) \mathbf{P}_{Y|Z}^\top \,.
\]
The statement $X \independent Y |Z$ also means that the following tensor product decomposition for the third order tensor describing the unconditional joint probability distribution of $X, Y,$ and $Z$ is valid:
\[
 \mathbfcal{P}_{X,Y,Z} =   \summing{z} \mathbb{P}(Z\!=\!z) \boldsymbol{\pi}_X^{\smallsuper{Z=z}} \otimes \boldsymbol{\pi}_Y^{\smallsuper{Z=z}} \otimes \unitvector{z} \,.
\]
Moreover if $X$ and $Y$ are just unconditionally independent, then their joint distribution is clearly encoded by the tensor product of their probability distribution vectors, i.e. $\mathbf{P}_{X,Y} = \boldsymbol{\pi}_X \otimes \boldsymbol{\pi}_Y$.\\

The above example, where $X$ and $Y$ are conditionally independent given $Z$, corresponds to a latent variable model where $Z$ is the latent variable secretly influencing $X$ and $Y$. (It also corresponds to a simple case of mixture model.) So the inherent suitability of non-negative tensors for representing the distributions of conditionally independent random variables makes them relevant for the analysis of latent variable models. For example, the paper \cite{10.1145/1102351.1102451} applies non-negative tensor factorization (NTF) to latent class analysis, and argues that this approach is superior to the EM method more commonly used for that problem. The authors of \cite{Kim_2013} attempt to provide a unified theoretical framework for other alternating algorithms used for NTF (cf. section \ref{sec:altern-least-squar}). See also the paper\cite{10.5555/1776684.1776708} and the related book\cite{amari_nonnegative} for more details about specific algorithms used for NTF.\\ 

The authors of \cite{10.1145/1102351.1102451} also note how the formulation of the latent class analysis problem in terms of non-negative tensor factorization also exposes its algebraic aspects. Such observations, for this and related problems, are part of the motivation for the field of algebraic statistics, which applies methods from algebraic geometry to better understand statistical problems. The algebraic interpretation of EM, for example, is discussed in detail in the paper \cite{Kubjas_2015} as well as in the introductory survey book \cite{algebraic_statistics}. The book \cite{algebraic_statistics} also discusses numerous applications of such methods to problems in computational biology, such as phylogenetic analysis. Methods from algebraic statistics have proven especially helpful for the analysis of latent variable and graphical models, see for example \cite{Garcia2005}\cite{Seigal2018}, even for the analysis of such models in causal inference \cite{Weihs_2018}. For other introductions to the use of such methods in e.g. the analysis of latent variable models and phylogenetic analysis, consult either of the books \cite{sullivant2018algebraic} or \cite{zwiernik2016semialgebraic}. Compare also with section \ref{sec:odeco-tensors}.\\

Methods from algebraic geometry and semi-algebraic geometry\footnote{Roughly speaking, semi-algebraic geometry is the study of polynomial inequalities, as opposed to algebraic geometry, which can be considered the study of polynomial equalities or equations.} are also useful for the study of non-negative rank. A CP decomposition can be called non-negative if all of the vectors defining the elementary tensors in the CP decomposition are non-negative, i.e. all of their entries are non-negative. Then the non-negative rank of a non-negative tensor is the smallest possible number of elementary tensor summands among all non-negative CP decompositions of the tensor. Even for matrices, calculating the non-negative rank is NP-hard in the best case\cite{wang2013nonnegative}\cite{gillis_nmf}, and that remains true for higher-order tensors. Various properties of the non-negative rank of tensors have been identified in \cite{semialg_nonneg} using methods from semi-algebraic geometry. The analysis of non-negative tensors of non-negative rank $2$ was used in \cite{Allman2015} to study binary latent class models, a special case of the aforementioned latent variable models, using semialgebraic statistics.

%%% Local Variables:
%%% mode: latex
%%% TeX-master: "../new_notes_draft"
%%% End:

\subsection{Symmetric Tensors and Antisymmetric Tensors}
\label{sec:symm-antisymm-tens}

Roughly speaking, symmetric tensors are those tensors which are invariant under permutations of their modes. (Also loosely speaking, this invariance under permutations of modes corresponds to the commutativity of polynomial multiplication.) Analogously\footnote{Interchanging any two modes, a symmetric tensor does not change, while an antisymmetric tensor changes sign. The sign of a permutation describes how it can be written in terms of such interchanges, usually called transpositions.}, antisymmetric tensors are those tensors which are invariant under even permutations of their modes and change signs under odd permutations of their modes. There are deep connections between antisymmetric tensors and determinants, basic concepts of multivariable calculus (e.g. cross products and differential forms), mathematical physics, and differential geometry, all of which are of course outside the scope of this review. See e.g. \cite{winitzki2010linear}, \cite{Hestenes1984}, or sections 14.5-14.7 of \cite{silva_hla} for details about these connections. Heuristically anti-symmetric tensors correspond to ``anti-commutative polynomials'', i.e. ``polynomials'' which obey all of the rules of regular polynomials except that $yx \not= xy$, but rather $yx = -xy$. In particular $x^2=0$ for any variable $x$ in the so-called exterior, or Grassmann, algebra. However it isn't the case for every anti-commutative polynomial $f$ that necessarily $f^2 = 0$. That is necessarily true only for anti-commutative polynomials ``of degree one''. A necessary consequence of the definitions\footnote{Note also that some references refer to symmetric tensors as ``supersymmetric tensors'' or antisymmetric tensors as ``totally antisymmetirc tensors'', in order to facilitate discussion of weaker forms of (anti)symmetry.} is that all symmetric tensors and all antisymmetric tensors are cubical tensors.\\

The Waring decomposition of a symmetric tensor is a CP decomposition such that all of the factor matrices are identical\cite{landsberg}, or equivalently that all of the rank-one/elementary tensor summands are of the form $\mathbf{v}^{\otimes O}$ for some vector $\mathbf{v}$. There is a natural correspondence between symmetric tensors and homogeneous polynomials (see e.g. section 2.3 of \cite{comon_2002} or sections 14.5-14.7 of \cite{silva_hla}), and the Waring decomposition of symmetric tensors corresponds to a decomposition of homogeneous polynomials with the same name. As a result of this one-to-one correspondence, the study of symmetric tensors is very amenable to methods from algebraic geometry, see e.g. \cite{Comon_symmetric}.  Just like finding the CP decomposition of a general tensor, finding the Waring decomposition of a general symmetric tensor turns out to be an NP-hard problem\cite{hillar_lim}. The best\footnote{Here ``best'' is always with respect to the metric induced by the Frobenius norm.} approximation of a symmetric tensor by a rank-one tensor (not necessarily symmetric) is the same as its best approximation by a symmetric rank-one tensor, i.e. the best rank-one approximate CP decomposition is always the same as the best rank-one approximate Waring decomposition\cite{hillar_lim}. The symmetric rank (also known as Waring rank) of a symmetric tensor is the smallest possible number of summands in an exact Waring decomposition (i.e. the same as the definition of tensor rank when ``CP'' is replaced with ``Waring'')\cite{seigal_thesis}. Comon's Conjecture states that the symmetric rank of any symmetric tensor is equal to its tensor rank. A counterexample for the conjecture was found for complex-valued tensors in \cite{Shitov2018}, but for real tensors whose tensor rank is less than their order it was found in \cite{Zhang2016} that the symmetric rank always equals the tensor rank because in that case every CP decomposition can be shown to equal a Waring decomposition. As of the beginning of 2020, Comon's Conjecture for real-valued tensors may still be unresolved\cite{seigal_thesis}. See also section 5.3 of \cite{seigal_thesis} as well as \cite{Bernardi2011}.

%%% Local Variables:
%%% mode: latex
%%% TeX-master: "../new_notes_draft"
%%% End:

\subsection{Odeco Tensors}
\label{sec:odeco-tensors}

So-called odeco tensors, or \textbf{o}rthogonally \textbf{deco}mposable tensors, are those tensors for which an \mbox{\textit{exact}} Tucker decomposition with a hyper-diagonal core tensor and orthogonal factor matrices exists. In other words, odeco tensors consist of all tensors which admit a Tucker decomposition analogous to the singular value decomposition (SVD), i.e. they are precisely the ``SVD-izable'' tensors. Note that some sources call these tensors ``orthogonal tensors'' and the corresponding decomposition an ``orthogonal tensor decomposition'', e.g. \cite{Janzamin2019}, but at the same time other sources reserve the term ``orthogonal tensor decomposition'' for a strictly weaker term and refer to the aforementioned decomposition as a ``completely orthogonal tensor decomposition''\cite{Kolda2001}. So in the language of \cite{Kolda2001}, odeco tensors are the same as the ``completely orthogonally decomposable tensors''. The existence of the singular value decomposition shows that all tensors of order two are orthogonally decomposable (in the strong sense of the term as used in \cite{Janzamin2019}), but for higher-order tensors the subset of odeco tensors is a low-dimensional subspace (precisely, a low-dimensional algebraic variety) inside of the relevant coordinate space\cite{robeva_thesis}.  A principal reason odeco tensors are important is that power iteration methods can find exact\footnote{``Exact'' in the same weak sense that any iterative method produces exact solutions.} CP decompositions for them in polynomial time\cite{Zhang2001}\cite{learning_latent_variable_models}, compare section \ref{sec:tens-power-iter}. This also enables one to find singular vectors and singular values for such tensors in polynomial time, since they can be written in terms of the vectors in the orthogonal tensor decomposition\cite{Robeva2017}. Further algorithms giving exact decompositions for odeco tensors have been developed since \cite{Zhang2001}, for example the online algorithm from \cite{pmlr-v40-Ge15}.\\

Symmetrically odeco tensors can be defined as tensors which are both symmetric and odeco. What turns out to be an equivalent definition\cite{robeva_thesis} is the class of tensors for which an \textit{exact} Tucker decomposition exists with a hyper-diagonal core tensor and such that \textit{all} of the factor matrices are the same orthogonal matrix. In other words, they consist of those tensors which admit a Tucker decomposition analogous to the eigenvalue decomposition (EVD) for symmetric matrices, i.e. they are precisely the ``EVD-izable'' tensors. Obviously then such tensors are necessarily cubical. The existence of the eigenvalue decomposition for any symmetric matrix shows that all symmetric tensors of order two are symmetrically odeco, but again for higher-order symmetric tensors the set of symmetrically odeco tensors is a low-dimensional algebraic variety\cite{Boralevi2017}. Similar to the case for general odeco tensors, exact Waring decompositions for symmetrically odeco tensors can also be found in polynomial time\cite{1503.01375}\cite{learning_latent_variable_models}. Moreover, since all eigenvectors of a symmetrically odeco tensor can be written in terms of the vectors in its orthogonal Waring decomposition, which are themselves eigenvectors, finding the eigenvectors and eigenvalues of symmetrically odeco tensors also turns out to be a substantially easier problem than it is for general symmetric tensors\cite{Robeva2016}.\\

The higher order moment and cumulant tensors of important families of statistical models turn out to be odeco tensors, which has been another way in which tensor-based methods and algebraic statistics have proven useful for the analysis of latent variable models (cf. the discussion in section \ref{sec:non-negative-tensors}). The survey \cite{learning_latent_variable_models} gives a good overview of the applications of tensor decomposition methods to these problems. (The seminar notes \cite{1711.10781} give a concise summary of \cite{learning_latent_variable_models}.) The fourth-order cumulants approach to the independent components analysis (ICA) problem (cf. e.g. \cite{DeLathauwer2000}\cite{amari_nonnegative}\cite{Cardoso}\cite{Comon1990} for an overview) also leads to tensors which are naturally odeco\cite{Zhang2001}\cite{pmlr-v40-Anandkumar15}. It is worth noting explicitly that cumulant tensors are always symmetric\cite{mccullagh}. (So are the tensors consisting of (central) moments, compare section \ref{sec:moments-cumulants}.) Thus methods specifically designed for symmetrically odeco tensors, e.g. \cite{1503.01375}, can be applied to both of these problems. Section 3.4 of \cite{Janzamin2019} gives an overview of previous work which explains how the whitening technique can be used to extend methods for symmetrically odeco tensors to a wider class of problems than would be expected, as does section 4.1.3 of \cite{1711.10781}. Applications of the aforementioned whitening technique include learning mixtures of high-dimensional Gaussians\cite{10.1145/2746539.2746616} and robust tensor decomposition\cite{pmlr-v51-anandkumar16}.

%%% Local Variables:
%%% mode: latex
%%% TeX-master: "../new_notes_draft"
%%% End:

\subsection{Tensor Networks}
\label{sec:tensor-networks}

Tensor (hyper)networks are (hyper)graphs (see section \ref{sec:appl-netw-with}) where the nodes correspond to tensors and (hyper)edges correspond tensor contractions\cite{hypergraph-duality}. Tensor networks are a natural way to describe arbitrarily complicated tensor decompositions generalizing the Tucker decomposition\footnote{While the Tucker decomposition can be described using a tensor network (using a graph), without using ``tricks'' the CP decomposition requires a tensor hypernetwork (using a hypergraph) to be described, see \cite{hypergraph-duality}.}\cite{uschmajew_pde}. Tensor networks\footnote{Henceforth ``tensor network'' will be used to mean both the given (hyper)graph as well as the corresponding tensor decomposition, with the distinction usually being either irrelevant or determined by context.} have already been studied for a long time by physicists before coming to the attention of data scientists\cite{hypergraph-duality}\cite{tensor_network_ranks}, and other terms which have been used for tensor networks by the physics community include ``Penrose graphical notation'' or ``tensor diagram notation''. Tensor networks are a substantial topic in their own right deserving of their own review, and accordingly there are many, see e.g. \cite{cichocki_tensor_networks}\cite{Grasedyck2013}\cite{uschmajew_pde}.  The most commonly used decompositions based on tensor networks (besides the Tucker decomposition) seem to be (cf. \cite{NIPS2019_8429}\cite{Grasedyck2013}\cite{sidiropoulos_review}\cite{cichocki_tensor_networks}\cite{Huggins_2019}) the tensor train format\cite{Oseledets2011} (also known as MPS, matrix product states amongst physicists\cite{hypergraph-duality}\cite{tensor_network_ranks}) and the tree tensor format\cite{Oseledets2009}\cite{Oseledets2009b}\cite{Rohwedder2013}\cite{Grasedyck2010} (also known as the hierarchical Tucker decomposition\cite{hierarchical_tucker}\cite{Grasedyck2011}). Even more general classes of tensor decompositions can be defined (see e.g. equations 2.5 and 2.7 of \cite{uschmajew_pde}), but such a level of generality seems to be rarely, if ever, used in practice. Analogous to the various notions of rank corresponding to Tucker decompositions, one can also define notions of rank corresponding to other tensor networks\cite{tensor_network_ranks}. The notion of tensor network rank from\cite{tensor_network_ranks} helps to provide a theoretical explanation for the phenomenon frequently observed in practice that a tensor which can not be compactly described using one type of tensor decomposition can nevertheless be described compactly using another type\cite{Oseledets2009}\cite{uschmajew_pde}. This phenomenon motivates much of the research into tensor train and tree tensor decompositions, since in many instances they have led to better results than what was possible using either CP or Tucker decompositions, see e.g. \cite{perros}. A generalization of the popular HOSVD\cite{HOSVD} for Tucker decompositions to hierarchical Tucker decompositions was presented in \cite{Grasedyck2010}, the hierarchical SVD (HSVD) algorithm. In \cite{hypergraph-duality}, a dual correspondence between tensor (hyper)networks of non-negative tensors and probabilistic graphical models is explained, as mentioned in section \ref{sec:non-negative-tensors}. Other papers exploring the close relationship between tensor networks and graphical models include \cite{pmlr-v32-novikov14}\cite{10.5555/2230976.2230988}\cite{UAI2019-319}\cite{8812695}\cite{NIPS2019_8429}\cite{pmlr-v28-song13}\cite{1609.09230}. Tensor networks have also proven relevant not only to quantum mechanics or computer science individually, but also to the interdisciplinary field of quantum computing, see e.g. \cite{10.5555/2230976.2230988} or \cite{Huggins_2019}.

%%% Local Variables:
%%% mode: latex
%%% TeX-master: "../new_notes_draft"
%%% End:

\subsection{Applications of Tensors to Networks with Higher-Order Interactions}
\label{sec:appl-netw-with}

As mentioned implicitly in sections \ref{sec:tensor-networks}, hypergraphs are one possible higher-order generalizations of graphs\cite{hypergraph-duality}. Whereas any given edge inside of a graph can connect at most two distinct nodes and therefore can only encode a pairwise interaction, a hyperedge in a general hypergraph can connect any number of distinct nodes and thus is capable of encoding higher-order interactions\cite{1909.06503}\cite{Klamt2009}. As for graphs, the precise definition of hypergraph depends on issues such as whether multiple hyperedges between the same set of nodes are allowed, whether ``self-loops'' are allowed, or whether hyperedges are weighted. For more details, see either of the introductions \cite{Bretto2013} or \cite{berge1976graphs}. Examples of networks where higher-order interactions are important, and thus for which hypergraphs may be more informative models than graphs, include social networks\cite{1909.06503} and biological networks\cite{Klamt2009}\cite{Mithani2009}\cite{1303205}\cite{Gaudelet2018}. Instances where tensor-based methods were applied to the analysis of such hypergraph models of biological networks include \cite{Michoel2012}, \cite{7524736}, and \cite{Shen2018}. \\

Spectral graph theory refers to methods which use the eigenvalues and eigenvectors of matrices associated to a graph (such as the adjacency matrix or the Laplacian) to draw conclusions about the graph. This often succeeds when quantities calculated from these eigenpairs can be shown to correspond to the solutions of (polynomial-time) relaxations of difficult (NP-hard) combinatorial problems associated with important properties of the graph\cite{bolla2013spectral}\cite{1608.04845}. Notions of spectral hypergraph theory, which seek to generalize results from spectral graph theory to hypergraphs, generally take one of two approaches\cite{8887197}: (1) define certain matrices to associate with a hypergraph and study their (regular) eigenpairs\cite{Rodrguez2009}\cite{YueGao2013}\cite{Jost2019}, or (2) define certain tensors to associate with a hypergraph and study their $H$- and/or $Z$-eigenpairs\cite{pmlr-v37-ghoshdastidar15}\cite{1909.06503}\cite{8887197} (see section \ref{sec:gener-eigenv-eigenv} regarding terminology). (There have also been intermediate approaches, such as associating certain tensors with a hypergraph but then studying the (regular) eigenpairs of matrices derived from those tensors\cite{5871640}, or associating both a tensor and a matrix to a hypergraph and studying both\cite{pmlr-v84-chien18a}.)\\

When defining spectral hypergraph theory in terms of tensors associated to a hypergraph, usually tensors are chosen that generalize a matrix used in spectral graph theory. The most prevalent examples are adjacency tensors meant to generalize some notion of (weighted) adjacency matrix\cite{1909.06503}\cite{pmlr-v84-chien18a}\cite{Cooper2012}\cite{banerjee_spectral}\cite{pmlr-v37-ghoshdastidar15}\cite{Pearson2013}\cite{8887197}, and/or Laplacian tensors meant to generalize some notion of Laplacian matrix \cite{banerjee_spectral}\cite{Hu2011}\cite{Li2013}\cite{8887197}. Just like in graph theory, the precise definitions often vary between authors or depend upon the assumption that some subclass of hypergraphs with additional structure is being considered. Tensor decompositions\cite{8887197} and spectral information (eigenvalues/eigenvectors) of these tensors have been shown to be useful for analysis of hypergraphs. Both H-eigenvalues\cite{banerjee_spectral}\cite{Cooper2012}\cite{Pearson2013} and Z-eigenvalues\cite{banerjee_spectral}\cite{pmlr-v37-ghoshdastidar15}\cite{7524736}\cite{Pearson2013}\cite{Hu2011}\cite{Li2013}\cite{Shen2018} seem to be employed for studying these tensors, although one possible exception following the original paper\cite{LekHengLim} is that generalizations of the Perron-Frobenius theorem to hypergraphs seem to exclusively consider a non-negative tensor's H-eigenvalues, see e.g. any of  \cite{chang2008}, \cite{Michoel2012}, or \cite{Friedland2013} for further examples. That H-eigenvalues have so much applicability to the study of tensor generalizations of the Laplacian might be somewhat surprising because only the Z-eigenvalue problem is invariant under orthogonal transformations (since only $Z$-eigenvalues are defined in terms of an $\ell^2$ variational problem), a property which spectral graph theory uses to establish connections with Riemannian geometry when studying the Laplacian matrix\cite{Hu2011}. A comprehensive overview of many of these developments can be found in chapters 3 and 4 of the book \cite{liqun2017tensor}.\\

These methods could possibly have a close relationship with topological data analysis\cite{carlsson}\cite{doi:10.1146/annurev-statistics-031017-100045}, which can be used for instance in the study of genomic and other biological data (cf. the book \cite{Rabad_n_2019}). Like the use of hypergraphs for modelling biological and other complex networks, one of the main goals of topological data analysis also is to discover higher-order interactions in data, and the analysis of simplicial complexes derived from the data is one of the principal ways this is done\cite{Giusti2016}\cite{ghristTDA}. It turns out that simplicial complexes can be formulated as a special type\footnote{One that is ``hereditary''\cite{Bretto2013} or ``downwards closed'', e.g. the existence of a hyperedge encoding a $3$-way interaction implies the existence of $3$ edges encoding all of the $3$ corresponding pairwise (sub)interactions.} of hypergraph\cite{Bretto2013}, which suggests that tensor-based approaches to the analysis of hypergraphs could possibly be modified for use in topological data analysis. For instance, the motif tensors defined in \cite{7524736} might be useful for encoding $k$-simplexes and other important sub-complexes.\\

Hypergraphs, as a model of multi-way interactions, should not be conflated with multilayer networks. Multilayer networks are normal graphs, but the nodes are partitioned into distinct layers, with the nodes in each layer understood to be qualitatively different. One then makes a distinction between intralayer links and interlayer links, with each type of link representing qualitatively different information. Paths which traverse one node of each layer can then be understood to represent multiway interactions between qualitatively different objects, with an ``adjacency tensor'' (conceptually similar to the motif tensors of \cite{7524736}) recording each such multiway interaction\cite{8068245}. Other definitions of ``adjacency tensor'' are possible for multilayer networks\cite{Wang_multilayer2017}\cite{Wu2019}, with a possibly more common one being a tensor whose slices are the adjacency matrices of the subgraphs corresponding to each layer\cite{2002.04457}\cite{Chen2019}\cite{PhysRevE.95.042317}. Regardless of definitions, the utility of tensor decompositions for the analysis of multilayer networks has also been demonstrated multiple times \cite{Wu2019}\cite{2002.04457}\cite{Chen2019}\cite{8068245}\cite{PhysRevE.95.042317}\cite{10.1093/biomet/asz068}\cite{Wang_multilayer2017}, suggesting that tensor methods are relevant for the analysis of higher-oder interactions regardless of the precise model used to describe them.

%%% Local Variables:
%%% mode: latex
%%% TeX-master: "../new_notes_draft"
%%% End:

\subsection{Applications of Tensors to Neural Networks}
\label{sec:appl-neur-netw}

The discovery that tensors can be used to both analyze and describe neural networks has recently caused an increased amount of interest into tensors, particularly among computer scientists (and has also managed to provide the inspiration for the name of one popular programming library for building neural networks). For instance, comparisons of the optimization problems underlying tensor decompositions and the backpropagation algorithm for neural networks have led to insights regarding both\cite{1506.08473}\cite{1506.07540}. Tensor fields have been used to design neural networks which correctly do not make distinctions between rotated versions of the same $3$-dimensional point cloud\cite{1802.08219}, and tensor contractions\cite{8014977} as well as tensor regression\cite{kossaifi2018tensor} have both been used to create new types of layers for neural networks. A correspondence between certain shallow network architectures and CP decompositions as well as between certain deep network architectures and hierarchical Tucker decompositions (see section \ref{sec:tensor-networks}) is established in \cite{pmlr-v49-cohen16} with the goal of explaining the expressive ability of this class of convolutional neural networks as well as (to some extent) the observed utility in practice of general deep neural networks. Follow-up work\cite{khrulkov2018expressive} showed an analogous correspondence between certain classes of recurrent neural networks and tensor train decompositions (see section \ref{sec:tensor-networks}) and thus was also able to show that deep networks with this architecture are exponentially more efficient than shallow ones. Similar to some of the benefits mentioned in sections \ref{CP-tensor-predictor}, \ref{CP-tensor-response}, and \ref{tucker-tensor-predictor}, the ability of tensor decompositions to substantially decrease the number of parameters to be estimated has been a major reason for their use with neural networks. More specifically, using tensor decompositions to ``compress'' neural network models has been shown in some cases to provide significant increases in computational efficiency while leading to small or negligible decreases in accuracy. For a small and likely non-representative example of such applications, see e.g. any of \cite{10.5555/2969239.2969289}, \cite{8489213}, or \cite{ijcai2018-88}. Tensor decompositions have also been used to provide a new method for training recurrent neural networks\cite{1603.00954} as well as for applying deep neural networks to multi-task representation learning\cite{DBLP:conf/iclr/YangH17}.

%%% Local Variables:
%%% mode: latex
%%% TeX-master: "../new_notes_draft"
%%% End:

\subsection{Tensor Regression and Applications of Tensors to Imaging}
\label{sec:applications-imaging}

As was mentioned already above in sections \ref{CP-tensor-predictor}, \ref{CP-tensor-response}, \ref{tucker-tensor-predictor}, and \ref{tucker-tensor-response}, tensors are a natural format to store neuroimaging data\cite{1910.09499}. Tensors are well-suited to describing many types of images, in particular multispectral images or those with multiple color channels, those which are time dependent, or those which are $3$-dimensional (or higher), since such data is inherently multi-modal. For examples, the authors of \cite{Zhong2015} applied a novel tensor regression method to colorimetric sensor array data describing the volatile metabolites produced by various strains of bacteria, and in \cite{6138863} the authors develop several tensor completion algorithms for application to missing value problems in visual images, videos, and even MRI data. As was described earlier in sections \ref{tucker-tensor-predictor} and \ref{tucker-tensor-response}, Tucker decomposition methods have found wide popularity in applications to the problem of tensor regression, see for example \cite{1910.09499}\cite{Qibin_Zhao_2013}\cite{pmlr-v48-yu16}\cite{NIPS2016_6302}\cite{Hoff_2015}\cite{Ding_2015}\cite{kossaifi2018tensor}, although the use of CP decompositions\cite{Weiwei_Guo_2012} (see sections \ref{CP-tensor-predictor} and \ref{CP-tensor-response}) and even esoteric approaches, e.g. using Kronecker products\cite{Li_2010}, have also been investigated. Tensor regression is often applied to problems in neuroscience, not only as was already described in sections \ref{CP-tensor-predictor}, \ref{CP-tensor-response}, \ref{tucker-tensor-predictor}, and \ref{tucker-tensor-response}, but for example tensor regression has also been used for combining both electroencephalography and fMRI data\cite{Deshpande_2017} and for calibration of Brain Computer Interface systems \cite{Eliseyev_2013}. Tensor regression has even found applications to neural networks \cite{kossaifi2018tensor}\cite{1712.09520}. Analogous to how CP decompositions have been useful for studying simpler mixture models (see section \ref{sec:non-negative-tensors}), they have also been used to study mixtures of generalized linear models (GLMs) \cite{DBLP:conf/aistats/SedghiJA16}. The assumption that the residuals have a separable Kronecker covariance (which was found in section \ref{tucker-tensor-response}) is also a common assumption, consult for example \cite{Hoff2011}\cite{Manceur_2013}\cite{NIPS2015_5920}\cite{Fosdick_2014}\cite{Hoff_2015}. For a detailed investigation of the theory surrounding regularization methods for tensor regression see \cite{Raskutti_2019}.

%%% Local Variables:
%%% mode: latex
%%% TeX-master: "../new_notes_draft"
%%% End:

\subsection{Tensor Clustering}
\label{sec:tensor-clustering}

Tensor-based methods have been the source for several new clustering techniques, with many applications to biological data. In \cite{Seigal2019}, polynomial inequalities are used to provide interpretable clusters of experimental results from assays measuring the effects of various ligands on mutated cell lines, with the hope of identifying mechanisms that coud lead to the onset of breast cancer. The authors in \cite{1702.07449} use a power iteration method (see section \ref{sec:tens-power-iter}) to develop a variation of CP decomposition intended to generalize PCA, called tensor PCA, which they then use to cluster gene expression data from the brain. A similar approach to tensor PCA, also involving a variation of CP decomposition and the use of power iteration methods, was carried out in \cite{10.5555/2969033.2969150}. The article \cite{pmlr-v65-anandkumar17a} analyzes the convergence of power iteration methods for tensor PCA and recommends improved implementations. The study in \cite{Sun2016} used a truncated power iteration method to create provably sparse clusters of tensor-valued data, and applied it successfully to both advertising and genetics data. Finally, following up on the previous method, the authors in \cite{Sun2019} developed techniques to account for fusion structures from time-varying tensor data, which they were able to apply successfully to the analysis of the results from an fMRI study of autism spectrum disorder. As explained in \cite{Wang_2019}, where a new clustering method based on semi-nonnegative tensor factorization is developed and applied to gene expression data, tensor clustering methods are often substantially more useful for datasets which naturally have multiple modes compared to alternative methods. Intuitively this is to be expected since clusters for such datasets need to incorporate information from higher-order/multi-way interactions between the modes in order to have useful interpretations.

%%% Local Variables:
%%% mode: latex
%%% TeX-master: "../new_notes_draft"
%%% End:

\subsection{Related Surveys}
\label{sec:related-surveys}

Other works have surveyed material similar to this review. The survey by \cite{kolda_bader} is probably one of the most well-known, and focuses largely on CP and Tucker decompositions from the perspective of applied mathematicians. The older review \cite{comon_2002} focuses on applications to signal processing (in particular independent components analysis) as well as aspects of tensor-based methods which are relevant to numerical analysis. The more recent survey paper \cite{Cichocki_2015} also focuses on the applications to signal processing, and \cite{de_Almeida_2015} discusses at length the relevance of these tensor-decomposition-based tools from signal processing to communication problems. Similarly \cite{Comon_2014} also focuses on signal processing applications, but examines tensors in a manner similar to a physicist, in other words as objects whose coordinates transform in a predictable way (cf. appendices \ref{sec:coord-isom-tens} and \ref{sec:tens-in-tens-out}). The review \cite{qi_survey} focuses on these topics from the perspective of numerical analysis and addresses several topics which are often not discussed in other reviews, such as tensor eigenvalues and applications to optimization. The survey \cite{Acar2009} is oriented towards the perspective of computer scientists, and while covering material very similar to \cite{kolda_bader} it also emphasizes applications in computer science where tensor methods have recently found prominence, such as computer vision, social network analysis, text mining, and real-time process monitoring. Another survey from the perspective of computer science is \cite{Lu_2011}, which focuses on the applications of tensor decompositions to dimensionality reduction problems. The overview \cite{Grasedyck2013} focuses on applications of tensor decompositions, particularly the tensor train and hierarchical Tucker decompositions (see section \ref{sec:tensor-networks}) to problems in numerical analysis and scientific computing, especially those related to functional analysis. The survey \cite{uschmajew_pde} takes an approach similar to \cite{Grasedyck2013}, but focuses even more specifically on problems related to the solution of partial differential equations. Another article focusing on applications to scientific computing is \cite{Vervliet_2014}, but is arguably more widely accessible than \cite{Grasedyck2013} or \cite{uschmajew_pde}, and it also draws some attention to statistical applications of tensor decompositions (for ``big data''). As mentioned previously in section \ref{sec:odeco-tensors}, \cite{learning_latent_variable_models} is a survey which limits its focus to the application of tensor decompositions to the analysis of latent variable models and closely related machine learning problems. It is notable, among other reasons, for drawing attention towards odeco tensors and the fact that these problems are a focus of study in algebraic statistics. The seminar notes \cite{1711.10781} provide a succinct summary of most of the important parts of both \cite{learning_latent_variable_models} and \cite{kolda_bader}. The paper \cite{sidiropoulos_review} aims to give an updated overview of the field compared to \cite{kolda_bader} (in particular introducing the now common tensor train and hierarchical Tucker decompositions), while focusing primarily on the specifics of algorithms and applications to ``data mining'' and ``big data''. The more recent survey \cite{cichocki_tensor_networks} focuses on the applications of tensor networks (in particular the tensor train format) to the analysis of extremely large data sets, very high-dimensional data, and optimization problems. The earlier preprint\cite{1403.2048} by one of the authors of \cite{cichocki_tensor_networks} covers much of the same material and shares the same emphasis on applications to big data, but is much more concise and so might be a better starting point for those unfamiliar with most of the topics.\\

There have also been several books published about these topics, which tend to have widely varying emphases, and have very inconsistent assumptions regarding the background knowledge of the reader. For example, \cite{landsberg} focuses on the connections with problems in algebraic geometry and the use of algebraic geometric tools (e.g. as evidenced by its focus on complex-valued tensors and frequent formulation of problems in terms of projective spaces). The book \cite{hackbusch} is very comprehensive, focuses on applications to numerical analysis and scientific computing, and is perhaps daunting for some beginners (due to its at times abstract and algebraic nature and its frequent formulation of problems in terms of infinite-dimensional spaces, where distinctions that are irrelevant in the finite-dimensional case become important, and the accompanying heavy utilization of functional analysis). In contrast, the book \cite{amari_nonnegative} is less mathematically daunting, but seems to be geared towards experts in signal processing and computer science and focuses exclusively on the factorization of non-negative tensors and the connections these methods have with also very popular non-negative matrix factorization techniques (cf. sections \ref{sec:spec-case:-matr} and \ref{tucker-cnmf}). The book \cite{kroonenberg2008applied} seems to be geared towards applied statisticians whose primary training is in other fields like the social or environmental sciences, and uses terminology quite different from that found in most papers on these topics published by researchers in the fields of machine learning or computer science. Similar to \cite{kroonenberg2008applied}, the textbook \cite{chemometrics} seems to orient itself towards applications relevant for readers whose primary interest is chemometrics or similar applied statistics problems. In contrast, the recent book \cite{Janzamin2019} is oriented towards researchers whose backgrounds are in computer science or (more theoretical) statistics and focuses almost exclusively on applications to machine learning (similar to the related review article \cite{learning_latent_variable_models}). Books discussing the use of tensor-based techniques in mathematical physics and differential geometry go back to at least the beginning of the 20th century\cite{Qi2005}, and the notation and concepts used, while actually closely related, are usually wildly different from those in the sources above. Appendices \ref{sec:coord-isom-tens} and \ref{sec:tens-in-tens-out} attempt to bridge part of this gap. The book \cite{mccullagh}, which focuses exclusively on the applications of tensors to theoretical statistics, uses much of this notation and perspective on tensors from the physics community. As mentioned before, the article \cite{Comon_2014} takes a similar perspective on tensors, and \cite{hypergraph-duality} attempts to highlight the close mathematical connections between recent work on graphical models and tensor networks in the field of machine learning to concepts from quantum mechanics and related areas.

%%% Local Variables:
%%% mode: latex
%%% TeX-master: "../new_notes_draft"
%%% End:

%%% Local Variables:
%%% mode: latex
%%% TeX-master: "new_notes_draft"
%%% End:

\clearpage
\break

\addcontentsline{toc}{section}{References}
\bibliographystyle{plain}
\bibliography{main.bib,regression.bib,tensor_product/algebraic_geometry.bib,tensor_product/bibliography.bib,tensor_product/physics_notation.bib,candecomp/matrix_factorizations.bib,candecomp/nmf.bib,candecomp/rank.bib,tucker/applications.bib,tucker/fourier.bib,tucker/sparse_vectors.bib,tucker/linear_equations.bib,tucker/factorizations.bib,tensor_networks.bib,appendices/random.bib,coredump.bib}

\break
\clearpage

\appendix
\appendixpage

\addtocontents{toc}{\protect\setcounter{tocdepth}{1}}
\setcounter{section}{0}
\renewcommand{\theHsection}{A.\arabic{section}}
\renewcommand{\thesection}{A.\arabic{section}}

\section{Conventions Used in this Review}
\label{sec:append-assumpt-this}

\subsection{Notation}
\label{sec:notation}

\begin{itemize}
\item Analogous to the convention used in \cite{kolda_bader}, in most cases bound
  variables will be denoted by lower-case Latin letters, and their
  corresponding ``maximum possible value'' (which is a free variable) will
  be denoted by the corresponding Latin letter. For example,
  $o \in [O]$, $m \in [M]$, $w \in [W]$, $h \in [H]$, $k \in [K]$, or $m_o \in [M_o]$.
  
\item  In cases where the free variable will later be used to define the bound variable for another expression, the free variable will usually be denoted with a lower-case Latin letter, and the bound variable will be denoted with the corresponding lower-case Greek letter.\\

  For example, in the expression:
  \[ 
 \unittensor{\m} (\mltidx{\mu}) \defequals
    \begin{cases}
      0 &\mltidx{\mu} \not= \m\\
      1 & \mltidx{\mu} = \m
    \end{cases} \,,
\]
or in the expression
\[   
\genMltidx{M}{[O]\setminus o}  \defequals \bigtimes_{\omega\not=o}^O [M_{\omega}] \,.  
 \]

\item Tensors will be denoted with boldface letters, while scalars\footnote{To those following the convention that scalars are ``$0$th order tensors'', then the above reads ``tensors of positive order will be denoted with boldface letters, while $0$th order tensors will be denoted with non-boldface letters''.} will be denoted with non-boldface letters. In particular, the elements of any tensor will be denoted with non-boldface letters. Tensors of order $1$, i.e. vectors, will be denoted with lower-case letters, while tensors of order $\ge 2$ will be denoted with upper-case letters\footnote{This is the same convention as used in e.g. \cite{Comon_symmetric}, see the second paragraph of the second section.}\endnote{This conflicts unfortunately with the common convention that random scalars and random vectors be denoted with upper-case letters. Thus in this review the reader must memorize whether the symbol denotes a random or a deterministic quantity. One could be perverse and argue that this is a non-issue, inasmuch as technically deterministic quantities are just degenerate random quantities (i.e. corresponding to Dirac measures).}. Tensors of order $\ge 3$ (so-called ``higher-order tensors'') will be denoted with calligraphic letters, and lower-order tensors will be denoted with non-calligraphic letters. If the order of a tensor is unspecified, i.e. it could be $1$, $2$, or $\ge 3$, then the notational conventions for higher-order tensors will be used by default. The elements of a tensor, being scalars, will always be denoted with non-boldface letters, but will otherwise share the notation of the tensor of which they are elements.

\item Linear functions will generally be denoted by upper-case, non-boldface, calligraphic letters. This conflicts somewhat with the notation for elements of a higher-order tensor, but while the subscripts for linear functions will generally be a single index (usually corresponding to a given mode of a tensor), the subscripts in the notation of elements of a higher-order tensor will always be multi-indices. Moreover, the calligraphic letter used for a linear function will usually either be $\mathcal{L}$ ($\mathcal{L}$ for \textit{l}inear) or $\mathcal{K}$ (since $\mathcal{K}$ is the letter in the Latin alphabet before $\mathcal{L}$).

\end{itemize}

\subsection{Simplifying Assumptions}
\label{sec:simpl-assumpt}

The following are a list of simplifying assumptions generally applied in this review, so as to prevent obfuscation by issues not deemed essential. This list aims to be but may not be comprehensive.

\begin{enumerate}
\item Only finite dimensional vector spaces are considered. In particular, this removes the need to consider the distinction between ``purely algebraic'' notions and ``topologically compatible'' notions, e.g. algebraic dual vs. topological dual, algebraic tensor product vs. topological tensor product, etc., which are considered in detail in e.g. \cite{hackbusch}. All of those distinctions are relevant only for the case of infinite dimensional vector spaces.
\item Only vector spaces over the field of real numbers are considered, since this is what is most used and familiar in practice. Thus issues brought up by the algebraic peculiarities of fields like $\mathbb{C}$ or $\mathbb{F}_2$, and the distraction they represent, can be avoided.
\item For every vector space, some basis is chosen, and the corresponding standard linear isomorphism (depending on the choice of basis) with $\mathbb{R}^d$ (where $d$ is the dimension of the vector space) is applied. Compare the reasoning behind this convention with \cite{lim_hla}: ``Tensors can be represented as hypermatrices by choosing a basis. Given a set of bases, the essential information about a tensor $T$ is captured by the coordinates $a_{j_1 \cdots j_d}$'s''. Compare this also with section 2.1 of \cite{uschmajew_pde}. Up to linear space isomorphisms, and even up to Hilbert Space isomorphisms, no loss of generality is incurred by restricting discussion to the specific case of tensors which are ``hypermatrices'' or ``multidimensional arrays''. See appendix \ref{sec:coord-vector-spaces}.
\item For any $\R^M$, the basis chosen is always the standard basis of indicator functions, $(1, 0, 0, \dots, 0)$, $(0,1,0,\dots, 0)$, $\dots$, $(0, \dots, 0, 1)$. Correspondingly the basis for any $\R^{\M}$ is always chosen to be the unit tensors, which are the tensor products of the standard basis vectors for each mode. This has the effect that the standard isomorphism generated by this basis is always the identity, rather than some other automorphism of $\R^M$ or $\R^{\M}$. (This is perhaps somewhat tautological or circular inasmuch as the standard isomorphism is defined in terms of the standard basis, but in any case it is definitely easier for humans to reason with.) There is also no loss of generality here, since to use another basis, one just needs to first apply the automorphism sending that basis to the standard basis (this is usually described as ``using the change of basis formula''), and then follow the discussion in this review as before. The specifics of this assumption are discussed in much greater detail in appendix \ref{sec:coord-vector-spaces} and appendix \ref{sec:coord-isom-tens}.
\item The chosen identification of/linear isomorphism between $\R^M$ and its dual $(\R^M)^\ast$, or between $\R^{\M}$ and its dual $(\R^{\M})^\ast$, will always be the unique isomorphism which sends the standard basis to the dual basis of the standard basis. Identifications of a vector space with its dual amount to defining a non-degenerate bilinear product on the vector space. In this case, the chosen identification amounts to the standard inner product (i.e. the operation which returns the sum all of the entries of the entrywise a.k.a. Hadamard product).

\item Only contiguous set partitions will be used, rather than arbitrary ordered set partitions. This is because all ordered set partitions (even when distinguishing between partitions based on the order of the elements within each block) can be generated from a contiguous set partition and a suitable permutation. It is more straightforward to create arbitrary ordered set partitions from the simpler building blocks of contiguous set partitions and permutations. (Compare this convention with that used in \cite{Hitchcock_1927} with regards to ``multipartite indices''.)
\end{enumerate}

%%% Local Variables:
%%% mode: latex
%%% TeX-master: "../new_notes_draft"
%%% End:

\section{Partitions of Finite Sets}
\label{sec:part-finite-sets}

A \textbf{set partition} of $[N]$ is a set of subsets of $[N]$ such that (i) their union is equal to $[N]$ and (ii) they are all disjoint, i.e. all of their intersections with each other are empty\cite{Wang2017}. Furthermore, a \textbf{contiguous set partition} of $[N]$ is a set partition such that if $n_1, n_2$ are in the same ``block'', i.e. element of the partition (which is a subset of $[N]$), then $n_1 \le n_3 \le n_2$ implies that $n_3$ is also in the same block. An \textbf{ordered set partition} is a set partition where the order of the blocks is considered important, i.e. a ``tuple'' of subsets of $[N]$ rather than just a set of subsets of $[N]$.\\

There is a bijection between surjective functions $[N] \to [B]$ and ordered set partitions of $[N]$ into $B$ blocks. There is also a bijection between \textit{monotone} and surjective functions $[N] \to [B]$ and contiguous set partitions\footnote{In turn there is a bijection between contiguous set partitions of $[N]$ into $B$ blocks and ordered (integer) partitions, also known as compositions, of $N$ of length $B$, see e.g. \cite{stanley} for definitions of these terms.} of $[N]$ into $B$ blocks.\\

Given contiguous set partitions $P_1$ of $[N]$ into $B_1$ blocks and $Q$ of $[B_1]$ into $B_2$ blocks, the \textbf{coarsening}/\textbf{fattening}/\textbf{merging} of $P_1$ along $Q$, denoted $P_1 \blacktriangleright Q$, will be defined as the unique contiguous set partition of $[N]$ into $B_2$ blocks corresponding to the surjective and monotone function:
\[
  [N] \xrightarrow{P_1 \blacktriangleright  Q} [B_2] \defequals [N] \xrightarrow{Q \circ P_1} [B_2] \,.
 \]
The same operation is also defined for ordered set partitions by dropping the assumption that $P_1$ and $Q$ correspond to monotone functions and only requiring them to be surjective functions.\\

As an example, let $N=6$, and consider the contiguous set partition:
\[ 
P_1 \defequals \{ \{1,2,3\}, \{4\}, \{5,6\} \} \,,
 \]
denoted using ``stars and bars'' notation (without the stars) as
\[ 
 P_1 = 123|4|56 \,. 
\]
This corresponds to the function $[6] \xrightarrow{P_1} [3]$ given by $P_1(1)=1$, $P_1(2)=1$, $P_1(3)=1$, $P_1(4)=2$, $P_1(5)=3$, $P_1(6)=3$. This function is both monotone and surjective. Given an element of $[6]$ the function states which of the three blocks the element has been placed into by the partition $P_1$. Note that here $B_1=3$. Consider also the contiguous set partition $Q$ of $[3]$ into $B_2=2$ blocks given by $12|3$, which corresponds to the function $[3] \xrightarrow{Q} [2]$ given by $Q(1)=1$, $Q(2)=1$, $Q(3)=2$, which again is surjective and monotone.\\  

Then in this case $P_1 \blacktriangleright Q$ is the contiguous set partition of $[6]$ into $2$ blocks given by
\[ 
 1234|56 \,, 
\]
since $Q \circ P_1 (1) =1$, $Q \circ P_1(2)=1$, $Q \circ P_1(3)=1$, $Q \circ P_1(4)=1$, $Q\circ P_1(5) =2$, $Q \circ P_1(6) =2$. Notice how $Q \circ P_1$ itself is a surjective and monotone function, since surjective functions are closed under composition and monotone functions are closed under composition, so functions which are both surjective and monotone are also closed under composition. The operation $P_1 \blacktriangleright Q$ basically says to merge the first two blocks of $P_1$ (since the first block of $Q$ is $\{1,2\}$) and to leave the $3$rd block unchanged (since $\{3\}$ is its own block in $Q$).\\

A contiguous set partition $P_2$ of $[N]$ into $B_2$ blocks is said to be \textbf{coarser} than the contiguous set parition $P_1$ of $[N]$ into $B_1$ blocks (necessarily $B_2 \le B_1$) when for any $n_1, n_2 \in[N]$ one has that $P_1(n_1)=P_1(n_2)$ implies that $P_2(n_1)=P_2(n_2)$ (even though the converse need not be true). The same definition also works for general ordered set partitions. If $P_2$ is coarser than $P_1$, then $P_1$ is said to be \textbf{finer} than $P_2$, for both contiguous set partitions and for ordered set partitions\cite{Wang2017}.\\

Given any contiguous set partition $P_2$ of $[N]$ into $B_2$ blocks which is coarser than the contiguous set partition $P_1$ of $[N]$ into $B_1$ blocks, a unique contiguous set partition $Q$ of $[B_1]$ into $B_2$ blocks such that $P_2 = P_1 \blacktriangleright Q$ exists, denoted $\frac{P_2}{P_1} \defequals Q$, and which will be called the \textbf{refinement quotient} of $P_2$ by $P_1$. The analogous notion exists and is well-defined for general ordered set partitions also. All of these notions exist and are well-defined because of the following elementary lemmas:
\begin{itemize}
\item For surjective functions $f:X \to Y$, $h:X \to Z$, if $h = g \circ f$, then $g$ is surjective, since for each $z\!\in\! Z$ there is an $x \!\in\! X$ with $g(f(x))=h(x) =z$, so $g(y)=z$ for $y \defequals f(x)$.
\item For a monotone\footnote{The assumption is that there are total orders on $X$, $Y$, and $Z$, not just partial orders. Then the contrapositive of the claimed statement $f(x_1) \prec f(x_2) \implies x_1 \preceq x_2$ can be seen to be valid for a monotonic function $f$.} function $h: X \to Z$, and a monotone and surjective function $f: X \to Y$, if $h = g \circ f$, then $g$ is also monotone. Given $y_1, y_2 \in Y$ with $y_1 \prec y_2$, there exist $x_1, x_2 \in X$ with $x_1 \preceq x_2$ with $f(x_1) = y_1$ and $f(x_2)=y_2$, since $f$ is monotone and surjective. Since $h$ is monotone, $g(y_1) =g(f(x_1)) = h(x_1) \preceq h(x_2) = g(f(x_2)) = g(y_2)$. If $y_1 = y_2$, then clearly $g(y_1) = g(y_2) \preceq g(y_2)$, so in both cases $y_1 \preceq y_2$ means that $g(y_1) \preceq g(y_2)$.
\item For a surjective function $f:X\to Y$, and any function $h: X \to Z$ such that $f(x_1) = f(x_2)$ implies $h(x_1) = h(x_2)$, there exists a function $g$ such that $h = g \circ f$. Namely, for any $y \in Y$, choose any $x \in X$ such that $f(x)=y$, and then define $g(y)=h(x)$. This is well-defined, since for any $x_1, x_2 \in X$ with $f(x_1)=y=f(x_2)$, by the assumptions on $h$ one also that $h(x_1)=h(x_2)$. Clearly one has that $g(f(x))=h(x)$ for all $x\in X$.
\item For a surjective $f:X \to Y$, and any $h:X \to Z$, if $g:Y\to Z$ is such that $h = g\circ f$, then $g$ is unique with this property, since given any $\tilde{g}$ such that $\tilde{g} \circ f =h$ and any $y \in Y$, one has for any $x \in X$ with $f(x)=y$ that $\tilde{g}(y) = \tilde{g}(f(x))=h(x)=g(f(x))=g(y)$, i.e. $\tilde{g} = g$. 
\end{itemize}

As discussed implicitly in appendix \ref{sec:simpl-assumpt}, any ordered partition of the set $[N]$ can be derived from a unique pair consisting of (i) a permutation of the elements of $[N]$ and (ii) a contiguous set partition of $[N]$ (corresponding to a composition a.k.a. ordered integer partition \cite{stanley} of the integer $N$). It is straightforward to see this in the case where both the order of the blocks of the partition of $[N]$ as well as the order of the elements inside of each block are considered important. In the case where only the order of the blocks of the partition of $[N]$ is considered important, but not the order of the elements inside of each block, one can without loss of generality assume that the elements inside of each block are in ascending order, and in that way still associate every such ordered partition with a unique pair of permutation of $[N]$ and contiguous partition of $[N]$ -- see section 2.2 of \cite{Huang_2018} for a more formal explanation of this argument. Therefore, when also allowing for permutations of the elements of $[N]$, there is no loss of generality in restricting attention only to contiguous partitions, rather than directly considering all possible ordered set partitions.  (Compare the above convention with the convention given for ``multipartite indices'' in \cite{Hitchcock_1927}.)

%%% Local Variables:
%%% mode: latex
%%% TeX-master: "../new_notes_draft"
%%% End:

\section{Orderings of Multi-Index Sets}
\label{sec:order-multi-index}

Here some basic facts allowing one to better understand vectorization of tensors are presented.

\subsection{Equivalent Characterization of Total Orders for Finite Sets}
\label{sec:equiv-char-total}

Let $X$ be a finite set, and define $N \defequals |X|$, the number of elements in $X$.\\

Then there is a one-to-one correspondence between total orders (orderings) on the set $X$ and bijections $[N] \overset{\phi}{\to} X$, allowing one to consider the two notions equivalent.\\

Given a bijection $[N] \overset{\phi}{\to} X$, one can define a total order $\preceq$ on $X$ via
\[
x_1 \preceq x_2 \iff \phi^{-1}(x_1) \le \phi^{-1}(x_2)\text{ for all }x_1, x_2 \in X \,.
\]
This is the total order generated by requiring the function $\phi$ to be monotone, i.e. order-preserving.\\

Conversely, given a total order $\preceq$ on $X$, one can define a bijection with $[N]$ via the rule:
\[
  \begin{array}{rcl}
    \phi(1) & \defequals & \text{smallest element of }X\text{ with respect to }\preceq \\
    \phi(2) & \defequals & \text{second smallest element of }X\text{ with respect to }\preceq\\
            & \vdots & \\
    \phi(N-1) & \defequals & \text{second largest element of }X\text{ with respect to }\preceq \\
       \phi(N) & \defequals & \text{largest element of }X\text{ with respect to }\preceq
  \end{array}
\]
Given this equivalence, elsewhere in the review there will in general be no explicit distinction made between a total order on a finite set and a bijection between that set and the index set corresponding to its cardinality. In particular, given a multi-index set $\M \defequals \bigtimes_{o=1}^O [M_o]$, a total order on $\M$ can equivalently be characterized via a choice of bijection $[M_1\dots M_o \dots M_O] \to \M$.\\

\subsection{Equivalent Characterizations of Permutations of Finite Sets}
\label{sec:equiv-char-perm}

There are two equivalent ways of thinking of permutations of a finite set $X$: (1) as orderings of the set $X$, i.e. bijections $[N] \xrightarrow{\phi} X$ (where $|X|=N$), or (2) as (when $X$ already has a specified ordering) re-orderings of the set $X$, i.e. (self-)bijections $X \xrightarrow{\pi} X$.\\

The elements of the set $X$ can be interpreted as ``labels'', and the elements of the set $[N]$ can be thought of as ``rankings''. So a bijection $[N] \xrightarrow{\phi} X$ can be interpreted as a way of assigning each ranking (i.e. position) a unique label, such that each label is also assigned a unique ranking.\\ 

One often sees the notation e.g. of the form $(312)$ for the permutation of the elements of a $3$-element set. This can be interpreted as specifying a bijection $[3] \xrightarrow{\phi} X$ which specifies that the ranking of one, i.e. the first position, belongs to the element of the set $X$ with label ``$3$'', the ranking two belongs to the element of the set $X$ with label ``$1$'', and the ranking three belongs to the element of the set $X$ with label ``$2$''. (In other words, it \textit{does} not say the opposite, e.g. that the element of the set $X$ with label $1$ gets sent to position $3$; in this case that element actually gets sent to position $2$, \textit{not} position $3$.) The above notation for a permutation can also be interpreted as specifying a (self-)bijection $[3] \xrightarrow{\pi} [3]$ such that $\pi(1)=3$, $\pi(2)=1$, $\pi(3)=2$. \\

Again, above $\pi(o_1)=o_2$ says that (after the permutation) the $o_1$'th position now belongs to $o_2$, and it does \textit{not} say that the position of $o_1$ after the permutation is $o_2$ (which would be the opposite statement, i.e. correspond to $\pi^{-1}$). In other words, this review uses the convention that permutations, in both of their equivalent formulations, are specified via bijections:
\[ 
\text{rankings} \quad \longrightarrow \quad \text{labels}  \,,
 \]
even though it is also possible to specify orderings using the inverse functions ${\text{labels} \to \text{rankings}}$.\\

Given a specified ordering $[N] \xrightarrow{\psi} X$ of the finite set $X$, e.g. the elements of $X$ have been named such that $x_1 \defequals \psi(1)$, \dots, $x_N \defequals \psi(N)$, one would like to consider how to re-arrange the elements of $X$ so that they have the ordering corresponding to the bijection $[N] \xrightarrow{\pi} X$. In other words, given an element $x\in X$, which has rank/position $\psi^{-1}(x)$ under the old ordering, one wants to know which element has that position under the new ordering, $\phi (\psi^{-1}(x)) = (\phi \circ \psi^{-1})(x)$. In other words, the new ordering $[N] \xrightarrow{\phi} X$ corresponds to the (self-)bijection $\phi \circ \psi^{-1}$ of $X$:
\[  
X \xrightarrow{\pi} X \defequals X \xrightarrow{\psi^{-1}} [N] \xrightarrow{\phi} X \,.
  \]
Conversely, starting with a self-bijection $X \xrightarrow{\pi} X$, and some specified (``standard'') ordering of $X$ specified via the bijection $[N] \xrightarrow{\psi} X$, one gets a new ordering $\phi$ of $X$ via the rule:
\[  
[N] \xrightarrow{\phi} X \defequals [N] \xrightarrow{\psi} X \xrightarrow{\pi} X \,. 
 \]
In other words, if the element $x$ is in the $n$'th position under the old order ($\psi(n)=x$), after the re-ordering specified by $\pi$, the element $\pi(x)$ is now in the $n$'th position, i.e. $\phi(n) = \pi(x)$.\\

Thus these two distinct ways of thinking about permutations can be seen to be equivalent, since (given a ``standard ordering'' on the set $X$) one permutation of the first type specifies a unique permutation of the second type, and vice versa. According to \cite{cameron}, the more straightforward interpretation as\footnote{Technically this is being sloppy and implicitly using the identification of orderings of $X$ with bijections $[N] \to X$ described in section \ref{sec:equiv-char-total} above. Strictly speaking the elementary interpretation corresponds to total orders of $X$.} bijections $[N] \to X$ is older (and can be referred to for the sake of unambiguity as ``passive permutations''), while the more abstract and modern interpretation as bijections $X \to X$ used to be referred to as ``substitutions'' (and can be referred to for the sake of unambiguity as ``active permutations''). Of course, in the case that the finite set $X$ is itself an index set $[N]$ there is mathematically speaking no distinction between the two. Since that is the most common scenario occurring in this review (usually the index set is the mode index set), no explicit distinction will be made between ``passive permutations'' and ``active permutations'' in this review. (The ``standard ordering'' in this case can be assumed to be the identity ${[N] \xrightarrow{\Id_{[N]}} [N]}$.)

\subsection{Lexicographical and Colexicographical Order}
\label{sec:lexic-colex-order}

Given a multi-index set $\M = \bigtimes_{o=1}^O [M_o]$, for each $o \in [O]$ the corresponding mode $[M_o]$ has a canonical order, namely the numerical order of the first $M_o$ integers $1, 2, \dots, M_o$. Given $O$ totally ordered sets, there is a standard partial order $\preceq$ defined on their Cartesian product, called the \textbf{product order}. For the multi-index set $\M$, the product order $\preceq$ amounts to the following:
\[ 
\forall \m, \n[O] \in \M\,, \quad  \m \preceq \n[O]  \iff m_1 \le n_1\,, m_2 \le n_2 \,, \dots \,, m_o \le n_o \,, \dots \,, m_O \le n_O \,. 
\]
This is not a total order unless $O=1$ because it leaves many, and usually most, pairs of multi-indices incomparable. For example, when $\M = [2] \times [2]$, it is neither the case that $(1,2) \preceq (2,1)$ nor the case that $(2,1) \preceq (1,2)$. All that can be said is that with respect to the product order $\preceq$, there is no way to compare $(1,2)$ and $(2,1)$. Thus more needs to be done to find a total order.\\

A total order $\le$ is said to be a \textbf{linear extension}\footnote{This definition generalizes to the linear extension of any partial order, and the name comes from the fact that an alternative name for total order is linear order, since the numbers on the number line are totally ordered.} of the product order $\preceq$ if $\le$ is compatible with $\preceq$ in the sense that $\m \preceq \n[O]$ \textit{always} implies that $\m \le \n[O]$. Of course the converse, $\m \le \n[O]$ implies $\m \preceq \n[O]$, is usually false, since again the partial order leaves most pairs of multi-indices incomparable. Thus the task of a linear extension is to compare those pairs left untouched by the product order $\preceq$ in a way that preserves transitivity ($a \le b$ and $b \le c$ implies $a \le c$).\\

The requirement that the total order for the multi-index set be a linear extension of the product order is a sensible one, since any order in which, for example, the following was possible:
\[  
(2081, 496, 1287) < (1,2,3) \,, 
 \]
would seem and feel bizarre to most people. However, for most multi-index sets it still leaves far too many options available. For instance, for multi-index sets of the form either $[2] \times [M]$ or $[M] \times [2]$ with $M \ge 2$, the number of linear extensions of the product order is equal to the $M$'th Catalan number, which grows super-exponentially with $M$. The number is of course even higher for multi-index sets with either a larger number of modes or modes of greater sizes.\\

Thus to limit the options available to a reasonable amount, a second restriction on the total order can be imposed, namely that it be defined via a process\footnote{\label{nicht-einmal-falsch}Of course this is a sufficiently imprecise statement as to be non-falsifiable, i.e. ``nicht einmal falsch'', and therefore at best somewhat misleading. What is an ``ordering process'' or ``ordering scheme'' which ``works for multiple sets''?} which works for \textit{all} non-degenerate multi-index sets with more than one mode of any size. The smallest such multi-index set is $[2] \times [2]$, and thankfully there are only two linear extensions of its product order, each corresponding to one of the two possible ways to choose how to resolve the question $(1,2) \overset{?}{\lessgtr} (2,1)$ mentioned above:
\[ 
(1,1) < (1,2) < (2,1) < (2,2) \quad \text{and} \quad (1,1) < (2,1) < (1,2) < (2,2) \,. 
\]
The first generalizes\footnote{Both probably also generalize to other ordering schemes too. So more precise restrictions seem to be required to conclude that these are the ``only possible'' generalizations of the total orders for $[2] \times [2]$. Compare footnote \ref{nicht-einmal-falsch}.} to a process definable on all multi-index sets called \textbf{lexicographical order}; the second generalizes to a process definable on all multi-index sets called \textbf{colexicographical order}. Since lexicographical order and colexicographical order are the ``only'' general ``ordering schemes'' which satisfy both of these sensible requirements, they are the only ones which will be used in this review. Not coincidentally, they are also the most widely used in practice.\\

\textbf{Lexicographical ordering} is defined such that for any $\m, \n[O] \in \M$, $\m \not= \n[O]$, one has:
\begin{itemize}
\item if $m_1 < n_1$ then $\m < \n[O]$, if $m_1 > n_1$ then $\m > \n[O]$, otherwise
\item if $m_2 < n_2$ then $\m < \n[O]$, if $m_2 > n_2$ then $\m > \n[O]$, otherwise
\item $\vdots$
\item if $m_{O-1} < n_{O-1}$ then $\m < \n[O]$, if $m_{O-1} > n_{O-1}$ then $\m > \n[O]$, otherwise
\item if $m_O < n_O$ then $\m < \n[O]$, otherwise $\m > \n[O]$.
\end{itemize}

\textbf{Colexicographical ordering} is defined such that for any $m, \n[O] \in \M$, $\m \not= \n[O]$, one has:
\begin{itemize}
\item if $m_O < n_O$ then $\m < \n[O]$, if $m_O > n_O$ then $\m > \n[O]$, otherwise,
\item if $m_{O-1} < n_{O-1}$ then $\m < \n[O]$, if $m_{O-1} > n_{O-1}$ then $\m > \n[O]$, otherwise,
\item $\vdots$
\item if $m_2 < n_2$ then $\m < \n[O]$, if $m_2 > n_2$ then $\m > \n[O]$, otherwise
\item if $m_1 < n_1$ then $\m < \n[O]$, otherwise $\m > \n[O]$.
\end{itemize}

Notice the recursive nature of both definitions, specifically that for lexicographical ordering:
\begin{itemize}
\item if $m_1 < n_1$ then $\m < \n[O]$, if $m_1 > n_1$ then $\m > \n[O]$, otherwise
\item if $\genmltidx{m}{[O]\setminus 1} < \genmltidx{n}{[O]\setminus 1}$ in lexicographical order then $\m < \n[O]$, otherwise $\m > \n[O]$.
\end{itemize}
Similarly for colexicographical ordering:
\begin{itemize}
\item  if $m_O < n_O$ then $\m < \n[O]$, if $m_O > n_O$ then $\m > \n[O]$, otherwise
\item if $\genmltidx{m}{[O]\setminus O} < \genmltidx{n}{[O]\setminus O}$ in colexicographical order then $\m < \n[O]$, otherwise $\m > \n[O]$.
\end{itemize}

%%% Local Variables:
%%% mode: latex
%%% TeX-master: "../new_notes_draft"
%%% End:

\section{The Categories of Multi-Index Sets}
\label{sec:categ-multi-index}

Explicitly defining categories is helpful for identifying the specific mathematical structure of interest. Working generally with the category of (unstructured) multi-index sets, or in special cases with the categories of structured multi-index sets, enables one to identify those issues in tensor algebra which are inherently combinatorial in nature and isolate them. The tensor shape functor then enables the transfer of combinatorial observations from the multi-index categories back to tensors, where they can subsequently be composed with algebraic observations.

\subsection{The (Unfactorized) Category of Multi-Index Sets}
\label{sec:unstr-categ-multi}

The category of multi-index sets $\multiindex$ is a\textbf{ full subcategory}\endnote{A full subcategory contains only a subset of the objects of its parent category, yet all of the morphisms between those objects which exist in the parent category. For example, the category of real vector spaces $\mathsf{Vect}_\R$ is \textit{not} a full subcategory of the category of sets $\sets$, because the morphisms between vector spaces are restricted to be linear functions in the category of real vector spaces  $\mathsf{Vect}_\R$, whereas in the parent category of sets $\sets$ all possible arbitrary set functions between any two vector spaces are also present. So while the function ${f: \R^2 \to \R^2}$ given by $(x,y) \mapsto (x^2, y^2)$ is in $\morphisms_\sets(\R^2, \R^2)$, it is \textit{not} in $\morphisms_{\mathsf{Vect}_\R}(\R^2, \R^2)$ since it is not a linear function.} of the category of finite sets $\setfin$, which in turn is a full subcategory of the category of sets $\sets$. The category $\sets$ has all sets as its objects and all functions between sets as its morphisms, while the category of finite sets $\setfin$ has all \textit{finite} sets as its objects and all functions between those sets as its morphisms.\\

Thus the objects of $\multiindex$ are \textit{all} multi-index sets, i.e. all finite sets of the form:
\[  
 [M_1] \times [M_2] \times [M_3] \times \dots \times [M_O]\,, \quad \quad O, M_1, M_2, \dots, M_O \in \mathbb{N} \,, 
 \]
and the morphisms of $\multiindex$ are \textit{all} functions between multi-index sets.

\subsection{Cartesian Product in the Category of Multi-Index Sets}
\label{product-pedantry}

The Cartesian product defines a functor $\multiindex \times \multiindex \to \multiindex$.

\subsubsection{Definition of the Cartesian Product on Objects}
\label{sec:defin-cart-prod-obj}

The product operation defined for $\multiindex$ is different from the Cartesian product construction that is typically the product in subcategories of $\sets$. The idea is to exploit the highly specific structure of multi-index sets to define a modification of the typical Cartesian product which:
\begin{itemize}
\item is associative up to equality, instead of only being associative up to isomorphism,
\item still satisfies the universal property of the (categorical) product,
\item returns another multi-index as output when given two multi-index as input.
\end{itemize}

For conceptual simplicity, since both product operations lead to sets which are uniquely isomorphic to each other in $\setfin$ and $\sets$\footnote{Note that it would not be correct to say that they are also isomorphic in $\multiindex$, even though it is a full subcategory of $\setfin$, since the typical Cartesian product of two multi-index sets is usually not even an object of $\multiindex$.}, and they correspond to functors which are naturally equivalent\footnote{For insight about why the distinct products are nevertheless naturally equivalent functors, compare the discussion in \ref{sec:tens-prod-funct}. Here the universal property satisfied by both functors is the universal property of the (categorical) product.}, and since many might not even notice the distinction at all without it being pointed to them explicitly, the modified product operation in $\multiindex$ will also be referred to as the \textbf{Cartesian product}, even though it is distinct from the construction which is more typically called ``Cartesian product''.\\

Given two objects of $\multiindex$ (where below $\bigtimes$ refers to the Cartesian product of $\sets$): 
\[ 
\MI = \bigtimes_{o=1}^{O_1} [M_o^{\smallone}] \,, \quad \MII = \bigtimes_{o=1}^{O_2} [M_o^{\smalltwo}] \,,
\]
their product $\times_{\multiindex}$ in $\multiindex$ is defined to be the set:
\[  
 \MI \times_{\multiindex} \MII \defequals \bigtimes_{o=1}^{O_1 + O_2} [M_o] \,, \quad M_o =
  \begin{cases}
    M_o^{\smallone} & \text{when } 1 \le o \le O_1 \\
M^{\smalltwo}_{o-O_1} & \text{when } O_1 + 1 \le o \le O_1 + O_2  
  \end{cases} \,. 
\]

For example, if $\M[2] = [2] \times [2]$, and $\N[1] = [2]$, then one has that:
\[
  \begin{array}{rcl}
 \M[2] \times_{\multiindex} \N[1]    & = &  [2] \times_{\sets} [2] \times_{\sets} [2]  \\
& = & \{ (1,1,1), (1,1,2), (1,2,1), (1,2,2),\\
&& \phantom{\ \ }(2,1,1), (2,1,2), (2,2,1), (2,2,2)  \} \,.
  \end{array}
 \]
Note the single level of parentheses, as well as that this is a multi-index set. In contrast:
\[
  \begin{array}{rcl}
 \M[2] \times_{\sets} \N[1]    & = &  ([2] \times [2]) \times [2]  \\
& = & \{ ((1,1),1), ((1,1),2), ((1,2),1), ((1,2),2),\\
&& \phantom{\ \ }((2,1),1), ((2,1),2), ((2,2),1), ((2,2),2)  \} \,.
  \end{array} 
\]
Note the two levels of parentheses, and that this is not a Cartesian product of three index sets, but a Cartesian product of one multi-index set and one index set, and thus not a multi-index set itself.\\

Observe that by design these two constructions coincide for index sets: 
\[ 
[M] \times_{\sets} [N] = [M] \times_{\multiindex} [N] \,, \quad \forall M,N \in \mathbb{N} \,. 
\]
Observe also that the product in $\multiindex$ is associative up to equality, whereas the typical Cartesian product is only associative up to isomorphism. For example, for $\times_{\sets}$ one has that:
\[  
\begin{array}{rcl}
 ([2] \times_{\sets} [2]) \times_{\sets} [2]  & = & \{ ((1,1),1), ((1,1),2), ((1,2),1), ((1,2),2),\\
&& \phantom{\ \ }((2,1),1), ((2,1),2), ((2,2),1), ((2,2),2)  \}  \\
& \not= & \{ (1,(1,1)), (1,(1,2)), (1,(2,1)), (1,(2,2)),\\
&& \phantom{\ \ }(2,(1,1)), (2,(1,2)), (2,(2,1)), (2,(2,2)) \}  \\
& = & [2] \times_{\sets} ([2] \times_{\sets} [2]) \,.
 \end{array} 
\]
Even though there is an obvious bijection between the two sets above, such that they are isomorphic, they are nevertheless not equal. In contrast, for $\times_{\multiindex}$ one has that:
\[ 
([2] \times_{\multiindex} [2]) \times_{\multiindex} [ 2] = [2] \times_{\multiindex} [2] \times_{\multiindex} [2] = [2] \times_{\multiindex} ([2] \times_{\multiindex} [2]) \,,  
\]
where by definition $ [2] \times_{\multiindex} [2] \times_{\multiindex} [2] =  [2] \times_{\sets} [2] \times_{\sets} [2]$. No isomorphisms are required to identify any of the three sets, instead all three of them are already equal.\\

More generally, one only has for the Cartesian product in $\sets$ that:
\[
\begin{array}{rcccl}
  (X \times_{\sets} Y) \times_{\sets} Z  & \not= & X \times_{\sets} Y \times_{\sets} Z & \not= &  X \times_{\sets} (Y \times_{\sets}) Z\,, \\
(X \times_{\sets} Y) \times_{\sets} Z & \cong_{\sets} & X \times_{\sets} Y \times_{\sets} Z & \cong_{\sets} & X \times_{\sets} (Y \times_{\sets}) Z \,,
  \end{array}
\]
since for arbitrary elements $x \in X$, $y \in Y$, $z \in Z$, elements of $(X \times_{\sets} Y) \times_{\sets} Z $ are of the form $((x,y),z)$, elements of $ X \times_{\sets} Y \times_{\sets} Z $ are of the form $(x,y,z)$, and elements of $X \times_{\sets} (Y \times_{\sets}) Z $ are of the form $(x,(y,z))$, none of which are equal. Here $\cong_{\sets}$ means equivalent up to isomorphism in the category $\sets$, i.e. that there exists a bijection between the sets\footnote{The relation $\cong_{\sets}$ of equivalence up to isomorphism is a transitive relation, while technically $\not=$ is not. Nevertheless it also true that  $(X \times_{\sets} Y) \times_{\sets} Z \not=  X \times_{\sets} (Y \times_{\sets}) Z$, even if it did not say so explicitly above.}.  \\

In contrast, for the product operation $\times_{\multiindex}$ in the category $\multiindex$, one has that:
\[ 
 ( \M[O_1]  \times_{\multiindex} \N ) \times_{\multiindex} \P  = \M[O_1] \times_{\multiindex} \N \times_{\multiindex} \P = \M[O_1] \times_{\multiindex} ( \N \times_{\multiindex} \P) \,,
 \]
for \textit{all} multi-index sets $\M[O_1]$, $\N$, and $\P$ in $\multiindex$. This is associativity up to equality.\\

For two given multi-index sets ${\M[O_1] = \bigtimes_{o=1}^{O_1} [M_{o}]}$ and ${\N = \bigtimes_{o=1}^{O_2}[N_o]}$, there always exists a canonical\footnote{\label{canonical-bijection-comment}It is unique among all bijections in transforming the projection maps of one set to the projection maps of the other, i.e. in being compatible with the universal property of the product satisfied by both sets.} bijection $\phi: \M[O_1] \times_{\sets} \N \to \M[O_1] \times_{\multiindex} \N$. Here is its rule of assignment:
\[  
  \begin{array}{rlcrl}
  & (\m[O_1] , \n) =    &&& \\
\phi:&  ((m_1, \dotsc, m_{O_1}), (n_1, \dotsc, n_{O_2})) & \mapsto &  (m_1, \dotsc, m_{O_1}, n_1, \dotsc, n_{O_2})  &\defequals \\
&&& \m[O_1] \times_{\multiindex} \n \phantom{\,.}& \defequals \\
&&&\m[O_1]\n \,.&
  \end{array}
\]
For example, taking $\M[2] = [2] \times [2]$ and $\N[1] = [2]$ as in the example above, one has that:
\[   
\phi :  ((1,1),2)  \mapsto (1,1,2) \,, \quad \phi :((2,2),1) ) \mapsto (2,2,1) \,, \quad \phi:  ((1,2),1 )  \mapsto (1,2,1) \,, \quad \text{etc.} 
 \]

This definition of product in $\multiindex$ also makes the definition of the Segre outer product $\otimes$ much less unwieldy compared to a comparable definition using the tensor product of functions $\fotimes$ defined in section \ref{sec:tensor-product}. In particular, one has for tensors $\mathbfcal{T}: \M[O_1] \to \R$ and $\mathbfcal{U}: \N \to \R$ that: 
\[ 
 \mathbfcal{T} \otimes \mathbfcal{U} \defequals (\mathbfcal{T} \fotimes \mathbfcal{U}) \circ \phi^{-1} \,, 
 \]
since, comparing domains and codomains, one sees that:
\[ 
  \mathbfcal{T} \otimes \mathbfcal{U} : \M[O_1] \times_{\multiindex} \N \to \R \,, \quad \mathbfcal{T} \fotimes \mathbfcal{U}: \M[O_1] \times_{\sets} \N \to \R \,.
  \]

One example of why this is useful is because it allows one to say for any three tensors $\mathbfcal{S}, \mathbfcal{T}, \mathbfcal{U}$:
\[ 
{(\mathbfcal{S} \otimes \mathbfcal{T}) \otimes \mathbfcal{U}} =  \mathbfcal{S} \otimes \mathbfcal{T} \otimes \mathbfcal{U}  = \mathbfcal{S} \otimes (\mathbfcal{T} \otimes \mathbfcal{U}) \,,
 \]
i.e. that all three tensors are the same up to \textit{equality}, not merely up to isomorphism\footnote{Which strictly speaking is all one can do using the typical tensor product construction found e.g. in Chapter 2 of \cite{atiyah-macdonald}.}. However, the analogous statement would be false using the definition $\fotimes$: the three functions ${(\mathbfcal{S} \fotimes \mathbfcal{T}) \fotimes \mathbfcal{U}}$, ${ \mathbfcal{S} \fotimes \mathbfcal{T} \fotimes \mathbfcal{U} }$, and $\mathbfcal{S} \fotimes (\mathbfcal{T} \fotimes \mathbfcal{U})$ don't even have the same domain and therefore certainly aren't equal. Moreover, in general \textit{none} of the functions ${(\mathbfcal{S} \fotimes \mathbfcal{T}) \fotimes \mathbfcal{U}}$, ${ \mathbfcal{S} \fotimes \mathbfcal{T} \fotimes \mathbfcal{U} }$, and $\mathbfcal{S} \fotimes (\mathbfcal{T} \fotimes \mathbfcal{U})$ are tensors, since their domains are not multi-index sets, only Cartesian products of multi-index sets. Thus the definition for the product in $\multiindex$ ensures that the Segre outer product returns a tensor as output when given tensors as input.\\

Throughout this review, whenever $\bigtimes$ or $\times$ is written to denote some product of multi-index sets, one can safely assume that the product in question is the product operation $\times_{\multiindex}$ and not $\times_{\sets}$.

\subsubsection{Definition of the Cartesian Product on Morphisms}
\label{sec:defin-cart-prod-mor}

One has the same issues for functions that one has for sets with $\times_{\sets}$. Specifically, given functions:
\[
 f_1: X_1 \to Y_1 \,, \quad f_2: X_2 \to Y_2 \,, \quad f_3: X_3 \to Y_3 \,,  
\]
none of the following three functions are equal:
\[
  \begin{array}{rccl}
(f_1 \times_{\sets} f_2) \times_{\sets} f_3 :   & (X_1 \times_{\sets} X_2) \times_{\sets} X_3  & \to  & (Y_1 \times_{\sets} Y_2) \times_{\sets} Y_3   \,, \\
f_1 \times_{\sets} f_2 \times_{\sets} f_3:  & X_1 \times_{\sets} X_2 \times_{\sets} X_3  & \to & Y_1 \times_{\sets} Y_2 \times_{\sets} Y_3 \,, \\
 f_1 \times_{\sets} (f_2 \times_{\sets} f_3):  &  X_1 \times_{\sets} (X_2 \times_{\sets} X_3) & \to & Y_1 \times_{\sets} (Y_2 \times_{\sets} Y_3) \,.
  \end{array}  
 \]
This is clearly true since none of their domains are the same, and none of their codomains are the same. They all take different types of inputs and return different types of outputs:
\[
  \begin{array}{rccl}
(f_1 \times_{\sets} f_2) \times_{\sets} f_3 :   & (( x_1 , x_2), x_3)  & \mapsto  & ( (f_1(x_1), f_2(x_2)), f_3(x_3) )  \,, \\
f_1 \times_{\sets} f_2 \times_{\sets} f_3:  & (x_1, x_2, x_3) & \mapsto &   ( f_1(x_1), f_2(x_2), f_3(x_3)  )  \,, \\
 f_1 \times_{\sets} (f_2 \times_{\sets} f_3):  & (x_1, (x_2, x_3)  )& \mapsto & ( f_1(x_1), (f_2(x_2), f_3(x_3))  )\,.
  \end{array}  
 \]
To avoid problems like this, the Cartesian product of functions in $\multiindex$ is defined differently. (It needs to be defined differently in any case, since the Cartesian product in $\sets$ of two functions $f_1: \MI \to \NI$ and $f_2: \MII \to \NII$ is the function:
 \[ 
f_1 \times_{\sets} f_2: \MI \times_{\sets} \MII \to \NI \times_{\sets} \NII \,,
\] 
for which neither the domain nor codomain are multi-index sets in general. Moreover, the Cartesian product functor in $\multiindex$ applied to the domains of $f_1$ and $f_2$ yields ${ \MI \times_{\multiindex} \MII}$, and applied to their codomains yields ${\NI \times_{\multiindex} \NII}$, so those must be the domain and codomain respectively for any definition of $f_1 \times_{\multiindex} f_2$.) Specifically, given functions\\ 
$f_1: \MI \to \NI$ and $f_2: \MII \to \NII$, their Cartesian product in $\multiindex$ set is
 \[
f_1 \times_{\multiindex} f_2: \MI \times_{\multiindex} \MII \to \NI \times_{\multiindex} \NII 
\]
 which has the rule of assignment:
\[ 
f_1 \times_{\multiindex} f_2: \mi \times_{\multiindex} \mii \mapsto f_1(\mi) \times_{\multiindex} f_2(\mii) \,. 
 \]
More precisely, if $\phi$ is the canonical\footnote{See footnote \ref{canonical-bijection-comment}.} bijection $\phi: \MI \times_{\sets} \MII \to \MI \times_{\multiindex} \MII$ and $\psi$ is the canonical bijection $\psi: \NI \times_{\sets} \NII \to \NI \times_{\multiindex} \NII$, then one has:
\[
 f_1 \times_{\multiindex} f_2 \defequals \psi \circ (f_1 \times_{\sets} f_2)  \circ \phi^{-1} \,. 
 \]
The bijections $\phi, \psi$ basically ``remove parentheses'', and the bijections $\phi^{-1}, \psi^{-1}$ ``add parentheses''.\\

Note that this definition of Cartesian product, in addition to being the ``correct choice'' by virtue of being the morphism induced by the universal property of the product, also has the virtue of being associative up to equality, not just isomorphism. In other words, given any morphisms of $\multiindex$:
\[
 f_1: \MI \to \NI \,, \quad f_2: \MII \to \NII \,, \quad f_3: \Mltidx[O_3]{M^{\smallsuper{3}}} \to \Mltidx[D_3]{N^{\smallsuper{3}}} \,, 
 \]
one always has the following relationships:
\[ 
 (f_1 \times_{\multiindex} f_2 ) \times_{\multiindex} f_3 = f_1 \times_{\multiindex} f_2 \times_{\multiindex} f_3 = f_1 \times_{\multiindex} ( f_2 \times_{\multiindex} f_3) \,. 
 \]
Therefore the order and number of parentheses do not affect the result.\\

For an example of the definition of Cartesian product for functions in $\multiindex$, consider
\[
 f_1: [4] \to [2] \times [2] \,, \quad f_1(1) = (1,1)\,, f_1(2) = (2,1) \,, f_1(3)=(1,2) \,, f_1(4) = (2,2)\,, 
 \]
and
\[ 
f_2: [2] \times [2] \to [4] \,, \quad f_2(1,1) = 1\,, f_2(1,2) = 2 \,, f_2(2,1) = 3 \,, f_2(2,2) = 4 \,. 
\]
Then for the values of the Cartesian product function of $\multiindex$: 
\[
f_1 \times_{\multiindex} f_2: [4] \times_{\multiindex} ([2] \times [2]) = [4] \times [2] \times [2] \to [2] \times [2] \times [4] = ([2] \times [2]) \times_{\multiindex} [4] 
\]
one has the following:
%\[  
\begin{longtable}{CCC}
%\begin{array}{rcl}
  f_1 \times_{\multiindex} f_2:  (1,1,1)  &\mapsto& (1,1,1)  \\
f_1 \times_{\multiindex} f_2:  (1,1,2) & \mapsto& (1,1,2)\\
f_1 \times_{\multiindex} f_2:  (1,2,1) &\mapsto& (1,1,3)\\
f_1 \times_{\multiindex} f_2:  (1,2,2) &\mapsto& (1,1,4) \\
f_1 \times_{\multiindex} f_2:  (2,1,1) &\mapsto& (2,1,1) \\
f_1 \times_{\multiindex} f_2:  (2,1,2) &\mapsto& (2,1,2) \\
f_1 \times_{\multiindex} f_2:  (2,2,1) &\mapsto& (2,1,3) \\
f_1 \times_{\multiindex} f_2:  (2,2,2) &\mapsto& (2,1,4) \\
f_1 \times_{\multiindex} f_2:  (3,1,1) &\mapsto& (1,2,1) \\
f_1 \times_{\multiindex} f_2:  (3,1,2) &\mapsto& (1,2,2) \\
f_1 \times_{\multiindex} f_2:  (3,2,1) &\mapsto& (1,2,3) \\
f_1 \times_{\multiindex} f_2:  (3,2,2) &\mapsto& (1,2,4) \\
f_1 \times_{\multiindex} f_2:  (4,1,1) &\mapsto& ( 2,2,1) \\
f_1 \times_{\multiindex} f_2:  (4,1,2) &\mapsto& (2,2, 2)\\
f_1 \times_{\multiindex} f_2:  (4,2,1) &\mapsto& ( 2,2,3)\\
f_1 \times_{\multiindex} f_2:  (4,2,2) &\mapsto& (2,2,4)  \,.
\end{longtable}
%  \end{array}
%\]
In contrast, for the values of the Cartesian product function of $\sets$:
\[ 
 f_1 \times_{\sets} f_2 : [4] \times_{\sets} ([2] \times [2]) \to ([2] \times [2]) \times_{\sets} [4] \,,
  \]
one has the following instead:
%\[ 
% \begin{array}{rcl}
\begin{longtable}{CCC}
  f_1 \times_{\sets} f_2:  (1,(1,1))  &\mapsto& ((1,1),1)  \\
f_1 \times_{\sets} f_2:  (1,(1,2)) & \mapsto& ((1,1),2)\\
f_1 \times_{\sets} f_2:  (1,(2,1)) &\mapsto& ((1,1),3)\\
f_1 \times_{\sets} f_2:  (1,(2,2)) &\mapsto& ((1,1),4) \\
f_1 \times_{\sets} f_2:  (2,(1,1)) &\mapsto& ((2,1),1) \\
f_1 \times_{\sets} f_2:  (2,(1,2)) &\mapsto& ((2,1),2) \\
f_1 \times_{\sets} f_2:  (2,(2,1)) &\mapsto& ((2,1),3) \\
f_1 \times_{\sets} f_2:  (2,(2,2)) &\mapsto& ((2,1),4) \\
f_1 \times_{\sets} f_2:  (3,(1,1)) &\mapsto& ((1,2),1) \\
f_1 \times_{\sets} f_2:  (3,(1,2)) &\mapsto& ((1,2),2) \\
f_1 \times_{\sets} f_2:  (3,(2,1)) &\mapsto& ((1,2),3) \\
f_1 \times_{\sets} f_2:  (3,(2,2)) &\mapsto& ((1,2),4) \\
f_1 \times_{\sets} f_2:  (4,(1,1)) &\mapsto& ( (2,2),1) \\
f_1 \times_{\sets} f_2:  (4,(1,2)) &\mapsto& ((2,2), 2)\\
f_1 \times_{\sets} f_2:  (4,(2,1)) &\mapsto& ( (2,2),3)\\
f_1 \times_{\sets} f_2:  (4,(2,2)) &\mapsto& ((2,2),4)  \,.
%  \end{array}
%\]
\end{longtable}
Clearly these functions are very similar, and in fact they are related via canonical bijections induced by the universal property of the product, but only the first one is a morphism in $\multiindex$.

%%% Local Variables:
%%% mode: latex
%%% TeX-master: "../new_notes_draft"
%%% End:

\subsection{The $D$-Factorized Categories of Multi-Index Sets}
\label{sec:k-struct-categ}

The definition of the $D$-factorized categories of multi-index sets $\multiindex_D$ initially requires some explanation, but becomes straightforward to understand after thinking through some examples.

\subsubsection{Objects of the $D$-Factorized Category of Multi-Index Sets}
\label{sec:objects-multiindex_k}

For a given $D \ge 1$, the objects of $\multiindex_D$ each consist of two data: (i) a multi-index set with $O \ge D$ modes, and (ii) a chosen \textit{contiguous}\footnote{As explained elsewhere, only considering contiguous partitions does not actually come with any loss of generality, due to the possibility of permuting modes beforehand to ensure that one's desired partition becomes contiguous.} partition of those $O$ modes into $D$ blocks.\\

For an example of what this means, consider the $2$-factorized category of multi-index sets, $\multiindex_2$. Then for the following multi-index set with $3 \ge 2$ modes:
\[
 \M[3] \defequals [7] \times [5] \times [14]
  \]
there are $\binom{3-1}{2-1} = \binom{2}{1} = 2$ objects of $\multiindex_2$ associated with it, since $\binom{O-1}{D-1}$ counts the number of distinct ways to partition a set with $O$ elements into $D$ contiguous blocks.\\

Specifically, the two objects of $\multiindex_2$ associated with $\M[3] = [7] \times [5] \times [4]$ are:
\[ 
\M[3]_{12|3} \defequals ([7] \times [5] \times [14], \{ \{1,2\}, \{3\} \}), \quad \M[3]_{1|23} \defequals ([7]\times[5]\times[14],\{ \{1\}, \{2,3\} \}) \,.
 \]

In contrast, given a multi-index set with $2$ modes, there is exactly $\binom{2-1}{1-1}=\binom{1}{0}=1$ object of $\multiindex_2$ which is associated with it. For instance, the multi-index set:
\[
  [23] \times [17]  
 \]
is associated uniquely with the following object of $\multiindex_2$:
\[ 
 ([23] \times [17], \{ \{1\}, \{2\} \}) \,. 
 \]
The objects of $\multiindex_D$ can be thought of as consisting of two distinct strata: (i) all multi-index sets with exactly $D$ modes, which are endowed with no extra structure, and (ii) multi-index sets with $O > D$ modes endowed with a given contiguous partition of their $O$ modes into $D$ blocks (with the result that each such multi-index set is associated with $\binom{O-1}{D-1}$ distinct objects of $\multiindex_D$).\\

The idea is that the objects of $\multiindex_D$ should have ``$D$-fold Cartesian product structure''.\\

While there is clearly only one way that a multi-index set with exactly $D$ modes can be considered to have such $D$-fold Cartesian product structure, for multi-index sets with $O>D$ modes one needs to be specific about how the multi-index set is being considered as having $D$-fold Cartesian product structure, since there are $\binom{O-1}{D-1}$ distinct possible groupings of the $O$ modes which would enable such an interpretation, each corresponding to a different contiguous partition of $[O]$.\\

Also notice that, since for every $O \ge 1$ there is exactly $\binom{O-1}{1-1}=\binom{O-1}{0}=1$ way to partition $O$ modes into one block\footnote{Since there is only one block, the restriction that the partition be contiguous is redundant.}, the objects of $\multiindex_1$ are essentially the same as those of $\multiindex$.

\subsubsection{Morphisms of the $D$-Factorized Category of Multi-Index Sets}
\label{sec:morph-k-part}

Given two objects  of $\multiindex_D$, the morphisms between them consist of all $D$-fold Cartesian products of functions between their $D$ corresponding blocks of modes.\\

For an example of what this definition means when $D=2$, consider the objects $\M[3]_{12|3}$ and $[23] \times [17]$ of $\multiindex_2$.  Given any two functions (equivalently morphisms in $\multiindex$):
\[ 
 f_1: [7] \times [5] \to [23] \,, \quad f_2: [14] \to [17]\,, 
 \]
then the Cartesian product of those two functions:
\[
  f_1 \times f_2 : [7] \times [5] \times [14] \to [23] \times [17] 
 \]
is a morphism in $\multiindex_2$ from  $\M[3]_{12|3}$ to $[23] \times [17]$.\\

Conversely, by definition \textit{every} morphism in $\multiindex_2$ starting from $\M[3]_{12|3}$ and going to $[23] \times [17]$ is of the form $f_1 \times f_2$ for some functions $f_1: [7] \times [5] \to [23]$ and $f_2: [14] \to [17]$.\\

Similarly, the morphisms in $\multiindex_2$ starting from $\M[3]_{1|23}$ (note the different partition) and going to $[23] \times [17]$ are the set of all functions which are of the form:
\[  
g_1 \times g_2 : [7] \times [5] \times [14] \to [23] \times [17] \,, 
 \]
for \textit{any} functions $g_1$ and $g_2$ such that:
\[  
g_1: [7] \to [23] \,, \quad g_2: [5] \times [14] \to [17] \,. 
  \]
The idea is that the morphisms in $\multiindex_D$ should preserve the specified $D$-fold Cartesian product structures of their domain and codomain, and thus have $D$-fold Cartesian product structure as well.\\

Note that for functions $g_1: [7] \to [23]$ and $g_2: [5] \times [14] \to [17]$, the function $g_1 \times g_2$ is \textit{not} a morphism in $\multiindex_2$ from $\M[3]_{12|3}$ to $[23] \times [17]$, even though the function is a morphism from $\M[3]_{1|23}$ to $[23] \times [17]$ in $\multiindex_2$, as well as a morphism from $\M[3]$ to $[23] \times [17]$ in $\multiindex$.\\

Similarly, for functions $f_1: [7] \times [5] \to [23]$ and $f_2: [14] \to [17]$, the function $f_1 \times f_2$ is \textit{not} a morphism in $\multiindex_2$ from $\M[3]_{1|23}$ to $[23] \times [17]$, even though the function is a morphism from $\M[3]_{12|3}$ to $[23] \times [17]$ in $\multiindex_2$, as well as a morphism from $\M[3]$ to $[23] \times [17]$ in $\multiindex$.\\

Thus the principal effect of attaching the extra structure of a partition of modes to the objects in $\multiindex_D$ is to limit the number of functions between those objects which are considered valid morphisms. Moreover, when different partitions of modes into $D$ blocks are attached to the same multi-index set, the answer to which functions are valid morphisms in $\multiindex_D$ changes.\\

Also notice that, since the $1$-fold Cartesian product of a function $f$ is just the function $f$ itself, the morphisms of $\multiindex_1$ are essentially the same as those of $\multiindex$.

\subsubsection{Remarks about the $D$-Factorized Categories of Multi-Index Sets}
\label{sec:final-remarks-about}

Since both the objects of $\multiindex_1$ and the morphisms of $\multiindex_1$ are essentially the same as their counterparts in $\multiindex$, the two categories $\multiindex_1$ and $\multiindex$ are for all intents and purposes the same category. Therefore, from now on $\multiindex_1$ will be redefined so that $\multiindex_1 = \multiindex$.\\

Notice that for any $O > D$, the objects of $\multiindex$ which are associated with the objects of $\multiindex_O$ (namely those multi-index sets with $O$ or more modes) are a strict subset of the objects of $\multiindex$ which are associated with objects of $\multiindex_D$ (those multi-index sets with $D$ or more modes).\\

Moreover, for $O > D$, one can define $\binom{O-1}{D-1}$ distinct forgetful functors from $\multiindex_O$ to $\multiindex_D$. I.e. there are $\binom{O-1}{D-1}$ distinct ways for a multi-index set to ``forget'' its $O$-fold Cartesian product structure, and to remember only a coarser $D$-fold Cartesian product structure instead. Each given contiguous partition of $[O]$ into $D$ blocks provides a pattern for merging the $O$ blocks of modes of the objects in $\multiindex_O$ into $D$ blocks of modes, and thus selects associated objects of $\multiindex_D$.\\

For example, there are $\binom{3-1}{2-1} = 2$ distinct forgetful functors from $\multiindex_3$ to $\multiindex_2$.\\

The first such functor, denoted $\mathscr{F}_{1|23}$, corresponding to the contiguous partition $\{ \{1\}, \{2,3\} \}$, allows an object of $\multiindex_3$ to remember its first block of modes, but makes that object of $\multiindex_3$ forget that its second and third blocks of modes were distinct and thus merges them into one block of modes. Here are some examples of the behavior of $\mathscr{F}_{1|23}$, where $\N[4] \defequals [5] \times [37] \times [29] \times [41]$:
\[
\begin{array}{lcl}
   
 \mathscr{F}_{1|23}(\M[3]) & = & \M[3]_{1|23} \\
\mathscr{F}_{1|23}(\N[4]_{12|3|4}) & = & \N[4]_{12|34} \\ 
                                                         
  \end{array}
\]

The second such functor, denoted $\mathscr{F}_{12|3}$, corresponding to the contiguous partition $\{  \{1,2\}, \{3\}\}$, makes an object of $\multiindex_3$ forget that its first and second of modes were distinct and thus merges them into one block of modes, but allows that object of $\multiindex_3$ to remember its third block of modes. Here are some examples of the behavior of $\mathscr{F}_{12|3}$:
\[
\begin{array}{lcl}
    
\mathscr{F}_{12|3}(\M[3]_{1|2|3}) & = & \M[3]_{12|3} \\
\mathscr{F}_{12|3}(\N[4]_{12|3|4}) & = & \N[4]_{123|4}  \\
                                                          
\end{array}
\]

Observe that these forgetful functors do not modify any of the morphisms in $\multiindex_O$. If a function (i.e. morphism in $\multiindex$) $f$ is a valid morphism in $\multiindex_O$ from some $\MI_{P_1}$ to $\MII_{P_2}$, then for any contiguous partition $Q$ of $[O]$ into $D$ blocks, $f$ will also be a valid morphism in $\multiindex_D$ from\footnote{Here $P_1$ denotes a contiguous partition of $[O_1]$ into $O$ blocks, $P_2$ denotes a contiguous partition of $[O_2]$ into $O$ blocks. See appendix \ref{sec:part-finite-sets} for an explanation of the notation $\blacktriangleright$.} $\mathscr{F}_Q(\MI_{P_1}) = \MI_{P_1 \blacktriangleright Q}$ to $\mathscr{F}_Q(\MII_{P_2}) = \MII_{P_2\blacktriangleright Q} $. In other words, for any $O>D$, any function preserving $O$-fold Cartesian product structure will also preserve any coarser $D$-fold Cartesian product structure (and thus also be a morphism in $\multiindex_D$).\\

For instance, consider the function:
\[ 
g = g_1 \times g_2 \times g_3:  \M[3] \to \N[4] \,, 
\]
where the functions $g_1$, $g_2$, and $g_3$ have the following domains and codomains:
\[ 
 g_1: [7] \to [5]\times [37]\,, \quad g_2: [5] \to [29]\,, \quad g_3:[14] \to [41]\,. 
\]
Then $g$ is a morphism in $\multiindex_3$:
\[ 
\M[3]_{1|2|3} \quad \to \quad \N[4]_{12|3|4} \,, 
\]
a morphism in $\multiindex_2$ after applying the second forgetful functor $\mathscr{F}_{12|3}:\multiindex_3 \to \multiindex_2$:
\[  
 \mathscr{F}_{12|3}(\M[3]_{1|2|3}) = \M[3]_{12|3} \quad \to \quad \N[4]_{123|4} = \mathscr{F}_{12|3}(  \N[4]_{12|3|4} )  
 \,,\]
and a morphism in $\multiindex_2$ after applying the first forgetful functor $\mathscr{F}_{1|23}:\multiindex_3 \to \multiindex_2$:
\[
\mathscr{F}_{1|23}(\M[3]_{1|2|3}) = \M[3]_{1|23} \quad \to \quad \N[4]_{12|34} = \mathscr{F}_{1|23}(  \N[4]_{12|3|4} ) \,.
\]

However, the converse is false. In other words, there will always be morphisms in $\multiindex_D$ from $\mathscr{F}_Q (\MI_{P_1})$ to $\mathscr{F}_Q (\MII_{P_2})$ which do not preserve $O$-fold Cartesian product structure, i.e. which are not morphisms in $\multiindex_O$ from $\MI_{P_1}$ to $\MII_{P_2}$. Such morphisms in $\multiindex_D$ are not the image of any morphism in $\multiindex_O$ under any forgetful functor $\multiindex_O \to \multiindex_D$. Thus the image of any of these forgetful functors always fails to be a full subcategory of $\multiindex_D$.\\

As an example, while the function
\[
f = f_1 \times f_2 : \M[3] \to \N[4] \,,
 \]
where the functions $f_1$ and $f_2$ have the following domains and codomains:
\[
f_1: [7] \times [5] \to  [5] \times [37] \times [29]\,, \quad f_2: [14] \to [41]\,, 
\] 
\textit{is} a morphism in $\multiindex_2$:
\[
\mathscr{F}_{12|3}(\M[3]) = \M[3]_{12|3} \quad \to \quad \N[4]_{123|4} =\mathscr{F}_{12|3}(\N[4]_{12|3|4} )  \,, 
\]
in general it is \textit{not} a morphism:
\[  
 \M[3]_{1|2|3} \quad \to \quad  \N[4]_{12|3|4}  
\]
in $\multiindex_3$ (except whenever $f_1 = h_1 \times h_2$ for some $h_1: [7] \to [5] \times [37]$ and $h_2: [5] \to [29]$).\\

Note finally that, for each $D \ge 1$, the category $\multiindex_D$ is equivalent to the image of the $D$-fold Cartesian product functor defined above for the category $\multiindex$ in section \ref{product-pedantry}:
\[ 
 \bigtimes_{d=1}^D \underbrace{\multiindex \times \cdots \times \multiindex}_{D\text{ times}} \to \multiindex \,. 
 \]
In other words, every object of $\multiindex_D$ has isomorphic $D$-fold Cartesian product structure to some object $\bigtimes_{d=1}^D \Mltidx[O_d]{M^{\smallsuper{d}}}$ in the image of the $D$-fold Cartesian product functor.

%%% Local Variables:
%%% mode: latex
%%% TeX-master: "../new_notes_draft"
%%% End:

\section{Identification of $\R^L$ with $\R^{[L]}$}
\label{sec:vector-identification}

It can be verified for each $\ell \in [L]$, the projection functions $\pi_{\ell}: \mathbb{R}^{[L]} \to \mathbb{R}$ given by $\mathbf{v} \mapsto \mathbf{v}(\ell) \defequals v_{\ell}$ are linear, and that $\mathbb{R}^{[L]}$ together with the $L$ projection functions $\pi_{\ell}$ satisfy the universal property of the $L$-fold product for $\R$ in the category of finite-dimensional real vector spaces.\\

One may also verify that the $L$-fold Cartesian product of copies of $\mathbb{R}$, $\bigtimes_{\ell=1}^L \mathbb{R}$, whose elements correspond to ordered lists, or tuples, of real numbers of length $L$, $(v_1, \dots, v_{\ell}, \dots, v_L)$, is also a real vector space, and that $\bigtimes_{\ell=1}^L \mathbb{R}$ together with the $L$ linear projection functions $p_{\ell} : \bigtimes_{\ell=1}^L \mathbb{R} \to \mathbb{R}$ (which, given such a tuple $(v_1, \dots, v_{\ell}, \dots, v_{L})$ return the $\ell$'th element $v_{\ell}$) also satisfy the universal property of the $L$-fold product of $\R$ in the category of finite-dimensional real vector spaces.\\

Therefore, there is a \textit{canonical} linear isomorphism\footnote{It is the unique isomorphism compatible with the specified projection functions $\pi_{\ell}$ and $p_{\ell}$.} between $\mathbb{R}^{[L]}$ and $\bigtimes_{\ell=1}^L \mathbb{R}$, such that they are ``essentially the same'' real vector space. In light of this, in what follows $\mathbb{R}^{[L]}$ and $\bigtimes_{\ell=1}^L \mathbb{R}$ will generally not be distinguished between explicitly. Instead the notation\footnote{Which is more commonly reserved for $\bigtimes_{\ell=1}^L \mathbb{R}$ only.} $\mathbb{R}^L$ will be used to denote both spaces and the notation $(v_1, \dots, v_{\ell}, \dots, v_L)$ to denote elements of either space.

%%% Local Variables:
%%% mode: latex
%%% TeX-master: "../new_notes_draft"
%%% End:

\section{Coordinate Isomorphisms}
\label{sec:coord-isom}

Every object $\V$ in $\finvect$ is isomorphic as a vector space to $\R^M$ for $M= \dim(V)$. Similarly, every object $\bT$ (with ``tensor factors'' $\V_1$, $\V_2$, \dots, $\V_D$) of $\fintensor{D}$ is isomorphic as a $D$-fold tensor space to $\R^{\M[D]}$ where $\M[D] = \bigtimes_{d=1}^D [M_d]$ with $M_d = \dim(\mathbb{V}_d)$ for all $d \in \mathbb{V}_d$. Combining these two statements, one could reach the conclusion that (for finite-dimensional spaces with real scalars) all problems in linear algebra and all problems in multilinear algebra can be reduced to the study of coordinate spaces $\R^{\M[D]}$. This is the reason why such spaces are the focus of this review.

\subsection{Coordinate Isomorphisms for Vector Spaces}
\label{sec:coord-vector-spaces}

Given a vector space $\V$ with $\dim(\V) = M$, choose some basis $\B = \{ \mathbf{b}_1, \dots, \mathbf{b}_M \}$, which then uniquely defines $M$ linear functions in $\L(\V;\R) \defequals \dual{\V}$ via the formulae:
\[  
 \dualfunc{m}{\B} : \mathbf{b}_{\mu} \mapsto \delta_{m\mu} \,,
\] 
where $\delta_{m\mu}$ denotes the Kronecker delta ($\delta_{m\mu} = 1$ if $m=\mu$ and $0$ otherwise). Then with respect to the chosen basis $\B$, any $\mathbf{v} \in \V$ may be written in the form:
\[ 
\mathbf{v} = \dualfunc{1}{\B}(\mathbf{v} )\mathbf{b}_1 + \dualfunc{2}{\B}(\mathbf{v}) \mathbf{b}_2 + \dots + \dualfunc{M}{\B}(\mathbf{v})\mathbf{b}_M \,.
 \]
For any vector $\mathbf{v} \in V$ the $M$ linear functions $\dualfunc{m}{\B}$ give the scalar coefficients in the unique linear combination writing $\mathbf{v}$ in terms of the elements of the basis $\B$. In other words, the $M$ linear functions $\dualfunc{m}{\B}$ define, for every vector $\mathbf{v} \in \V$, the coordinates of $\mathbf{v}$ with respect to the basis $\B$.\\

For any chosen basis $\B$, one can verify that all of the coordinate functions $\dualfunc{1}{\B}$ together collectively form a basis for $\dual{\V}$, denoted $\dual{\B}$. The basis $\dual{\B}$ is called the \textbf{dual basis} of $\B$. \\

As an example, if $\V = \{ a \sin(x) + b \sin(2x): a,b \in \R \}$, then for $\mathbf{v} = \sin(x) + \sin(2x)$, for 
\[
\B = \{\sin(x), \sin(2x)\} \,,
\] 
then the values of the coordinate functions are $\dualfunc{1}{\B} (\mathbf{v}) = 1$ and $\dualfunc{2}{\B}(\mathbf{v}) = 1$, and for 
\[
\mathscr{C} = \left\{  \frac{1}{2} \sin (x) + \frac{1}{2} \sin(2x), \frac{1}{2} \sin(x) - \frac{1}{2} \sin(2x) \right\} \,,
\] 
the values of the coordinate functions are $\dualfunc{1}{\mathscr{C}}(\mathbf{v}) = 2$ and $\dualfunc{2}{\mathscr{C}}(\mathbf{v}) = 0$.\\ 

As another example, the dual basis corresponding to the standard basis of $\R^L$ consists of the projection functions $\pi_{\ell}$ or equivalently $p_{\ell}$ defined in section \ref{sec:vector-identification}. For each $\ell \in [L]$, one has that $\dualfunc{\ell}{L} (\mathbf{v}) = \pi_{\ell}(\mathbf{v}) = v_{\ell}$, i.e. it just returns the $\ell$'th entry of the vector $\mathbf{v}$. It is also true for higher-order tensors that the dual basis of the standard basis is very simple, consisting of coordinate functions which return the corresponding entry of the tensor.\\

For any basis $\B$ of any $M$-dimensional real vector space $\V$, the dual basis $\dual{\B}$ construction allows one to define an isomorphism identifying $\V$ with $\R^M$:
\[ 
\isodual{\B} : \V \to \R^M 
\]
called the \textbf{coordinate isomorphism} with respect to $\B$, using the rule of assignment:
\[
 \isodual{\B}: \mathbf{v} \mapsto \dualfunc{1}{\B}(\mathbf{v}) \unitvector[M]{1} + \dualfunc{2}{\B}(\mathbf{v})\unitvector[M]{2} + \cdots + \dualfunc{M}{\B} (\mathbf{v}) \unitvector{M} \,.
 \]
Therefore everything that can be said about $\V$ as a finite-dimensional real vector space can be explored by choosing a basis for $\B$ and subsequently working in the coordinate space $\R^M$.

\subsection{Coordinate Isomorphisms for Tensor Spaces}
\label{sec:coord-isom-tens}

This is largely a straightforward extension\footnote{Compare what follows with section 2 of \cite{uschmajew_pde} as well as Facts 12 and 13 from section 15-8 of \cite{lim_hla}.} of the coordinate isomorphism construction in $\finvect$, just modified to explicitly capture the $D$-fold tensor product structure for an object $\bT$ in $\fintensor{D}$. Specifically, given such a $\bT$ (cf. section \ref{sec:objects-categ-finite}), consider its associated $D$-tuple of vector spaces, $(\V_1, \dots, \V_D)$, and for each $d \in [D]$ define $M_d = \dim(\V_d)$. Choose, for each $d \in [D]$, a basis $\B_d$ for $\V_d$; this in turn defines $D$ dual basises $\dual{\B_d} = \{ \dualfunc{1}{\B_d}, \dots, \dualfunc{M_d}{\B_d} \}$, one for each $d \in [D]$.\\

Note that, since each $\B_d$ is a basis for $\V_d$, the set:  
\[  
\B \defequals \{ \mathbfcal{B}_{\m[D]} \defequals \mathbf{b}^{\smallone}_{m_1} \otimes_{\bT} \mathbf{b}^{\smalltwo}_{m_2} \otimes_{\bT} \cdots \otimes_{\bT} \mathbf{b}^{\smallsuper{D}}_{m_D}  : \mathbf{b}^{\smallone}_{m_1} \in \B_1, \dots, \mathbf{b}^{\smallsuper{D}}_{m_D} \in \B_d   \} 
 \]
is always a basis for $\bT$ \cite{hackbusch}\footnote{Specifically, cf. Lemma 3.11 of \cite{hackbusch}.}. In the above  expression, observe that $\otimes_{\bT}$ need not denote the Segre outer product, and refers to whatever the multilinear map:
\[ 
\V_1 \times \V_2 \times \cdots \times \V_D \overset{\otimes_{\bT}}{\longrightarrow} \bT
 \]
is which endows $\bT$ with its $D$-fold tensor product structure, cf. section \ref{sec:objects-categ-finite}.\\ 

Therefore, one may define a dual basis for $\dual{\bT}$ as 
\[ 
 \dual{\B} \defequals \{  \dualfunc{\m[D]}{\B}: \bT \to \R \,, \m[D] \in \M[D]  \} 
\]
where, to clarify, $\M[D] \defequals \bigtimes_{d=1}^D [M_d]$, and for each $\m[D] \in \M[D]$, the corresponding element of $\dual{\B}$ is defined via the formula $\dualfunc{\m[D]}{\B} (\mathbfcal{B}_{\mltidx[D]{\mu}}) = \delta_{\m[D] \mltidx[D]{\mu} }$, which again denotes the Kronecker delta, which equals $1$ if and only if $\m[D] = \mltidx[D]{\mu}$ and $0$ otherwise.\\

Observe that, if $\otimes_{func}$ denotes the tensor product of functions defined in section \ref{sec:tensor-product}, then for each multi-index $\m[D] \defequals (m_1, \dots, m_D)$, the corresponding dual basis element $\dualfunc{\m[D]}{\B}$ is the unique linear function $\bT \to \R$ guaranteed to exist by the universal property of the tensor product such that 
\[   
\dualfunc{m_1}{\B_1} \otimes_{func} \dualfunc{m_2}{\B_2} \otimes_{func} \cdots \otimes_{func} \dualfunc{m_D}{\B_D} = \dualfunc{\m[D]}{\B} \circ \otimes_{\bT} \,. 
  \]

Note that as before this dual basis gives one the coordinates of any tensor $\T \in \bT$ with respect to the basis $\B$. Specifically, one has for every $\T \in \bT$ the following:
\[  
\T = \sum_{\m[D]\in \M[D] } \dualfunc{\m[D]}{\B}(\T) \mathbfcal{B}_{\m[D]} = \sum_{\m[D]\in \M[D]} \dualfunc{\m[D]}{\B}(\T) \mathbf{b}_{m_1}^{\smallone} \otimes_{\bT} \cdots \otimes_{\bT} \mathbf{b}_{m_D}^{\smallsuper{D}}\,.
 \]
Note further that one can exploit multilinearity such that for any elementary tensor:
\[ 
\T = \mathbf{v}_1 \otimes_{\bT} \mathbf{v}_2 \otimes_{\bT} \cdots \otimes_{\bT} \mathbf{v}_D \,,
 \]
one has the following relationship:
\[  
\dualfunc{\m[D]}{\B}(\T) = \dualfunc{m_1}{\B_1}(\mathbf{v}_1)\dualfunc{m_2}{\B_2}(\mathbf{v}_2) \cdots \dualfunc{m_D}{\B_D}(\mathbf{v}_D) \,,
 \]
and therefore:
\[ 
 \T = \sum_{\m[D] \in \M[D] } \dualfunc{\m[D]}{\B}(\T) \mathbfcal{B}_{\m[D]} = \sum_{\m[D] \in \M[D]}  \left(  \prod_{d=1}^D \dualfunc{m_d}{\B_d}(\mathbf{v}_d)  \right) \mathbfcal{B}_{\m[D]} \,. 
 \]
Thus, for \textit{any} tensor $\T \in \bT$, given any outer product decomposition of $\T$:
\[ 
 \T = \sum_{k=1}^K \v{k}_1 \otimes_{\bT} \v{k}_2 \otimes_{\bT} \cdots \otimes_{\bT} \v{k}_D \,,
 \]
one has the relationship:
\[ 
\dualfunc{\m[D]}{\B}(\T) = \sum_{k=1}^K \left(  \prod_{d=1}^D \dualfunc{m_d}{\B_d}( \v{k}_d  ) \right)  \,,
 \]
so that finally:
\[ 
 \T = \sum_{\m[D] \in \M[D]} \left( \sum_{k=1}^K  \left(  \prod_{d=1}^D \dualfunc{m_d}{\B_d}( \v{k}_d   ) \right)  \right) \mathbf{b}_{m_1}^{\smallone} \otimes_{\bT} \mathbf{b}_{m_2}^{\smalltwo} \otimes_{\bT} \cdots \otimes_{\bT} \mathbf{b}_{m_D}^{\smallsuper{D}} \,.
  \]
The result does not depend on the particular outer product decomposition of $\T$ chosen due to the universal property of the $D$-fold tensor product satisfied (by assumption) by $\bT$ using $\otimes_{\bT}$.\\

Analogous to before, the existence of a dual basis $\dual{B}$ for the chosen basis $\B$ of $\bT$, which allows one to write the coordinates of any $\T \in \bT$ with respect to $\B$ using the elements of $\dual{B}$, also enables one to define a $D$-fold tensor space isomorphism:
\[  
\isodual{\B}: \bT \to \R^{\M[D]} \,, 
 \]
called again the \textbf{coordinate isomorphism}, via the rule of assignment:
\[ 
 \isodual{\B}: \T \mapsto \sum_{\m[D] \in \M[D]  } \dualfunc{\m[D]}{\B}(\T) \unittensor{\m[D]} = \sum_{ \m[D]  \in \M[D] } \dualfunc{\m[D]}{\B}(\T) \unitvector{m_1} \otimes \cdots \otimes \unitvector{m_D} \,,
  \]
where as usual $\unittensor{\m[D]}$ denotes the unit tensor (see section \ref{sec:tensors}) of $\R^{\M[D]}$ corresponding to the multi-index $\m[D]$, and $\otimes$ denotes the Segre outer product.\\

Observe that this $D$-fold tensor space coordinate isomorphism $\isodual{\B}$ is already completely determined by the $D$ respective vector space coordinate isomorphisms defined in section \ref{sec:coord-vector-spaces}
\[ 
 \isodual{\B_1}:\V_1 \to \R^{M_1} \,, \dots \,, \isodual{\B_d}: \V_d \to \R^{M_d} \,, \dots, \,, \isodual{\B_D}: \V_D \to \R^{M_D} \,,
 \]
since given any outer product decomposition of any tensor $\T \in \bT$, one has that:
\[ 
\begin{array}{rcl}
\isodual{\B}(\T)     &=&\displaystyle \sum_{ \m[D]  \in \M[D] } \left( \sum_{k=1}^K \left( \prod_{d=1}^D \dualfunc{m_d}{\B_d} ( \v{k}_d  ) \right) \right) \unitvector{m_1} \otimes \cdots \otimes \unitvector{m_D} \\
& = & \displaystyle\sum_{ \m[D]  \in \M[D] } \left( \sum_{k=1}^K \left( \prod_{d=1}^D \dualfunc{m_d}{\B_d} (  \v{k}_d  ) \right) \right) \unittensor{\m[D]} \\
& = &  \displaystyle\sum_{ \m[D]  \in \M[D] } \left( \sum_{k=1}^K \left( \prod_{d=1}^D  \left(\isodual{\B_d}(  \v{k}_d  ) \right)_{m_d}  \right) \right) \unittensor{\m[D]}  \,.
  \end{array} 
 \]
In fact, one even has that $\isodual{\B}$ is the unique linear function $\bT \to \R^{\M[D]}$ such that
\[ 
 \isodual{\B} \circ \otimes_{\bT} = \otimes \circ \left(  \isodual{\B_1} \times \cdots \times \isodual{\B_D}  \right) 
\]
which is guaranteed to exist by the universal property of the $D$-fold tensor product. Thus everything that can ever be said about $\bT$ as a $D$-fold tensor space can be explored by choosing basises for each of the $\V_1$, \dots, $\V_D$, and subsequently working in the coordinate space $\R^{\M[D]}$.

\section{Representing Dual Spaces with Coordinates}
\label{sec:repr-dual-spac}

Given a finite-dimensional real vector space $\V$ with $\dim(\V) = M$, one has that $\dim(\dual{\V}) = M$ (recall that $\dual{\V} \defequals \L(\V; \R)$), thus $\V$ and $\dual{\V}$ are isomorphic as real vector spaces. The most common choices of isomorphism used to identify them choose a basis $\B = \{ \mathbf{b}_1, \dots, \mathbf{b}_M \}$ for $\V$ and then associate each element $\mathbf{b}_m$ of $\B$ with the corresponding element $\dualfunc{m}{\B}$ of the dual basis $\dual{\B}$ of $\dual{\V}$. Specifically, the isomorphism $\idendual{\B}: \dual{\V} \to \V$ has the rule of assignment:
\[  
\idendual{\B}: \phi \mapsto \phi(\mathbf{b}_1)\mathbf{b}_1 + \phi(\mathbf{b}_2)\mathbf{b}_2 + \cdots + \phi(\mathbf{b}_M) \mathbf{b}_M
 \]
and has inverse $(\idendual{\B})^{-1}: \V \to \dual{\V}$ with rule of assignment:
\[ 
(\idendual{\B})^{-1}: \mathbf{v} \mapsto \dualfunc{1}{\B}(\mathbf{v})\dualfunc{1}{\B} + \dualfunc{2}{\B}(\mathbf{v})\dualfunc{2}{\B} + \cdots + \dualfunc{M}{\B}(\mathbf{v})\dualfunc{M}{\B} \,. 
\]
The fact that $\idendual{\B}$ and $(\idendual{\B})^{-1}$ as defined above are inverses can be seen using the fact that $\V$ is naturally isomorphic to its double dual $\dual{\dual{\V}}$ via the isomorphism $\mathbf{v} \mapsto (\phi \mapsto \phi(\mathbf{v}))$. (This isomorphism does \textit{not} depend on any choice of basis for $\V$, which is why it is possible for it to be natural\footnote{\label{natural_footnote}This is precisely defined using the concept of natural transformation, cf. footnote \ref{sec:repr-dual-spac} of appendix \ref{sec:tens-prod-funct}.} in $\V$.) The basis $\B$ is effectively also a dual basis for $\dual{\B}$, since for any $\phi \in \dual{\V}$:
\[ 
\phi = \phi(\mathbf{b}_1) \dualfunc{1}{\B} + \phi(\mathbf{b}_2)\dualfunc{2}{\B} + \cdots + \phi(\mathbf{b}_M)\dualfunc{M}{\B} \,, 
 \]
the functions $\phi \mapsto \phi(\mathbf{b}_m)$ give the coordinates of $\phi$ with respect to the basis $\dual{\B}$ for any $\phi \in \dual{\V}$.\\

Observe that the given isomorphism $\dual{\V} \overset{\idendual{\B}}{\underset{(\idendual{\B})^{-1}}{\rightleftarrows}} \V$ depends on a choice of basis for $\V$ (or equivalently a choice of basis for $\dual{\V}$), and thus is not natural\footnote{In fact there is no natural isomorphism between $\V$ and $\dual{\V}$, although discussing that observation in the detail it deserves would derail the discussion too far afield.} in $\V$. \\

Observe also that a choice of isomorphism $\dual{\V} \overset{\idendual{\B}}{\underset{(\idendual{\B})^{-1}}{\rightleftarrows}} \V$ is equivalent to choosing a bilinear function $\V \times \V \to \R$, in the sense that there is a one-to-one correspondence between identifications $\psi$ of $\dual{\V}$ with $\V$ and bilinear functions $\langle \cdot, \cdot \rangle_{\psi}: \V \times \V \to \R$ given by
\[ 
 \psi^{-1} : \mathbf{v} \mapsto \langle \mathbf{v}, \cdot \rangle_{\psi} 
 \]
and
\[ 
 \langle \cdot , \cdot \rangle_{\psi}: (\mathbf{v}_1, \mathbf{v}_2) \mapsto \left(  \psi^{-1}(\mathbf{v}_1) \right)(\mathbf{v}_2) \,.
 \]
This fact is related to the Riesz Representation theorem and the statement that Hilbert spaces are isomorphic to their duals, although discussing this in the detail it deserves would also derail the discussion too far afield. Note at the very least that the identification between $\L (\V; \dual{V} )$ and $\L(\V, \V; \R)$ is not just a bijection, but also a linear isomorphism.\\

Usually the vector space $\V$ under discussion will be $\R^M$, with the chosen basis the standard basis $\B = \{ \unitvector[M]{1}, \dots, \unitvector{M} \}$. Note that the dual basis to the standard basis consists of the coordinate projection functions $\pi_m$ as defined in section \ref{sec:vector-identification}. For each $m \in [M]$, the element $\dualfunc{m}{M}$ of the dual basis to the standard basis corresponding to $\unitvector{m}$  gives the $m$'th entry of the vector, $\dualfunc{m}{M}(\mathbf{v}) = v_m$. The identification of $\R^M$ with $\dual{\R^M}$ via the identification of the standard basis $\B$ of $\R^M$ with its dual basis $\dual{\B}$ in $\dual{\R^M}$ also corresponds to the standard dot product $\bullet$ on $\R^M$.\\ 

Let $\dualfunc{1}{M}, \dots, \dualfunc{M}{M}$ denote the dual basis $\dual{\B}$ of $\dual{\R^M}$ corresponding to the standard basis\\ 
${\B = \{ \unitvector[M]{1}, \dots, \unitvector{M} \} }$ of $\R^M$, i.e. for all $m \in [M]$:
\[ 
\dualfunc{m}{M} (\unitvector[M]{\mu}) = \delta_{m\mu} \,.
 \]
Denote the isomorphism $\dual{\R^M} \to \R^M$ defined via sending the standard dual basis $\dual{\B}$ of $\dual{\R^M}$ to the standard basis $\B$ of $\R^M$ by $\idendual{M}$, so that for all $m \in [M]$:
\[ 
\idendual{M}: \dualfunc{m}{M} \mapsto \unitvector{m} \,. 
 \]
Then one has that, for every $\mathbf{v}_1, \mathbf{v}_2 \in \R^M$:
\[ 
 \mathbf{v}_1 \bullet \mathbf{v}_2 = \left(  \left( \idendual{M} \right)^{-1} (\mathbf{v}_1) \right)(\mathbf{v}_2) \,, \quad \left( \idendual{M}  \right)^{-1} : \mathbf{v} \mapsto ( \mathbf{w} \mapsto \mathbf{v} \bullet \mathbf{w} ) \,.
 \]

Given a vector space $\V$ and a choice of basis $\B$ for $\V$, thus also a choice of basis $\dual{\B}$ for $\dual{\V}$, one can define a ``dual coordinate isomorphism'' $\dual{\isodual{\B}} : \dual{\V} \to \dual{\R^M}$ such that for all $m \in [M]$:
\[
  \dual{\isodual{\B}}: \dualfunc{m}{\B} \mapsto \dualfunc{m}{M} \,.
 \]
For any basis $\B$ of any $M$-dimensional real vector space $\V$ the following is also true:
\[ 
 \idendual{M} \circ \dual{\isodual{\B}} = \isodual{\B} \circ \idendual{\B} \,,
 \]
and since all functions in question are invertible:
\[
\begin{array}{rcl}
\isodual{\B}    & = & \idendual{M} \circ \dual{\isodual{\B}} \circ \left(  \idendual{\B} \right)^{-1} \,, \\
\dual{\isodual{\B}} & = & \left( \idendual{M}  \right)^{-1} \circ \isodual{\B} \circ \idendual{\B} \,, \\
\idendual{M} & = & \isodual{\B} \circ \idendual{\B} \circ \left( \dual{\isodual{\B}} \right)^{-1} \,, \\
\idendual{\B} & = & \left(  \isodual{\B} \right)^{-1} \circ \idendual{M} \circ \dual{\isodual{\B}} \,. 
  \end{array}
\]
Therefore all questions regarding the identification of any $M$-dimensional vector space $\V$ with its dual $\dual{\V}$, using any choice of basis $\B$ for $\V$, can always be reduced to the study of the identification of $\R^M$ with its dual $\dual{\R^M}$ using its standard basis. Thus, in what follows the discussion will be restricted exclusively to the identification of $\R^M$ with $\dual{\R^M}$ via the standard basis of $\R^M$ and its dual basis, with it implicit that the restricted discussion can readily be generalized from coordinate spaces $\R^M$ with the standard basises to arbitrary finite-dimensional real vector spaces with arbitrary basises via the use of the coordinate isomorphisms and dual coordinate isomorphisms.

\section{Representing Linear Functions with Coordinates}
\label{sec:repr-line-funct}

Given the standard identification $\idendual{M}: \dual{\R^M} \to \R^M$ from appendix \ref{sec:repr-dual-spac}, it is possible not only to identify $\L (\R^M; \R) = \dual{\R^M}$ with $\R^M$, it is also possible to identify the space of all linear functions $\R^M \to \R^N$, denoted $\L(\R^M; \R^N)$, with $\R^{[N] \times [M]}$. Specifically, given a linear function $\L : \R^M \to \R^N$, one may focus attention onto the values it takes for the standard basis of $\R^M$, i.e. $\L(\unitvector{m}) \in \R^N$ for all $m \in [M]$. Doing this one then discovers the following for any $\mathbf{v} \in \R^M$:
\[  
\L : \mathbf{v} \mapsto \sum_{m=1}^M \dualfunc{m}{M} (\mathbf{v}) \L (\unitvector{m}) = \sum_{m=1}^M v_m \L (\unitvector{m}) \,. 
\]
One can then show that $\L(\R^M; \R^N)$ satisfies the universal property of the tensor product for $\left(\R^N, \dual{\R^M}\right)$ by using the tensor product defined for each $\mathbf{w} \in \R^N$, $\phi \in \dual{\R^M}$ via the rule:
\[ 
 \bigotimes_{\L(\R^M; \R^N)} : (\mathbf{w}, \phi) \mapsto ( \mathbf{v} \mapsto \phi(\mathbf{v}) \mathbf{w} ) \,.  
\]
Hence one usually chooses to identify $\L \in \L(\R^M; \R^N)$ with the following tensor $\mathbf{L} \in \R^{N \times M}$:
\[  
\mathbf{L} = \sum_{m=1}^M \L (\unitvector{m}) \otimes \idendual{M}( \dualfunc{m}{M} ) = \sum_{m=1}^M \L (\unitvector{m} ) \otimes \unitvector{m} \,,
 \]
where in the above equation $\otimes$ denotes the Segre outer product.\\ 

The above identification is a tensor product of linear maps as defined in appendix \ref{sec:morph-categ-finite}:
\[ 
\operatorname{Id}_{\R^N} \otimes \idendual{M}: \L(\R^M; \R^N) = \R^N \underset{{\L(\R^M; \R^N)}}{\otimes} \dual{\R^M} \to \R^N \otimes \R^M = \R^{N \times M} \,,
  \]
which is an isomorphism in $\fintensor{2}$ as it is the tensor product of  two vector space isomorphisms.\\

$\mathbf{L}$ is the matrix whose $m$'th column  $\mathbf{L}_{:m}$ is $\L(\unitvector{m})$ represented with respect to the standard basis of $\R^N$. This leads to the typical\footnote{E.g. as opposed to say something unconventional like Cracovian notation. For more about Cracovians, see \cite{cracovian}.} definition of matrix multiplication: for any $\mathbf{v} \in \R^M$,
\[ 
  \sum_{m=1}^M v_m \L (\unitvector{m}) = \sum_{m=1}^M v_m \mathbf{L}_{:m}  = \mathbf{L} \bullet_2 \mathbf{v} = \L (\mathbf{v}) \,.
 \]

\subsection{Linear Functions with Vector Inputs and Tensor Outputs}
\label{sec:vec-in-tens-out}

Clearly one can extend this identification further to linear functions $\L \in \L(\R^M; \R^{\N[O]})$, since analogous to the relationship above one has for every $\mathbf{v} \in \R^M$ that
\[ 
 \L : \mathbf{v} \mapsto \sum_{m=1}^M \dualfunc{m}{M}(\mathbf{v}) \L(\unitvector{m}) = \sum_{m=1}^M v_m \L(\unitvector{m}) \,, 
\] 
Using $\idendual{M}$ one may identify $\L \in \L(\R^M; \R^{\N[O]})$ with an order $(O+1)$ tensor:
\[  
\LL = \sum_{m=1}^M \L(\unitvector{m}) \otimes \idendual{M}( \dualfunc{m}{M}) = \sum_{m=1}^M \L (\unitvector{m}) \otimes  \unitvector{m} \,,
\]
where as before $\otimes$ denotes the Segre outer product in the above expression. $\LL$ is a tensor whose $m$'th hyperslice of shape $\N[O]$ (corresponding to the first $O$ modes) $\LL_{:_{[O]}m}$ is the value of $\L(\unitvector{m})$ represented with respect to the standard basis of unit tensors $\unittensor{\n[O]}$ of $\R^{\N[O]}$.\\ 

Once more analogous to the relationship above, the value of $\L$ applied to any vector $\mathbf{v} \in \R^M$ can be expressed via contraction with the final (in this case $(O+1)$'th) mode of the tensor $\LL$:
\[ 
\summing{m} v_m \L(\unitvector{m} ) = \summing{m} v_m \LL_{:_{[O]}m}  = \LL \bullet_{O+1} \mathbf{v} = \L(\mathbf{v}) \,. \]
And also again analogous to the above, $\L(\R^M; \R^{\N[O]})$ satisfies the universal property of the tensor product for the pair of vector spaces $\left(\R^{\N[O]}, \dual{\R^M} \right)$, when using the tensor product defined for each $\mathbfcal{U} \in \R^{\N[O]}$ and $\phi \in \dual{\R^M}$ via the rule of assignment:
\[  
\bigotimes_{\L(\R^M; \R^{\N[O]})} : (\mathbfcal{U}, \phi) \mapsto (\mathbf{v} \mapsto \phi(\mathbf{v}) \mathbfcal{U} ) \,. 
\]
The above identification of $\L(\R^M; \R^{\N[O]})$ with $\R^{\N[O] \times [M]}$ is a tensor product of linear maps:
\[ 
 \operatorname{Id}_{\R^{\N[O]}} \otimes \idendual{M} : \L( \R^M; \R^{\N[O]}) = \R^{\N[O]} \underset{\L(\R^M; \R^{\N[O]})}{\otimes} \dual{\R^M} \to \R^{\N[O]} \otimes \R^M = \R^{\N[O] \times [M]} \,, 
\]
which again is an isomorphism in $\fintensor{2}$ as a result of being the tensor product of two vector space isomorphisms as defined in appendix \ref{sec:morph-categ-finite}.

\subsection{Linear Functions with Tensor Inputs and Tensor Outputs}
\label{sec:tens-in-tens-out}

Moreover, this procedure extends yet further to identify linear functions $\L (\R^{\M[D]} ; \R^{\N[O]})$ with tensors in $\R^{\N[O] \times \M[D]}$ once one establishes a standard way of identifying $\dual{\R^{\M[D]}}$ with $\R^{\M[D]}$, since $\L ( \R^{\M[D]}; \R^{\N[O]})$ satisfies the universal property of the tensor product (and thus is an object of $\fintensor{2}$) for the pair of vector spaces $\left(\R^{\N[O]} ,\dual{\R^{\M[D]}}\right)$ when using the tensor product defined for every $\mathbfcal{U} \in \R^{\N[O]}$ and $\phi \in \dual{\R^{\M[D]}}$ via the relationship:
\[
 \bigotimes_{\L(\R^{\M[D]}; \R^{\N[O]})} : (\mathbfcal{U}, \phi) \mapsto (\T \mapsto \phi(\T) \mathbfcal{U}) \,.
 \]
Note that the object of $\fintensor{2}$ which is most commonly mentioned as the representative vector space of those which satisfy the universal property of the tensor product for the pair\endnote{Or technically speaking rather usually as the object of $\fintensor{D+1}$ satisfying the universal property of the tensor product for the $(D+1)$-tuple of vector spaces ${(\R^{\N[O]}, \dual{\R^{M_1}}, \dual{\R^{M_2}}, \dots, \dual{\R^{M_D}}  )}$, or as the object of $\fintensor{O+D}$ satisfying the universal property of the tensor product for the $(O+D)$-tuple of vector spaces\\ ${ (  \R^{N_1}, \dots, \R^{N_O}, \dual{\R^{M_1}}, \dots, \dual{\R^{M_D} }   )}$.  } $\R^{\N[O]}$ and $\dual{\R^{\M[D]}}$ is usually $\L(\R^{M_1}, \dots, \R^{M_D};\R^{\N[O]})$, the space of multilinear functions:
\[ 
 \R^{M_1} \times \cdots \times \R^{M_D} \to \R^{\N[O]} \,,
 \]
cf. Chapter 8 of \cite{vinberg}. However, the space of linear functions $\L(\R^{\M[D]}; \R^{\N[O]})$ and the space of multilinear functions $\L(\R^{M_1}, \dots, \R^{M_D};\R^{\N[O]})$ are isomorphic vector spaces, with isomorphism:
\[ 
 \Upsilon: \L(\R^{M_1}, \dots, \R^{M_D};\R^{\N[O]}) \to \L (\R^{\M[D]}; \R^{\N[O]}) \,, \quad \Upsilon: \mathscr{M} \mapsto \L \,,
 \]
where for a given multilinear function $\mathscr{M} \in \L(\R^{M_1}, \dots, \R^{M_D};\R^{\N[O]})$, $\L$ denotes the unique linear function $\L: \R^{\M[D]} \to \R^{\N[O]}$ such that $\mathscr{M} = \L \circ \bigotimes$ (where $\bigotimes$ denotes the Segre outer product $\R^{M_1} \times \cdots \times \R^{M_D} \to \R^{\M}$) which is guaranteed to exist due to the fact that $\R^{\M}$, together with the Segre outer product, satisfies the universal property of the tensor product for the $D$-tuple of vector spaces $(\R^{M_1} ,\dots, \R^{M_D})$.\\ 

It is therefore just as legitimate\endnote{Recall that for any two vector spaces $\bT_1$, $\bT_2$ which are isomorphic (as vector spaces) via the isomorphism $\Upsilon: \bT_2 \to \bT_1$, and for which $\bT_1$ satisfies the universal property of the $D$-fold tensor product for vector spaces $(\V_1, \dots, \V_D)$ using the tensor product $\bigotimes_{\bT_1}$, it is always the case that $\bT_2$ then also satisfies the universal property of the $D$-fold tensor product for $(\V_1, \dots, \V_D)$ using the tensor product $\bigotimes_{\bT_2} \defequals \Upsilon^{-1} \circ \bigotimes_{\bT_1}$. Cf. endnote \ref{tensor-universal-property-uniqueness} of section \ref{sec:tensor-product}.} to define tensors of mixed variance $(O,D)$ as being linear functions $\L : \R^{\M[D]} \to \R^{\N[O]}$ as it is to use the more common definition of tensors of mixed variance $(O,D)$, namely as being multilinear functions $\mathscr{M} \in \L(\R^{M_1}, \dots, \R^{M_D};\R^{\N[O]})$.\\

Recall from section \ref{sec:coord-isom-tens} that\footnote{Technically that section discussed a more general situation. What is discussed here is the special case where one has that $\bT = \R^{\M[D]}$, $\V_1 = \R^{M_1}$, \dots, $\V_D = \R^{M_D}$.} that the standard basis of unit tensors $\B$ for $\R^{\M[D]}$ is
\[  
\B = \{  \unittensor{\m[D]} = \unitvector{m_1} \otimes \cdots \otimes \unitvector{m_D}: m_1 \in [M_1] , \dots, m_D \in [M_D] \} \,,
 \]
in other words the standard basis of unit tensors for $\R^{\M[D]}$ consists of all possible combinations of the $D$-fold Segre outer product applied to the standard basis vectors of $\R^{M_1}$, \dots, $\R^{M_D}$, respectively.\\ 

Note that for the dual basis $\dual{\B} \subset \dual{\R^{\M[D]}}$ of $\B$, each element is simply the linear function $\R^{\M[D]} \to \R$ which returns the corresponding entry of the tensor: for each $\m[D] \in \M[D]$, for every $\T \in \R^{\M[D]}$ one has that $\dualfunc{\m[D]}{\M[D]}(\T) = \mathcal{T}_{\m[D]}$, the $\m[D]$'th entry of $\T$. Moreover, the following is also true for the dual basis $\dual{\B}$ of $\dual{\R^{\M[D]}}$ corresponding to the standard basis $\B$ of $\R^{\M[D]}$: for any tensor $\T \in \R^{\M[D]}$, and any outer product decomposition
\[ 
\T = \summing{k}  \v{k}_1 \otimes \cdots \otimes \v{k}_D \,, \quad \text{where for all }k\in[K], \v{k}_1 \in \R^{M_1}\,, \dots\,, \v{k}_D \in \R^{M_D} \,, 
\]
then for each multi-index $\m[D] \in \M[D]$ the following is true of $\dualfunc{\m[D]}{\M[D]}$:
\[  
\dualfunc{\m[D]}{\M[D]}(\T) = \mathcal{T}_{\m[D]} = \summing{k} \left( \prod_{d=1}^D \dualfunc{m_d}{M_d} ( \v{k}_d) \right) = \summing{k} \left( \prod_{d=1}^D ( \v{k}_d )_{m_d}  \right) \,.  
\]
As before, $\dualfunc{\m[D]}{\M[D]}$ denotes the element of the dual basis for $\dual{\R^{\M[D]}}$ corresponding to the unit tensor $\unittensor{\m[D]} \in \R^{\M[D]} = \unitvector{m_1} \otimes \cdots \otimes \unitvector{m_D}$ in the standard basis for $\R^{\M[D]}$, and $\dualfunc{m_d}{M_d}$ denotes the element of the standard dual basis for $\dual{\R^{M_d}}$ corresponding to the unit vector $\unitvector[M_d]{m_d}$ in the standard basis for $\R^{M_d}$. The above result does not depend on the chosen outer product decomposition for $\T$ due to the universal property of the $D$-fold tensor product.\\

Thus, given any linear function $\L \in \L(\R^{\M[D]}; \R^{\N[O]})$, one has that for every $\T \in \R^{\M[D]}$:
\[ 
\L : \T \mapsto \sum_{\m[D] \in \M[D]} \dualfunc{\m[D]}{\M[D]}(\T) \L (\unittensor{\m[D]}) = \sum_{\m[D] \in \M[D] } \mathcal{T}_{\m[D]} \L (\unittensor{\m[D]})\,. 
\]
Moreover, for every outer product decomposition of $\T$:
\[  
\T = \summing{k} \v{k}_1 \otimes \cdots \otimes \v{k}_D \,, 
\]
one can write additionally that:
\[ 
 \L : \T \mapsto \sum_{\m[D] \in \M[D]} \left(  \summing{k} \left( \prod_{d=1}^D \dualfunc{m_d}{M_d}( \v{k}_d   ) \right)  \right) \L(\unittensor{\m[D]}) = \sum_{\m[D] \in \M[D]} \left( \summing{k} \left( \prod_{d=1}^D ( \v{k}_d   )_{m_d} \right) \right) \L (\unittensor{\m[D]}) \,. 
\]
As a result, one then generally chooses to identify $\L$ with the following tensor $\LL \in \R^{\N[O] \times \M[D]}$:
\[ 
\LL = \sum_{\m[D] \in \M[D]} \L(\unittensor{\m[D]}) \otimes \idendual{\M[D]} ( \dualfunc{\m[D]}{\M[D]}) = \sum_{\m[D] \in \M[D]} \L(\unittensor{\m[D]}) \otimes \unittensor{\m[D]} \,, 
 \]
where $\idendual{\M[D]}: \dual{\R^{\M[D]}} \to \R^{\M[D]}$ denotes the identification of $\R^{\M[D]}$ with its dual $\dual{\R^{\M[D]}}$ which sends each $\dualfunc{\m[D]}{\M[D]} \in \dual{\B}$ of the standard dual basis of $\dual{\R^{\M[D]}}$ to the corresponding element $\unittensor{\m[D]} \in \B$ of the standard basis  of unit tensors of $\R^{\M[D]}$, analogous to the identification defined in appendix \ref{sec:repr-dual-spac} above. Here $\otimes$ also denotes the Segre outer product.\\

$\LL$ is the order $(O+D)$ tensor whose $\m[D]$'th hypersplice of shape $\N[O]$, $\LL_{:_{[O]} \m[D] } $, is the value of $\L$ applied to the $\m[D]$'th unit tensor $\unittensor{\m[D]}$. Therefore one has that, for any $\T \in \R^{\M[D]}$:
\[  
\sum_{ \m[D]\in \M[D]} \mathcal{T}_{\m[D]} \L(\unittensor{\m[D]}) = \sum_{\m[D]\in \M[D]} \mathcal{T}_{\m[D]} \LL_{:_{[O]} \m[D] } = \LL \bullet_{[O+D]\setminus [O]} \T = \L (\T) \,. 
 \]
Given any outer product decomposition $\T = \summing{k} \v{k}_1 \otimes \cdots \otimes \v{k}_D$ of $\T$, the above also equals\footnote{Compare this calculation with Fact 12 of section 15-1 in \cite{lim_hla}.}
\[ 
 \summing{k} \LL \bullet_{O+1} \v{k}_1 \bullet_{O+2} \v{k}_2 \cdots \bullet_{O+D} \v{k}_D \,.
 \]

As seen above, tensor contractions amount to applying linear functions between coordinate spaces. For example, using the notation from \cite{mccullagh}, which is usually called Einstein notation, the symbol:
\[ 
 \mathcal{T}^{m_1m_2}_{m_3m_4m_5}
 \]
corresponds to the standard contravariant representation (as defined above) of some linear function $\R^{[M_3] \times [M_4] \times [M_5]} \to \R^{[M_1] \times [M_2]}$. The contraction written in Einstein notation as
\[  
\mathcal{T}^{m_1m_2}_{m_3m_4m_5} \mathcal{U}^{m_3 m_4 m_5}
 \]
corresponds to the result of applying the aforementioned linear function to some $\mathbfcal{U} \in \R^{[M_3] \times [M_4] \times [M_5]}$. Finally, the contraction written in Einstein notation as:
\[  
\mathcal{T}^{m_1m_2}_{m_3m_4m_5} \mathcal{S}^{m_3 m_4 m_5}_{m_6 m_7} 
\]
corresponds to composing another linear function in $\L(\R^{[M_6] \times [M_7]} ; \R^{[M_3] \times [M_4] \times [M_5]})$ with the aforementioned linear function in $\L(\R^{[M_3] \times [M_4] \times [M_5]}; \R^{[M_1] \times [M_2]})$. Of course the above three examples don't provide a direct interpretation for all possible types of contractions which can be written using Einstein notation, e.g. the very existence of the notion of tensor network implies the existence of far more complicated patterns, but it is a start.\\

Observe that, analogous to previous examples, the above identification of $\L(\R^{\M[D]}; \R^{\N[O]})$ with $\R^{\N[O] \times \M[D]}$ corresponds to the tensor product of linear maps:
\[ 
\operatorname{Id}_{\R^{\N[O]}} \otimes \idendual{\M[D]} : \L (\R^{\M[D]};\R^{\N[O]}) = \R^{\N[O]} \underset{\L(\R^{\M[D]};\R^{\N[O]})}{\otimes} \dual{\R^{\M[D]}} \to \R^{\N[O]} \otimes \R^{\M[D]} = \R^{\N[O] \times \M[D]} \,,
 \]
which again is a morphism in $\fintensor{2}$ whose existence and unique definition can be inferred from the universal property of the tensor product.

%%% Local Variables:
%%% mode: latex
%%% TeX-master: "../new_notes_draft"
%%% End:

\section{Kronecker Products and Coordinate Representation of the Tensor Product of Linear Transformations}
\label{sec:kron-prod-coord}

Recall from section \ref{sec:tens-prod-line} the definition of the tensor product
\[ 
\bigotimes: \L(\R^{\MI}; \R^{\NI}) \times \L (\R^{\MII};\R^{\NII}) \to \L( \R^{\MI \times \MII}; \R^{\NI \times \NII} )  
  \]
given the rule of assignment:
\[ 
 \bigotimes: (\L_1, \L_2) \mapsto
  \begin{array}{l}
\left( \T     =  \displaystyle\summing{k} \v{k}_1 \otimes \cdots \otimes \v{k}_{O_1} \otimes \v{k}_{O_1 +1} \otimes \cdots \otimes \v{k}_{O_1 + O_2}  \right.\\
\left.\phantom{(()\T = }\mapsto\displaystyle \summing{k} \L_1 (\v{k}_1 \otimes \cdots \otimes \v{k}_{O_1}) \otimes \L_2 (\v{k}_{O_1+1} \otimes \cdots \otimes \v{k}_{O_1+O_2}) \right) \,.
  \end{array}
 \]

In section \ref{sec:repr-line-funct} it was explained that the standard way to represent ${\L_1 \in \L(\R^{\MI}; \R^{\NI} )}$ as a contravariant tensor is as the following tensor $\LL_1 \in \R^{\NI \times \MI}$:
\[ 
\LL_1 =  \sum_{ \mi \in \MI} \L_1 ( \unittensor{\mi}) \otimes \unittensor{\mi} \,,   
\] 
and the standard way to represent $\L_2 \in \L(\R^{\MII}; \R^{\NII})$ as a contravariant tensor is as the following tensor $\LL_2 \in \R^{\NII \times \MII}$:
\[  
\LL_2 = \sum_{\mii \in \MII} \L_2(\unittensor{\mii}) \otimes \unittensor{\mii} \,. 
\]
Thus one might be tempted to conclude that $\LL_1 \otimes \LL_2$ would be the contravariant representation of $\L_1 \otimes \L_2$, but that is not true. Most noticeably, while $\LL_1 \otimes \LL_2$ is a tensor in $\R^{\NI \times \MI \times \NII \times \MII}$, the pattern in section \ref{sec:repr-line-funct} gives the contravariant representation of $\L_1 \otimes \L_2$ as the following tensor $\LL$ in $\R^{\NI \times \NII \times \MI \times \MII}$:
\[
  \begin{array}{rcl}
   \LL & = &\displaystyle \sum_{\mi  \mii \in \MI \times \MII} (\L_1 \otimes \L_2) ( \mathbfcal{I}_{\mi  \mii}) \otimes  \mathbfcal{I}_{\mi  \mii} \\
& = &\displaystyle \sum_{\mi  \mii \in \MI \times \MII} (\L_1 \otimes \L_2) ( \unittensor{\mi} \otimes \unittensor{\mii} ) \otimes \unittensor{\mi} \otimes \unittensor{\mii} \\
& = &\displaystyle \sum_{\mi  \mii \in \MI \times \MII}  \L_1 (\unittensor{\mi}) \otimes \L_2 ( \unittensor{\mii}) \otimes \unittensor{\mi} \otimes \unittensor{\mii} \,.
  \end{array}
\]
In contrast, one has that $\LL_1 \otimes \LL_2$ equals
\[ 
\LL_1 \otimes \LL_2 = \sum_{\mi  \mii \in \MI \times \MII}  \L_1 (\unittensor{\mi}) \otimes \unittensor{\mi} \otimes \L_2 ( \unittensor{\mii}) \otimes \unittensor{\mii} \,. 
\]
Comparing the above two outer product representations, it is clear that, although $\LL \not = \LL_1 \otimes \LL_2$, all of the information necessary to compute $\LL$ is still present in $\LL_1$ and $\LL_2$. There exists a well-defined bilinear function which, given $\LL_1$ and $\LL_2$, will return the corresponding $\LL$:
\[ 
\bigotimes_Z : \R^{\NI \times \MI} \times \R^{\NII \times \MII} \to \R^{\NI \times \NII \times \MI \times \MII} \,,
 \]
which can be seen to differ from the usual Segre outer product, and yet also be recovered from the Segre outer product by applying a suitable permutation of the modes.\\

Given two arbitrary outer product decompositions of $\T^{\smallone} \in \R^{\NI \times \MI}$ and $\T^{\smalltwo} \in \R^{\NII \times \MII}$:
\[
  \begin{array}{rcl}
   \T^{\smallone} & = & \displaystyle \summing{k_1} (\w{k_1}_1 \otimes \cdots \otimes \w{k_1}_{D_1}   ) \otimes ( \v{k_1}_1 \otimes \cdots \otimes \v{k_1}_{O_1} )\\
\smallskip\T^{\smalltwo} & =  & \displaystyle \summing{k_2} ( \w{k_2}_1 \otimes \cdots \otimes \w{k_2}_{D_2}  ) \otimes (\v{k_2}_1 \otimes \cdots \otimes \v{k_2}_{O_2} ) \,,
  \end{array}
\]
and contiguous partitions\footnote{Which are already implicitly specified by writing the multi-index sets this way, but to be unambiguous, $\{  \{ N_1^{\smallone}, \dots, N_{D_1}^{\smallone} \}, \{ M_1^{\smallone}, \dots, M_{O_1}^{\smallone} \} \}$ and $\{ \{ N_1^{\smalltwo}, \dots, N_{D_2}^{\smalltwo}  \}, \{ M_1^{\smalltwo}, \dots, M_{O_2}^{\smalltwo}  \}  \}$.} of the mode index sets of $\NI \times \MI$ and $\NII \times \MII$, then this bilinear function, to be called the \textbf{Zehfuss product}\footnote{Johann Georg Zehfuss described the Kronecker product of matrices in 1858, the first person known to do so.}, applied to $\T^{\smallone}$ and $\T^{\smalltwo}$ has the value:
\[ 
 \bigotimes_Z ( \T^{\smallone}, \T^{\smalltwo} ) = \summing{k_1} \summing{k_2} (\w{k_1}_1 \otimes \cdots \otimes \w{k_1}_{D_1}) \otimes (\w{k_2}_1 \otimes \cdots \otimes \w{k_2}_{D_2}) \otimes (\v{k_1}_1 \otimes \cdots \otimes \v{k_1}_{O_1}) \otimes (\v{k_2}_1 \otimes \cdots \otimes \v{k_2}_{O_2}) \,, 
\]
whereas the standard Segre outer product would return the following tensor:
\[ 
\T^{\smallone} \otimes  \T^{\smalltwo}  = \summing{k_1} \summing{k_2} (\w{k_1}_1 \otimes \cdots \otimes \w{k_1}_{D_1}) \otimes (\v{k_1}_1 \otimes \cdots \otimes \v{k_1}_{O_1}) \otimes (\w{k_2}_1 \otimes \cdots \otimes \w{k_2}_{D_2})  \otimes (\v{k_2}_1 \otimes \cdots \otimes \v{k_2}_{O_2}) \,.
 \]
Although the contiguous partitions of the modes of $\T^{\smallone}$ and $\T^{\smalltwo}$ have been suppressed from the notation for the Zehfuss product given in the above equations, the contiguous partitions \textit{must} be specified somehow before the Zehfuss product can be well-defined.\\ 

In the case that both tensors are the standard contravariant representation of linear functions between coordinate spaces, as in the examples above, the contiguous partitions of the modes into two blocks can be considered implicitly specified by the requirements that (i) the first block of modes corresponds to the function's codomain and (ii) the second block of modes corresponds to the function's domain. Yet even in that case it is necessary to make explicit \textit{which} linear function between coordinate spaces the tensor is intended to represent.\\

For example, already for an order-$2$ tensor (matrix) $\mathbf{L} \in \R^{H \times W}$ can be considered, in an equally valid manner\footnote{Note that switching between any of these interpretations corresponds to the operations of ``raising indices'' and ``lowering indices'' in Einstein tensor notation, i.e. that used in e.g. \cite{mccullagh} or \cite{amari}.}, the contravariant representation of any of the following linear functions:
\begin{enumerate}
\item a linear function $\mathscr{L}_{|12}:\R \to \R^{H \times W}$ such that $\L_{|12} (1) = \mathbf{L}$, 
\item a linear function $\L_{1|2}: \R^W \to \R^H$ such that $\L_{1|2} (\mathbf{v}) = \mathbf{L} \bullet_2 \mathbf{v}$,
\item a linear function\footnote{This is the unique linear function $\L: \R^{H \times W} \to \R$ such that $\L \circ \otimes = \mathscr{M}$, where $\mathscr{M}$ is the bilinear function(/form) $\R^{H} \times \R^W \to \R$ such that $\mathscr{M} (\mathbf{v}, \mathbf{w}) = \mathbf{L} \bullet_1 \mathbf{v} \bullet_2 \mathbf{w}$. Cf. the discussion in section \ref{sec:tens-in-tens-out}, i.e. $\L = \Upsilon(\mathscr{M})$.} $\L_{12|}:\R^{H \times W} \to \R$ such that $\L_{12|}(\mathbf{T}) = \langle \mathbf{L}, \mathbf{T} \rangle$.
\end{enumerate}
In general, an order $O$ tensor can be interpreted, in an equally valid manner, as the contravariant representation of $(O+1)$ distinct linear functions, even after having chosen a single ``standard'' way of identifying linear functions with any domain or codomain as tensors. Thus simply stating that two tensors represent linear functions is by itself insufficient to implicitly specify the contiguous partitions of modes needed to make the Zehfuss product well-defined; the domains and codomains of those linear functions must also be specified.\\

Observe that the map $\L(\R^{\MI \times \MII} ; \R^{\NI \times \NII})$ to $\R^{\NI \times \NII \times \MI \times \MII}$ which returns the standard contravariant representation of a linear function corresponds to the morphism in $\fintensor{2}$ which is the tensor product of (i) the standard contravariant representation of $\L(\R^{\MI}; \R^{\NI})$ with tensors in $\R^{\NI \times \MI}$ and (ii) the standard contravariant representation of $\L(\R^{\MII}; \R^{\NII})$ with tensors in $\R^{\NII \times \MII}$, \textit{as long as} the chosen tensor product:
\[  
\R^{\NI \times \MI} \times \R^{\NII \times \MII} \to \R^{\NI \times \NII \times \MI \times \MII} 
 \]
is the corresponding Zehfuss product, and \textit{not} the standard Segre outer product. When using $\zeta$ to denote identifications of linear functions with corresponding contravariant tensors, one has the following identity (cf. definition of tensor product of linear functions in section \ref{sec:morph-categ-finite}):
\[  
\bigotimes_Z \circ ( \zeta_{\MI ; \NI} \times \zeta_{\MII ; \NII}  ) = \zeta_{\MI \times \MII ; \NI \times \NII } \circ \bigotimes_{\L(\R^{\MI \times \MII}; \R^{\NI \times \NII})} \,. 
 \]

The definition of Zehfuss product can be extended to more than $2$ tensors (each with a specified contiguous partitions of its modes) in a straightforward way. The most common example in the practice is the case of $O$ order $2$ tensors, each being the contravariant representation $\mathbf{L}_o$ of a linear function in $\L_o \in \L(\R^{M_o}; \R^{N_o})$. Then the $O$-fold Zehfuss product of these tensors is the order $2O$ tensor given by the outer product decomposition:
\[
  \begin{array}{rcl}
   \displaystyle\bigotimes_Z (\mathbf{L}_1, \dots, \mathbf{L}_O) & = &\displaystyle \summing{m_1}\cdots \summing{m_O} \L_1 (\unitvector{m_1}) \otimes \cdots \otimes \L_o (\unitvector{m_O}) \otimes \unitvector{m_1} \otimes \cdots \otimes \unitvector{m_O} \\
& = &\smallskip \displaystyle \summing{m_1}\cdots \summing{m_O} (\mathbf{L}_1)_{:m_1} \otimes \cdots \otimes (\mathbf{L}_O)_{:m_O} \otimes \unitvector{m_1} \otimes \cdots \otimes \unitvector{m_O} \,,
  \end{array}
\]
since this is the (standard) contravariant representation of the linear transformation $\L \defequals \bigotimes_{o=1}^O \L_o$.\\

Defining $\M = \bigtimes_{o=1}^O [M_o]$, the lexicographical matricization of the above Zehfuss product $\bigotimes_Z (\mathbf{L}_1, \dots, \mathbf{L}_O)$ representing the linear function $\L \defequals \bigotimes_{o=1}^O \L_o$ is the matrix:
\[  
\begin{array}{cl} 
& \displaystyle \sum_{\m \in \M} \lex ( \L_1 (\unitvector{m_1}) \otimes \cdots \otimes \L_o (\unitvector{m_O})) \otimes \lex (\unitvector{m_1} \otimes \cdots \otimes \unitvector{m_O}) \\
= &\smallskip \displaystyle  \sum_{\m \in \M}  \lex ((\mathbf{L}_1)_{:m_1} \otimes \cdots \otimes (\mathbf{L}_O)_{:m_O} )\otimes \lex ( \unitvector{m_1} \otimes \cdots \otimes \unitvector{m_O} ) \\
 = & \smallskip \mathbf{L}_1 \kronecker \mathbf{L}_2 \kronecker \cdots \kronecker \mathbf{L}_O \,,   \end{array} 
 \]
and the colexicographical matricization is the matrix:
\[
  \begin{array}{cl} 
&\displaystyle\sum_{\m \in \M} \vectorize{ \L_1 (\unitvector{m_1}) \otimes \cdots \otimes \L_o (\unitvector{m_O})} \otimes \vectorize{\unitvector{m_1} \otimes \cdots \otimes \unitvector{m_O}} \\
= &\smallskip \displaystyle  \sum_{\m \in \M}  \vectorize{(\mathbf{L}_1)_{:m_1} \otimes \cdots \otimes (\mathbf{L}_O)_{:m_O} }\otimes \vectorize{ \unitvector{m_1} \otimes \cdots \otimes \unitvector{m_O} } \\
 = & \smallskip \mathbf{L}_O \kronecker \mathbf{L}_{O-1} \kronecker \cdots \kronecker \mathbf{L}_1 \,,   \end{array} 
 \]
where $\kronecker$ denotes the Kronecker product of matrices. The above formulae explain the reason why the Kronecker product occurs so frequently in the discussion of tensors: it can be used to represent matricized forms of the contravariant representations of linear functions between spaces of tensors.\\

In addition to the $O$-fold Zehfuss product of two tensors with contiguous partitions of modes which have $2$ blocks, one can also define the $2$-fold Zehfuss product of two tensors, $\T \in \R^{\M[D_1]}$ and $\mathbfcal{U} \in \R^{\N[D_2]}$, each with contiguous partitions of modes which have $O$ blocks:
\[ 
\begin{array}{rcl} \displaystyle
 \bigotimes_Z  (\T , \mathbfcal{U}) & =& \displaystyle \summing{k_1} \summing{k_2} (\v{k_1}_1 \otimes \cdots \otimes \v{k_1}_{B_1}) \otimes (\w{k_2}_1 \otimes \cdots \otimes \w {k_2}_{C_1} ) \otimes \cdots \\
&& \displaystyle \phantom{\summing{k_1}\summing{k_2}}\otimes (\v{k_1}_{D_1 - B_O +1} \otimes \cdots \otimes \v{k_1}_{D_1}) \otimes (\w{k_2}_{D_2 - C_O +1} \otimes \cdots \otimes \w{k_2}_{D_2}) \,, 
\end{array}
 \]
where $\summing{k_1} \v{k_1}_1 \otimes \cdots \otimes \v{k_1}_{D_1}$ and $\summing{k_2} \w{k_2}_1 \otimes \cdots \otimes \w{k_2}_{D_2}$ are arbitrary tensor product decompositions of $\T$ and $\mathbfcal{U}$ respectively, and $B_1$ and $C_1$ are the lengths of the first blocks of the contiguous partitions of the mode index sets $[D_1]$ and $[D_2]$ respectively, $B_2$ and $C_2$ are the lengths of the second blocks of the contiguous partitions of $[D_1]$ and $[D_2]$ respectively, \dots, and $B_O$ and $C_O$ are the lengths of the $O$'th blocks of the contiguous partitions of $[D_1]$ and $[D_2]$. Compare this with the definition of covariance tensor from \cite{Hoff2011} discussed below in appendix \ref{sec:separ-kron-covar}.\\

Obviously the above definitions can be extended even further to the $O_1$-fold Zehfuss product of $O_1$ tensors each of which has contiguous partitions of their modes into $O_2$ blocks. 

%%% Local Variables:
%%% mode: latex
%%% TeX-master: "../new_notes_draft"
%%% End:

\section{Categories of Vector Spaces and Tensor Spaces}
\label{sec:categ-vect-spac}

The category $\finvect$ of finite-dimensional real vector spaces has as objects (unsurprisingly) finite-dimensional real vector spaces\footnote{The detailed definition of a finite-dimensional real vector space is outside the scope of this review, and assumed to be prerequisite knowledge for the review. Consider consulting e.g. \cite{meckes} for an introduction before continuing further.}, and as morphisms all linear functions between them. \\

However the category of finite-dimensional real $D$-factorized tensor spaces $\fintensor{D}$ is assumed to possibly be unfamiliar to the reader. Compare what follows with section 3.2.5 of \cite{hackbusch}.

\subsection{The Tensor Product Functor}
\label{sec:tens-prod-funct}

Given $D$ finite-dimensional real vector spaces $\V_1, \dots, \V_D$, the pair consisting of the two data:

\begin{enumerate}[label=(\roman*)]
\item a vector space $\bT$,
\item a multilinear map $\bigotimes_{\bT}: \V_1 \times \dots \times \V_D \to \bT$,
\end{enumerate}

is said to satisfy the \textbf{universal property of the $D$-fold tensor product}\footnote{See section \ref{sec:tensor-product}, as well as section 14.2 of \cite{silva_hla}, section 15.2 of \cite{lim_hla}, Proposition 3.22 of \cite{hackbusch}, or chapter 2 of \cite{atiyah-macdonald}.} if for every other pair of a vector space $\W$ and multilinear map $\Phi: \V_1 \times \dots \times \V_D \to \W$, there exists a \textbf{\textit{unique linear}} map $\phi: \bT \to \W$ such that $\Phi = \phi \circ \bigotimes_{\bT}$. This property is useful inasmuch as it describes the desired behavior, but it is inadequate for specifying a construction which realizes that behavior. In other words, there is never\footnote{See endnote \ref{tensor-universal-property-uniqueness} from section \ref{sec:tensor-product} for an explanation of why this is true. Compare also proposition 3.21 of \cite{hackbusch}.} a unique $(\bT, \bigotimes_{\bT})$ pair satisfying this universal property.\\

However, given both a $(\bT, \bigotimes_{\bT})$ and a $(\tilde{\bT}, \tilde{\bigotimes}_{\bT})$ satisfying the universal property, a corollary of the universal property is that there is a \textbf{\textit{unique}}\footnote{Specifically, only unique among those isomorphisms which are compatible with both of the chosen tensor products, there are of course other isomorphisms between $\bT$ and $\tilde{\bT}$, those isomorphisms just aren't as ``good''.} linear isomorphism between $\bT$ and $\tilde{\bT}$. Thus to some extent there is no problem in making an arbitrary choice among the infinitely many pairs of $(\bT, \bigotimes_{\bT})$ satisfying the universal property for given $\V_1, \dots, \V_D$, since any other choice one might have made instead can still be recovered uniquely from the initial choice.\\

Because of the aforementioned indeterminacy when referring to ``the'' tensor product of vector spaces, the correct definition in terms of the universal property is often overlooked in favor of a construction which can be defined and guaranteed to exist for every possible choice of $\V_1$, \dots, $\V_D$, the idea being to make the same arbitrary choice for all vector spaces rather than making one arbitrary choice for some class of vector spaces and making a distinct arbitrary choice for another class of vector spaces, compare chapter 2 of \cite{atiyah-macdonald} or section 3.2.1 of \cite{hackbusch}.\\

However, in this review the preferred tensor product construction for the finite-dimensional real vector spaces $\R^{\M}$ (for all possible multi-index sets $\M$) will always be the Segre outer product, even though the Segre outer product is not a construction which is defined for all possible finite-dimensional real vector spaces. Therefore this review falls into the group of expositions which wish to make one arbitrary choice of tensor product for one class of vector spaces (Segre outer product for the $\R^{\M}$'s), while making a distinct arbitrary choice for all other vector spaces.\\

Thus the approach pursued in chapter 2 of \cite{atiyah-macdonald} or section 3.2.1 of \cite{hackbusch}, of \textit{always} defining the tensor product space as the suitable quotient of the (massive) free vector space\endnote{Actually the notion of ``free vector space'' opens its own separate can of worms, since it is also defined via a universal property which only specifies a vector space up to isomorphism, and does not give priority to any particular construction over another. The approach taken in section 3.1.2 of \cite{hackbusch} for constructing the free vector space is very similar to the usual construction of taking the direct sum of the vector spaces $\R^{s}$ for every element $s \in S$ for an arbitrary set $S$. This would seem to not open yet another can of worms since the direct sum refers to a specific construction, although that construction is of interest primarily because it satisfies the universal property of the coproduct.} generated by all of the vectors in the vector space $\V_1 \times \cdots \times \V_D$, is not suitable here.\\

This causes some difficulty, however, when trying to define a tensor product functor:
\[  
\bigotimes: \underbrace{\finvect \times \cdots \times \finvect}_{D\text{ times}} \to \finvect \,, 
 \]
since using the Segre outer product leads to this functor not automatically being defined for all $D$-tuples of finite-dimensional real vector spaces (in contrast to the situation where the ``one arbitrary choice for all vector spaces'' approach had been utilized instead), since the Segre outer product only defines the functor for those $D$-tuples which consist exclusively of $\R^{\M}$'s.\\

So the concern which needs to be addressed is the following: will different arbitrary choices of vector space satisfying the universal property of the tensor product to use as the value of the tensor product functor for all remaining $D$-tuples lead to (meaningfully) different tensor product functors? Will the study of one type of tensor product functor defined partially in terms of the Segre outer product lead to different conclusions compared to the tensor product functor defined on objects by always using the aforementioned quotient space construction?\\

First an interlude which will soon prove to be relevant. In order to define a tensor product functor, it actually suffices to define it only on objects, doing so by choosing for each $D$-tuple of finite-dimensional real vector spaces an element of the equivalence class of vector spaces satisfying the universal property for them. This is because the universal property can then be used to automatically define the tensor product functor on morphisms.\\

More concretely, for linear functions $\V_1 \xrightarrow{f_1} \W_1$, \dots, $\V_D \xrightarrow{f_D} \W_D$, what one wants to do is have a choice of linear function from the chosen value $\bigotimes_{d=1}^D \V_d$ of the tensor product functor for $\V_1, \dots, \V_D$ to the chosen value $\bigotimes_{d=1}^D \W_d$ of the tensor product functor for $\W_1, \dots, \W_D$, so that the chosen linear function can then represent the chosen value $\bigotimes_{d=1}^D f_d$ for the tensor product functor on the morphisms $f_1, \dots, f_D$. The universal property automatically makes such a choice.\\

Consider the Cartesian product of linear functions (which itself is linear):
\[ 
  \V_1 \times \cdots \times \V_D \xrightarrow{f_1 \times \cdots \times f_D} \W_1 \times \cdots \times \W_D \,,
  \]
given via the rule of assignment $(f_1 \times \cdots \times f_D)(\mathbf{v}_1, \dots, \mathbf{v}_D) = (f_1(\mathbf{v}_1), \dots, f_D(\mathbf{v}_D))$. Then
 \[
\bigotimes_{\W} \circ (f_1 \times \cdots \times f_D): \V_1 \times \cdots \times \V_D \to \bigotimes_{d=1}^D \W_d
\] 
is multilinear (since the function $\bigotimes_{\W}: \W_1 \times \cdots \times \W_D \to \bigotimes_{d=1}^D \W_d$ is multilinear). Therefore the universal property of the tensor product for $(\bigotimes_{d=1}^D\V_d, \bigotimes_{\V})$ guarantees the existence of a \textit{unique} linear function $h: \bigotimes_{d=1}^D\V_d \to \bigotimes_{d=1}^D\W_d$ such that $h \circ \bigotimes_{\V} = \bigotimes_{\W} \circ (f_1 \times \cdots \times f_D)$.\\

It turns out that, given the choices of $(\bigotimes_{d=1}^D \V_d, \bigotimes_{\V})$ for the value of the tensor product functor on $\V_1$, \dots, $\V_D$, and $(\bigotimes_{d=1}^D \W_d, \bigotimes_{\W})$ for the value of the tensor product functor on $\W_1$, \dots, $\W_D$, then the linear function $h: \bigotimes_{d=1}^D \V_d \to \bigotimes_{d=1}^D \W_d$ defined uniquely via the universal property of the tensor product is the natural choice for the value of the tensor product functor on $f_1, \dots, f_D$, i.e. the natural choice is given by defining $\bigotimes_{d=1}^D  f_d \defequals h$.\\

The choice is natural in the following sense: given one tensor product functor $\bigotimes^{\smallone}$ defined on objects by making one class of arbitrary choices of vector spaces satisfying the universal property, and defined on morphisms by using the arbitrary choices on objects and the universal property in the manner explained above, and another tensor product functor $\bigotimes^{\smalltwo}$ defined on objects by making another class of arbitrary choices of vector spaces satisfying the universal property, and also defined on morphisms by using the universal property as explained above to extend the arbitrary choices on objects, then the two functors $\bigotimes^{\smallone}$ and $\bigotimes^{\smalltwo}$ will always be naturally isomorphic\footnote{\label{second_natural_footnote}``Natural isomorphism'' is a precisely defined notion of category theory, cf. footnote \ref{natural_footnote} of appendix \ref{sec:repr-dual-spac}.}, or in other words they will be for all intents and purposes entirely equivalent. This is because of the unique isomorphisms which, due to the universal property, exist between the arbitrary choices made on objects by the first functor and the arbitrary choices made on objects by the second functor, and because both functors were defined on morphisms by using the unique linear functions found using the universal property in the way which was described above.\\

Note that the tensor product of linear transformations as it has been defined in section \ref{sec:tens-prod-line} is the same definition one gets by using the Segre outer product to define the tensor product functor on $\R^{\M}$'s and then using the universal property to define the values of the tensor product functor on morphisms. So even though the tensor product functor as defined in this review is slightly different, at least on $\R^{\M}$'s and linear functions between them, then the tensor product functor more commonly defined elsewhere, both functors are still naturally isomorphic to one another, and this remains true regardless of which arbitrary choices of vector spaces satisfying the universal property are used to define the tensor product functor on $D$-tuples of vector spaces for which not all of the $\V_d$'s are equal to $\R^{\M}$ for some $\M$. Therefore, for the sake of simplicity, when at least one of the vector spaces $\V_d$ is not equal to some $\R^{\M}$, for this review define the $D$-fold tensor product of those vector spaces (and associated $D$-linear map) via the usual construction as found e.g. in chapter 2 of \cite{atiyah-macdonald}. Moreover, for this review the value of the tensor product functor on morphisms will always use the linear maps generated by the universal product as described above.

%%% Local Variables:
%%% mode: latex
%%% TeX-master: "../new_notes_draft"
%%% End:

\subsection{Objects of the Category of Finite-Dimensional Real $D$-Factorized \\Tensor Spaces}
\label{sec:objects-categ-finite}

For a given $D \ge 1$, the objects of $\fintensor{D}$ consist of three data: 
\begin{enumerate}[label=(\roman*)]
\item a finite-dimensional real vector space $\bT$,
\item a $D$-tuple of finite-dimensional real vectors spaces $(\V_1, \dots, \V_D)$,
\item  a multilinear map
\[  
\bigotimes: \V_1 \times \V_2 \times \cdots \times \V_D \to \bT
 \]
which satisfies the universal property of the tensor product.
\end{enumerate}

The standard example will always be (where ${\M[D] \defequals \bigtimes_{d=1}^D [M_d] }$) the special case when\\ 
$\bT = \R^{\M[D]}$, $\V_1 = \R^{[M_1]}$, $\V_2 = \R^{[M_2]}$, \dots, $\V_D = \R^{[M_D]}$, and $\bigotimes$ denotes the Segre outer product.

\subsection{Morphisms of the Category of Finite-Dimensional Real $D$-Factorized Tensor Spaces}
\label{sec:morph-categ-finite}

Given two objects $\mathbb{V}$ and $\mathbb{W}$ of $\fintensor{D}$, a morphism $\mathbb{V} \to \mathbb{W}$ is any linear function of the form:
\[ 
 \bigotimes_{d=1}^D f_d : \mathbb{V} \to \mathbb{W}\,, \quad\quad f_1: \mathbb{V}_1 \to \mathbb{W}_1 \,, \dots \,, f_d: \mathbb{V}_d \to \mathbb{W}_d \,, \dots \,, f_D: \mathbb{V}_D \to \mathbb{W}_D \,. 
\]
Here $\bigotimes_{d=1}^D f_d$ refers to the unique linear function $\mathbb{V} \to \mathbb{W}$ such that
\[ 
 \left( \bigotimes_{d=1}^D f_d \right) \circ \bigotimes_{\mathbb{V}} = \bigotimes_{\mathbb{W}} \circ (f_1 \times \cdots \times f_D)  
 \]
which exists since both $\mathbb{V}$ and $\mathbb{W}$ satisfy the universal property of the $D$-fold tensor product, cf. the discussion section \ref{sec:tens-prod-funct} above. It can be thought of as the unique linear function $\mathbb{V} \to \mathbb{W}$ corresponding to the multilinear function $\bigotimes_{\mathbb{W}} \circ (f_1 \times \cdots \times f_D)$ which is compatible with the tensor product structure on $\mathbb{V}$ (which is encapsulated via $\bigotimes_{\mathbb{V}}$).\\

In this review the standard example will always be functions of the form:
\[  
\bigotimes_{d=1}^D f_d: \R^{\M[D]} \to \R^{\N[D]} \,, \quad \bigotimes_{d=1}^D f_d: \mathbf{v}_1 \otimes \cdots \otimes \mathbf{v}_D \mapsto f_1(\mathbf{v}_1) \otimes \cdots \otimes f_D(\mathbf{v}_D) \,,  
\]
where it suffices to define the function only for elementary tensors due to the universal property of the tensor product, and where $\otimes$ refers to the Segre outer product; compare this with section \ref{sec:tens-prod-line}.\\

Tensor spaces $\mathbb{V}$, $\mathbb{W}$ are \textbf{isomorphic} (as $D$-factorized tensor spaces) if there are morphisms in $\fintensor{D}$:
\[
 f: \mathbb{V}\to \mathbb{W}  \,, \quad g:\mathbb{W} \to \mathbb{V}  \,,
 \]
such that $g \circ f = \operatorname{Id}_{\mathbb{V}}$ and $f \circ g = \operatorname{Id}_{\mathbb{W}}$.

\subsection{Remarks about the Category of Finite-Dimensional Real $D$-Factorized Tensor Spaces}
\label{sec:remarks-about-categ}

Analogous to $\multiindex$ and $\multiindex_1$, both the objects and morphisms of $\fintensor{1}$ and $\finvect$ are essentially the same, making both categories for all intents and purposes essentially the same. Thus, from now $\fintensor{1}$ will be redefined so that $\fintensor{1} = \finvect$.\\

Also analogous to the multi-index categories, for any $O > D$ one also always has $\binom{O-1}{D-1}$ distinct forgetful functors from $\fintensor{O}$ to $\fintensor{D}$, with each distinct forgetful functor corresponding to a different contiguous partition of the $O$ vector spaces $\mathbb{V}_1, \dots, \mathbb{V}_O$.\\

As an example, consider the case when $O=3$, $D=2$, $\mathbb{V} = \R^{[24] \times [24]}$, $\bigotimes_{\mathbb{V}} = \kronecker$ (the Kronecker matrix product), $\mathbb{V}_1 = \R^{[4] \times [2] }$, $\mathbb{V}_2 = \R^{[3] \times [3]}$, and $\mathbb{V}_3 =\R^{[2] \times [4]}$. Then $\mathscr{F}_{12|3}(\mathbb{V})$ corresponds to $\mathbb{V}_1 = \R^{[12]\times[6]}$ and $\mathbb{V}_2 = \R^{[2] \times [4]}$, while $\mathscr{F}_{1|23}(\mathbb{V})$ corresponds to $\mathbb{V}_1 = \R^{[4] \times [2]}$ and $\mathbb{V}_2 = \R^{[6] \times [12]}$.\\

Also, as before, any morphism in $\fintensor{O}$ between $\mathbb{V}$ and $\mathbb{W}$ will also be a morphism in $\fintensor{D}$ between $\mathscr{F}_Q(\mathbb{V})$ and $\mathscr{F}_Q(\mathbb{W})$ for any contiguous partition $Q$ of $[O]$ into $D$ blocks, but there will also always be morphisms between $\mathscr{F}_Q(\mathbb{V})$ and $\mathscr{F}_Q(\mathbb{W})$ in $\fintensor{D}$ which are not morphisms in $\fintensor{O}$. Preserving $O$-fold tensor product structure is sufficient to preserve any coarser $D$-fold tensor product structure, but preserving $D$-fold tensor product structure is \textit{not} enough to guarantee also preserving $O$-fold tensor product structure for $O > D$.\\

Note that, for each $D \ge 1$, at least as long as the $D$-fold tensor product functor is defined in any of the naturally isomorphic ways described above in section \ref{sec:tens-prod-funct}, the category $\fintensor{D}$ is equivalent to the image of the $D$-fold tensor product functor:
\[ 
 \bigotimes_{d=1}^D: \underbrace{\finvect \times \cdots \times \finvect}_{D\text{ times}} \to \finvect \,, 
 \]
in the sense that every object of $\fintensor{D}$ is isomorphic as a $D$-factorized tensor space (isomorphic in $\fintensor{D}$) to an object $\bigotimes_{d=1}^D \mathbb{V}_d$ in the image of the $D$-fold tensor product functor $\bigotimes_{d=1}^D$.\\

Note also that, by choosing coordinates, every object of $\fintensor{D}$ is isomorphic as a $D$-factorized tensor space to $\R^{\M[D]}$ for some multi-index set $\M[D]$ of order $D$. More specifically, given an object $\bT$ of $\fintensor{D}$ which has the $D$-tuple of vector spaces $(\V_1, \dots, \V_D)$ as its ``tensor factors'', and if $M_d \defequals \dim (\V_d)$ for all $d \in [D]$, then the aforementioned $\M[D]$ equals  $\bigtimes_{d=1}^D [M_d]$. One construction of such an isomorphism is described in appendix \ref{sec:coord-isom-tens}.\\

Thus, even when considering $D$-factorized tensor spaces and $D$-fold tensor product structure, there is still is no loss of generality in restricting attention only to coordinate spaces $\R^{\M[D]}$ with the Segre outer product as their tensor product, analogous to the situation for vector spaces.

%%% Local Variables:
%%% mode: latex
%%% TeX-master: "../new_notes_draft"
%%% End:

\section{Tensor Shape Functors}
\label{sec:tens-shape-funct-1}

The following sections describe the interpretation of the dual tensor shape functor as well as how to relate general linear functions to those in the image of $\shape$.

\subsection{Interpretation of Dual Tensor Shape Functor}
\label{sec:interpr-dual-tens}

By construction, the tensor shape functor $\multiindex \to \finvect$ can be considered the restriction of a free functor $\shape: \setfin \to \finvect$, since $\shape ( \M) = \R^{\M}$ satisfies the universal property of the free real vector space generated by $\M$, cf. endnote \ref{free-functor-endnote} of section \ref{sec:tens-shape-funct} or section 3.1.2 of \cite{hackbusch}. \\

Any functor has a corresponding opposite functor which behaves the same on objects and morphisms as the original functor. The tensor shape functor in particular has an opposite functor:
\[ 
 \shape^{\operatorname{op}}: \multiindex^{\operatorname{op}} \to \finvect^{\operatorname{op}}  \,. 
\]
The universal property of the free real vector space implies that, for any finite-dimensional real vector space $\V$, the following functors $\multiindex^{\operatorname{op}} \to \finvect$ are naturally equivalent:
\[  
 \hom_{\finvect}( \shape^{op}(\cdot) , \V) \cong \hom_{\sets}(\cdot, \mathscr{F}(\V)) = \hom_{\sets}(\cdot, \V)   \,,
 \]
where $\mathscr{F}$ denotes the forgetful functor $\finvect \to \sets$. Note that $\hom_{\sets} (\M, \V) \defequals \V^{\M}$ is a finite-dimensional vector space (for any multi-index set $\M$ and finite-dimensional real vector space $\V$, where addition and scalar multiplication are defined ``pointwise'' for the functions, and the dimension as a real vector space is equal to the product of the cardinality of $\M$ and the dimension of $\V$ as a real vector space, i.e. $\dim(\V^{\M}) = |\M| \dim(\V)$).\\

When $\V = \R$, this gives the following natural equivalence of functors $\multiindex^{\operatorname{op}} \to \finvect$:
\[  
 \hom_{\finvect}( \shape^{op}(\cdot) , \R) \cong \hom_{\sets}(\cdot, \R) \,.
 \]
By definition, the dual tensor shape functor $\dualshape: \multiindex^{\operatorname{op}} \to \finvect $ is the same as the functor in the right-hand side of the natural equivalence written above:
\[
 \dualshape \defequals \hom_{\sets}(\cdot, \R)  \,,
 \]
so the above equivalence of functors also implies the following:
\[ 
 \hom_{\finvect}( \shape^{op}(\cdot) , \R) \cong \dualshape \,.
 \]
Similarly, definitionally for the dual functor $\mathfrak{D}: \finvect^{op} \to \finvect$ one has that:
\[  \mathfrak{D} \defequals \hom_{\finvect} (\cdot, \R) \,.  \]
Therefore one can rewrite left-hand side of the above natural equivalence:
\[ 
 \hom_{\finvect}( \shape^{op}(\cdot) , \R)  = \hom_{\finvect}( \cdot , \R) \circ \shape^{\operatorname{op}} = \mathfrak{D} \circ \shape^{\operatorname{op}}  \,. 
 \]
Combing the above, the universal property of the free vector space finally leds to the equivalence:
\[ 
 \mathfrak{D}\circ \shape^{\operatorname{op}} \cong \dualshape \,,
  \]
i.e. the dual tensor shape functor $\dualshape$ is a naturally equivalent functor to the functor resulting from composing the opposite functor $\shape^{\operatorname{op}}$ of the tensor shape functor $\shape$  with the dual functor $\mathfrak{D}$. This is the justification for the name ``dual tensor shape functor'' and also explains why $\dualshape(f)$ always corresponds to the adjoint linear transformation of $\shape(f)$ for any morphism $f$ in $\multiindex$.

\subsection{Combinatorial Linear Algebra}
\label{sec:comb-line-algebra}

The sub-category of $\finvect$ which is the image\footnote{
That the image of a functor in this case is a bona fide subcategory, i.e. closed under composition of morphisms, can be derived as a corollary of the fact that the functor is injective on objects.
} of the tensor shape functor $\shape$ can be thought of as a subset of (multi)linear algebra which is secretly entirely combinatorial in nature, in the sense that all problems can be reduced to problems about multi-index sets and functions between them via the tensor shape functor. Such problems include (but are not limited to) permuting modes of a tensor and reshaping a tensor. The objects of this category are all of the coordinate spaces, and the morphisms are a heavily restricted subset of the possible linear functions between them, namely linear functions of the form $\shape (f)$ for some function between multi-index sets $f$.\\

However, it is possible to use this very limited subset of linear functions to recover all non-zero linear functions\footnote{Which is the same thing as all linear functions with rank greater than or equal to one.} by combining them with changes of basis\footnote{In more algebraic terms: linear automorphisms, or linear self-isomorphisms, of vector spaces.} of the coordinate spaces. Specifically, \textit{every} non-zero linear function $\L$ between coordinate spaces may be factorized as:
\[  
\left(\R^{\M[O_1]} \xrightarrow{\L} \R^{\N} \right)= \left(\R^{\M[O_1]} \xrightarrow{\mathscr{A}_S} \R^{\M[O_1]} \xrightarrow{\mathscr{C}} \R^{\N} \xrightarrow{\mathscr{A}_T} \R^{\N} \right)\,, 
 \]
where $\mathscr{A}_S$ is a change of basis for $\R^{\M[O_1]}$ (i.e. an \textbf{a}utomorphism of the \textbf{s}ource vector space), $\mathscr{C}$ is equal to $\shape (f)$ for some function $f: \M[O_1] \to \N$ (i.e. a ``\textbf{c}ombinatorial linear map''), and $\mathscr{A}_T$ is a change of basis for $\R^{\N}$ (i.e. an \textbf{a}utomorphism of the \textbf{t}arget space).\\

The general idea is always to choose a basis $\mathscr{B}_S$ for $\R^{\M[O_1]}$ and a basis $\mathscr{B}_T$ for $\R^{\N}$ such that the linear function $\L$ sends every element of $\mathscr{B}_S$ either to an element of $\mathscr{B}_T$ or to zero. (While for any basis of $\R^{\M[O_1]}$ it is true that $\L$ will send it to a finite subset of $\R^{\N}$, for a general basis of $\R^{\M[O_1]}$ it need not be the case that this subset of vectors is linearly independent excluding the zero vector.) That such a choice can always be made can be seen, for example, as consequence of the existence of a singular value decomposition\endnote{\label{vectorize-endnote}
In the case that any of $\M[O_1]$ or $\N$ are ``proper multi-index sets'', i.e. of order $2$ or greater, then one has to apply the SVD to the linear function $\L_{\operatorname{vec}} : \R^\mathbf{M} \to \R^\mathbf{N}$, where $\mathbf{M} \defequals \prod_{o_1=1}^{O_1} M_{o_1}$, $\mathbf{N} \defequals \prod_{o_2=1}^{O_2} N_{o_2}$, and
\[ 
\L_{\operatorname{vec}} = \operatorname{vec}_N \circ \L \circ \operatorname{vec}_M^{-1} \,, \quad \L = \operatorname{vec}_N^{-1} \circ \L_{\operatorname{vec}} \circ \operatorname{vec}_M \,,  
\]
with $\operatorname{vec}_M$ and $\operatorname{vec}_N$ any arbitrary isomorphisms $\R^{\M[O_1]} \to \R^\mathbf{M}$ and $\R^{\N} \to \R^\mathbf{N}$ respectively.\\ 

From the left singular vectors $\mathbf{u}$ of $\L_{\operatorname{vec}}$ one then gets ``left singular tensors'' $\mathbfcal{U}$ via $\mathbfcal{U} = \operatorname{vec}_N^{-1}(\mathbf{u})$, and from the right singular vectors $\mathbf{v}$ of $\L_{\operatorname{vec}}$ one gets ``right singular tensors'' $\mathbfcal{V}$ via $\mathbfcal{V} = \operatorname{vec}_M^{-1}(\mathbf{v})$. Then one has $\L(\mathbfcal{U}) = \sigma \mathbfcal{V}$ for every pair of ``singular tensors'' and their corresponding singular value, allowing one to factorize $\L$ as a combinatorial linear map preceded and followed by changes of basis in the manner described above. \\

Of course this will in general destroy whatever additional structure $\L$ might have as a tensor product of other linear functions $\L = \bigotimes_{d=1}^D \L_d$ (clearly $D \le \min\{O_1,O_2\}$). In that case it would be best to first separately perform SVD's in the manner described above for each of the linear functions $\L_d$ in the tensor product factorization, and then define ``singular tensor'' basises for each of $\R^{\M[O_1]}$ and $\R^{\N}$ via the tensor products of the respective ``singular tensors'' of the linear functions $\L_d$, i.e. $\mathbfcal{U} = \mathbfcal{U}_1 \otimes \cdots \otimes \mathbfcal{U}_D$ and $\mathbfcal{V} = \mathbfcal{V}_1 \otimes \cdots \otimes \mathbfcal{V}_D$. Then it can be verified that
\[ 
\L(\mathbfcal{V}) = \L_1(\mathbfcal{V}_1) \otimes \cdots \otimes \L_D(\mathbfcal{V}_D) = \left( \prod_{d=1}^D \sigma_d \right) \mathbfcal{U}_1 \otimes \cdots \otimes \mathbfcal{U}_D = \left( \prod_{d=1}^D \sigma_d \right) \mathbfcal{U} \,,
 \]
meaning that this choice of basises for $\R^{\M[O_1]}$ and $\R^{\N}$ satisfies the conditions above while also better reflecting the tensor product structure of $\L$ than the result one would get by vectorizing $\R^{\M}$ and $\R^{\N}$. (Note that in general it will still be necessary to vectorize the domains and codomains of the respective $\L_d$'s however.)
}. If the rank is at least one, then a trick can be employed to make everything work when the null space of $\L$ is non-zero (see section \ref{sec:non-zero-null}).\\

However, the fact that the zero linear map can not factorize this way, and that an ugly trick is required whenever the null space is non-zero, correctly suggests that $\multiindex$ is not the ``correct'' category to which one should attempt to reduce all multilinear algebra. The problem is that $\multiindex$ does not have any ``zero object'' and consequently no notions of ``zero morphisms'' or ``sending an element to zero'', whereas $\finvect$ does. Thus these aspects of $\finvect$ cannot be completely or directly captured using $\multiindex$ and the tensor shape functor $\shape$, even after changing basises.\\

This can be fairly easily rectified however by replacing $\multiindex$ with the category of \textbf{pointed} multi-index sets\footnote{The idea is analogous to the ``four subspace'' diagrams of Gilbert Strang (when the multi-index sets are index sets).} $\multiindex^{\star}$, i.e. multi-index sets with an extra ``point''/element $\star$ appended to represent a notion of ``zero object'', and functions $f: \M[O_1] \to \N$ limited to those which send the zero points $\star_M$ of their domains to the zero points $\star_N$ of their codomains. This requires modifying the Cartesian product operation to be the so-called ``smash product'' which ensures that the product of pointed multi-index sets is still a pointed multi-index sets (i.e. only has \textit{one} zero point, not multiple), and some other relatively straightforward changes. In particular, the category $\multiindex$ can be ``embedded'' as a subcategory of $\multiindex^\star$ in a fairly obvious way, so there is no ``loss of information''. The details of these constructions are tedious and outside the scope of this review.

\subsection{Non-Zero Null Space Trick}
\label{sec:non-zero-null}

As explained in endnote \ref{vectorize-endnote}, without loss of generality one may assume that the multi-index sets $\M[O_1]$ and $\N$ are index sets $[M]$ and $[N]$ respectively. Then one possibly way of choosing $\mathscr{B}_S$ and $\mathscr{B}_T$ consists of making the right singular vectors the elements of $\mathscr{B}_S$ and 
\[  
 \tau_i \mathbf{u}_i  \quad \text{where }\mathbf{u}_i \text{ is a left singular vector and }\quad \tau_i =
  \begin{cases}
    \sigma_i & \sigma_i > 0 \\
1 & \sigma_i = 0
  \end{cases} 
\]
($\sigma_i$ denote singular values) the first elements of $\mathscr{B}_T$, with $\mathscr{B}_T$ itself being an arbitrary completion of the $\tau_i \mathbf{u}_i$ to a basis of $\R^N$. Then clearly $\L$ sends every element of $\mathscr{B}_S$ either to an element of $\mathscr{B}_T$ (when the corresponding $\sigma_i$ is non-zero) or to $\mathbf{0}$.\\

Starting from this, the first attempt at choosing $\mathscr{A}_S$ sends the elements of $\mathscr{B}_S$ to the standard basis of $\R^M$, and the first attempt at choosing $\mathscr{A}_T$ sends the elements of $\mathscr{B}_T$ to the standard basis of $\R^M$.\\ 

This works without any problems in the case that $\L$ is injective (i.e. all of the singular values are non-zero, or equivalently the null space is zero), since in this case $\mathscr{C} $ can be chosen to be $\shape(f)$ for $f: [M] \to [N]$ which embeds $[M]$ into $[N]$. (Note that $\L: \R^M \to \R^N$ being injective necessarily implies that $M < N$.) However $\mathscr{A}_S$ must be modified when $\L$ is not injective.\\

The idea of the trick is that ``sending $k$ dimensions to zero'' is the same as ``sending $k+1$ dimensions to the same (one) dimension'', and thus requires the assumption that there is at least one dimension in the image of $\L$, i.e. that $\L$ has rank of at least one.\\

Assume without loss of generality that $\mathscr{A}_S^{\smallone}$ sends a right singular vector corresponding to a non-zero singular value to the first standard basis vector, and that $\mathscr{A}_S^{\smallone}$ sends all of the right singular vectors corresponding to zero singular values to the $2$nd through $(k+1)$'st standard basis vectors. (If there are any remaining right singular vectors corresponding to non-zero singular values, $\mathscr{A}_S^{\smallone}$ sends them to the remaining standard basis vectors in a one-to-one manner\footnote{Basically $\mathscr{A}_S^{\smallone}$ is the same as the function $\mathscr{A}_S$ described before up to reordering of the elements of $\mathscr{B}_S$.}.) Then the goal is to find a second change of basis $\mathscr{A}_S^{\smalltwo}$ such that $\mathscr{C} \circ \mathscr{A}_S^{\smalltwo}$ will send the $2$nd through $(k+1)$'st standard basis vectors of $\R^M$ to zero and the first standard basis vector of $\R^M$ to the first standard basis vector of $\R^N$, when $\mathscr{C} = \shape(f)$ for an $f:[M] \to [N]$ such that $f(m) = 1$ for all $1 \le m \le k+1$, with the result that $\mathscr{C}$ sends all of the $k+1$ first standard basis vectors of $\R^M$ to the first standard basis vector of $\R^N$. (Also it is assumed that $f(m) = m-k$ for all $k+2 \le m \le M$, so that $\mathscr{C}$ sends all of the $m$'th standard basis vectors of $\R^M$ to the $(m-k)$'th standard basis vectors of $\R^N$ for $k+2 \le m \le M$, assuming that $\mathscr{A}_S^{\smalltwo}$ sends the $m$'th standard basis vector of $\R^M$ to itself for all $k+2 \le m \le M$. This is much easier to understand by drawing a picture, e.g. when $k=1$.) \\

One possible choice of $\mathscr{A}_S^{\smalltwo}$ which accomplishes this goal:
\[ 
\begin{array}{rcll}
  \mathscr{A}_S^{\smalltwo}:  & \unitvector[M]{1} & \mapsto &  \displaystyle\frac{1}{k+1} \sum_{i=1}^{k+1} \unitvector[M]{i}  \\
\mathscr{A}_S^{\smalltwo}: & \unitvector[M]{2}  & \mapsto& \displaystyle\frac{1}{2} \unitvector[M]{2} - \frac{1}{2k}\left( \sum_{i \not =2 , 1 \le i \le k+1} \unitvector[M]{i}  \right)  \\
& \vdots && \\
\mathscr{A}_S^{\smalltwo}: & \unitvector[M]{j} & \mapsto & \displaystyle\frac{1}{2} \unitvector[M]{j} - \frac{1}{2k}\left( \sum_{i \not =j , 1 \le i \le k+1} \unitvector[M]{i}  \right) \\
& \vdots && \\
\mathscr{A}_S^{\smalltwo}: & \unitvector[M]{k+1}  & \mapsto & \displaystyle \frac{1}{2} \unitvector[M]{k+1} - \frac{1}{2k} \left( \sum_{i=1}^k \unitvector[M]{i} \right) \\
\mathscr{A}_S^{\smalltwo}: & \unitvector[M]{k+2} & \mapsto & \unitvector[M]{k+2} \\
& \vdots &&\\
\mathscr{A}_S^{\smalltwo} : & \unitvector{M} & \mapsto & \unitvector{M}
  \end{array}
\]
With such a choice of $\mathscr{A}_S^{\smalltwo}$, one then has that $\L = \mathscr{A}_T \circ \mathscr{C} \circ \mathscr{A}_S$ where $\mathscr{A}_S \defequals \mathscr{A}_S^{\smalltwo} \circ \mathscr{A}_S^{\smallone}$.

%%% Local Variables:
%%% mode: latex
%%% TeX-master: "../new_notes_draft"
%%% End:

%%% Local Variables:
%%% mode: latex
%%% TeX-master: "../new_notes_draft"
%%% End:

\section{Non-Commutative Polynomials}
\label{sec:append-non-comm}

Informally speaking, the equivalence between tensors and non-commutative polynomials can be understood by applying the following rules.\\

\textbf{Rule 1:} Only tensors of the exact same shape can be added. All other addition is formal addition.\\

(One way the notion of formal addition can be made rigorous is by taking the direct sum of vector spaces, in this case all of the $\R^{\Mltidx{I}}$ for all possible shapes $\Mltidx{I}$. Compare with e.g. \cite{vinberg}.) \\

\textbf{Example of non-formal addition:} $\mathbf{x}_1 \otimes(\sqrt{2} \mathbf{y}_3) \otimes \mathbf{z}_2 + \pi\mathbf{x}_1 \otimes \mathbf{y}_3 \otimes \mathbf{z}_5 = \mathbf{x}_1 \otimes \mathbf{y}_3 \otimes (\sqrt{2}\mathbf{z}_2 + \pi\mathbf{z}_5)$.\\

\textbf{Example of formal addition:} $3\mathbf{x}_2 \otimes \mathbf{y}_4 \otimes \mathbf{z}_1 + 5\mathbf{v}_8 \otimes \mathbf{w}_3$ can never be simplified further.\\

\textbf{Rule 2:} For each given shape $\Mltidx{I} = [I_1] \times [I_2] \times \dots \times [I_O]$, choose a basis for $\R^{I_o}$ for every ${o \in [O]}$.\\

This leads to $O$ basises, $\mathscr{B}_1= \{ \x{1}_1, \dots, \x{1}_{I_1} \} \subset \mathbb{R}^{I_1}, \dots, \mathscr{B}_O = \{  \x{O}_1, \dots, \x{O}_{I_O}  \} \subset \R^{I_O}$.\\

\textbf{Example:} For the given shape $\Mltidx{I} = [3] \times [2]$, one must choose a basis $\mathscr{B}_1$ for $\R^3$ and a basis $\mathscr{B}_2$ for $\R^2$. For example, one might choose $\mathscr{B}_1 = \{ \unitvector[3]{1}, \unitvector[3]{2} , \unitvector{3}  \}$ and $\mathscr{B}_2 = \{ \unitvector[2]{1},  \unitvector{2} \}$.\\
 
\textbf{Rule 3:} Every tensor of a given shape $\Mltidx{I}$ must be written as a sum of elementary (i.e. rank-one) tensors such that each factor belongs to the corresponding basis chosen in the step above.\\

It is always possible to write a tensor this way, and to do so uniquely, since given basises\\
${\mathscr{B}_1 =\{ \x{1}_1, \dots, \x{1}_{I_1} \} \subset \mathbb{R}^{I_1}}$, $\dots, \mathscr{B}_O = \{  \x{O}_1, \dots, \x{O}_{I_O}  \} \subset \R^{I_O}$, the set:
\[ 
 \mathscr{B} \defequals \{   \x{1}_{i_1} \otimes \x{2}_{i_2} \otimes \cdots \otimes \x{O}_{i_O} : \x{1}_{i_1} \in \mathscr{B}_1, \x{2}_{i_2} \in \mathscr{B}_2, \dots, \x{O}_{i_O} \in \mathscr{B}_O   \} 
 \]
is always a basis for $\R^{\Mltidx{I}}$\cite{hackbusch}\footnote{Specifically, cf. Lemma 3.11 of \cite{hackbusch}.}. (As a corollary, the unit tensors of $\R^{\Mltidx{I}}$ are always a basis.)\\

After applying rule 3, the chosen vectors $\x{1}_1, \dots, \x{1}_{I_1}, \x{2}_1, \dots, \x{O}_{I_O}$ are the variables of the non-commutative polynomials corresponding to tensors of shape $\Mltidx{I}$.\\

\textbf{Example:} repeating from the last example, for the given shape $\Mltidx{I} = [3] \times [2]$, choose $\x{1}_1 = \unitvector[3]{1} $, $\x{1}_2 = \unitvector[3]{2}$ , $\x{1}_3 = \unitvector{3}$ , $\x{2}_1 = \unitvector[2]{1}$ , and $\x{2}_2 = \unitvector{2}$. Then the tensor
\[
  \begin{bmatrix}
    3 & 2 \\
    6 & 0 \\
    0 & 4
  \end{bmatrix}
\]
is written as the non-commutative polynomial $3 \x{1}_1 \otimes \x{2}_1 + 2 \x{1}_1 \otimes \x{2}_2 + 6\x{1}_2 \otimes \x{2}_1 + 4\x{1}_3 \otimes \x{2}_2$ or even more suggestively as $3 \x{1}_1  \x{2}_1 + 2 \x{1}_1  \x{2}_2 + 6\x{1}_2 \x{2}_1 + 4\x{1}_3  \x{2}_2$.\\

The total number of variables can be reduced by applying:\\

\textbf{Rule 4 (Optional):} Always choose the same set to be the basis for $\R^I$ every time $[I]$ occurs as one of the modes of a tensor of a given shape.\\

\textbf{Example:} For $\Mltidx{I} = [3] \times [3] \times [2]$, one needs to choose a basis $\mathscr{B}_1$ for $\R^3$, another basis $\mathscr{B}_2$ for $\R^3$, and a basis $\mathscr{B}_3$ for $\R^2$. In general, $\mathscr{B}_1 \not=\mathscr{B}_2$, and the non-commutative polynomials of shape $\Mltidx{I}$ are written in terms of $3+3+2 = 8$ non-commuting variables (which are the distinct elements of $\mathscr{B}_1, \mathscr{B}_2,\mathscr{B}_3$). For example, one could choose
\[ 
\mathscr{B}_1 = \{ \unitvector[3]{1}, \unitvector[3]{2} , \unitvector{3} \}\,, \quad \mathscr{B}_2 = \{ (1,1,1), (1,1,0), (0,1,1) \}\,,\quad \mathscr{B}_3 = \{\unitvector[2]{1}, \unitvector{2}\} \,.
 \]
Then the $8$ non-commuting variables are $\mathbf{x}_1=(1,0,0),\mathbf{x}_2 =(0,1,0),\mathbf{x}_3 = (0,0,1)$,\\ ${\mathbf{y}_1=(1,1,1)}$, $\mathbf{y}_2=(1,1,0),\mathbf{y}_3=(0,1,1)$, $\mathbf{z}_1=(1,0)$, and $\mathbf{z}_2 = (0,1)$.\\

However, if one chooses the same basis for $\R^3$ twice, i.e. $\mathscr{B}_1 = \mathscr{B}_2$, then the number of non-commuting variables becomes $3+2 = 5$ (the distinct elements of $\mathscr{B}_1=\mathscr{B}_2$ and $\mathscr{B}_3$). For example, one could choose $\mathscr{B}_1 = \{ \unitvector[3]{1}, \unitvector[3]{2}, \unitvector{3}  \} = \mathscr{B}_2$ and $\mathscr{B}_3 = \{ \unitvector[2]{1}, \unitvector{2}  \}$. Then the $5$ non-commuting variables are $\mathbf{v}_1 = (1,0,0), \mathbf{v}_2=(0,1,0), \mathbf{v}_3=(0,0,1)$,$\mathbf{w}_1=(1,0)$, and $\mathbf{w}_2=(0,1)$.\\

In particular, when rule 4 is observed, the same variables are used for every mode of cubical tensors, which makes them easier to work with as non-commutative polynomials.\\

\textbf{Example:} Consider when $\Mltidx{I} = [2] \times [2] \times [2]$. When rule 4 is not observed, there could potentially be up to $6$ distinct non-commuting variables just for tensors of shape $\Mltidx{I}$ alone, not to mention tensors of other shapes. But when rule $4$ is observed, one only has $2$ non-commuting variables. Moreover, those $2$ non-commuting variables would also be the same as the $2$ non-commuting variables used for tensors of shape $[2] \times [2]$ or of shape $[2] \times [2] \times [2] \times [2]$, as well as the same as $2$ of the $5$ non-commuting variables used to describe tensors of shape $[3] \times [3] \times [2]$ and $[3] \times [2]$.\\

\textbf{Rule 5 (Optional):} Only consider cubical tensors, and of those only cubical tensors whose modes all have the same dimension, regardless of their order.\\

\textbf{Example:} If one restricts to tensors of shape $[2]$, $[2] \times [2]$, $[2] \times [2] \times [2]$, $[2] \times [2] \times [2] \times [2]$, $\dots$, and always uses the same basis for $\R^2$, then regardless of the order of the tensor, the number of non-commuting variables is always $2$. Thus these tensors correspond to bivariate non-commutative polynomials in a very straightforward way\footnote{Strictly speaking one also needs to take care to include formal sums with ``tensors or order zero'', i.e. scalars or constants, in order to be fully analogous with (commutative) polynomials.}. Similarly, tensors of shape $[3], [3]\times[3],[3]\times [3] \times [3]$,$[3] \times [3] \times [3] \times [3]$, $\dots$ can be made to correspond to trivariate non-commutative polynomials.\\

To go from the result of applying rule 5 to something isomorphic to regular, commutative polynomials, one needs to do at least both of the following:

\begin{itemize}
\item Restrict the subset of cubical tensors under consideration to only those which are symmetric,
\item Use a different product which preserves the symmetric property, i.e. a property under which symmetric tensors are closed.
\end{itemize}

This is because the regular tensor product does \textit{not} preserve the property of being a symmetric tensor, i.e. given two symmetric tensors, the value of their tensor product need not be symmetric.\\

\textbf{Example:} Letting as always $\unitvector[2]{1}, \unitvector{2}$ denote the standard basis of $\R^2$, then while $\unitvector[2]{1} \otimes \unitvector[2]{1}$ and $\unitvector{2} \otimes \unitvector{2}$ are both symmetric tensors, one also has that $(\unitvector[2]{1} \otimes \unitvector[2]{1} )\otimes ( \unitvector{2}  \otimes \unitvector{2})= \unitvector[2]{1}\otimes \unitvector[2]{1}\otimes \unitvector{2}\otimes \unitvector{2} $ is \textit{not} a symmetric tensor. Thus $\otimes$ did not preserve the property of being a symmetric tensor.\\

For details about how to rectify this and go from the symmetric but non-commutative polynomials above to regular, commutative polynomials\footnote{Or in other words how to go from the tensor algebra generated by $\R^I$ for some fixed integer $I$ to the symmetric algebra generated by $\R^I$ (which is isomorphic to the set of $I$-variate real-valued polynomials).}, consult for example \cite{silva_hla}.\\

Thus, in conclusion the dictionary between tensors and polynomials can be extended roughly as:\\

\begin{tabular}{|l|l|}
\hline
  \textbf{Polynomials}  & \textbf{Tensors} \\
  \hline
  degree & order \\
  \hline
  monomial & elementary tensor \\
  \hline
  commutative & non-commutative \\
  \hline
  variable & basis element \\
  \hline
  number of variables & dimension \\
  \hline
  \vtop{\hbox{\strut which and how many}\hbox{\strut variables are available}\hbox{\strut does \textit{not} depend}\hbox{\strut on degree or position}\hbox{\strut within product of factors}} &
\vtop{\hbox{\strut which and how many}\hbox{\strut variables are available}\hbox{\strut \textit{could} depend}\hbox{\strut on order or position}\hbox{\strut within product of factors}}\\
 \hline 
\end{tabular}\\

%%% Local Variables:
%%% mode: latex
%%% TeX-master: "../new_notes_draft"
%%% End:

\section{Separable Kronecker Covariance Structure}
\label{sec:separ-kron-covar}

Having a separable Kronecker covariance structure means that the covariance matrix of the vectorized\footnote{The paper \cite{tucker-response} uses the same unfolding conventions as \cite{kolda_bader}, which strongly suggests that the vectorization here is the colexicographical one. However, the type of vectorization used is never specified explicitly.} error tensors can be written as a sequence of Kronecker product factors:
\[
  \begin{array}{rcl}

    \boldsymbol{\Sigma}_O \kronecker \boldsymbol{\Sigma}_{O-1} \kronecker \cdots \kronecker \boldsymbol{\Sigma}_1
   & = &    \operatorname{Cov} ( \vectorize{ \mathbfcal{E}_i }  )\\
                                                            & = &\displaystyle \mathbb{E} \left[  (\vectorize{\mathbfcal{E}_i} - \mathbb{E}[\vectorize{\mathbfcal{E}_i}])  \otimes   (\vectorize{\mathbfcal{E}_i} - \mathbb{E}[\vectorize{\mathbfcal{E}_i}])     \right] \\

    &=&   \mathbb{E} [  \vectorize{ \mathbfcal{E}_i - \mathbb{E}[ \mathbfcal{E}_i ]  } \otimes \vectorize{ \mathbfcal{E}_i - \mathbb{E}[ \mathbfcal{E}_i ]  }    ]     \,.
  \end{array}
\]
The last equation follows from the linearity of the colexicographical vectorization function.\\

Represent each covariance matrix $\boldsymbol{\Sigma}_o$ in terms of its eigenvalue decomposition:
\[   
\boldsymbol{\Sigma}_o = \sum_{m_o=1}^{M_o} ( \sqrt{\lambda_o^{\smallsuper{m_o}}} \u{m_o}_o ) \otimes (\sqrt{\lambda_o^{\smallsuper{m_o}}} \u{m_o}_o) \defequals \sum_{m_o=1}^{M_o} \v{m_o}_o \otimes \v{m_o}_o \,, 
 \]
where for each $m_o \in [M_o]$, one has that $\lambda_o^{\smallsuper{m_o}} $ is the $m_o$'th eigenvalue of $\boldsymbol{\Sigma}_o$ and is non-negative since each $\boldsymbol{\Sigma}_o$ is PSD, that $\u{m_o}_o$ is the $m_o$'th eigenvector of $\boldsymbol{\Sigma}_o$, and by definition ${ \v{m_o}_o = \sqrt{\lambda_o^{\smallsuper{m_o}}} \u{m_o}_o  .}$ Then the Kronecker product of the $\boldsymbol{\Sigma}_o$'s is equal to the following matricization:
\[   
 \boldsymbol{\Sigma}_O \kronecker \boldsymbol{\Sigma}_{O-1} \kronecker \cdots \kronecker \boldsymbol{\Sigma}_1 = \sum_{\m \in \M} \vectorize{ \v{m_1}_1 \otimes \cdots \otimes \v{m_O}_O  } \otimes \vectorize{ \v{m_1}_1 \otimes \cdots \otimes \v{m_O}_O  } \,. 
 \]
Applying the linear transformation inverse to the matricization, the above means that for $\mathbfcal{E}_i$:
\[ 
 \mathbb{E} [ (\mathbfcal{E}_i - \mathbb{E}[\mathbfcal{E}_i]) \otimes (\mathbfcal{E}_i - \mathbb{E}[\mathbfcal{E}_i]) ] = \sum_{\m \in \M}  (\v{m_1}_1 \otimes \cdots \otimes \v{m_O}_O) \otimes (\v{m_1}_1 \otimes \cdots \otimes \v{m_O}_O)\,,
 \]
\textit{whenever} the residuals $\mathbfcal{E}_i$ have a separable Kronecker covariance structure. In particular, if one writes $\boldsymbol{\Sigma}=\boldsymbol{\Sigma}_O \kronecker \cdots \kronecker \boldsymbol{\Sigma}_1$, then the above equations imply\footnote{Since the product of two matrices is the sum of the outer products of the columns of the first matrix and the rows of the second matrix, and the equation below gives a standard column-wise outer product decomposition.} the following:
\[
  \begin{array}{rcl}
    \boldsymbol{\Sigma}^{1/2}    & = & \displaystyle \sum_{\m \in \M} \vectorize{ \v{m_1}_1 \otimes \cdots \otimes \v{m_O}_O  } \otimes \vectorize{ \unitvector{m_1} \otimes \cdots \otimes \unitvector{m_O} } \smallskip\\
&=& \smallskip\boldsymbol{\Sigma}_O^{1/2} \kronecker \boldsymbol{\Sigma}_{O-1}^{1/2} \kronecker \cdots \kronecker \boldsymbol{\Sigma}_1^{1/2} \,,
  \end{array} 
 \]
where for each $o \in [O]$, the square root has been defined using the eigenvalue decomposition as:
\[ 
 \boldsymbol{\Sigma}_o^{1/2} = \sum_{m_o=1}^{M_o} \v{m_o}_o \otimes \unitvector[M_o]{m_o} \defequals \sum_{m_o=1}^{M_o}  \sqrt{\lambda_o^{\smallsuper{m_o}}} \u{m_o}_o \otimes \unitvector[M_o]{m_o} \,. 
 \]
Thus one finds the following relationship implied by separable Kronecker covariance structure:
\[
\mathbb{E} [ (\mathbfcal{E}_i - \mathbb{E}[\mathbfcal{E}_i]) \otimes (\mathbfcal{E}_i - \mathbb{E}[\mathbfcal{E}_i]) ]     =\displaystyle \sum_{\m \in \M} \left( \unittensor{\m} \bullet_1 \boldsymbol{\Sigma}_1^{1/2} \cdots \bullet_O \boldsymbol{\Sigma}_O^{1/2}  \right) \otimes \left( \unittensor{\m} \bullet_1 \boldsymbol{\Sigma}_1^{1/2} \cdots \bullet_O \boldsymbol{\Sigma}_O^{1/2}  \right) \,.
\]
Define temporarily for convenience $\mathcal{S}: \mathbfcal{T} \mapsto \mathbfcal{T} \bullet_1 \boldsymbol{\Sigma}_1^{1/2} \cdots \bullet_O \boldsymbol{\Sigma}_O^{1/2}$, which is a linear function. Therefore its tensor product with itself, namely ${\mathcal{S} \otimes \mathcal{S}: \mathbfcal{T}_1 \otimes \mathbfcal{T}_2 \mapsto \mathcal{S}(\mathbfcal{T}_1) \otimes \mathcal{S}(\mathbfcal{T}_2)}$, is also a linear function. The terms in the above summation are all of the form ${\mathcal{S}(\unittensor{\m}) \otimes \mathcal{S}(\unittensor{\m}) } = (\mathcal{S} \otimes \mathcal{S})(\unittensor{\m} \otimes \unittensor{\m})$. Therefore the right hand side of the above equation also equals:
\[
  (\mathcal{S} \otimes \mathcal{S})\left( \sum_{\m \in \M} \unittensor{\m} \otimes \unittensor{\m}  \right) \,.  
\]
Note that the tensor $\sum_{\m \in \M} \unittensor{\m} \otimes \unittensor{\m} $ corresponds \textit{exactly} to the covariance tensor of a random tensor $\mathbfcal{U}_i$ whose entries are all uncorrelated and which all have unit variance.\\ 

Thus the covariance tensor of $\mathbfcal{E}_i$ is the same as the covariance tensor of the tensor:
\[   \mathbfcal{U}_i \bullet_1  \boldsymbol{\Sigma}_1^{1/2} \cdots \bullet_O \boldsymbol{\Sigma}_O^{1/2} = \mathcal{S}( \mathbfcal{U}_i) \,, \]
where again $\mathbfcal{U}_i$ is \textit{any} random tensor whose entries are uncorrelated and have unit variance. The above calculations should be compared with section 2.2 of \cite{Hoff2011}, noting that $k, K$ in \cite{Hoff2011} correspond to $o, O$ here, and in particular that the $\mathbf{A}_k$ of \cite{Hoff2011} essentially correspond to the $\boldsymbol{\Sigma}_o^{1/2}$ here so that the notation $\mathbf{Z} \times \mathbf{A}$ corresponds to $\mathcal{S}(\mathbfcal{U}_i)$. The definition of covariance tensor of the random tensor $\mathbfcal{E}_i$ (corresponding to $\mathbf{Y}$ of \cite{Hoff2011}) in \cite{Hoff2011} is a permutation of the modes of $\mathbb{E} [ (\mathbfcal{E}_i - \mathbb{E}[\mathbfcal{E}_i]) \otimes (\mathbfcal{E}_i - \mathbb{E}[\mathbfcal{E}_i]) ] $, i.e. essentially $\mathbb{E} [ (\mathbfcal{E}_i - \mathbb{E}[\mathbfcal{E}_i]) \otimes_Z (\mathbfcal{E}_i - \mathbb{E}[\mathbfcal{E}_i]) ]$ with $\otimes_Z$ denoting a Zehfuss product for the only possible contiguous partition of the modes of ${\mathbfcal{E}_i - \mathbb{E}[\mathbfcal{E}_i]}$ into $O$ blocks, see appendix \ref{sec:kron-prod-coord}.  \\

The definition of covariance tensor from \cite{Hoff2011} is also such that the following is true:
\[
  \begin{array}{rcl}
   \mathbb{E} [ (\mathbfcal{E}_i - \mathbb{E}[\mathbfcal{E}_i]) \otimes_Z (\mathbfcal{E}_i - \mathbb{E}[\mathbfcal{E}_i]) ]    & = & \displaystyle \sum_{\m \in \M}  (\v{m_1}_1 \otimes \v{m_1}_1) \otimes \cdots \otimes (\v{m_O}_O \otimes \v{m_O}_O) \smallskip\\
&=& \displaystyle \left( \summing{m_1} \v{m_1}_1 \otimes \v{m_1}_1 \right) \otimes \cdots \otimes \left( \summing{m_O} \v{m_O}_O \otimes \v{m_O}_O \right) \smallskip  \\
& = & \boldsymbol{\Sigma}_1 \otimes \boldsymbol{\Sigma}_2 \otimes \cdots \otimes \boldsymbol{\Sigma}_O \,.
  \end{array} 
 \]
Thus the intuition is that, for a separable Kronecker covariance structure, $\boldsymbol{\Sigma}_1$ is the ``covariance matrix for the first mode'', $\boldsymbol{\Sigma}_2$ is the ``covariance matrix for the second mode'', and so on.\\

In the case that all of the $\boldsymbol{\Sigma}_o$'s are positive definite, or equivalently in the case that all of the $\boldsymbol{\Sigma}^{1/2}$'s are invertible, one can also say that the random tensor with the following definition:
\[
\mathbfcal{U}_i \defequals \mathbfcal{E}_i \bullet_1 \boldsymbol{\Sigma}^{-1/2} \cdots \bullet_O \boldsymbol{\Sigma}^{-1/2} = \mathcal{S}^{-1}(\mathbfcal{E}_i)
\]
has all of its entries uncorrelated from one another and with unit variance.\\

When the residuals $\mathbfcal{E}_i$ have what is called a multilinear normal distribution\cite{multilinear_normal}, one \textup{can} go further in making statements about the distribution of $\mathbfcal{E}_i$ based on its covariance tensor alone. More specifically, as is partially explained in \cite{multilinear_normal}, under such an assumption one can decompose the vectorized form of the tensor, $\vectorize{\mathbfcal{E}_i}$, via the following relationship:
\[
  \begin{array}{rcl}
    \vectorize{ \mathbfcal{E}_i - \mathbb{E}[ \mathbfcal{E}_i ] } &=& \\
    \vectorize{\mathbfcal{E}_i}  -  \vectorize{ \mathbb{E}[ \mathbfcal{E}_i ] } & = & \boldsymbol{\Sigma}^{1/2} \vectorize{\mathbfcal{U}_i} \\
       & = &  (\boldsymbol{\Sigma}_O^{1/2} \kronecker \boldsymbol{\Sigma}_{O-1}^{1/2} \kronecker \cdots \kronecker \boldsymbol{\Sigma}_1^{1/2}) \vectorize{\mathbfcal{U}_i} \\
    & = & \vectorize{  \mathbfcal{U}_i \bullet_1 \boldsymbol{\Sigma}_1^{1/2} \cdots \bullet_O \boldsymbol{\Sigma}_O^{1/2}   } \,,
  \end{array}
\]
where by definition $\mathbfcal{U}_i$ is a tensor whose entries consist of independent and identically distributed ${\mathscr{N}(0,1)}$ random variables, and where the identity \cite{cumulant_book}\cite{kolda} from \ref{sec:spec-cases:-vect} giving the colexicographically vectorized form of the Tucker decomposition was used. Since $\operatorname{vec}_{\operatorname{col}}$ is an invertible and linear function, the above relationship means that assuming the residuals have a multilinear normal distribution is equivalent to implicitly defining them in terms of a Tucker decomposition:
\[  
\mathbfcal{E}_i = \mathbb{E}[ \mathbfcal{E}_i ]  + \mathbfcal{U}_i \bullet_1 \boldsymbol{\Sigma}^{1/2}_1 \cdots \bullet_O \boldsymbol{\Sigma}_O^{1/2} \,.
 \]
Note in particular that a commonly assumed case, where the entries of $\mathbfcal{E}_i$ are all i.i.d. standard normal random variables, is (a rather uninteresting) special case of the above. The above gives a sense in which the statement that the matrix $\boldsymbol{\Sigma}_o$ is ``the covariance matrix for the $o$-th mode of $\mathbfcal{E}_i$'' can be made precise for a multilinear normal distribution.

%%% Local Variables:
%%% mode: latex
%%% TeX-master: "../new_notes_draft"
%%% End:

\section{Relationships Between Eigen- and Singular Vectors and Values of Tensors}
\label{sec:relat-betw-eigen}

For a general non-cubical tensor, only singular values and singular vectors exist, so there is no need to wonder about their possible relationships with eigenpairs. However, starting already with arbitrary cubical tensors, one might begin to wonder whether certain relationships are guaranteed to exist. For example, the definition of eigenpair seems similar to the definition of singular value and vector tuple, and so one might wonder whether, given any eigenvector of a cubical tensor, the corresponding tuple of $O$ copies of that eigenvector is also a singular vector tuple. However, one needs to keep in mind that the condition of being a mode-$o$ eigenvector for some $o \in [O]$ for a general cubical tensor corresponds to nothing stronger than satisfying at least one of the $O$ necessary conditions for the corresponding tuple of $O$ copies of that vector to be a singular vector tuple, much weaker than satisfying all of them simultaneously with the same value of $\lambda$ or $\sigma$. In order to have its $O$ copies be a singular vector tuple, the given vector would need to not only be a mode-$o$ eigenvector simultaneously for all $O$ modes of the tensor, it would also need to correspond to the same eigenvalue for each mode. Since, as discussed previously, higher-order tensors need not have the same eigenvalues for all modes, those tensors for which this is the case cannot have any singular vector tuples consisting of $O$ copies (up to signs) of the same vector, and thus their eigenpairs and singular values and singular vector tuples will be unrelated.\\

Even when the various modes of a cubical tensor happen to have the same eigenvalues (e.g. as in the matrix case), they need not necessarily have the same eigenvectors, and even in this case any eigenvector which does not correspond to all modes of the tensor has no chance of corresponding to a singular vector tuple of $O$ copies (up to signs) of itself. It seems a direct consequence from the definitions that a vector has $O$ copies of itself (up to signs) as a singular vector tuple if and only if it is an eigenvector simultaneously for all modes corresponding to the same eigenvalue. In particular a tensor has a singular vector tuple consisting of $O$ copies (up to signs) of the same vector if and only if it has such an eigenvector. Thus the definitions of eigenpair and singular value and vector tuple for a cubical tensor are not as similar as lazily glancing would suggest. One might still wonder whether the remaining singular vector tuples with at least two distinct (up to signs) vectors have any weaker relationships with the eigenvectors for any of the modes, but already the case of matrices suggests that this is not the case. However, when the cubical tensor is symmetric, \textit{every} eigenvector automatically satisfies all of the $O$ necessary conditions with the same $\lambda$. So one can conclude that that every eigenpair of a \textit{symmetric} tensor corresponds (up to signs) to a singular value and a singular vector tuple consisting of $O$ copies (up to signs) of the same vector. This leads to the questions of whether all singular values and singular vector tuples of a higher-order symmetric tensor can be so characterized, and if not, what can be said about those singular vector tuples of a higher-order symmetric tensor consisting of distinct (even up to signs) vectors.\\

Like the case of order-two tensors (matrices), the largest $\ell^2$-singular value of a symmetric higher-order tensor equals the absolute value of its largest-in-magnitude $\ell^2$-eigenvalue, and (up to signs) the corresponding tuple of $\ell^2$-singular vectors is the same as the corresponding $\ell^2$-eigenvector\cite{hillar_lim}.  For symmetric matrices, the analogous statements are also true for the smaller singular values and smaller-in-magnitude eigenvalues. It is not immediately obvious whether this is also true for higher-order tensors. This is because the variational problem defining $\ell^2$-singular values for arbitrary tensors (not necessarily symmetric) is related to the optimization problem defining the best rank-one approximation in such a way that, if $\mathbf{v}_1 \otimes \cdots \otimes \mathbf{v}_O$ is the best rank-one approximation, then the largest $\ell^2$-singular value $\sigma_1 = \prod_{o=1}^O ||\mathbf{v}_o||_2$ and for each of the corresponding $\ell^2$-singular vectors $\mathbf{x}_o = \mathbf{v}_o / ||\mathbf{v}_o||_2$\cite{hillar_lim}\cite{LekHengLim}. (In fact this relationship allows one to prove the NP-hardness of finding a best rank-one approximation starting from the fact that finding the largest $\ell^2$-singular value is NP-hard\cite{hillar_lim}. Moreover just like for matrices the largest $\ell^2$-singular value equals, basically by definition, the spectral norm of the tensor, a fact utilized in \cite{Wang2017}.) In the case that the tensor is symmetric, a theorem by Banach guarantees that the corresponding optimization problem for the largest-in-magnitude $\ell^2$-eigenvalue coincides with the problem for finding the largest $\ell^2$-singular value (cf. the statement about best rank-one Waring decompositions made in section \ref{sec:symm-antisymm-tens}). \\

One might therefore suppose that the analogous statements for smaller $\ell^2$-singular values and smaller-in-magnitude $\ell^2$-eigenvalues are not true for higher-order tensors. The method of proof used for the largest $\ell^2$-singular value and largest-in-magnitude $\ell^2$-eigenvalue seems to break down since, for higher-order tensors, best rank-$r$ and rank-$(r+1)$ approximations need not exist, and even if they do exist they are generally unrelated to each other, unlike the matrix case. For the matrix case, one would use the fact that, given a matrix $\mathbf{T}$ and its best rank-one approximation $\mathbf{T}_{(1)}$, its best rank-two approximation would be $\mathbf{T}_{(1)} + \mathbf{T}_{(2)}$, where $\mathbf{T}_{(2)}$ denotes the best rank one approximation of $\mathbf{T} - \mathbf{T}_{(1)}$. For a given higher order tensor $\T$, even though a best rank-one approximation $\T_{(1)}$ to $\T$ and a best rank-one approximation $\T_{(2)}$ to $\T - \T_{(1)}$ will exist, in general $\T_{(1)} + \T_{(2)}$ is not the best rank-two approximation to $\T$ and in fact such an approximation need not even exist. On the other hand, even though $\T_{(1)} + \T_{(2)}$ is not the best rank-two approximation to $\T$, one might still hope that $||\T_{(2)}||_F$ is the second largest $\ell^2$-singular value and that the corresponding $\ell^2$-singular vector tuple is the same as the tensor product factors in $\T_{(2)}$ up to scaling. As mentioned before, in the case that $\T$ is symmetric, a theorem from Banach guarantees that $\T_{(1)}$ will be symmetric, and thus that also $\T - \T_{(1)}$ will be symmetric, which in turn by the theorem implies that $\T_{(2)}$ will be symmetric, and so one might hope in that case that $||\T_{(2)}||_F$ is also the absolute value of the second largest-in-magnitude $\ell^2$-eigenvalue of $\T$ and that the single repeated tensor product factor defining $\T_{(2)}$ will be (up to scaling and signs) the corresponding $\ell^2$-eigenvector.\\

However, if none of this is true, then one might also suppose that the $\ell^2$-singular vector tuple corresponding to the second largest $\ell^2$-singular value of a higher-order symmetric tensor could possibly (even up to signs) consist of distinct vectors, rather than $O$ copies of a related $\ell^2$-eigenvector. Even though it is fairly clear that the eigenpairs of a higher-order symmetric tensor have to correspond to singular tuples with $O$ copies (up to signs) of the eigenvector and the eigenvalue the same (up to signs) as the singular value, the converse relationship is much less clear except in the case of the $\ell^2$ definitions for the largest $\ell^2$-singular value only. It is less straightforward to discern whether every singular vector tuple of a higher-order symmetric tensor must consist of $O$ copies of the same (up to signs) vector (recall that when a singular vector tuple does, the repeated vector is automatically an eigenvector and the singular value is up to signs automatically an eigenvalue).

%%% Local Variables:
%%% mode: latex
%%% TeX-master: "../new_notes_draft"
%%% End:

\theendnotes

\end{document}